\documentclass{aa}

\usepackage{lscape}
\usepackage{graphicx}
\usepackage[varg]{txfonts}
\usepackage[colorlinks=true, allcolors=blue]{hyperref}

\usepackage{amsmath}
\usepackage{amssymb}
\usepackage{textcomp}
\usepackage{gensymb}
\usepackage{mathtools}
\usepackage{subcaption}
\usepackage{multirow}
\usepackage{colortbl}
\usepackage{xcolor}
\usepackage{arydshln}

\makeatletter
\renewcommand*\aa@pageof{, page \thepage{} of \pageref*{LastPage}}
\makeatother

\begin{document}

\title{A global view on star formation: The GLOSTAR Galactic plane survey}
\subtitle{II. Supernova remnants in the first quadrant of the Milky Way \thanks{Tables 2, 3, and 4 will be available in electronic form at the CDS via anonymous ftp to cdsarc.u-strasbg.fr (130.79.128.5) or via http://cdsweb.u-strasbg.fr/cgi-bin/qcat?J/A+A/}}
\titlerunning{GLOSTAR: Supernova Remnants in the first quadrant of the Milky Way}

\author{Rohit Dokara \inst{1}\thanks{Member of the International Max Planck Research School (IMPRS) for Astronomy and Astrophysics at the Universities of Bonn and Cologne},
        A.\,Brunthaler \inst{1},
        K.\,M.\,Menten \inst{1},
        S.\,A.\,Dzib \inst{1},
        W.\,Reich \inst{1},
        W.\,D.\,Cotton \inst{2,3},
        L.\,D.\,Anderson \inst{4},
        C.\,-H.\,R.\,Chen \inst{1},
        Y.\,Gong \inst{1},
        S.\,-N.\,X.\,Medina \inst{1},
        G.\,N.\,Ortiz-Le\'on \inst{1},
        M.\,Rugel \inst{1},
        J.\,S.\,Urquhart \inst{5},
        F.\,Wyrowski \inst{1},
        A.\,Y.\,Yang \inst{1},
        H.\,Beuther \inst{6},
        S.\,J.\,Billington \inst{5},
        T.\,Csengeri \inst{7},
        C.\,Carrasco-Gonz\'alez \inst{8}
        \and
        N.\,Roy \inst{9}
        }
\authorrunning{R. Dokara et al.}

\institute{Max-Planck-Institut f\"ur Radioastronomie (MPIfR), Auf dem H\"ugel 69, 53121 Bonn, Germany\\
           \email{rdokara@mpifr-bonn.mpg.de}
           \and
           National Radio Astronomy Observatory, 520 Edgemont Road, Charlottesville, VA 22903, USA
           \and
           South African Radio Astronomy Observatory, 2 Fir St, Black River Park, Observatory 7925, South Africa
           \and
           Department of Physics and Astronomy, West Virginia University, Morgantown WV 26506, USA
           \and
           Centre for Astrophysics and Planetary Science, University of Kent, Canterbury, CT2\,7NH, UK
           \and
           Max Planck Institute for Astronomy, K\"onigstuhl 17, 69117, Heidelberg, Germany
           \and
           Laboratoire d'astrophysique de Bordeaux, Univ. Bordeaux, CNRS, B18N, all\'ee Geoffroy Saint-Hilaire, 33615, Pessac, France
           \and
           Instituto de Radioastronom\'ia y Astrof\'isica (IRyA), Universidad Nacional Aut\'onoma de M\'exico Morelia, 58089, Mexico
           \and
           Department of Physics, Indian Institute of Science, Bengaluru, India 560012
           }

\date{Submitted on November 06, 2020; revised on March 01, 2021; accepted on March 09, 2021}


\abstract
{ The properties of the population of Galactic supernova remnants
(SNRs) are essential to our understanding of the dynamics of the interstellar medium (ISM) in the  Milky
Way.  However, the completeness of the catalog
of Galactic SNRs is expected to be only ${\sim}30\%$, with on order
700 SNRs yet to be detected.  Deep interferometric radio continuum surveys
of the Galactic plane help in rectifying this apparent deficiency by
identifying low surface brightness SNRs and compact SNRs that
have not been detected in previous surveys.  However, SNRs are routinely
confused with \ion{H}{ii}~regions, which can have similar radio morphologies.
Radio spectral index, polarization, and emission at mid-infrared (MIR)
wavelengths can help distinguish between SNRs and \ion{H}{ii}~regions.  }
{We aim to identify SNR candidates using continuum images from the
Karl G. Jansky Very Large Array GLObal view of the STAR formation in the
Milky Way (GLOSTAR) survey.}
{GLOSTAR is a C-band (4--8 GHz) radio wavelength survey of the Galactic
plane covering $358\degr \leq l \leq 60\degr, |b| \leq 1\degr$.  The
continuum images from this survey, which resulted from observations with
the most compact configuration of the array, have an angular resolution of
$18''$.  We searched for SNRs in these images to identify known SNRs,
previously identified SNR candidates, and new SNR candidates.  We study
these objects in MIR surveys and the GLOSTAR polarization data to
classify their emission as thermal or nonthermal. }
{ We identify 157 SNR candidates, of which 80 are new.
Polarization measurements provide evidence of nonthermal emission
from 9 of these candidates.  We find that two previously identified
candidates are filaments.  We
also detect emission from 91 of the 94 known SNRs in the survey region.
Four of these are reclassified as \ion{H}{ii}~regions following
detection in MIR surveys. }
{ The better sensitivity and resolution of the GLOSTAR data have led
to the identification of 157 SNR candidates, along with the
reclassification of several misidentified objects.  We show that
the polarization measurements can identify nonthermal emission,
despite the diffuse Galactic synchrotron emission.  These
results underscore the importance of higher resolution and higher
sensitivity radio continuum data in identifying and confirming SNRs.}


\maketitle

\section{Introduction}

Supernova remnants (SNRs) inject energy and material into the interstellar
medium (ISM) of the Galaxy; they produce and accelerate cosmic rays, drive
turbulence within molecular clouds, and impact the dynamics of the ISM
\citep{2017A&A...604A..70I,2020A&A...634A..59B}.  Based on the statistics
of massive stars, pulsars, supernova rates, and iron abundances, it was
estimated that the Milky Way should contain $\gtrsim 1000$ SNRs
\citep{1991ApJ...378...93L,1994ApJS...92..487T}.  However, the most
recent version of the
catalog of Galactic SNRs \citep{2019JApA...40...36G} contains fewer
than $300$ objects, pointing to a large inconsistency.  This is
thought to be the result of observational bias that hinders the
detection of low surface brightness SNRs in the one extreme, and the
small angular size of many SNRs in the other
\citep{2006ApJ...639L..25B,2017A&A...605A..58A}.

Radio surveys covering the Galactic plane have proved to be highly
effective in the identification of SNR candidates en masse
\citep[see][for a review including a historical perspective]{Dubner2015}.
While single-dish telescopes have repeatedly been used to study
Galactic SNRs \citep[see][for instance]{1976MNRAS.174..267C,2011A&A...536A..83S},
more recent efforts using interferometer arrays have identified more than
$170$ candidates in data from the following projects:
a 330~MHz survey conducted with the Very Large Array (VLA)
by \citet{2006ApJ...639L..25B},
the 1.4~GHz Multi-Array Galactic Plane Imaging Survey
\citep[MAGPIS;][]{2006AJ....131.2525H},
the 843~MHz Molonglo Galactic Plane Survey \citep[MGPS;][]{2014PASA...31...42G},
the 1--2~GHz \ion{H}{i}, OH, Recombination line survey of the Milky Way
\citep[THOR;][]{2017A&A...605A..58A},
and the 80--300~MHz GaLactic and Extragalactic All-sky Murchison Widefield
Array survey \citep[GLEAM;][]{2019PASA...36...45H}.

Generally, objects are labeled as SNR candidates when there is an
indication of nonthermal emission from only one or two studies, and
they may be "confirmed" when more evidence is demonstrated.  Of the
SNR candidates identified so far, only a fraction
have been confirmed as SNRs \citep[see][]{2019PASA...36...48H}.

\ion{H}{ii}~regions, like SNRs, are routinely observed at radio
wavelengths, and also have a morphology similar to that of SNRs in
the radio continuum.  This represents a major obstacle in the
confirmation of more candidates as SNRs.  In the earlier versions of
the catalog of Galactic SNRs by \citet{2019JApA...40...36G}, several
objects that were thought to be SNRs were subsequently reclassified as
\ion{H}{ii}~regions
\citep{2017A&A...605A..58A,2019PASA...36...48H,2019A&A...623A.105G}.
In addition, pulsar wind nebulae (PWNe), also referred to as plerion
or Crab-like remnants, are another source of confusion.  Their emission
is driven by the winds from the central pulsar and they may have
a spectral index\footnote{The spectral index,
$\alpha$, is defined as the slope of the linear fit to the log-log
plot of flux density, $S_\nu$, versus frequency, $\nu$:
$S_\nu \propto \nu^\alpha$} similar to that of
\ion{H}{ii}~regions \citep{2006ARA&A..44...17G}, although some
exceptions are known \citep[see][]{2008ApJ...687..516K}.
An extended radio object in the
Galaxy may be classified as a SNR or an \ion{H}{ii}~region based on
whether its emission is thermal or nonthermal.
\ion{H}{ii}~regions emit thermal Bremsstrahlung radiation, which is
unpolarized and has a spectral index of $-0.1 \lesssim \alpha \lesssim 2$
depending on the optical thickness \citep{2013tra..book.....W}.  These regions
also show strong mid-infrared (MIR) radiation emitted by warm dust
and polycyclic aromatic hydrocarbons
\citep{1986A&A...155..380C,2009PASP..121..213C}.  On the other hand,
the emission from SNRs is primarily nonthermal synchrotron radiation
emitted by relativistic electrons gyrating around magnetic field lines
in a magneto-ionic medium.  Synchrotron emission has a characteristic
falling spectral index ($\alpha \lesssim -0.5$), and is
generally linearly polarized \citep{2013tra..book.....W}.
PWNe may have a thermal-like spectral index ($\alpha \sim -0.1$),
but their emission is nevertheless linearly polarized.
SNRs are also generally quite faint or even not detected at MIR
wavelengths \citep{1987A&AS...71...63F}.
\citet{1996A&AS..118..329W} measured the ratio of $60~\mu$m MIR to
36~cm radio flux densities of SNRs and \ion{H}{ii}~regions to be
typically $\lesssim$ 50 and $\gtrsim 500$ respectively.

The above characteristics help to distinguish SNRs from
\ion{H}{ii}~regions. In particular, the presence or absence of MIR
wavelength emission has been widely used as a criterion by many of the
aforementioned surveys.  In this paper, we identify SNR candidates
using radio continuum data from the D-array data of the GLObal view
of STAR formation in the Milky Way survey that we conducted with the
Karl G. Jansky VLA
\citep[GLOSTAR-VLA,][]{2021arXiv210600377B,2019A&A...627A.175M}.
For these objects, along with already confirmed SNRs and previously
identified candidates, we have examined MIR surveys and the polarization
data of GLOSTAR-VLA in order to classify their emission as thermal or nonthermal.

This paper has the following structure: In \S \ref{sec:data}, we
discuss the GLOSTAR-VLA radio and the GLIMPSE and MIPSGAL MIR surveys,
as well as the catalogs of \ion{H}{ii}~regions and known and candidate
SNRs.  In \S \ref{sec:methods}, we describe the method we use to
identify new SNR candidates and the process of measuring the linearly
polarized and total flux densities of extended objects.  In \S
\ref{sec:results}, we present the results, consider their implications,
and discuss several individual objects.  \S \ref{sec:conclusions}
summarizes our work and concludes with remarks on future efforts.

\section{Data}
\label{sec:data}

\subsection{GLOSTAR survey}
\label{subsec:GLOSTARdata}

The GLOSTAR survey,
with a $4$--$8$ GHz frequency band, covers the Galactic center region
and the first quadrant of the Galactic plane up to a Galactic longitude of
$l=60\degr$ in a $\pm 1\degr$ wide band in latitude, $b$.  In addition,
the Cygnus X region was covered, but is not considered in the present study.  It was performed with the Jansky VLA
in the compact D-configuration, and in the more extended
B-configuration with the wideband receivers observing in full
polarization in a mixed setup of continuum and spectral lines.
For this paper, we used only the D-array continuum data
covering the region $358\degr \leq l \leq 60\degr, |b| \leq 1\degr$.
Calibration and imaging of the continuum data are done using standard
VLA procedures and calibrators.  Details of the data reduction of a
part of the survey ($28\degr \leq l \leq 36\degr, |b| \leq 1\degr$)
are described by \citet{2019A&A...627A.175M} and a full presentation
will be given by \citet{2021arXiv210600377B} and Medina et al. (in prep.).

The products of the continuum data-reduction process are mosaic FITS cubes
of Stokes $I$, $Q,$ and $U$, with each mosaic covering about 16 square
degrees.  Each cube has 11 planes containing the images for 9 frequency
intervals across the 4--8 GHz band with sections affected by man-made
radio frequency interference (RFI) discarded, an averaged image, and the
in-band spectral index map computed from the 9 frequency planes. The
averaged image is obtained by taking a mean of each pixel across the
9 planes weighted by the inverse of the square of the RMS noise.
This averaged image has an effective frequency of 5.8 GHz.
The images were smoothed to a common resolution of $18\arcsec$ after
the CLEAN process.  The RMS noise of the averaged Stokes $I$ images
typically ranges from $60$ to $150~\mu$Jy \citep{2019A&A...627A.175M}.

Although the largest scale that can be observed ($\sim \lambda/B_{min}$) is about
$2\arcmin$, mosaicking the pointings helped recover several larger angular scale
structures \citep[see][]{2019A&A...627A.175M}.  However, there still exists a
significant fraction of undetected flux density in objects larger than $1\arcmin$,
especially in the higher frequency images.  This ``missing'' flux density causes a
systematic reduction of spectral index.  Measuring the spectral index of an
extended structure  is only logical if the angular scales being probed are
roughly the same at all of the frequencies employed for its determination.
Within the 4--8 GHz band of GLOSTAR, the highest frequency
images are only sensitive to structures smaller than $\sim 1.5\arcmin$, and the
lowest frequency images to $\sim 3\arcmin$.  There is also no added single-dish
data, making the interpretation of spectral index of an extended object
quite uncertain.  Almost all the objects that we discuss in this paper have
sizes larger than $1\arcmin$, and we typically recover only a fraction of the flux density of extended objects (further discussed in \S\ref{subsubsec:missflux}).  Therefore, the GLOSTAR-VLA flux densities we report are only used as lower limits, and we measure spectral index only
in two cases: (i) if the size of the object is comparable to the beam, in which case the spectral index derived would be reliable,
and (ii) for deriving a lower limit of the spectral index by comparing the GLOSTAR-VLA flux density with
lower frequency data that have comparatively reliable flux density estimates, such as the 1.4 GHz
THOR+VGPS.  In the second case, a lower limit on the spectral index close to
zero is useful in identifying thermal emission and PWNe, because SNRs
(other than PWNe) do not have a spectral index $\gtrsim 0$.

\subsection{Other surveys covering the Galactic plane}

The Galactic Legacy Infrared Mid-Plane Survey Extraordinaire
(GLIMPSE) is a four-band ($3.6$--$8~\mu$m) survey of the Galactic plane by the
\textit{Spitzer} Space Telescope covering $|l|<65\degr$ and
$|b|<1\degr$ to $2\degr$, with a resolution $<2\arcsec$ \citep{2009PASP..121..213C}.  The MIPS
Galactic plane survey (MIPSGAL) is a complementary $24~\mu$m and
$70~\mu$m survey by \textit{Spitzer} with coverage overlapping
with that of the GLIMPSE survey \citep{2009PASP..121...76C}.
The resolutions of the $24~\mu$m and $70~\mu$m bands are $6\arcsec$ and $24\arcsec$,
respectively.
\citet{2014ApJS..212....1A} report that the sensitivity of these MIR
surveys is good enough to detect all the \ion{H}{ii}~regions present in the Milky Way.

We also studied the mosaics of recent radio surveys such as the
NRAO VLA Sky Survey \citep[NVSS; ][]{1998AJ....115.1693C}\footnote{\url{https://www.cv.nrao.edu/nvss/postage.shtml}} and the
TIFR GMRT Sky Survey \citep[TGSS; ][]{2017A&A...598A..78I}\footnote{\url{https://vo.astron.nl/tgssadr/q_fits/cutout/form}}.
The Galactic plane surveys MAGPIS \citep{2006AJ....131.2525H}\footnote{\url{https://third.ucllnl.org/cgi-bin/gpscutout}}
and THOR+VGPS \citep{2016A&A...595A..32B}\footnote{\url{https://www2.mpia-hd.mpg.de/thor/Overview.html}}
are particularly useful because they are at lower frequencies, but with surface
brightness sensitivity comparable to that of the GLOSTAR-VLA data.  As the 1.4 GHz
MAGPIS has a resolution of ${\sim}6''$, we convolved these
data to the beam size of GLOSTAR-VLA ($18''$) for our analysis.

\subsection{Lists of objects}
\label{listofobjects}

We search the literature for previously identified SNRs and \ion{H}{ii}~regions.
We find the catalogs by \citet{2019JApA...40...36G} and
\citet{2014ApJS..212....1A} to be the most authoritative compilations
of SNRs and \ion{H}{ii}~regions respectively.  A brief description of these catalogs,
along with some recent SNR candidate lists, is given below.

\subsubsection{The catalog of Galactic SNRs}

The D. Green catalog of Galactic SNRs is updated every few years with
additions of new SNRs and removals of misidentified objects
\citep[see][and references therein]{2019JApA...40...36G}.  The
most recent version contains $295$ SNRs, with 94 of these being found in the region
covered by the GLOSTAR survey.  Hereafter, we refer to these
objects as G19 SNRs.

\subsubsection{WISE catalog of H II regions}

\citet{2014ApJS..212....1A} produced the most complete catalog of
\ion{H}{ii}~regions in the Milky Way using data from the Wide-Field Infrared
Survey Explorer (WISE) satellite.  It contains about 6000 candidate
\ion{H}{ii}~regions (identified using their characteristic MIR morphology) and
about ${}2000$ confirmed \ion{H}{ii}~regions (with radio recombination line or
H$\alpha$ emission) spanning the entire Galaxy.  The GLOSTAR survey
region contains the positions of approximately ${}1000$ confirmed \ion{H}{ii}~regions in this
catalog.  We hereafter refer to these objects as A14 \ion{H}{ii}~regions.

\subsubsection{Previously discovered SNR candidates}
\label{subsubsec:prevcandslist}

We compiled a list of SNR candidates from the literature, especially
focusing on large-area surveys that have a significant overlap with
the GLOSTAR survey region.  These are summarized below.

\begin{itemize}
    \item{Using data from the THOR survey with the VGPS data added, \citet{2017A&A...605A..58A} identified 76 SNR candidates.  The GLOSTAR survey region covers 74 of these candidates.}
    \item{\citet{2006AJ....131.2525H} discovered 49 ``high-probability'' SNR candidates in the MAGPIS data,  all of which are located in the GLOSTAR survey region.}
    \item{The inner Galactic plane was observed at 330 MHz using the VLA by \citet{2006ApJ...639L..25B} and 35 SNR candidates were discovered, all of which are covered in the GLOSTAR survey region.}
    \item{\citet{2019PASA...36...45H} identified 27 SNR candidates using data from the GLEAM survey.  The GLOSTAR survey region covers 10 of these candidates.}
    \item{An internet-accessible version\footnote{\url{http://www.mrao.cam.ac.uk/surveys/snrs/snrs.info.html}} of the catalog of Galactic SNRs by \citet{2019JApA...40...36G} discusses several candidates identified across the electromagnetic spectrum.  A machine-readable list of 70 SNR candidates in the GLOSTAR survey region was provided by D. Green in a private communication.}
\end{itemize}

From these previously reported SNR candidates, we removed the
objects that have already been included in either the G19 SNR or the A14
\ion{H}{ii}~region catalogs, and those that were noted as
misidentifications, such as the candidates discussed by
\citet[Section 4.3]{2017A&A...605A..58A}.
The GLOSTAR-VLA data were then searched by eye at the positions of the
remaining objects.  From this search, we identify 77 previously reported
SNR candidates (presented in Table \ref{tab:prevcandSNRs}).
The process that was followed to discover {new} SNR candidates is
explained below in \S\ref{subsec:identifySNRs}.

\section{Methods}
\label{sec:methods}

\subsection{Identification of new candidate SNRs}
\label{subsec:identifySNRs}

\begin{figure}
    \centering
    \includegraphics[width=0.48\textwidth]{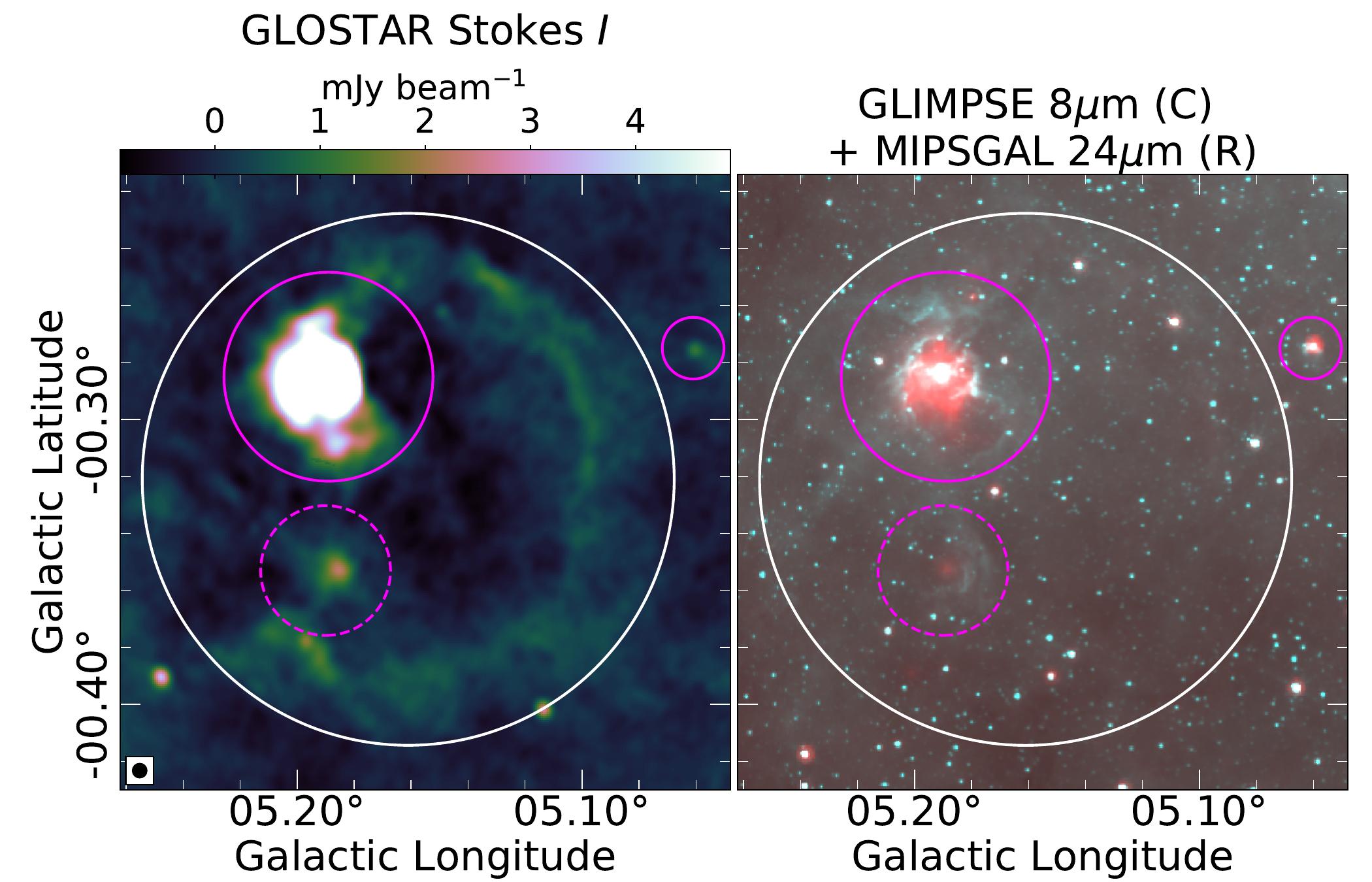}
    \caption{Example illustrating the criteria used to identify SNR candidates.  The SNR candidate G005.161$-$0.321 (encircled in white) has a bright confirmed \ion{H}{ii}~region G005.189$-$0.285 (large solid magenta circle) and a faint candidate \ion{H}{ii}~region G005.189$-$0.354 (dashed magenta circle) within its angular extent.  The data in the right panel are from MIPSGAL $24~\mu$m (red) and GLIMPSE 8~$\mu$m (cyan).  The GLOSTAR-VLA image is presented in the left panel with the synthesized beam shown in black in the bottom left corner.  This beam convention is followed throughout the paper.  The object-marking scheme for all the figures in this paper is as follows:  white circles mark newly identified SNR candidates, solid magenta circles mark confirmed \ion{H}{ii}~regions, dashed magenta circles mark candidate \ion{H}{ii}~regions, red circles mark G19 SNRs, and green circles mark previously identified SNR candidates described in \S \ref{subsubsec:prevcandslist}.  }
    \label{fig:example}
\end{figure}

In order to identify new SNR candidates, we follow an approach that is not
biased toward any particular morphology and is similar to the process followed by
\citet{2017A&A...605A..58A}.  First, we searched the GLOSTAR-VLA Stokes $I$
5.8~GHz integrated mosaics by eye
for extended emission regions that are absent from the list of previously identified
objects (including previously discovered SNR candidates; see \S \ref{listofobjects}).  We ignored regions where the negative
side lobes are as strong as the emission.
These radio emissions are then searched for in the
images of GLIMPSE $8~\mu$m and MIPSGAL $24~\mu$m MIR surveys, again visually.  At MIR
wavelengths, SNRs  usually emit little or no radiation.  In some cases,
such as SNR W49B, they may have significant MIR emission, but the MIR-to-radio
flux density ratio is still low \citep{1996A&AS..118..329W,2011AJ....142...47P}.
Conversely, \ion{H}{ii}~regions have strong
MIR emission and generally have a characteristic radio-MIR morphology:
coincident radio and $24~\mu$m emissions, which are surrounded by $8~\mu$m
emission \citep[see][for instance]{2009PASP..121..213C}.  Therefore,
the objects that  are associated with strong MIR emission in either of the GLIMPSE
$8~\mu$m or MIPSGAL $24~\mu$m images are removed from our list.  What
remains is a group of previously unclassified extended objects that emit at radio
wavelengths and have no associated MIR emission.  A circular region is
defined for each object such that it encompasses its radio emission, and
if only an arc or a partial shell is observed, then the curvature is followed.
Figure~\ref{fig:example} shows the GLOSTAR-VLA and MIR images of an example GLOSTAR
SNR candidate G005.161$-$0.321.  Although there is MIR emission from this
region, it is confined to the \ion{H}{ii}~regions G005.189$-$0.285 and G005.189$-$0.354,
but absent from the shell of the SNR candidate G005.161$-$0.321.

We do not make use of the GLOSTAR-VLA source catalogs by Medina et al. (in prep)
and \citet{2019A&A...627A.175M} because their method is optimized to identify
compact sources with high reliability.  These latter authors use {\tt{SExtractor}}
\citep{1996A&AS..117..393B} to create the background noise map, and then
{\tt{BLOBCAT}} \citep{2012MNRAS.425..979H} to perform the automated source
extraction.  A mesh size of 80$\times$80 pixels$^2$ and a detection threshold
of $5\sigma$ were used.  In addition,
during the visual inspection stage, these latter authors exclude the sources
with low signal-to-noise ratio if no counterparts are found in MIR
surveys \citep[see][for details]{2019A&A...627A.175M}.  This
process imparts a two-fold systematic bias against identifying SNR candidates.
Firstly, the constant mesh size is not suitable for identifying extended
emission as noise levels are overestimated\footnote{A mesh size of 80$\times$80
pixels$^2$ (${\sim}10{\times}10$ beam$^2$) is too small for
{\tt{SExtractor}}'s sigma-clipping algorithm to converge on a robust background
in a region with extended emission.}, and as we aim to identify candidates
that have not been detected before, we expect these objects to be quite faint
and possibly be judged by the software as
background noise.  Secondly, as the counterparts for these sources are
searched only in MIR surveys during the visual inspection,
SNRs are again excluded because they typically have no MIR emission.
For these reasons, we search the images visually.  An assessment
of our confidence level that an object is not an interferometric artefact
can be made by comparing the flux density with its uncertainty, whose
measurements are explained in the following section.

\subsection{Measuring flux densities}
\label{subsec:flux}

\begin{figure}
    \centering
    \includegraphics[width=0.32\textwidth]{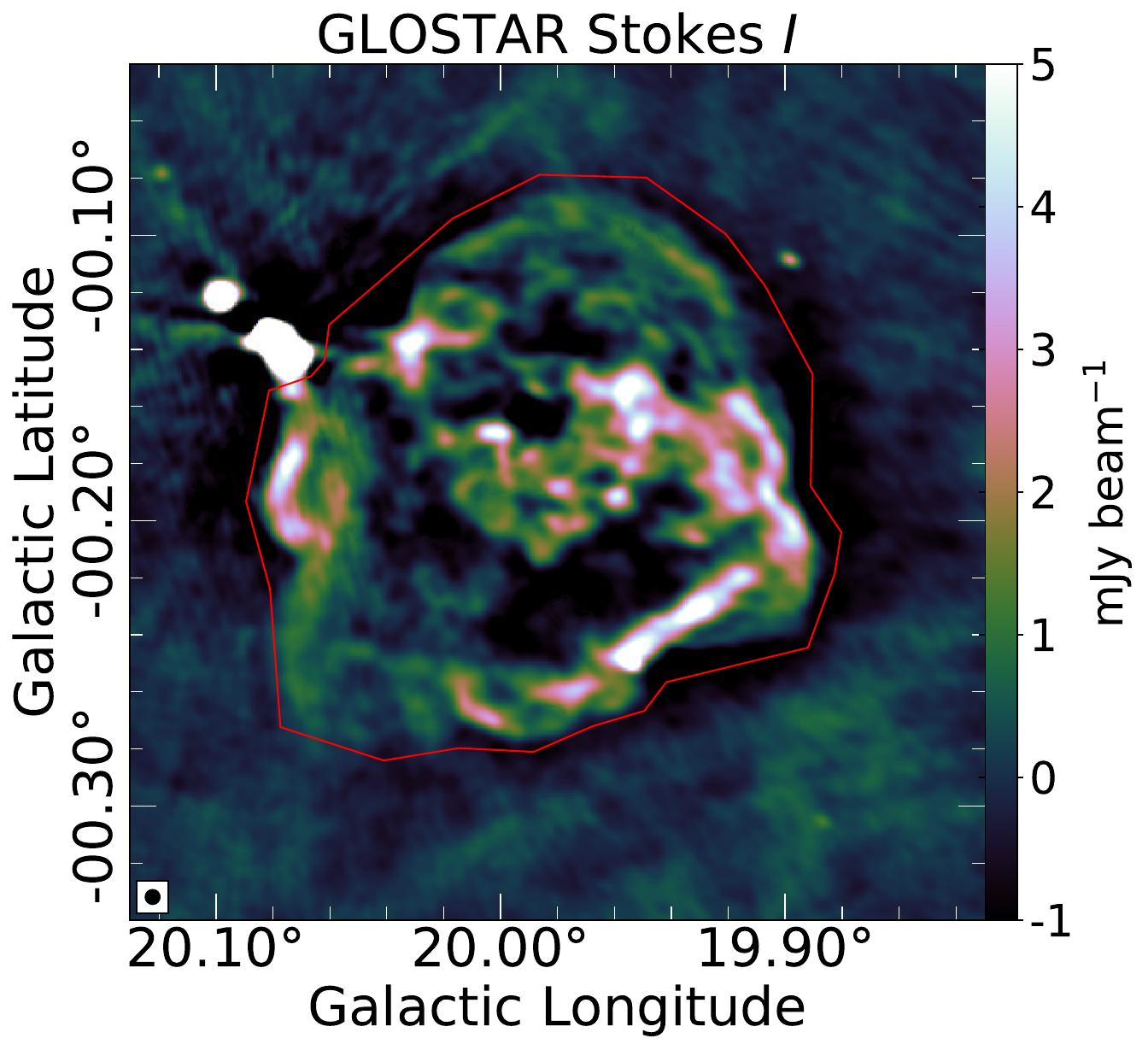}
    \caption{Measuring the integrated flux density of SNR G20.0$-$0.2.  The polygon used to select the region of the SNR is shown in red.  As we mask values below the $3\sigma_I$ level, choosing the negative bowls around real emission does not affect our measurement.  The regions with compact emission to the northeast of the SNR (near $l=20.09,~b=-0.13$) are A14 \ion{H}{ii}~regions, and hence excluded.  }
    \label{fig:measureflux}
\end{figure}

Because of confusion in the crowded Galactic plane, measuring the
integrated flux densities of extended objects is not
straightforward.  The lack of short-spacing data and the poor
$uv$-coverage of snapshot observations cause strong negative
side lobes near bright sources, further complicating the issue.
For these reasons, we decided to manually measure the flux densities
by drawing a polygon aperture around the emission that is clearly associated
with the object being analyzed.  An example illustrating this procedure is
shown in Fig.~\ref{fig:measureflux}.

The total flux density is measured using the integrated mosaic at an
effective frequency of $5.8$ GHz.  In order to solve the
negative side lobe problem, we masked all pixels below $3\sigma_I$,
where $\sigma_I$ is the background RMS in the surrounding regions
determined using an iterative
sigma-clipping algorithm.  The integrated flux density, $S$, of an
object and the error in this measurement, $\Delta S$, are given by
\begin{equation}
    S = \frac{\sum_i I_i}{A_{\mathrm{beam}}} ~~\mathrm{and}~~
    \Delta S = \sigma_I \sqrt{\frac{N_{\mathrm{src}}}{A_{\mathrm{beam}}}}
,\end{equation}
where $I_i$ is the value of the pixel, $A_{\mathrm{beam}}$ is the area of the beam in number of pixels,
$N_{\mathrm{src}}$ is the number of pixels in the aperture defined for
the source, and $i$ is the summation index that runs over the pixels with
values greater than the threshold ($I > 3\sigma_I$).

We do not attempt to measure the flux densities of partially observed or severely
confused objects, and of those that overlap with clear artefacts that arise
from imperfect CLEANing.  We note that the Stokes $I$ flux densities are
generally only lower limits due to missing short spacing data.
Nevertheless, we report these values because they can be used to derive
useful information such as
the degree of polarization and spectral index limits.


We measure linearly polarized flux densities, $L=\sqrt{Q^2+U^2}$, in
each frequency plane first and then use the weighted mean and variance of these
values to obtain the source integrated linearly polarized flux density
and its uncertainty.  This is done in order to take care of any
bandwidth de-polarization effects, which may cause significant
de-polarization in the ${\sim}4$ GHz wide bandwidth of our survey.  To
include only the statistically significant pixels, we applied two masks
to each plane of the  Stokes $Q$ and $U$ mosaics:  one based on the
Stokes $I$ ($>3\sigma_I$) and the other on the Stokes $Q$ or $U$
($>3\sigma_{Q,U}$, where $\sigma_{Q,U}$ is the local RMS
noise\footnote{$\sigma_{Q,U}$ generally varies from $40$ to $100~\mu$Jy.}).
This masking also somewhat helps in the removal of low-level noise, and
spurious polarization from artefacts.

Apart from artefacts due to bright, compact sources, we also observe
spurious polarization in regions without Stokes $I$ counterparts (see e.g., \S \ref{subsubsec:G22.00}), and
also from a few \ion{H}{ii}~regions.  This may be caused by diffuse Galactic
synchrotron emission, or a foreground intervening ionized medium with a
strong magnetic field generally known
as a Faraday screen \citep[see][for instance]{2007A&A...463..993S}.
Unlike Stokes $I$, which has mostly smooth structure, Stokes $Q$ and $U$ have fine-scale structure that is not filtered out by the interferometer.  As we limit
the polarization measurement to only the pixels above a $3\sigma_I$ level in
the Stokes $I$, the effects of differential filtering and diffuse
emissions are minimized.  However, we note that the degree of polarization,
$p=L/S$, is an overestimate because the filtering in Stokes $I$ is much more severe
compared to Stokes $Q$ and $U$ due to the small-scale structure mentioned above
\citep[also see Section 4.1 of][]{2001ApJ...549..959G}.

There exists a bias in the measured polarization because the
uncertainties in Stokes $Q$ and $U$ are added, which results in a
positive polarization measurement even if the true polarization is
null.  We corrected for this polarization bias in each pixel using
\begin{equation}
    L ~=~ \sqrt{Q^2 + U^2 - (1.2\sigma_{Q,U})^2}
,\end{equation}
where $\sigma_{Q,U}$ is the noise in Stokes $Q$ and $U$ maps
\citep{1974ApJ...194..249W}.  Above $3\sigma_{Q,U}$, all bias estimators
converge \citep[see Fig.~2 of][]{1985A&A...142..100S}, and we masked
all pixels below $3\sigma_{Q,U}$.  Hence, we do not expect the choice of
method of bias estimation to influence the measurement significantly.

\subsection{Rotation measures and electric field vectors}

When electromagnetic radiation passes through a plasma with a nonzero
magnetic field along the direction of propagation, the birefringence
property of the medium causes the polarization vector to rotate.  This
is known as the Faraday
effect.  The rotation of the electric vector position angle (EVPA;
$\chi = 0.5 \arctan (U/Q)$) can be measured using the relation
\begin{equation}
    \Delta \chi ~=~ \mathrm{RM} \cdot \lambda^2
,\end{equation}
where $\lambda$ is the wavelength and RM is the rotation measure.  The RM
is the strength of the magnetic field component parallel to the line of sight (l.o.s.), $B_{||}$, weighted by the
electron density, $n_e$, in the foreground medium integrated along the l.o.s.:
\begin{equation}
    \frac{\mathrm{RM}}{\mathrm{rad}~\mathrm{m}^{-2}} ~=~ 0.81 ~\int \frac{n_e(L)}{\mathrm{cm}^{-3}} ~ \frac{B_{||}(L)}{\mu\mathrm{G}} ~ \frac{dL}{\mathrm{pc}}
.\end{equation}

Although Faraday rotation measure synthesis is necessary to fully
disentangle the contribution of different ionized sources along the
l.o.s. to the RM \citep{2005A&A...441.1217B}, the reduced GLOSTAR-VLA data
are not suited for such an analysis because of the large width
(${\sim}1000$~rad~m$^{-2}$) of the RM transfer function
\citep[see equation 61 of][]{2005A&A...441.1217B}.  Therefore,
we estimate the RM from the slope of a simple linear fit of EVPA
versus wavelength-squared, where such a fit is possible.  The EVPA at
$\lambda=0$ is then estimated by extrapolating the linear fit, which is
plotted on the polarization maps such as Fig.~\ref{fig:G27.06}.  This
fitting and estimation of EVPAs at $\lambda=0$
are handled by the function \texttt{RMFit.Cube} of the software
\textit{Obit} \citep{2008PASP..120..439C}.  Given an input of Stokes $Q$
and $U$ cubes, \texttt{RMFit.Cube} produces maps of the RM and the EVPAs at $\lambda=0$.

\section{Results}
\label{sec:results}

\subsection{Degree of linear polarization as a measure of nonthermal emission}

\begin{figure*}
    \centering
    \includegraphics[width=0.8\textwidth]{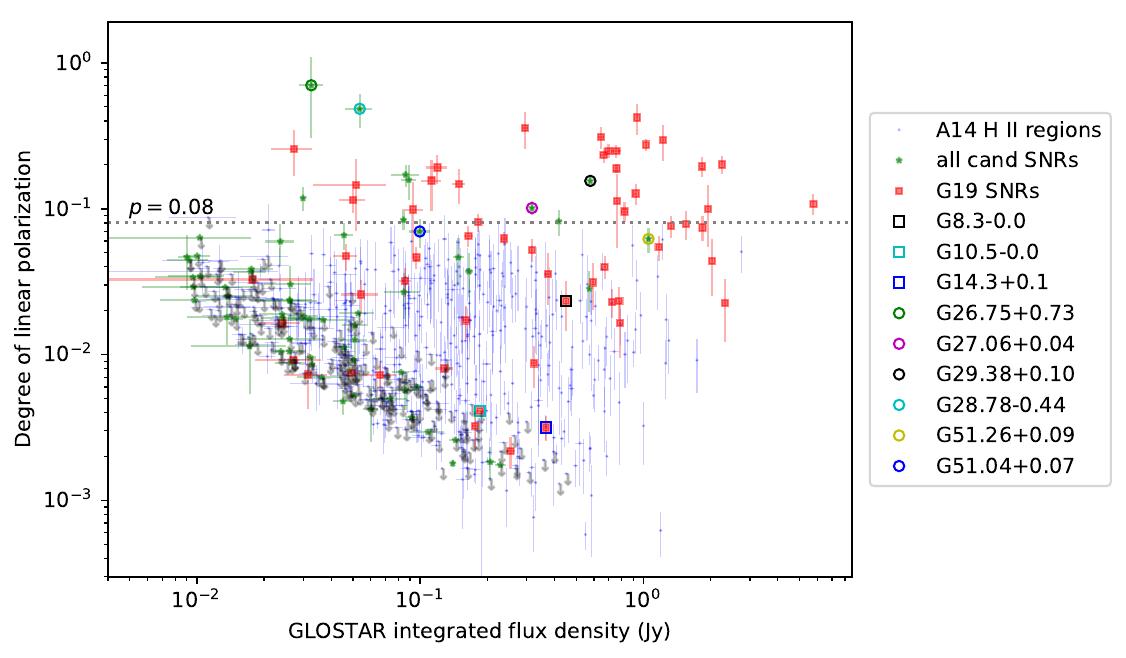}
    \caption{Degree of linear polarization $p$ plotted against flux density for A14 \ion{H}{ii}~regions, G19 SNRs, and SNR candidates brighter than 9~mJy.  Objects with upper limits on the degree of polarization are marked with gray arrows.   Some SNR candidates and G19 SNRs discussed in the later sections have been marked for the sake of comparison: open squares represent the misidentified G19 SNRs (see \S \ref{subsubsec:misG19}), and the open circles represent SNR candidates that we confirm as true SNRs (see \S \ref{subsec:candSNRs_pol}).  The lower end of the group of points follows an approximate linear relation due to the nearly constant detection threshold of linearly polarized flux density (${\sim}300$~$\mu$Jy). }
    \label{fig:polplot}
\end{figure*}

Synchrotron radiation, which is emitted by SNRs, is linearly
polarized with a degree of linear polarization ($p=L/S$) that
can be as high as 0.7 \citep{2013tra..book.....W}, although only a
few SNRs have been reported to have a degree polarization larger than
0.5 \citep{2017A&A...597A.116K}.  Thermal emission from
\ion{H}{ii}~regions on the other hand is inherently unpolarized.
However, because of the diffuse
Galactic synchrotron emission, \ion{H}{ii}~regions may also show an
apparent polarized emission.  To confirm that the SNRs have higher
degrees of polarization than the \ion{H}{ii}~regions, we plotted the
degree of linear polarization against the source integrated total
flux density for the three samples of A14 \ion{H}{ii}~regions,
G19 SNRs, and SNR candidates (Fig.~\ref{fig:polplot}).  The method
used to measure the flux densities is discussed in \S \ref{subsec:flux}.

We measure significant polarized emission from most SNRs, and also
some \ion{H}{ii}~regions.  The polarized emission from
\ion{H}{ii}~regions is probably from the diffuse Galactic synchrotron
emission.  Some SNRs have a low degree of polarization,
which is not uncommon \citep[see][]{2011A&A...536A..83S}.  However,
the majority of SNRs have a higher degree of polarization than
\ion{H}{ii}~regions.  As no \ion{H}{ii}~region brighter than 9~mJy
has $p>0.08$, we use this as the threshold to separate SNRs from
\ion{H}{ii}~regions and the diffuse Galactic synchrotron
emission.  Confirming the candidates with a lower degree of
polarization will require further observations at different
wavelengths.

\subsection{G19 SNRs}

We identify 91 out of 94 objects in the catalog of Galactic SNRs
\citep[G19 SNRs,][]{2019JApA...40...36G} covered in the GLOSTAR survey
region.  The SNRs G0.0$+$0.0 and G0.3$+$0.0 near the Galactic
center are in a  heavily confused region, and the radio emission
from the SNR G32.1$-$0.9 still remains undetected.  The positions of these
objects along with their measured flux densities are given in Table~\ref{table:G19SNRs}.
Studying the GLOSTAR-VLA images, we find that four G19 SNRs are actually
\ion{H}{ii}~regions, and four have
ambiguous radio emission.  We briefly discuss these objects in \S \ref{subsubsec:misG19} and \S \ref{subsubsec:ambigG19}.  Below, we discuss the flux densities of G19 SNRs as measured in the GLOSTAR-VLA data.

\subsubsection{Flux densities of G19 SNRs}
\label{subsubsec:missflux}

\begin{figure}
    \centering
    \includegraphics[width=0.48\textwidth]{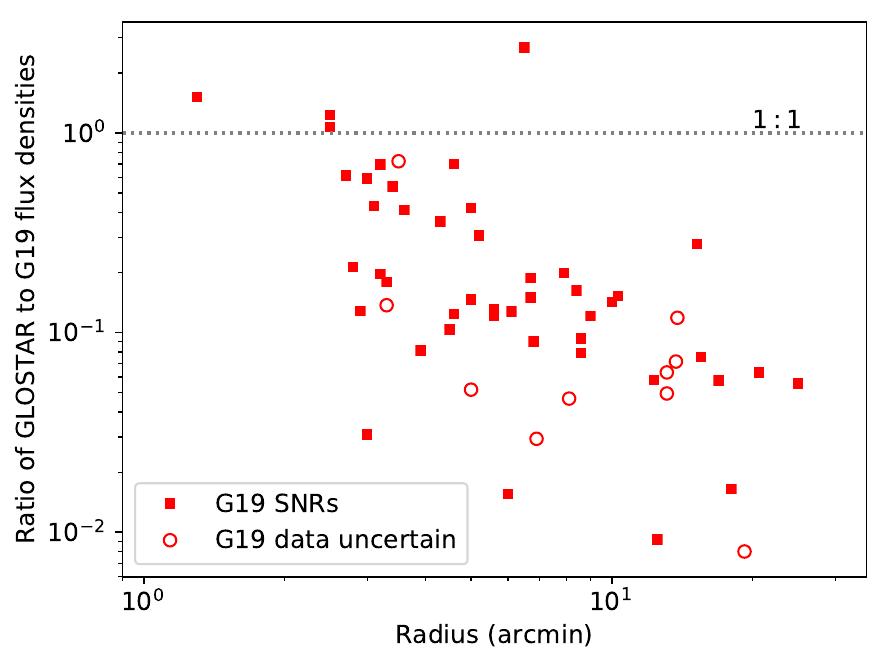}
    \caption{Comparing the flux densities of G19 SNRs measured in the GLOSTAR-VLA 5.8 GHz data and the flux densities reported in the G19 catalog scaled from 1 GHz to 5.8 GHz.  Those with uncertain flux densities or spectral indices in the G19 catalog are marked with circles. }
    \label{fig:missflux}
\end{figure}

Because of the missing short-spacing information, there is a significant amount
of undetected flux density in the GLOSTAR-VLA data.  The amount of flux density
recovered depends on the scale and structure of the emission,
although spectral index also plays a role because of the wide
bandwidth of the survey.  While it is true that a single power-law
model may not suffice for extrapolating over a wide range of
frequencies, one can still get a rough estimation of the undetected
flux density by measuring the flux densities of known extended
objects.  \citet{2019JApA...40...36G} report a flux density
and spectral index for most objects in their catalog
(albeit without uncertainties in the measurements).
In Fig.~\ref{fig:missflux}, we plot the ratio of the GLOSTAR-VLA
flux density to the G19 flux density, scaled to 5.8~GHz using their
individual spectral indices,
against the radius of each object in the G19 catalog.  The
median ratio is ${\sim}0.15$.  As expected, most objects fall
below the 1:1 line, and the amount of flux density not recovered increases
with the size of the object.  There are four G19 SNRs with
ratios $>1$, with the largest ratio being ${\sim}2.7$.  This is probably
because of uncertainties in the G19 flux density, and the fact that
SNRs need not follow a single power-law model.

\subsubsection{H II regions mistaken for SNRs}
\label{subsubsec:misG19}

\begin{figure*}
    \centering
    \includegraphics[width=0.48\textwidth]{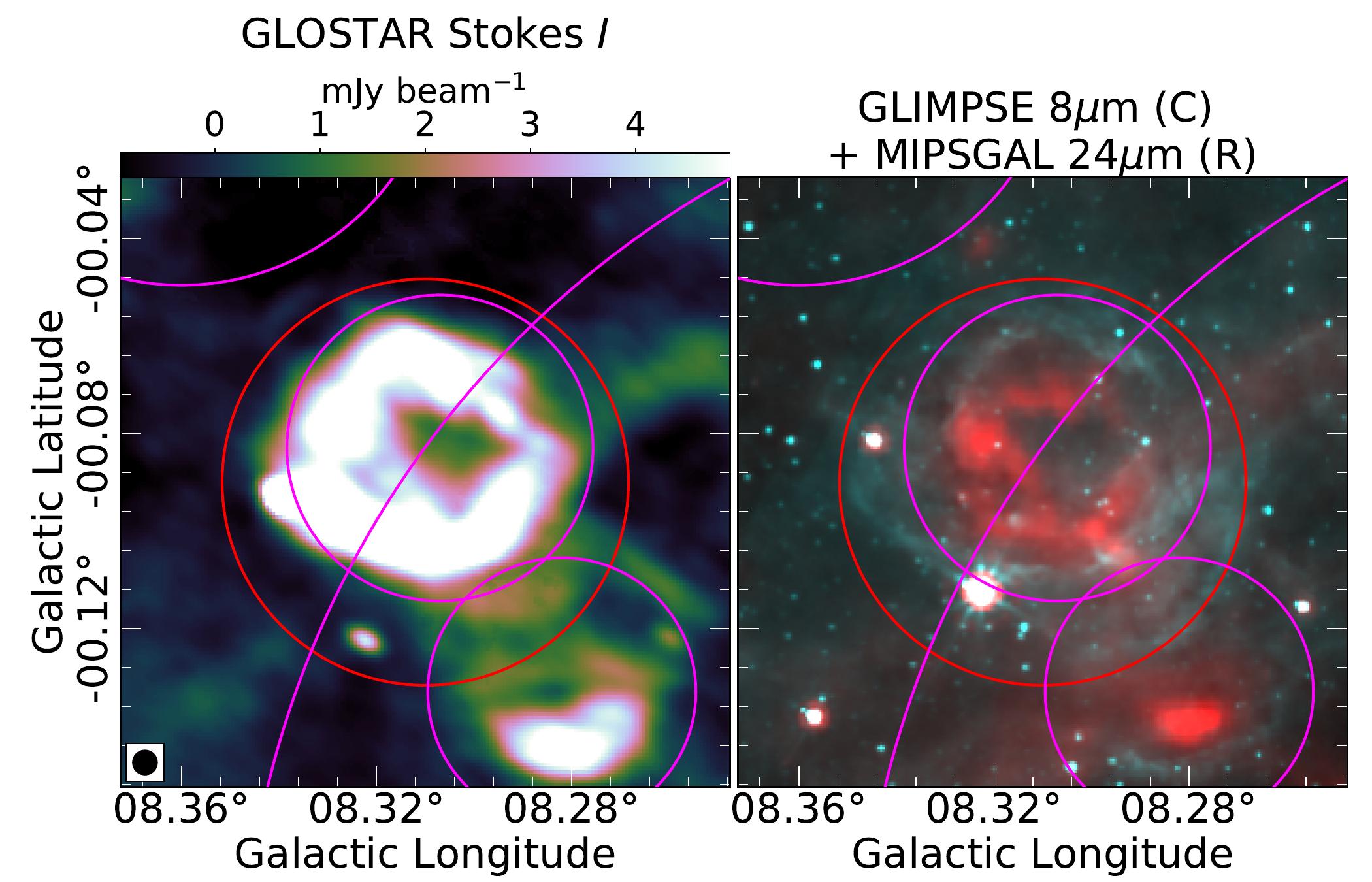}
    \includegraphics[width=0.48\textwidth]{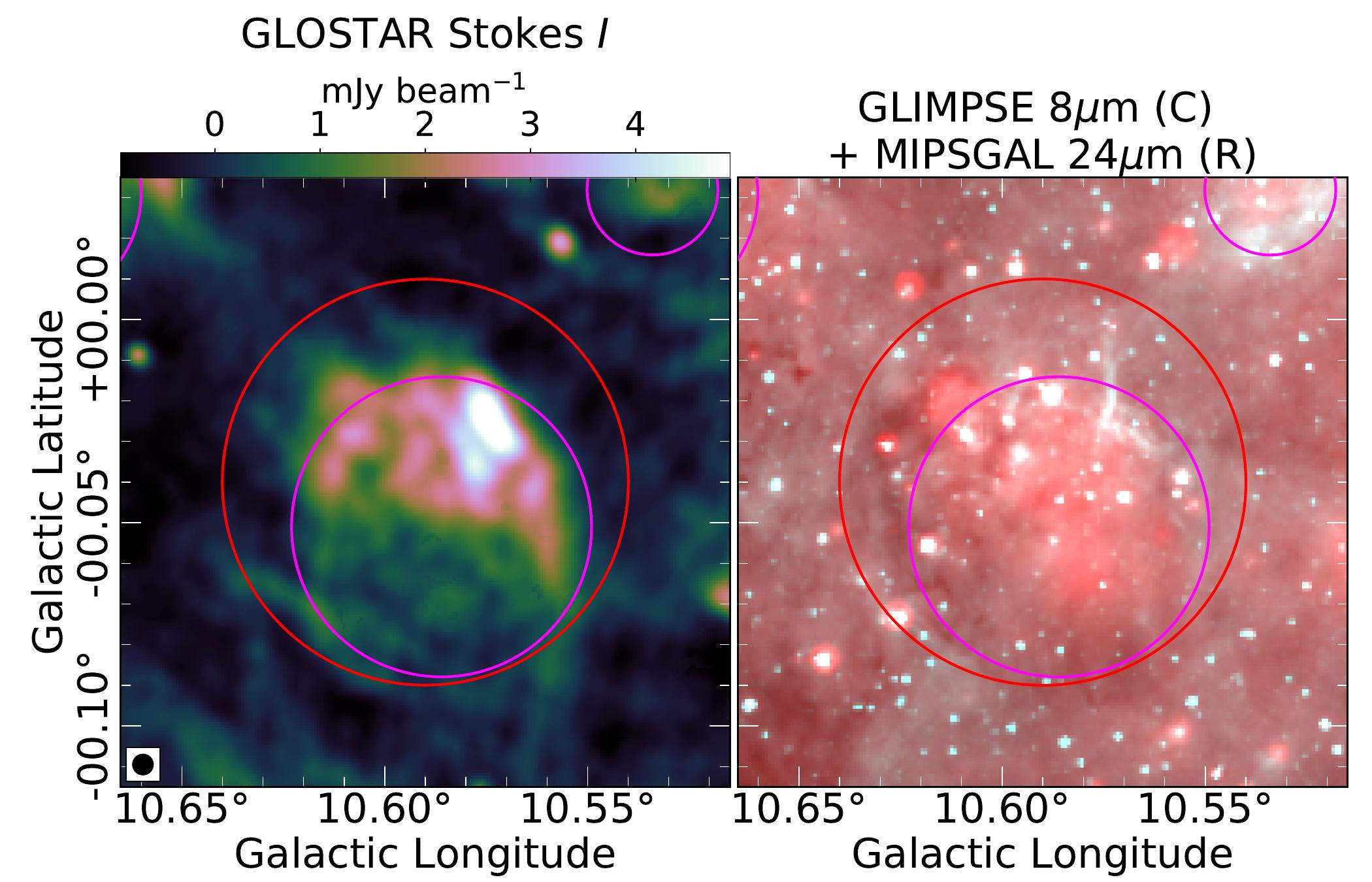}

    \includegraphics[width=0.48\textwidth]{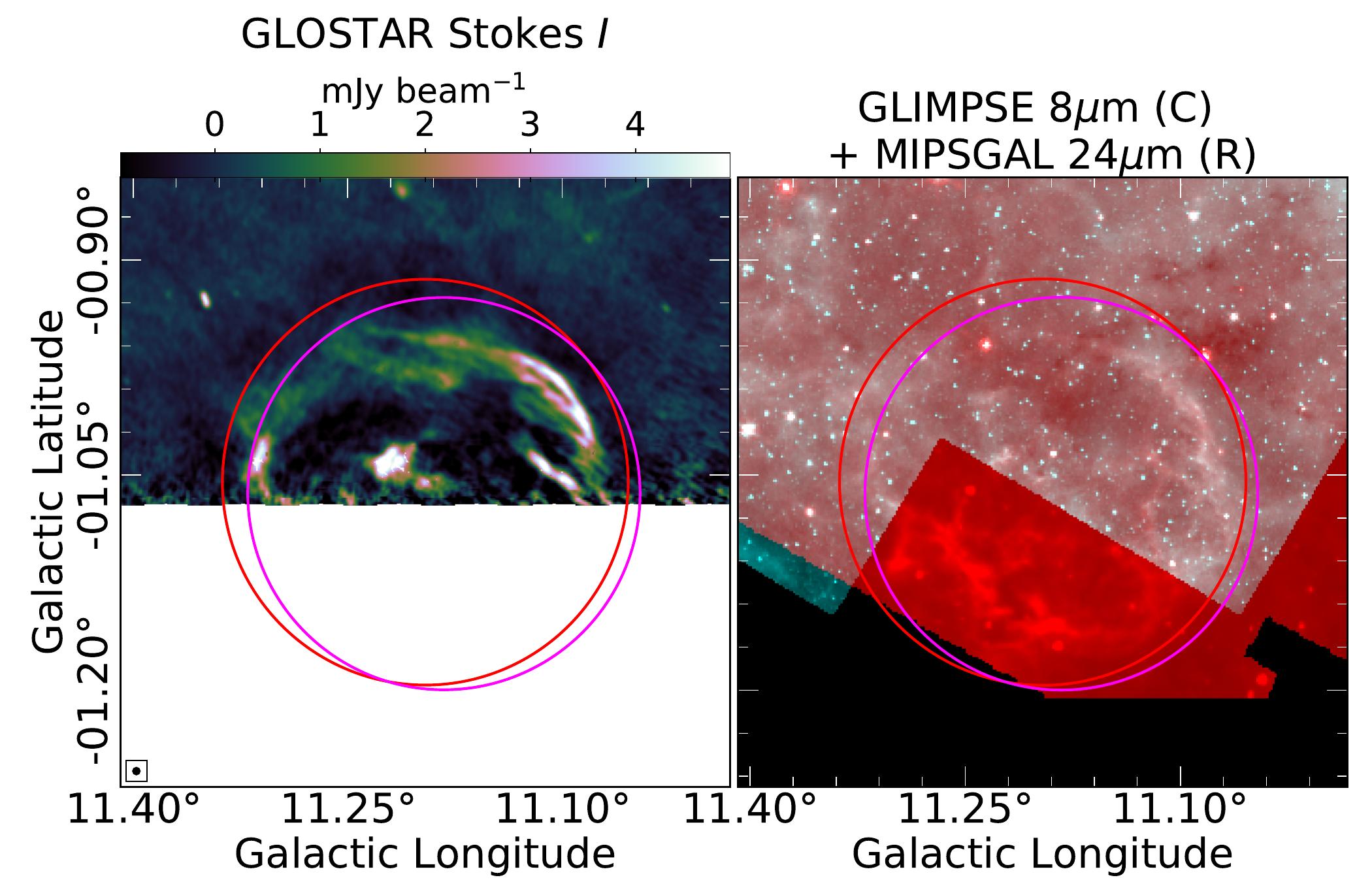}
    \includegraphics[width=0.48\textwidth]{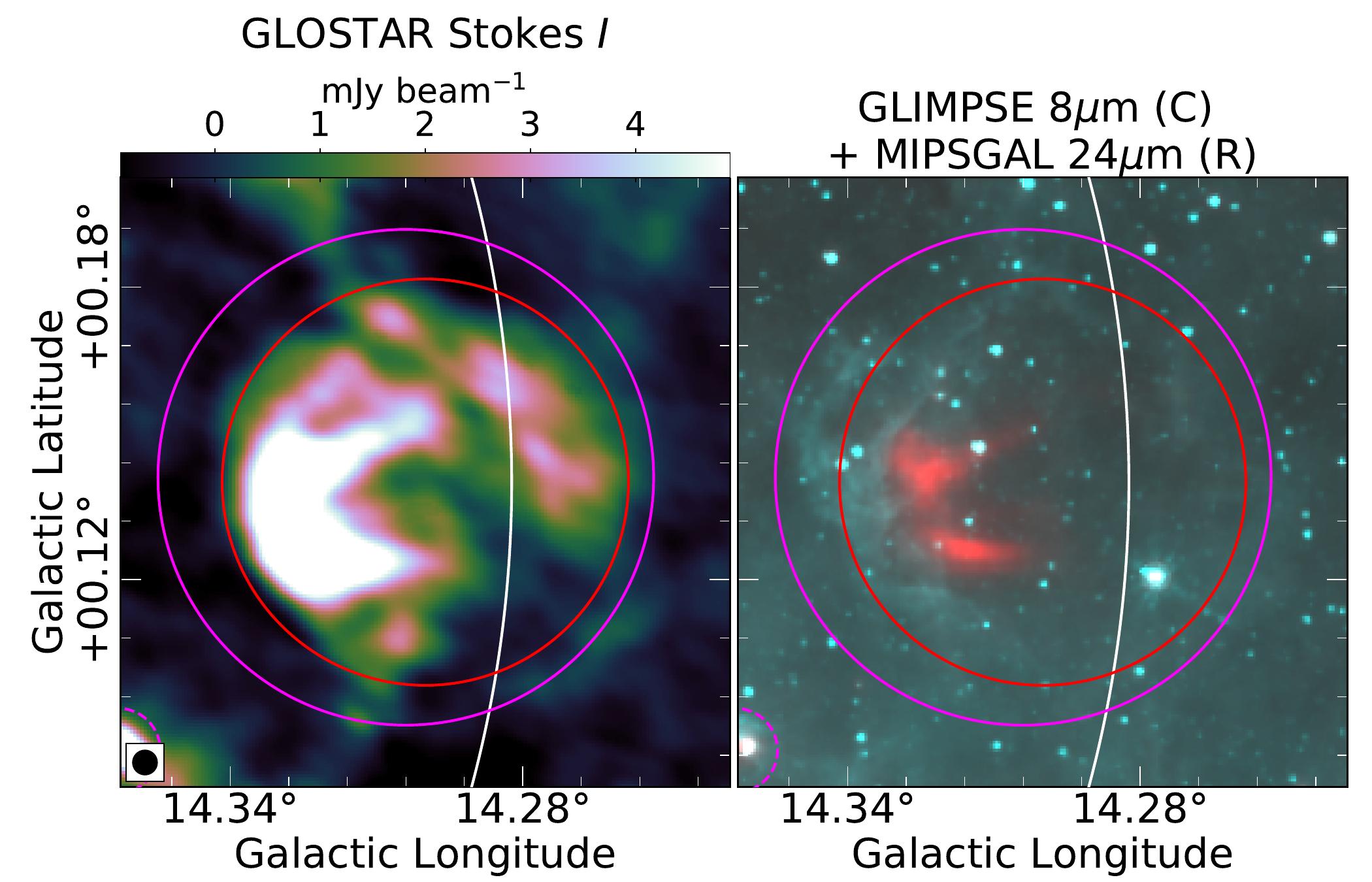}
    \caption{Objects in both G19 SNR and A14 \ion{H}{ii}~region catalogs: G8.3$-$0.0 (top left), G10.5$-$0.0 (top right), G11.1$-$1.0 (bottom left, partially covered) and G14.3$+$0.1 (bottom right).  The left panels are the GLOSTAR-VLA images and the right panels are MIR data: MIPSGAL 24~$\mu$m (red) and GLIMPSE 8~$\mu$m (cyan).  The marking scheme is explained in Fig.~\ref{fig:example}.}
    \label{fig:wrongSNRs}
\end{figure*}

Four G19 SNRs, G8.3$-$0.0, G10.5$-$0.0, G11.1$-$1.0, and G14.3$+$0.1, have
coincident MIR emission (Fig.~\ref{fig:wrongSNRs}) and are present in
the A14 \ion{H}{ii}~region catalog.  They were also noted as
\ion{H}{ii}~regions in earlier studies
\citep{1989ApJS...71..469L,1996ApJ...472..173L,2019A&A...623A.105G}.  We
detect no significant polarization from these objects, agreeing with
their identifications as \ion{H}{ii}~regions.

\subsubsection{G19 SNRs with ambiguous radio emission}
\label{subsubsec:ambigG19}

\begin{figure*}
    \sidecaption
    \includegraphics[width=0.7\textwidth]{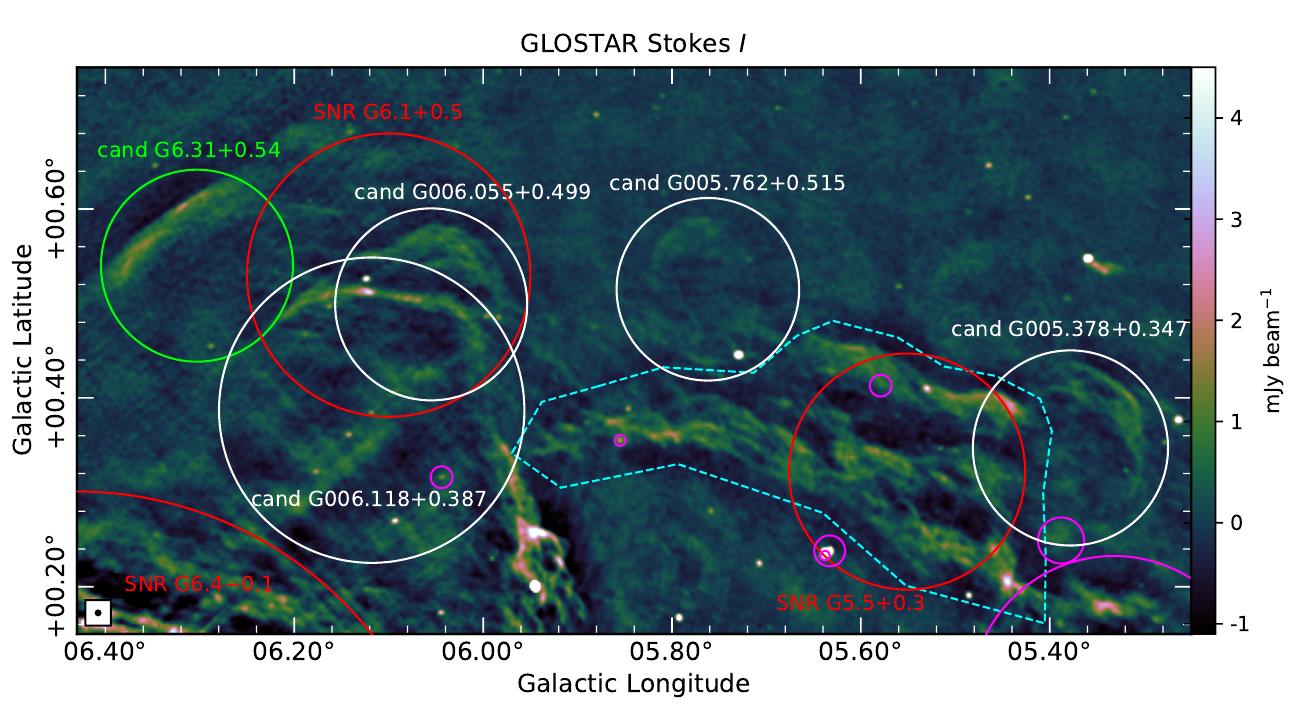}
    \caption{Field of G19 SNRs G5.5$+$0.3 and G6.1$+$0.5.  The region of G5.5$+$0.3 defined by \citet{2006ApJ...639L..25B} seems to be a part of a larger structure with no easily recognizable shape (marked with a dashed  cyan polygon). }
    \label{fig:G5.5_G6.1}
\end{figure*}

\indent\indent {\bf{G5.5$+$0.3}} was identified by \citet{2006ApJ...639L..25B}
as a class~II SNR candidate (class I being very likely to be a SNR, and
class III being least likely).  \citet{2009A&A...508.1331L} report strong CO
emission from the periphery of this SNR at $l=5.64,~b=0.23$; this is probably
associated with the presence of A14 \ion{H}{ii}~regions G005.633$+$00.238 and
G005.637$+$00.232.  \citet{2011MNRAS.414.2282S} studied G5.5+0.3 at optical
wavelengths, but could not confidently associate optical and radio emission.
\citet{2019JApA...40...36G} reports a spectral index of $-0.7$, a 1~GHz flux density
of $5.5$~Jy, and a size $15\arcmin\times12\arcmin$.  These values translate
to an expected average surface brightness of over ${\sim}1.1$ mJy beam$^{-1}$
in the GLOSTAR-VLA data, which is well above the local noise
(${\sim}0.1$~mJy~beam$^{-1}$).  Emission from this region is indeed detected
in the GLOSTAR-VLA data (Fig.~\ref{fig:G5.5_G6.1}), but it is part of a
much larger structure (marked with a dashed cyan  polygon in
Fig.~\ref{fig:G5.5_G6.1}) that is ${\sim}0.6\degr$ in angular extent.  We
also observe no significant polarization.  Considering the fact that
we recover only a small fraction of the flux density (see \S\ref{subsec:GLOSTARdata}
and \S\ref{subsubsec:missflux}), it is likely that the nonthermal emission from
the SNR is actually undetected.

{\bf{G6.1$+$0.5}}:  This appears to be a superposition of two
objects in the GLOSTAR-VLA data
(Fig.~\ref{fig:G5.5_G6.1}).  One object centered at $l=6.118\degr,b=0.387$
has a bright arc-shaped emission on its northern edge, whereas the rest of
its shell is faint.  This arc-shaped emission passes through another object,
centered at $l=6.055,b=0.499,$ which has a clear shell morphology.
These newly resolved shells from G6.1$+$0.5 are included in the list of
GLOSTAR SNR candidates (see Table~\ref{table:newcandSNRs}).

\begin{figure}
    \centering
    \includegraphics[width=0.48\textwidth]{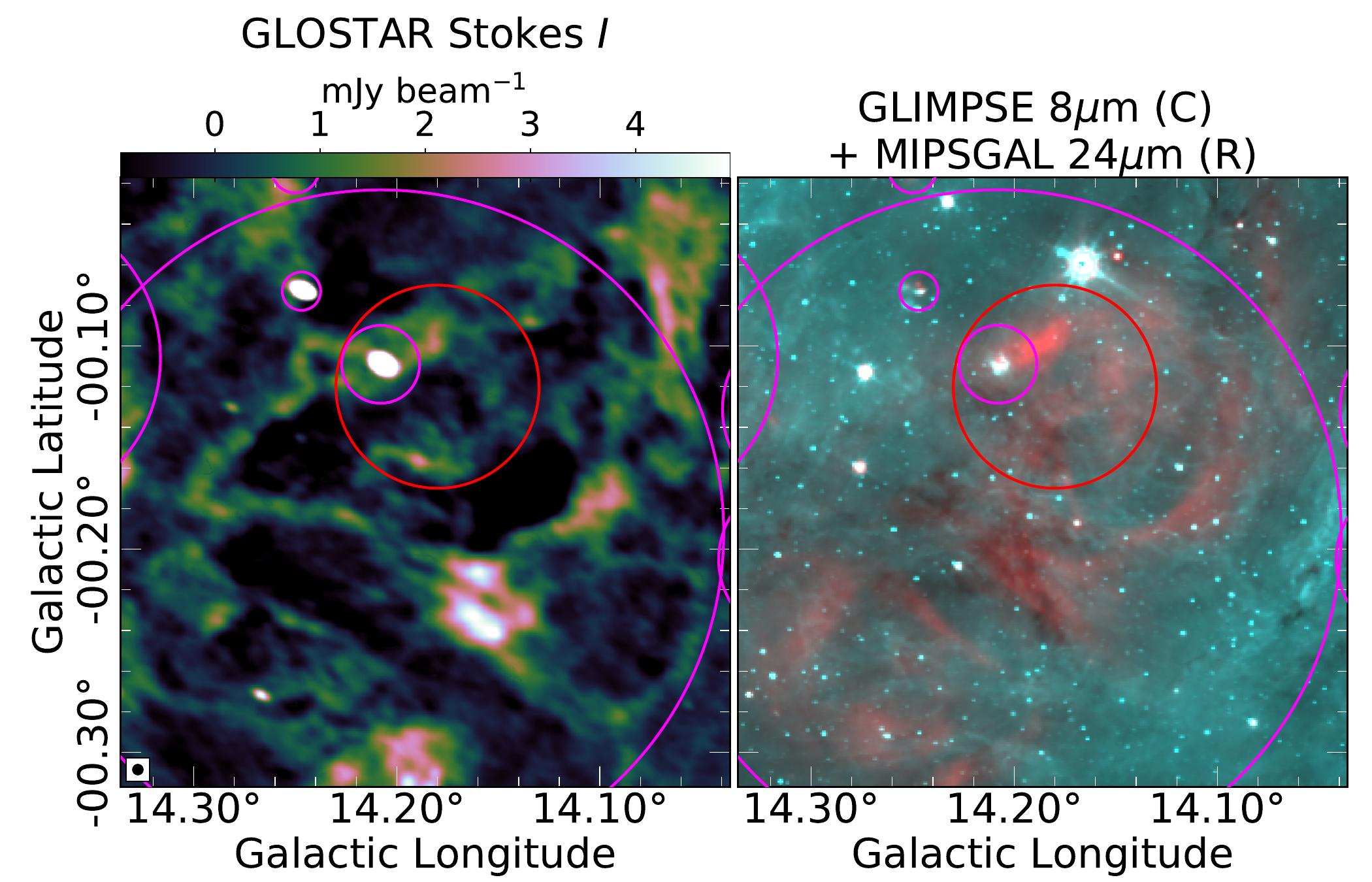}
    \caption{Environment of G19 SNR G14.1$-$0.1 (encircled in red).  The left panel shows the GLOSTAR-VLA data and the right panel shows MIPSGAL 24~$\mu$m data in cyan and GLIMPSE 8~$\mu$m in red.  The supposed shell of G14.1$-$0.1 is not clearly detected in the GLOSTAR-VLA data. }
    \label{fig:G14.1}
\end{figure}

{\bf{G14.1$-$0.1}}:  The emission from this object as seen in the
GLOSTAR-VLA data (Fig.~\ref{fig:G14.1}) is dominated by the bright
A14 \ion{H}{ii}~region G014.207$-$00.110
in the northeast, and these objects lie inside the large A14
\ion{H}{ii}~region G014.207$-$00.193.  Given the flux density
of 0.5~Jy, spectral index of $-0.6,$ and a size of $6\arcmin\times5\arcmin$
reported by \citet{2019JApA...40...36G}, the nonthermal emission from
G14.1$-$0.1 should have an average surface brightness of
${\sim}0.75$~mJy beam$^{-1}$ in the GLOSTAR-VLA data.  However, we cannot
positively identify an object distinct from the surrounding emission.
In addition, the northern part of G14.1$-$0.1 is detected in MIPSGAL.
These facts indicate that the emission we observe in the GLOSTAR-VLA
data may just be from the large \ion{H}{ii}~region G014.207$-$00.193,
and not the SNR G14.1$-$0.1.  Similar to G5.5$+$0.3, the
nonthermal emission from this SNR is also probably undetected.

\begin{figure}
    \centering
    \includegraphics[width=0.48\textwidth]{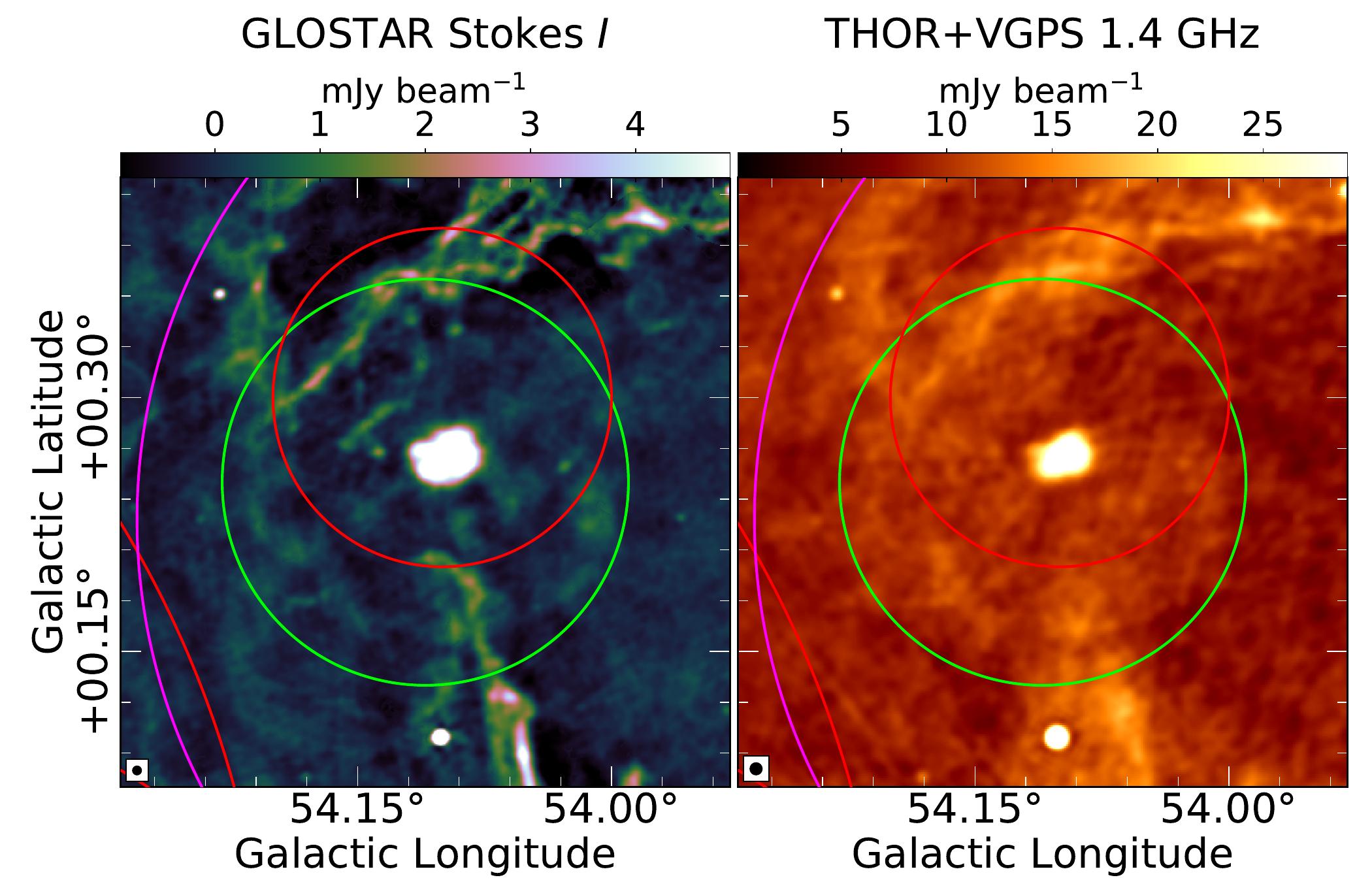}
    \caption{Supposed shell of G54.1$+$0.3 encircled in red, and THOR SNR candidate G54.11$+$0.25 encircled in green.  }
    \label{fig:G54.11}
\end{figure}

{\bf{G54.1$+$0.3}}: The pulsar wind nebula (PWN) G54.1$+$0.3 is
identified as a bright source in the GLOSTAR-VLA data,
with a faint shell surrounding this emission (Fig.~\ref{fig:G54.11}).
\citet{2010ApJ...709.1125L} and \citet{2017A&A...605A..58A} also
identify a shell with similar morphology,
but it is not yet known if it is associated with
the PWN.  \citet{2017A&A...605A..58A} included this shell in the
list of THOR SNR candidates, named G54.11$+$0.25.  Distance
measurements to the PWN have yielded inconsistent results
\citep{2018AJ....155..204R,2013ApJ...774....5K,2012JKAS...45..117L},
and a multiwavelength study by \citet{2018ApJ...860..133D} suggests
that the shell is unlikely to be a SNR, indicating that further
observational studies are required to fully disentangle the
emission from this region.

\subsection{SNR candidates}
\label{subsec:resultscands}

We identify 80 new candidate SNRs using the methodology described in \S
\ref{subsec:identifySNRs}.  We visually identify counterparts of
50 of these objects in either of the 20 cm (1400 MHz) THOR+VGPS, the 20/90 cm (1400/325 MHz)
MAGPIS, or the 150 cm (200 MHz) GLEAM data.  As SNRs are brighter at lower frequencies,
the detection of the SNR candidates in these lower frequency surveys
data can be used as an assessment of our confidence level in these
candidates.  The positions of
these objects, their sizes, and their flux densities are listed in
Table~\ref{table:newcandSNRs}.  The images of all these candidates are
presented in Appendix \ref{ims:newcandSNRs}.

In addition, we also identify 77 SNR candidates in the GLOSTAR-VLA data that were discovered in earlier
studies (see \S \ref{subsubsec:prevcandslist}).  The details of these candidates, along with references to the
studies that identified them, are presented in Table~\ref{tab:prevcandSNRs}; if they are found to have MIR emission in the MIPSGAL 24~$\mu$m and
GLIMPSE 8~$\mu$m images, they are marked as thermal in the remarks
column.

\subsubsection{Comparing the properties of SNR candidates with G19 SNRs and H II regions}

\begin{figure*}
    \centering
    \includegraphics[width=0.96\textwidth]{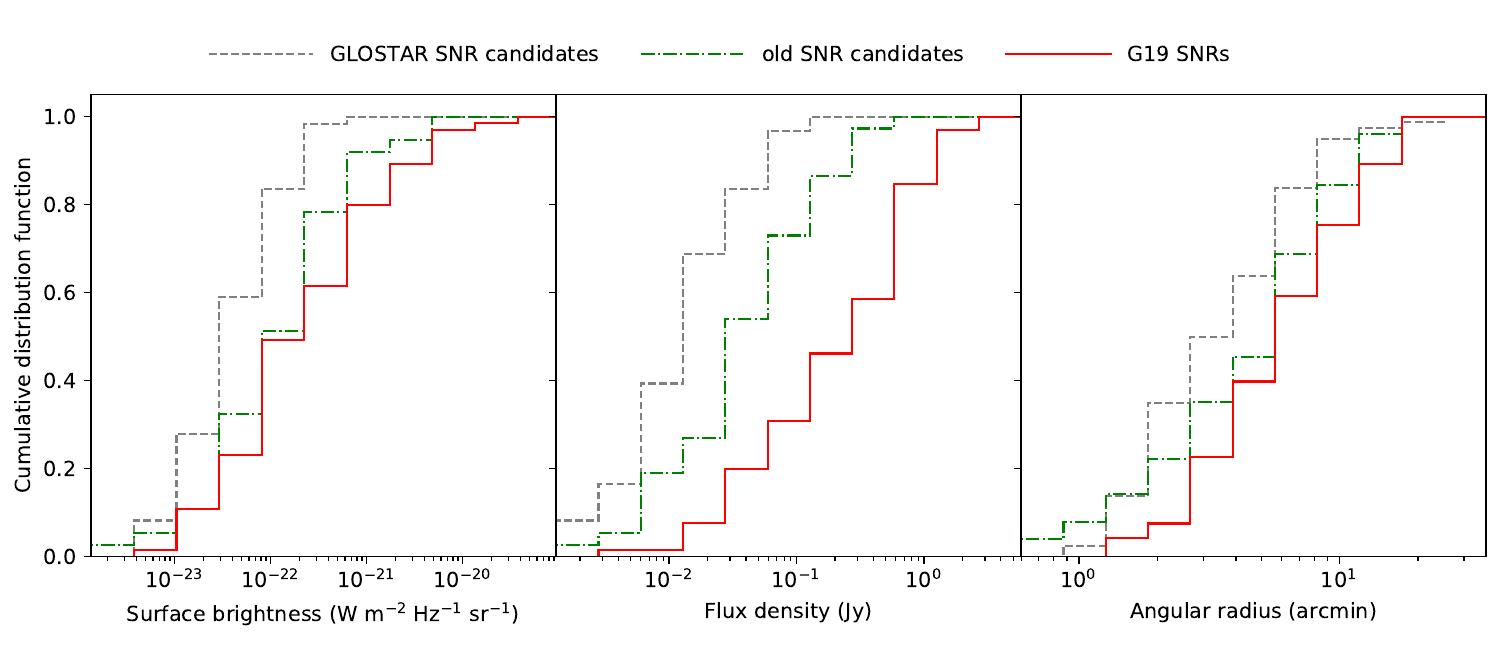}
    \caption{Cumulative distribution functions of average 5.8 GHz surface brightness (left), flux density (middle), and angular radius (right) of G19 SNRs (red), the SNR candidates discovered in earlier studies (green), and the new SNR candidates identified in the GLOSTAR-VLA survey (gray).  The average surface brightness is obtained by dividing the flux density by the angular area subtended by the object.  All the properties presented here are as measured in the GLOSTAR-VLA data.  }
    \label{fig:CDFs}
\end{figure*}

In Fig. \ref{fig:CDFs}, we present the cumulative distribution functions (CDFs) of 5.8 GHz surface brightness (defined as the ratio of flux density to the area subtended), GLOSTAR-VLA flux density, and radius of the three samples of G19 SNRs, the previously identified SNR candidates detected in the GLOSTAR-VLA data, and the newly discovered GLOSTAR-VLA SNR candidates.  The CDFs show that the new SNR candidates discovered in the GLOSTAR-VLA data are in general smaller and fainter than the other two samples; this is expected because of the survey's better surface brightness sensitivity and better resolution than many previous large-scale studies.  The sample of G19 SNRs consists of objects that were easily detected and well studied, and hence they are brighter and larger than the other two samples.  We note that the observed differences across the three samples in the 5.8 GHz surface brightness and flux density are not artefacts of the problem of flux density resolved out by the interferometer.  All the measurements presented in Fig. \ref{fig:CDFs} are from the GLOSTAR-VLA data, and the ``missing flux density'' problem affects all the measurements.  As G19 SNRs are in general larger than the two samples of SNR candidates, more flux density is resolved out from G19 SNRs than the other two.  Future addition of single-dish data (currently being collected with the Effelsberg 100 m telescope) is only expected to widen the differences in flux density and surface brightness of these samples.

\begin{figure}
    \centering
    \includegraphics[width=0.47\textwidth]{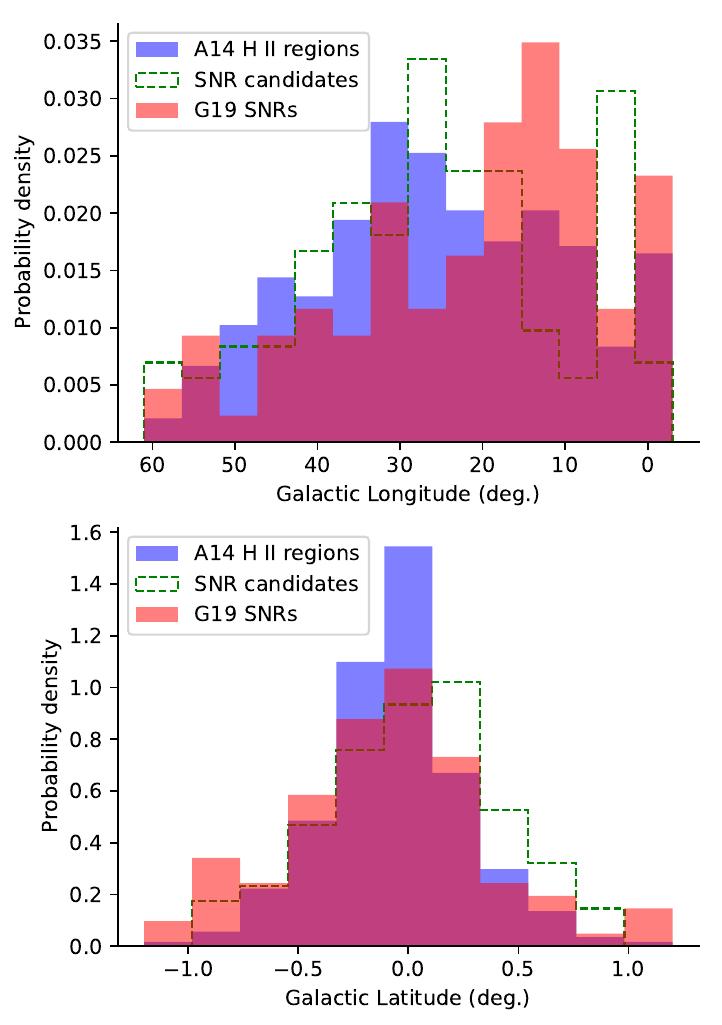}
    \caption{Probability distribution functions of the Galactic longitudes (top) and latitudes (bottom) for the three samples of G19 SNRs, A14 \ion{H}{ii}~regions, and SNR candidates detected in the GLOSTAR-VLA data (both previously discovered candidates and new GLOSTAR candidates together). }
    \label{fig:GLongGLat}
\end{figure}

The histograms of Galactic longitudes and latitudes of A14 \ion{H}{ii}~regions, G19 SNRs, and SNR candidates (both newly discovered and previously identified ones together) detected in the GLOSTAR-VLA survey are shown in Fig. \ref{fig:GLongGLat}.  \ion{H}{ii}~regions and SNRs trace recent massive star formation activity, and are generally expected to have similar distributions, although local discrepancies are common.  \citet{2017A&A...605A..58A} note that there could be physical reasons for the apparent differences, such as the progenitors of supernovae being both O- and B-stars, and \ion{H}{ii}~regions generally tracing only O-stars.  The survey by \citet{2006ApJ...639L..25B}, which covered the longitude range $22\degree>l>4.5\degree$, nearly doubled the number of confirmed SNRs in their survey region.  The number of SNR candidates in this region is also relatively small, and so the observed differences in Galactic longitude among the three samples (top panel, Fig. \ref{fig:GLongGLat}) seem to be a result of previous surveys focusing on selected regions.  We also observe a difference in the Galactic latitudes of SNR candidates compared with the other two samples.  Both G19 SNRs and A14 \ion{H}{ii}~regions peak at $b{\sim}0\degree$ and populate the $b<0\degree$ latitudes slightly more than $b>0\degree$, but the distribution of SNR candidates is quite asymmetric and skewed toward $b>0\degree$ with its peak at $b{\sim}0.2\degree$.  \citet{2017A&A...605A..58A} also report a similar shift toward positive latitudes for the THOR SNR candidates, although less apparent.  The reason for this unexpected shift seems unclear.  The current sample of G19 SNRs and SNR candidates may still not be representative of the overall Galactic SNR population.

\subsubsection{The number of SNRs in the Galaxy}
\label{subsubsec:nSNRs}

\citet{1989ApJ...341..151H} estimated that the Galaxy should contain ${\gtrsim}590$ SNRs by studying the distribution of approximately ${}155$ SNRs known at that time.  They arrived at this number by assuming that the surface density of Galactic SNRs in the longitude range $270\degree>l>90\degree$ provides a stringent lower limit to the actual surface density of SNRs across the Milky Way, and they used a linear gradient of SNR number density.  These latter authors also provide the total expected number of SNRs in different regions of the Galaxy which they named from A--M.  \citet{1991ApJ...378...93L} further analyzed the distribution of SNRs in a similar statistical manner, but they assumed a ``selection-free'' zone of 3 kpc around the Sun and used exponential disk and spiral arm scales to model the SNR number surface density.  These authors estimate that there must be ${\gtrsim}1000$ SNRs and also predict the number of SNRs in various regions of the Milky Way defined by \citet{1989ApJ...341..151H}.  The estimates given by \citet{1991ApJ...378...93L} outnumber the estimates of \citet{1989ApJ...341..151H} in most regions \citep[see Table 4 of][]{1991ApJ...378...93L}.  The total number of SNRs given by \citet[][]{1991ApJ...378...93L},  about ${}1000$, is in agreement with other studies \citep{1994ApJS...92..487T,2013JCAP...06..041M}.  A simple calculation involving the lifetimes of SNRs \citep[$\sim 60\,000$ years;][]{1994ApJ...437..781F} and a supernova rate of two per century \citep{1993A&A...273..383C,2013ApJ...778..164A} also gives a number upwards of 1000 SNRs.  With this context, below we discuss the distributions of SNRs observed in the GLOSTAR-VLA data.

\begin{table}
\caption{Comparing the distributions of Galactic longitudes of SNRs observed in the GLOSTAR-VLA survey with the predictions by \citet{1989ApJ...341..151H} and \citet{1991ApJ...378...93L}.}
\label{tab:SNR_comp}
\centering
\begin{tabular}{ l c c c }
\hline\hline
 & $0\degree-30\degree$ & $30\degree-45\degree$ & $45\degree-60\degree$  \\ \hline
\citet{1991ApJ...378...93L} & $316-380$ & $101$    & $30-36$               \\
\citet{1989ApJ...341..151H} & $146-176$ & $53$     & $27-32$               \\
SNRs in GLOSTAR             & $155$     & $58$     & $26$               \\
\hline
\end{tabular}
\tablefoot{The numbers presented against \citet{1991ApJ...378...93L} are for their 1000 SNRs model, and those against ``SNRs in GLOSTAR'' include the three samples of G19 SNRs, and previously and newly discovered SNR candidates in GLOSTAR.\\
The expected number of SNRs in the longitude range $0\degree<l<30\degree$ is obtained by assuming that SNRs in $0\degree<l<30\degree$ account for $50\%-60\%$ of the SNRs in the $|l|<30\degree$ range (regions F+G+H+I as defined by \citealt{1989ApJ...341..151H}; see their figure 8).  Similarly for $45\degree<l<60\degree$ ($33\%-40\%$ of SNRs in $45\degree<l<90\degree$, i.e., regions B+C).  }
\end{table}

In Table \ref{tab:SNR_comp}, we compare the results of our search with the expected numbers of SNRs given by \citet{1989ApJ...341..151H} and \citet{1991ApJ...378...93L}, assuming that all the SNR candidates are positive identifications.  Our results are a surprisingly good match with the predictions by \citet{1989ApJ...341..151H}, but fall well short of the numbers given by \citet{1991ApJ...378...93L}.  We believe that the agreement between our results and the predictions of \citet{1989ApJ...341..151H} is a coincidence.  Our survey is unlikely to be sensitive enough to detect all the SNRs in the survey region.  A simple way to test this would be to conduct deeper searches for SNRs in the longitude range $0\degree<l<30\degree$.  Comparing our results with the numbers estimated by \citet{1991ApJ...378...93L}, we find that over 150 SNRs remain to be detected in this region (see Table \ref{tab:SNR_comp}).  Therefore, if deeper surveys reveal more SNRs, then the good agreement with the expected number given by \citet{1989ApJ...341..151H} is purely a coincidence.  However, if no new SNR candidates were to be discovered in future deeper surveys---which we believe is unlikely---then we may need to rethink the distributions of SNRs and also possibly the total number of SNRs in the Milky Way.

\subsection{New SNR candidates with polarized emission}
\label{subsec:newcandpol}
\begin{figure*}
    \centering
    \includegraphics[width=0.96\textwidth]{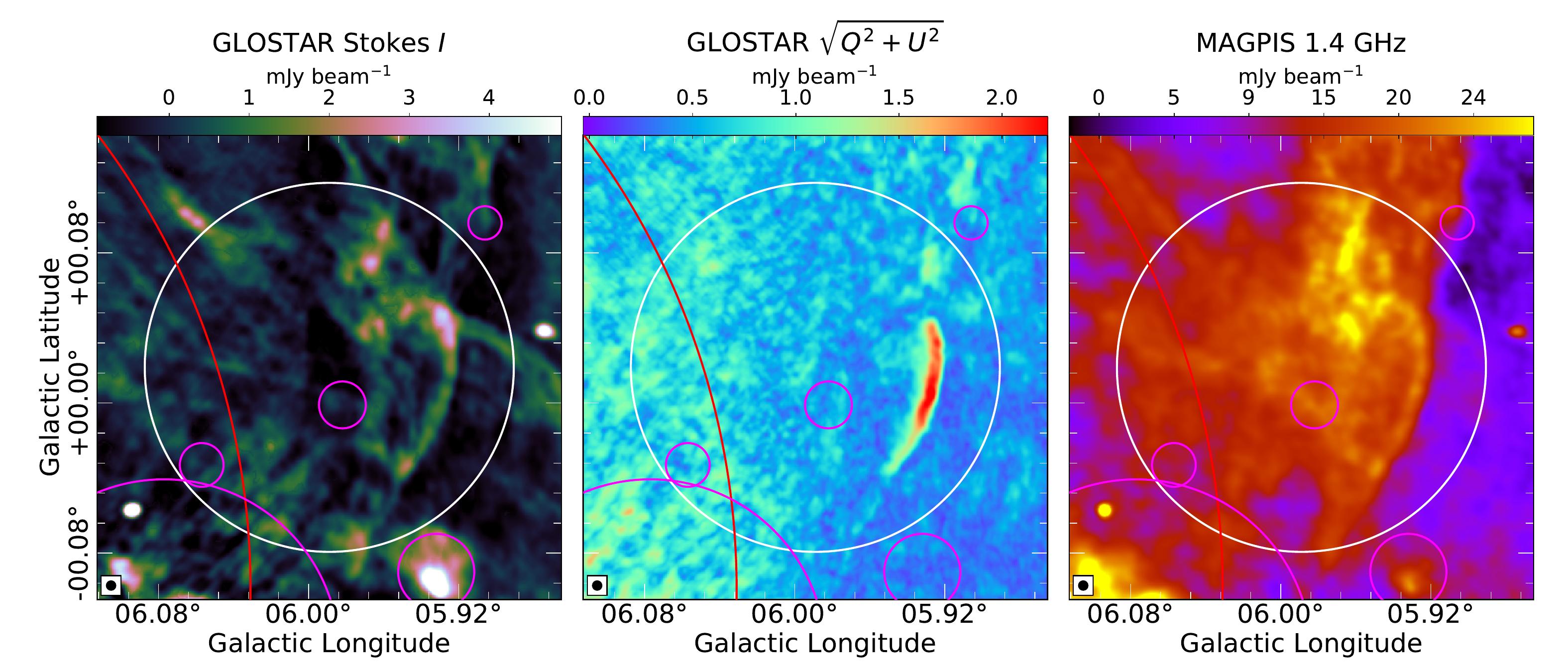}

    \includegraphics[width=0.96\textwidth]{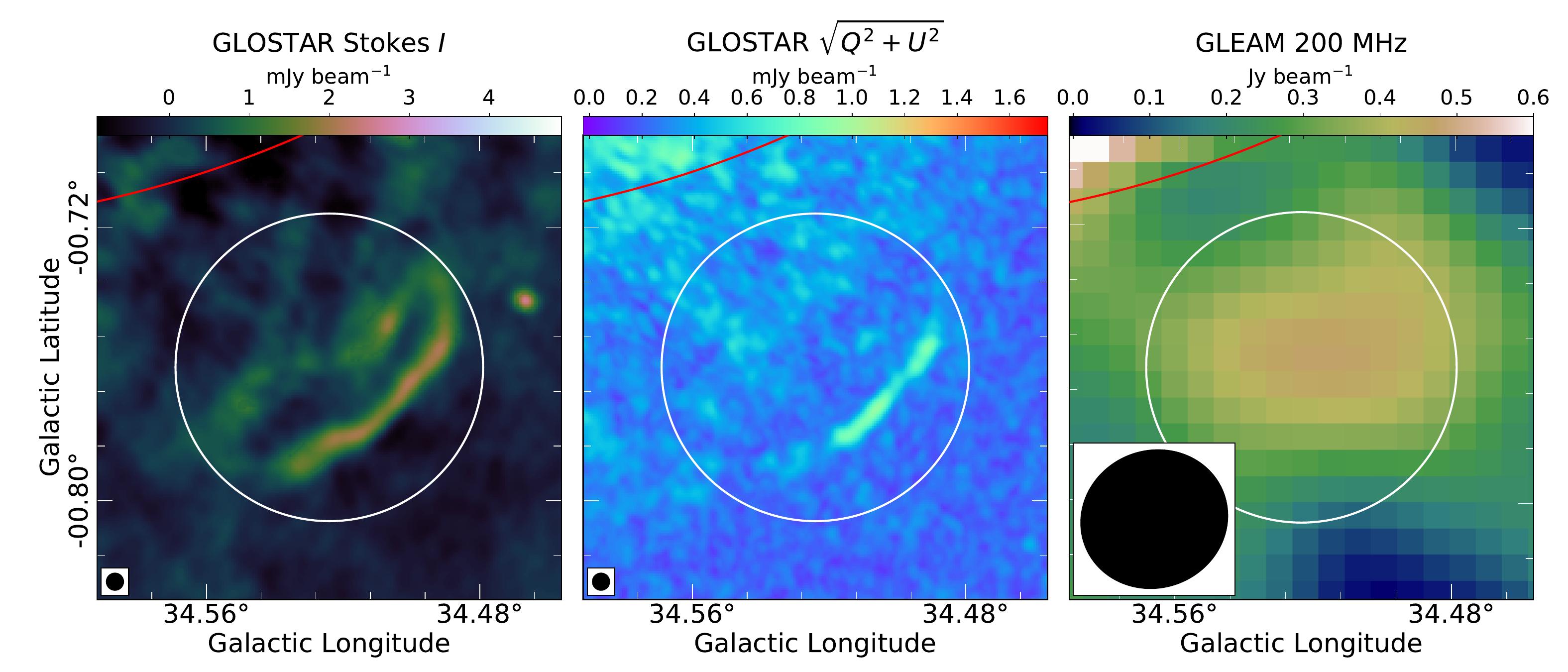}

    \includegraphics[width=0.96\textwidth]{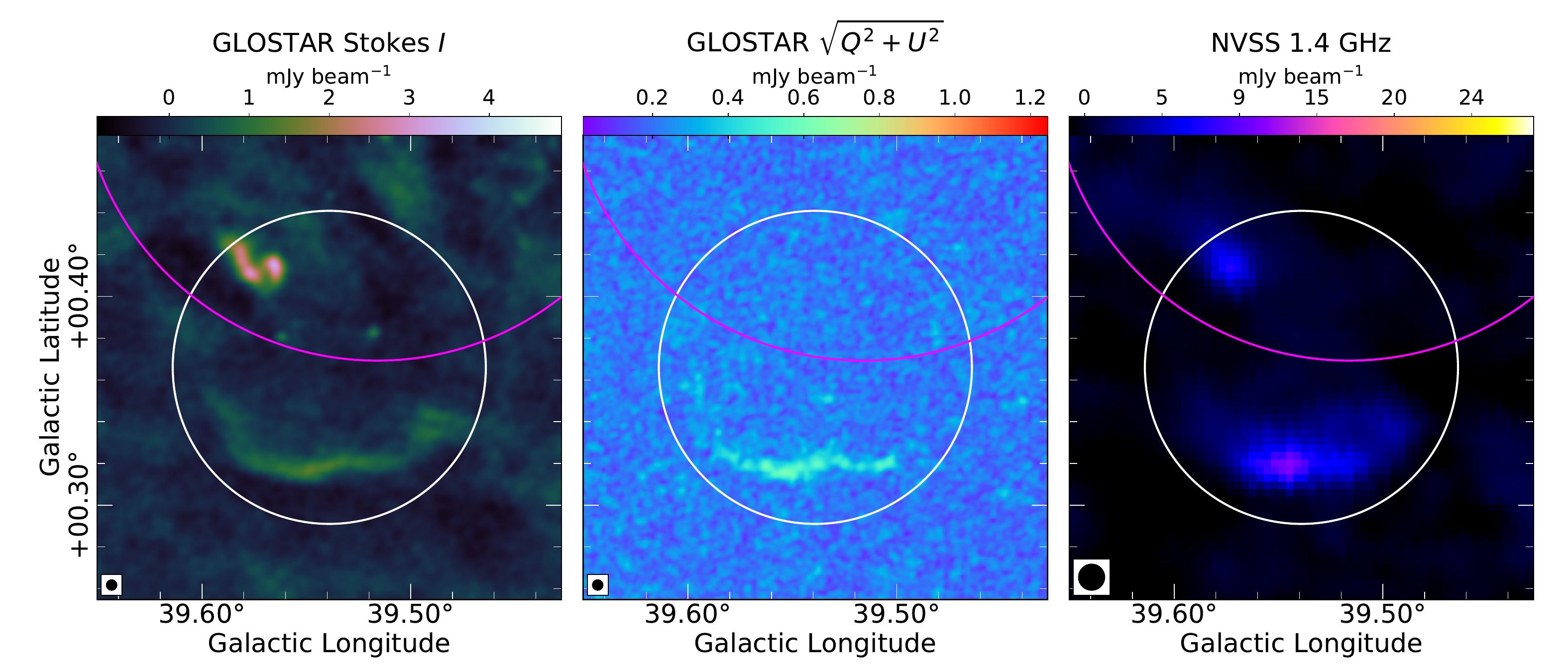}
    \caption{New SNR candidates identified in the GLOSTAR-VLA data with significant polarization: G005.989$+$0.019 (top panels), G034.524$-$0.761 (middle panels), and G039.539$+$0.366 (bottom panels).  Although we correct the linearly polarized flux density for Ricean bias (see \S \ref{subsubsec:polflux}), we present the polarization data in all the figures without bias correction so that the structures are clearly seen. }
    \label{fig:new_SNR_pol}
\end{figure*}

Three new SNR candidates have significant linearly polarized emission
clearly coming from their Stokes $I$ counterparts, namely
G005.989$+$0.019, G034.524$-$0.761, and
G039.539$+$0.366, with degrees of polarization $0.18 \pm 0.03$,
$0.07 \pm 0.02,$ and $0.06 \pm 0.02$, respectively.
All three have a lone arc morphology that is reminiscent of a shell (see
Fig.~\ref{fig:new_SNR_pol}).
These three candidates are also detected in lower frequency
surveys (shown in the right panels of Fig.~\ref{fig:new_SNR_pol}).  The other
newly discovered candidates with counterparts in lower frequency surveys are marked
in Table~\ref{table:newcandSNRs}.
The low number of detections in polarization may imply that a large portion
of our new candidates may in fact be \ion{H}{ii}~regions that are too faint to be detected
by the GLIMPSE and MIPSGAL surveys, although this is unlikely
\citep{2014ApJS..212....1A}.  Spectral index measurements can ascertain
the nature of these new candidates.

\subsection{Previously identified SNR candidates with polarized emission}
\label{subsec:candSNRs_pol}

\subsubsection{G26.75\texorpdfstring{$+$}{+}0.73}

\begin{figure*}
    \centering
    \includegraphics[width=0.96\textwidth]{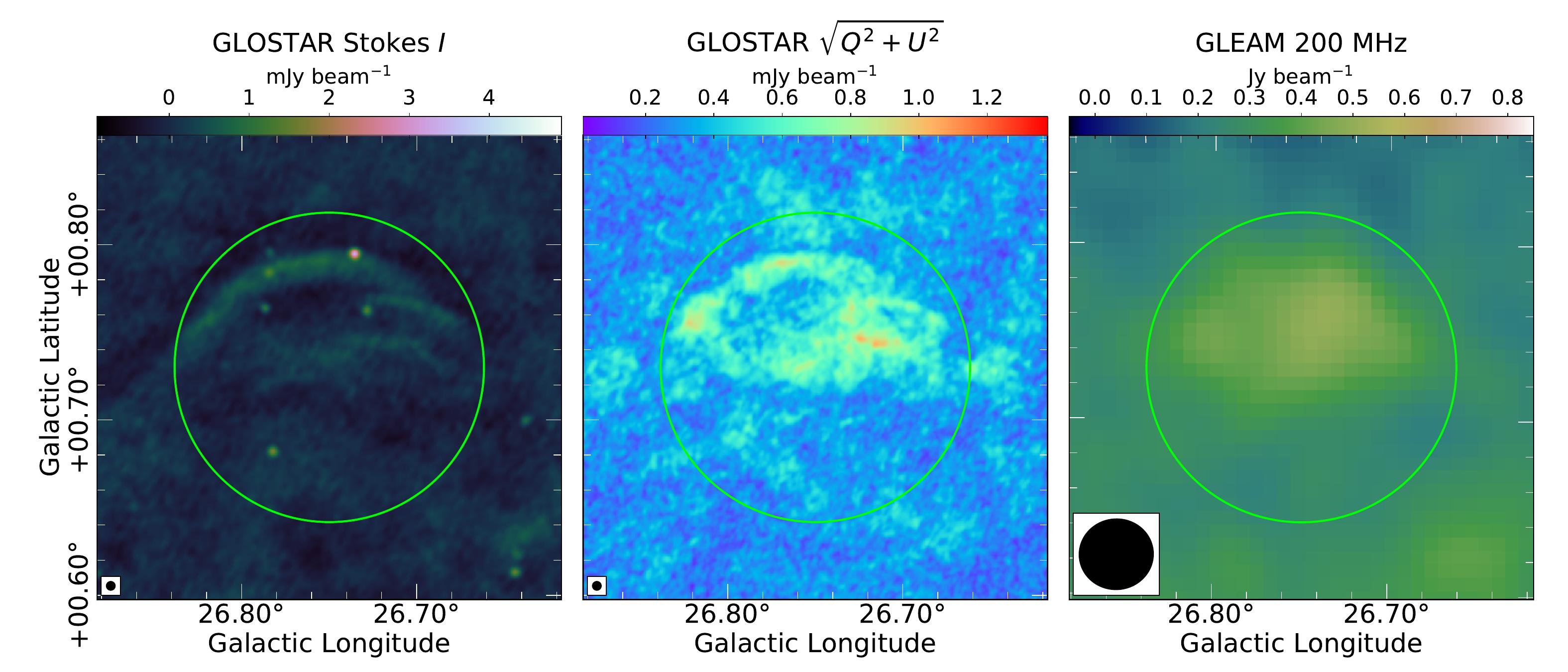}
    \caption{G26.75$+$0.73, encircled in green, as seen in the GLOSTAR-VLA data (left and middle panels) and the GLEAM 200 MHz data (right panel)}.
    \label{fig:G26.75}
\end{figure*}

Candidate SNR G26.75$+$0.73 was identified by \citet{2017A&A...605A..58A}
using data from the THOR survey.  We observe a partial shell morphology
in the GLOSTAR-VLA data similar to the THOR+VGPS data
(Fig.~\ref{fig:G26.75}).  We find
that it has a high degree of polarization, $0.70 \pm 0.40$, suggesting
that this shell-shaped object is a SNR.  We note that the degree of
polarization observed in the GLOSTAR-VLA data is an over-estimation
(see \S \ref{subsec:flux}).  In addition, we also find faint emission from
this object in the 200 MHz GLEAM
data\footnote{\url{http://gleam-vo.icrar.org/gleam_postage/q/form}}
\citep{2019PASA...36...47H}.  We measure its flux density as
$1.0 \pm 0.5$~Jy in the GLEAM data after subtracting the local background.
Comparing this to its THOR+VGPS flux density of $\sim 0.5$~Jy
\citep{2017A&A...605A..58A}, we measure a nonthermal
spectral index of $\sim -0.4$, agreeing with its identification as a SNR.

\subsubsection{G27.06\texorpdfstring{$+$}{+}0.04}

\begin{figure*}
    \centering
    \includegraphics[width=0.96\textwidth]{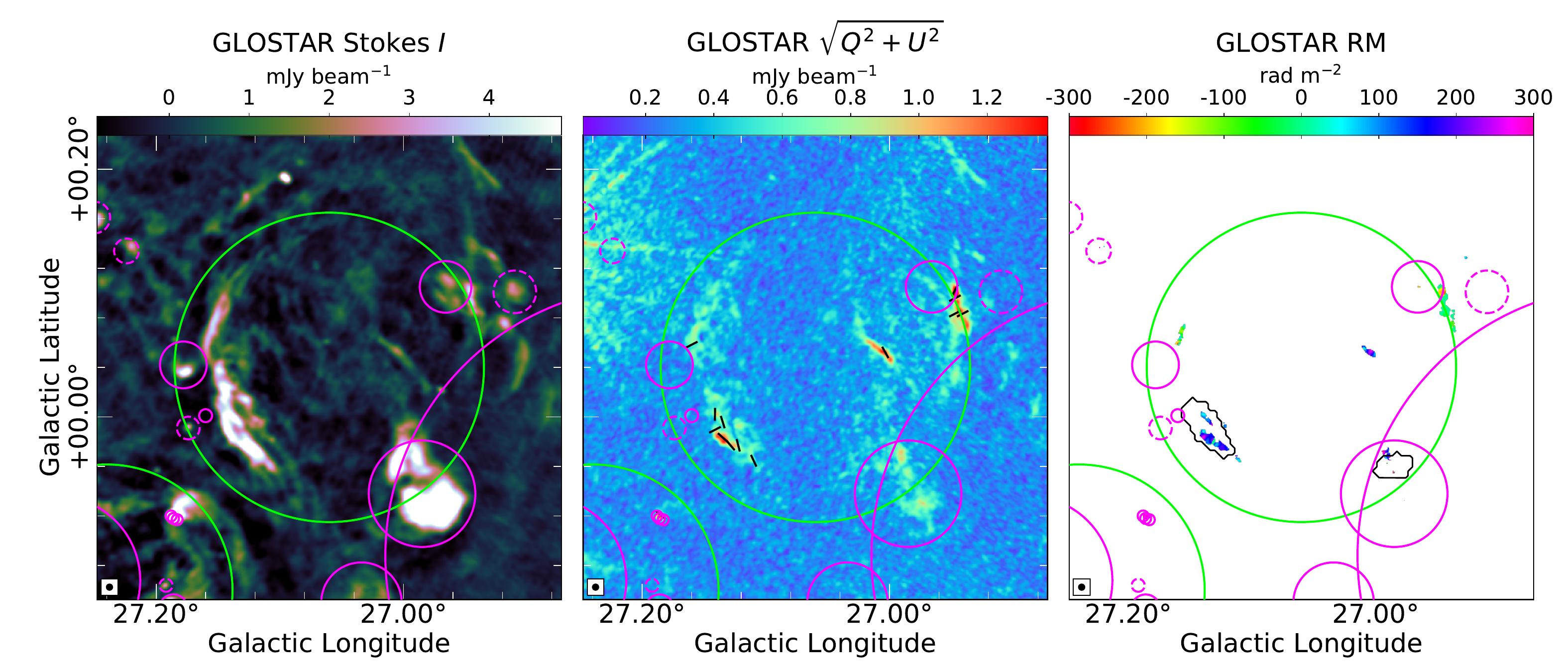}
    \caption{G27.06$+$0.04.  The RMs presented in the right-most panel typically have uncertainties of $30-40$~rad~m$^{-2}$.  The black contours on the rotation measure map show the regions for which the TIFR GMRT Sky Survey-NRAO VLA Sky Survey (TGSS-NVSS) spectral index was measured.  The eastern region (on the arc) and the western region (overlapping with two \ion{H}{ii}~regions) have a similar spectral index of ${\sim}-0.5$.  For this and all subsequent figures, the directions of the electric field vector position angles (after accounting for Faraday rotation) are plotted with black lines on the polarization map.}
    \label{fig:G27.06}
\end{figure*}

G27.06$+$0.04 is an arc-shaped SNR candidate detected
in MAGPIS, THOR, and GLEAM surveys
\citep{2006AJ....131.2525H,2017A&A...605A..58A,2019PASA...36...48H}.
We observe the same morphology in the GLOSTAR-VLA data (Fig.
\ref{fig:G27.06}).  For the arc, \citet{2019PASA...36...48H} report a
flux density of $4.9 \pm 0.1$~Jy at 200~MHz, while we measure its
flux density to be $1.4 \pm 0.3$~Jy in the THOR+VGPS data
\citep{2016A&A...595A..32B}.  This
implies that this arc has a spectral index of $-0.65 \pm 0.31$, consistent
with the value of $-0.53 \pm 0.22$ from the TGSS-NVSS spectral index
map\footnote{\url{http://tgssadr.strw.leidenuniv.nl/doku.php?id=spidx\#spectral_index_map}}
\citep{2018MNRAS.474.5008D}.  In the GLOSTAR-VLA data, we measure a
degree of polarization of $0.10 \pm 0.01$ for the arc.  We observe
different RMs for the northern ($\sim -100$~rad m$^{-2}$) and southern
($\sim +150$~rad m$^{-2}$) parts of the arc.  This is likely due to a
change in the magnetic field direction, or local Faraday screens.
Such large RMs and changes in RMs are not uncommon in SNRs
\citep[e.g.,][]{1974AuJPh..27..549M,1974IAUS...60..335M,2000ApJ...542..380G,2010ApJ...712.1157H}.
Further studies are necessary to fully analyze the emission from this
region.  Nonetheless, the polarization and spectral index measurements
provide sufficient evidence of the nonthermal nature of this object.

\subsubsection{G28.78\texorpdfstring{$-$}{-}0.44}

\begin{figure*}
    \centering
    \includegraphics[width=0.96\textwidth]{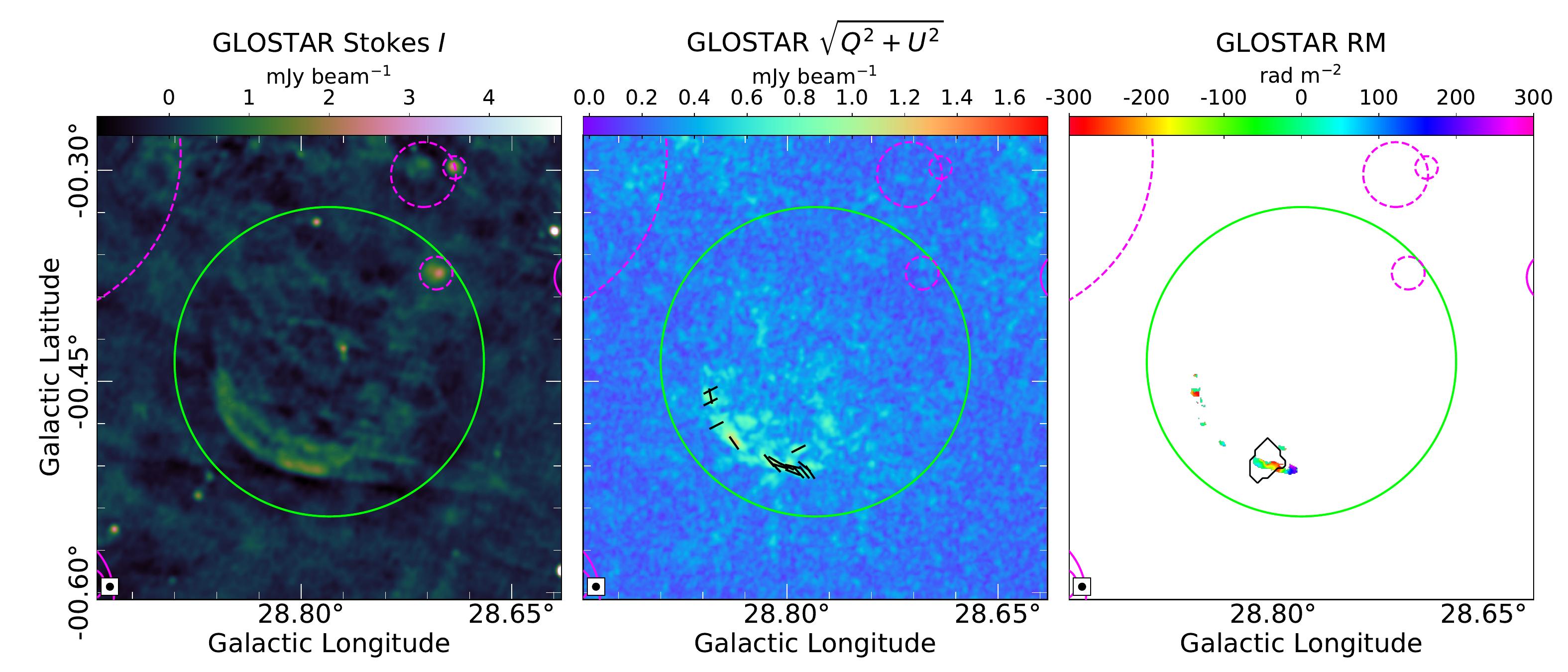}
    \caption{G28.78$-$0.44.  The black contour on the RM map shows the region for which the TGSS-NVSS spectral index ($\alpha = -0.75 \pm 0.22$) was measured.  The RMs have uncertainties of $25-35$~rad~m$^{-2}$.}
    \label{fig:G28.78}
\end{figure*}

G28.78$-$0.44 was first identified in the MAGPIS survey
\citep{2006AJ....131.2525H} and subsequently in the
THOR and GLEAM surveys \citep{2017A&A...605A..58A,2019PASA...36...48H}
as a near-complete shell.
A spectral index of $-0.79 \pm 0.12$ was derived by
\citet{2019PASA...36...48H}, which is consistent with the value of $-0.75 \pm 0.22$
for a part of the shell measured from the TGSS-NVSS spectral index data
\citep{2018MNRAS.474.5008D}.  In the GLOSTAR-VLA data, we find the
object to have a partial shell morphology along with
clear polarized emission ($p=0.49 \pm 0.12$, Fig.~\ref{fig:G28.78}).
The polarized emission from this object is further evidence
that this object is a SNR.
We find that the electric field vectors are generally tangential to the shell,
implying that the ambient magnetic field is either radial or nearly
parallel to the line of sight.  Radial magnetic fields are seen in
young shell-type SNRs, likely because of Rayleigh–Taylor instability
\citep{1987AuJPh..40..771M,1996ApJ...472..245J,2004mim..proc..141F}.

\subsubsection{G29.38\texorpdfstring{$+$}{+}0.10}
\label{subsubsec:G29.38}

\begin{figure*}
    \centering
    \includegraphics[width=0.96\textwidth]{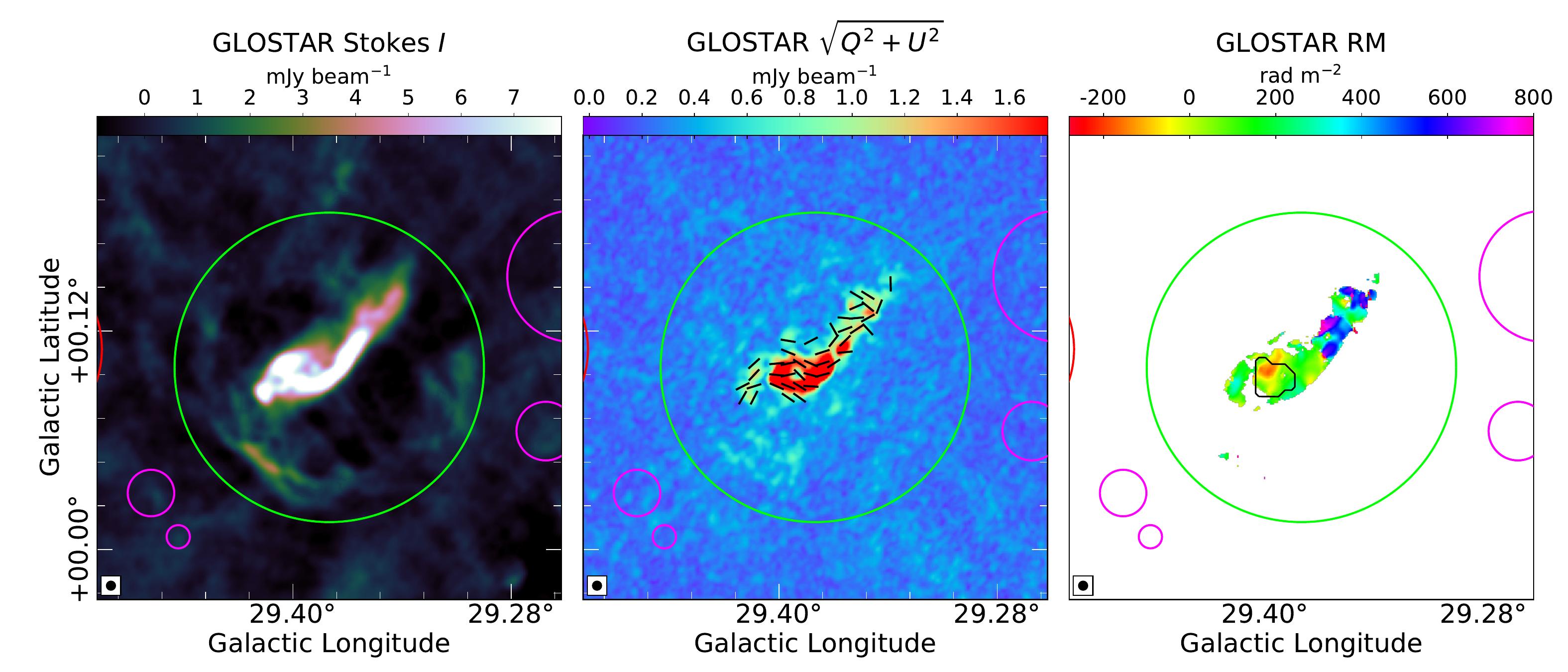}
    \caption{G29.38$+$0.10.  The black contour on the RM map shows the region for which the TGSS-NVSS spectral index ($\alpha = 0.17 \pm 0.06$) was measured.  The uncertainties in the RMs are $\sim 10-20$~rad~m$^{-2}$.}
    \label{fig:G29.38}
\end{figure*}

G29.38$+$0.10 was observed in the MAGPIS and THOR surveys as a
source with bright central compact emission inside a weakly emitting
shell \citep{2006AJ....131.2525H,2017A&A...605A..58A}.
\citet{2019PASA...36...48H} measured a spectral index of $0.09 \pm 0.14$,
noting it as a potential PWN.  A similar spectral
index is obtained from the TGSS-NVSS spectral index map
\citep[$0.17 \pm 0.06$][]{2018MNRAS.474.5008D}.  We observe the central
emission at higher resolution in the GLOSTAR-VLA data; it shows a faint
unresolved central object inside the bright elongated nebula, which
itself is surrounded by a weak shell, the remnant of the SN that had
the pulsar as its end product  (Fig.~\ref{fig:G29.38}).  An
ordered magnetic field can be inferred from the electric field
vectors.  The derived RMs range from $\sim -200$ to $\sim +600$ rad
m$^{-2}$.  Such a large spread is not typically seen in objects in the
Milky Way, and may be due to Faraday thick structures or a
superposition with sources unrelated to the PWN.  The PWN and its shell
have measured degrees of polarization of $0.17 \pm 0.02$ and
$0.02 \pm 0.01,$ respectively.

This region was analyzed across several spectral bands by
\citet{2017A&A...602A..31C} and \citet{2019A&A...626A..65P},
searching for evidence for an association of the radio detection with
the TeV source HESS J1844-030 \citep{2018A&A...612A...2H}.
No evidence for  pulsations was found in any band and
the S-shaped feature is suggested to be a radio galaxy.  It is argued
that the radio galaxy is responsible for only a part of the
observed emission and that this source most likely represents a
chance superposition of the radio galaxy, and a PWN and its remnant
shell.  This may explain the large variation in our RM measurements
near the tail.  The polarization and spectral index measurements,
combined with the analysis by \citet{2017A&A...602A..31C} and
\citet{2019A&A...626A..65P}, confirm
the status of this candidate as a SNR.

\subsubsection{G51.21\texorpdfstring{$+$}{+}0.11 complex: G51.04\texorpdfstring{$+$}{+}0.07 and G51.26\texorpdfstring{$+$}{+}0.11}

\begin{figure*}
    \centering
    \includegraphics[width=0.96\textwidth]{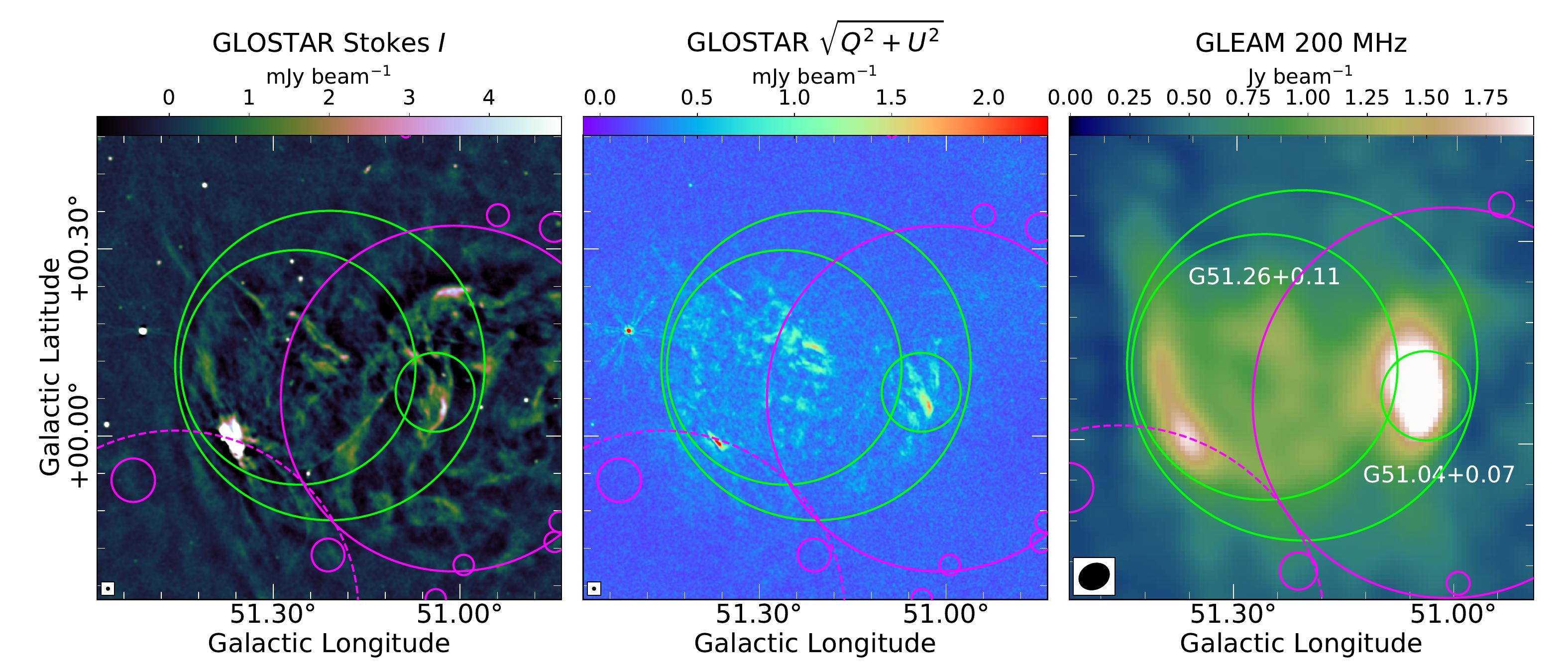}
    \caption{Complex G51.21$+$0.11 (largest green circle, defined by \citealt{2017A&A...605A..58A}) containing the SNRs G51.04$+$0.07 and G51.26$+$0.11 (smaller green circles, defined by \citealt{2018ApJ...866...61D}).}
    \label{fig:G51.21}
\end{figure*}

The candidate G51.21$+$0.11 was identified by \citet{2017A&A...605A..58A} in the
THOR survey.  Further studies by \citet{2018A&A...616A..98S}
and \citet{2018ApJ...866...61D} have shown evidence of nonthermal
emission arising from two distinct regions in this complex.
\citet{2018ApJ...866...61D} classify it as two separate SNRs,
G51.04$+$0.07 and G51.26$+$0.11.  Recently, \citet{2021arXiv210208851A} identified GeV emission from this region.  They rule out nearby star-forming regions and Bremsstrahlung radiation as the origin of this GeV emission and support the hypothesis that this emission is from at least one SNR.  In the GLOSTAR-VLA data, a morphology
similar to the one in THOR+VGPS data is observed (Fig.
\ref{fig:G51.21}).  We measure a degree of polarization of $0.07
\pm 0.01$ for G51.04$+$0.07 and $0.06
\pm 0.02$ for G51.26$+$0.11.
In the 1.4 GHz THOR+VGPS data and the 200 MHz GLEAM data, we subtracted
the local background and measured the flux densities of these objects.  The 200
MHz and 1.4 GHz flux densities of G51.04$+$0.07 are $6.3 \pm 2.1$~Jy and
$2.0 \pm 0.3$~Jy respectively, whereas G51.26$+$0.11 has $25.8 \pm 3.6$~Jy and
$12.4 \pm 0.6$~Jy, respectively.  These values imply spectral indices of
$\sim -0.6$ for G51.04$+$0.07 and $\sim -0.4$ for G51.26$+$0.11.  The above
polarization and spectral index measurements
further strengthen the case of these two objects as SNRs.

\subsection{Other observed SNR candidates}
\label{subsec:othercands}

\subsubsection{G15.51\texorpdfstring{$-$}{-}0.15}
\label{subsubsec:G15.51}

\begin{figure}
    \centering
    \includegraphics[width=0.48\textwidth]{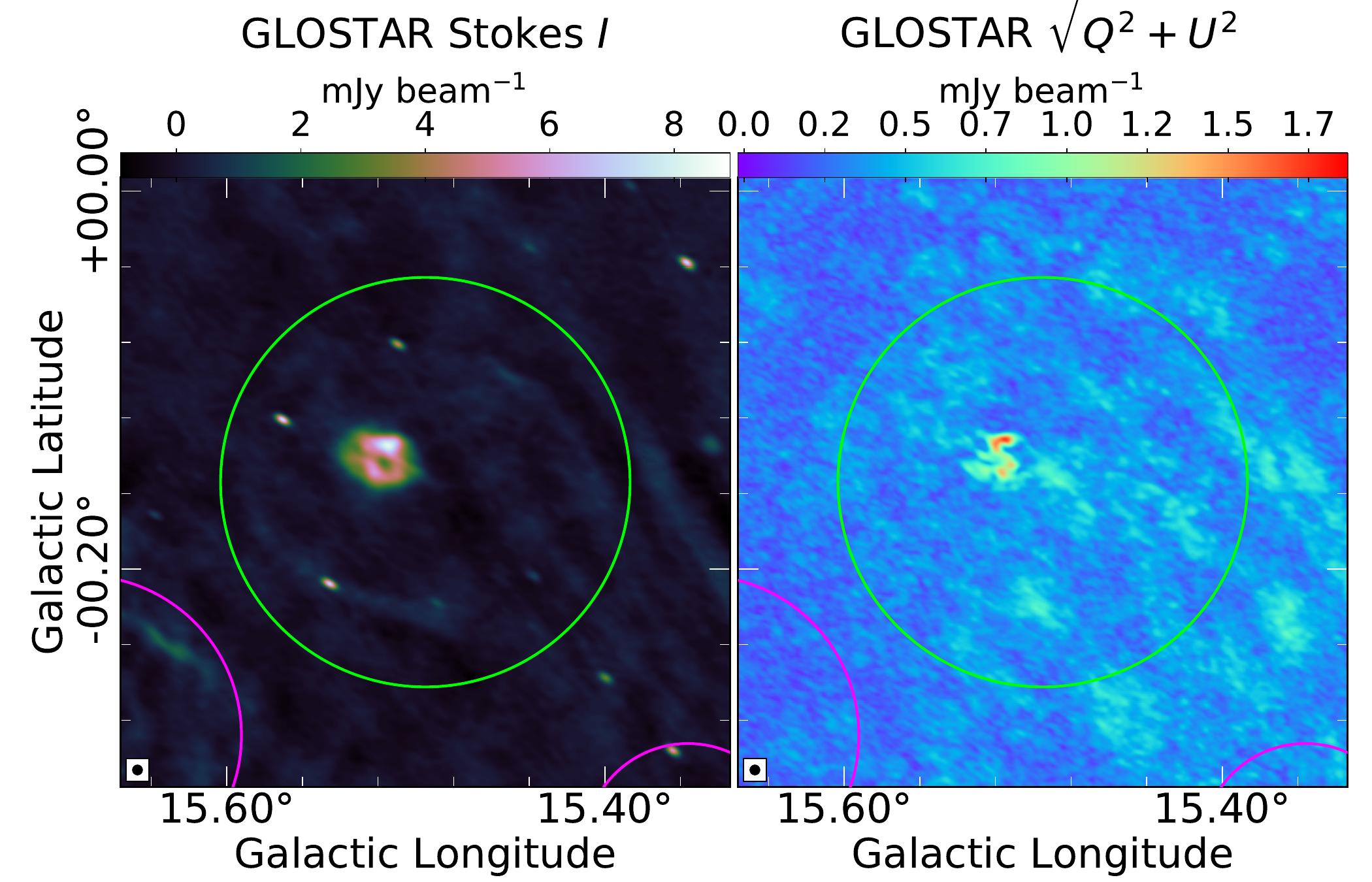}
    \caption{Candidate SNR G15.51$-$0.15.  The morphology and polarization from the central object imply the presence of a PWN at the center, but previous studies derive a spectral index of $\alpha{\sim}-0.5,$ which is not expected from PWNe. }
    \label{fig:G15.51}
\end{figure}

\citet{2006ApJ...639L..25B} identified G15.51$-$0.15 as a potential
shell-type remnant that is less likely to be a
SNR\footnote{\citet{2006ApJ...639L..25B} classified the candidate
G15.51$-$0.15 as a class~III-type shell 
in their Table~1 and have not discussed this candidate further in
the text.  However, the image of this candidate they showed in
their Plate 1 has a bright central object and
only a weak partial shell.}.  \citet{2019PASA...36...48H} studied
this object in GLEAM and the archival NRAO VLA Sky Survey data
\citep[NVSS;][]{1998AJ....115.1693C}.  They derive a spectral index of
${\sim}-0.6$ for both the central object and the surrounding
emission and speculate a common origin for both.

In the GLOSTAR-VLA data (Fig.~\ref{fig:G15.51}), we clearly resolve this
candidate into a shell that surrounds off-center compact emission.  This
morphology is indicative of a PWN.  However, PWNe generally have a
spectral index of $\alpha>-0.3$, although known exceptions exist
\citep[see][]{2008ApJ...687..516K}.  We note that the spectral
index calculations for the shell by \citet{2019PASA...36...48H} are
unlikely to be influenced significantly by the three point
sources on the shell.  In the GLOSTAR-VLA data, these point sources
have flux densities of 4--8 mJy and spectral indices close to
zero, implying that they would have similarly small flux densities in the
GLEAM band as well.  Comparing with the flux density of ${\sim}2.8$ Jy
derived by \citet{2019PASA...36...48H}, it can be seen that the contribution
of these three point sources to the flux density at 200 MHz, and hence to
their 200--1400 MHz spectral index calculation, would be negligible.
It is possible that
the shell and the central object are two separate SNRs, or it may be a
composite-type remnant.  Distance measurements are required to
study whether the shell and the central object are related.

In order to measure the polarization of the shell, we excluded the
three compact objects on the shell that are likely unrelated
sources.  The remaining part of the shell is faint and we could
only derive an upper limit on the degree of linear polarization,
$p<0.08$.  We measure a low degree of polarization of $0.03 \pm
0.01$ from the central object.

\subsubsection{G18.76\texorpdfstring{$-$}{-}0.07}

\begin{figure}
    \centering
    \includegraphics[width=0.48\textwidth]{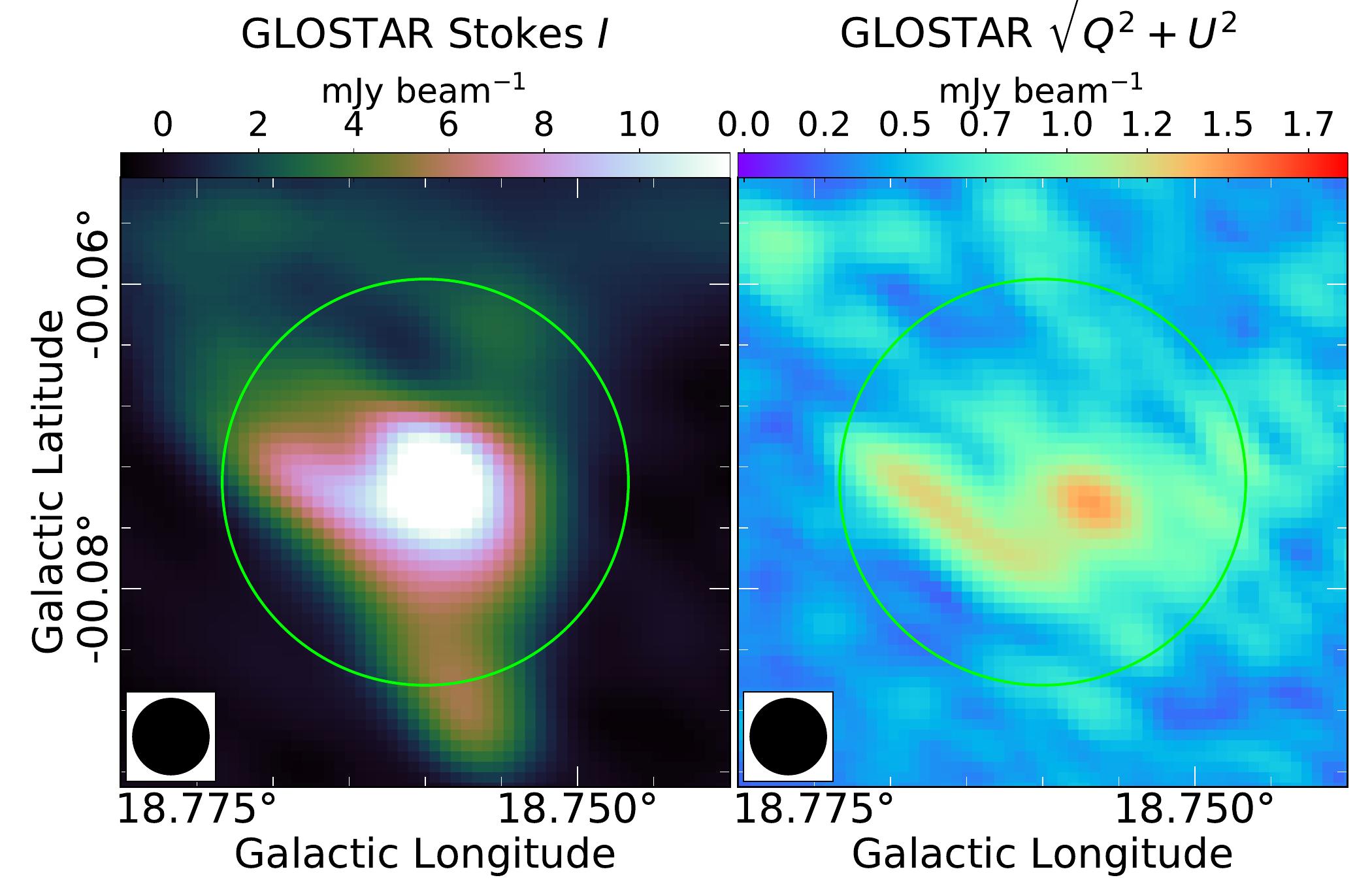}
    \caption{SNR candidate G18.76$-$0.07.  We hypothesize that this is a complex region with at least one extragalactic object.}
    \label{fig:G18.76}
\end{figure}

With a diameter of $96\arcsec$, G18.76$-$0.07 is one of the SNR candidates of the  smallest
angular size.  It was first discovered in the MAGPIS
survey \citep{2006AJ....131.2525H} and then in
the THOR survey \citep{2017A&A...605A..58A}.  We measure in the
GLOSTAR-VLA data a degree of polarization of $0.08 \pm 0.01$.  The
polarization map (Fig.~\ref{fig:G18.76}) shows a point source
and an elongated source in this region.  A large negative
spectral index of $\sim -1.8$ was measured from the GLEAM (${\sim}200$
MHz) and the NVSS data (1400 MHz) by \citet{2019PASA...36...48H}, and
the TGSS-NVSS spectral index map
\citep[150-1400 MHz;][]{2018MNRAS.474.5008D} also shows a similar
value ($\sim -1.2$).  Such values for SNRs have
been reported only at higher frequencies after a ``spectral break''
\citep[see][for instance]{2020MNRAS.496..723K}.  As such a spectral
index below L-band frequencies is generally seen only in extragalactic
objects, and because of the morphology of the linearly polarized emission,
we infer that at least one extragalactic object is located within
the angular extent of this candidate.

\subsubsection{G22.00\texorpdfstring{$+$}{+}0.00/G022.045\texorpdfstring{$-$}{-}0.028}
\label{subsubsec:G22.00}
\begin{figure*}
    \centering
    \includegraphics[width=0.96\textwidth]{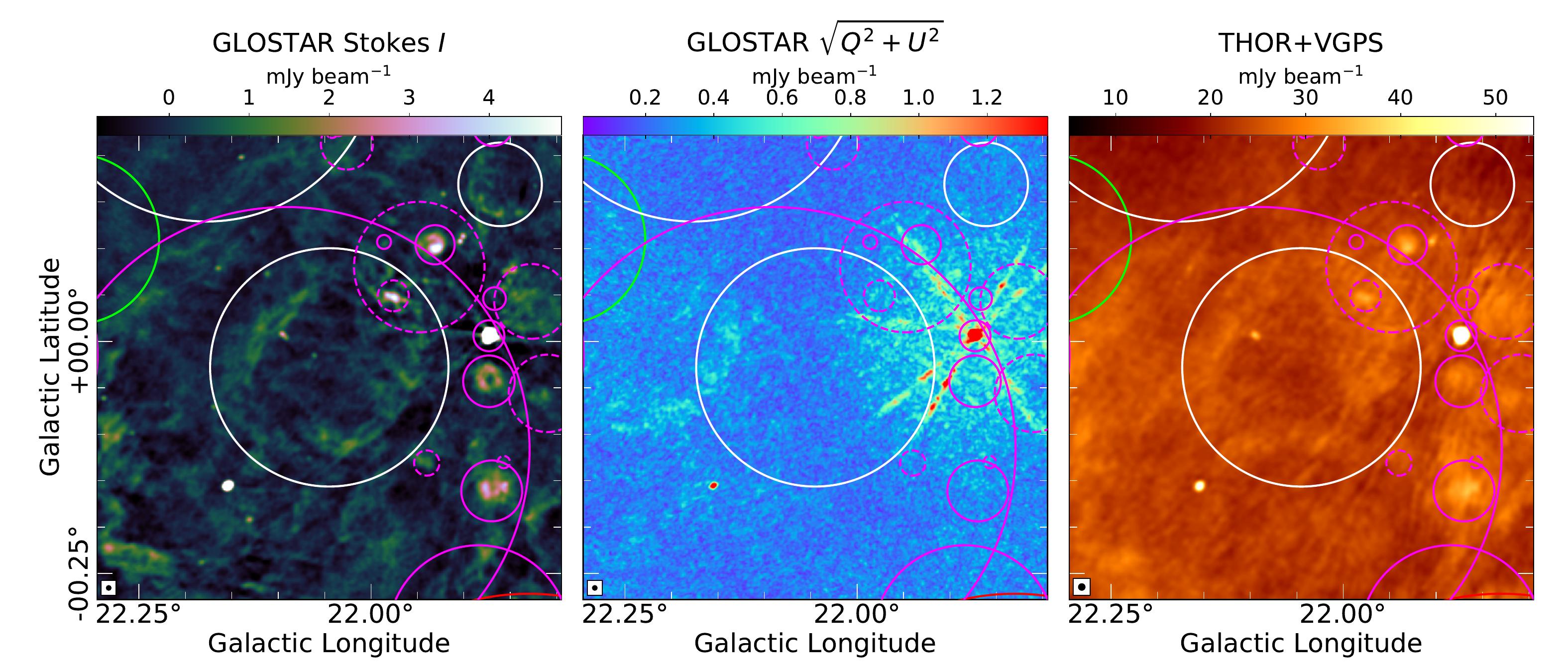}
    \caption{GLOSTAR SNR candidate G022.045$-$0.028: Shell-shaped object near the X-ray SNR candidate G22.00$+$0.00.  There is no clear association of the radio and the X-ray morphologies \citep[see Fig.~1 of][]{2006IAUS..230..333U}.}
    \label{fig:G22.00}
\end{figure*}

\citet{2006IAUS..230..333U} discovered G22.00$+$0.00 at X-ray energies,
noting that synchrotron X-ray emitting SNRs have low radio surface
brightness.  \citet{2016PASJ...68S...6Y} hint that this candidate
may be a PWN.  We do not find any PWN-like object, but we do identify a shell-like object in the GLOSTAR-VLA
and the THOR+VGPS data (Fig.~\ref{fig:G22.00}), overlapping with the diffuse X-ray emission
detected by \citet{2006IAUS..230..333U} and \citet{2016PASJ...68S...6Y}
at $l=22.00$, $b=0.00$.  We name this a GLOSTAR SNR candidate G022.045$-$0.028.
The spatial overlap indicates that this may be the shell corresponding to the
PWN suggested by \citet{2016PASJ...68S...6Y}.
We observe polarization in the GLOSTAR-VLA data from the eastern
part of the shell, but it appears to be from unrelated shell-shaped
emission extending further east without a Stokes $I$ counterpart
in either of the THOR+VGPS and GLOSTAR-VLA data (Fig.~\ref{fig:G22.00}).  It is as yet unclear whether the
radio shell is associated with the X-ray detection.

\subsubsection{G27.18\texorpdfstring{$+$}{+}0.30}

\begin{figure*}
    \centering
    \includegraphics[width=0.96\textwidth]{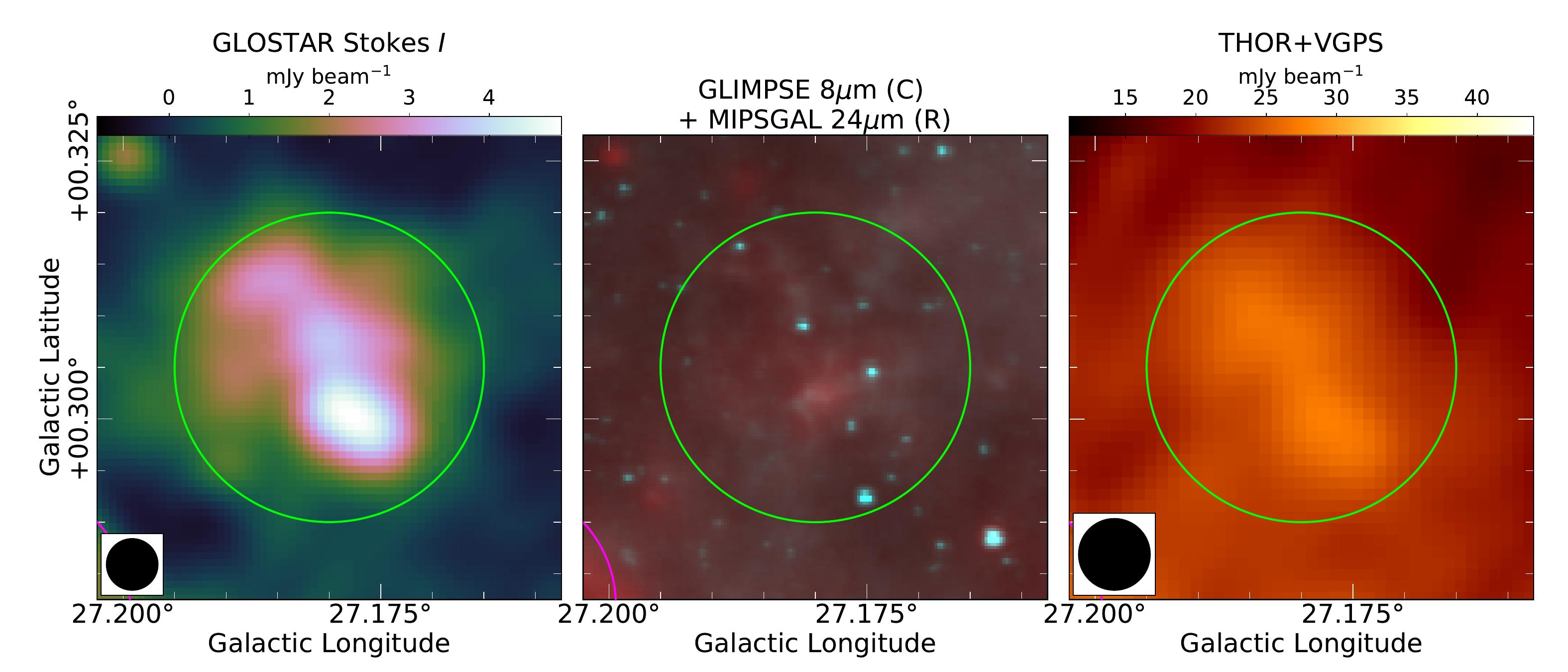}
    \caption{THOR SNR candidate G27.18$+$0.30, as seen in GLOSTAR Stokes $I$ (left), MIR (middle) and THOR+VGPS (right).}
    \label{fig:G27.18}
\end{figure*}

The THOR SNR candidate G27.18$+$0.30 is clearly seen in the GLOSTAR-VLA data
(Fig.~\ref{fig:G27.18}).  It seems to have faint counterparts in MIPSGAL
24~$\mu$m and GLIMPSE 8~$\mu$m images, but this may be just the diffuse MIR background
unrelated to the radio emission.  \citet{2017A&A...605A..58A} report a
flux density of 0.05$\pm$0.03~Jy in the THOR+VGPS data, similar to the GLOSTAR-VLA
data flux density of 0.048$\pm$0.001~Jy.  The object has a size of
${\sim}1.5\arcmin$, and so we take the GLOSTAR-VLA flux density as a lower limit
and estimate the lower limit of the spectral index of this candidate:
\begin{equation}
\begin{aligned}
    & \alpha_{\mathrm{low}} = \frac{\ln{S_{\mathrm{GLOSTAR-VLA}}} - \ln{S_{\mathrm{THOR+VGPS}}}}{\ln{5.8\mathrm{GHz}} - \ln{1.4\mathrm{GHz}}} \\
    & \implies -0.36 < \alpha_{\mathrm{low}} < 0.62.
\end{aligned}
\end{equation}
The lower limit of the spectral index implies that G27.18$+$0.30 may be a PWN,
although the morphology is atypical.

\subsubsection{G53.07\texorpdfstring{$+$}{+}0.49}

\begin{figure*}
    \centering
    \includegraphics[width=0.96\textwidth]{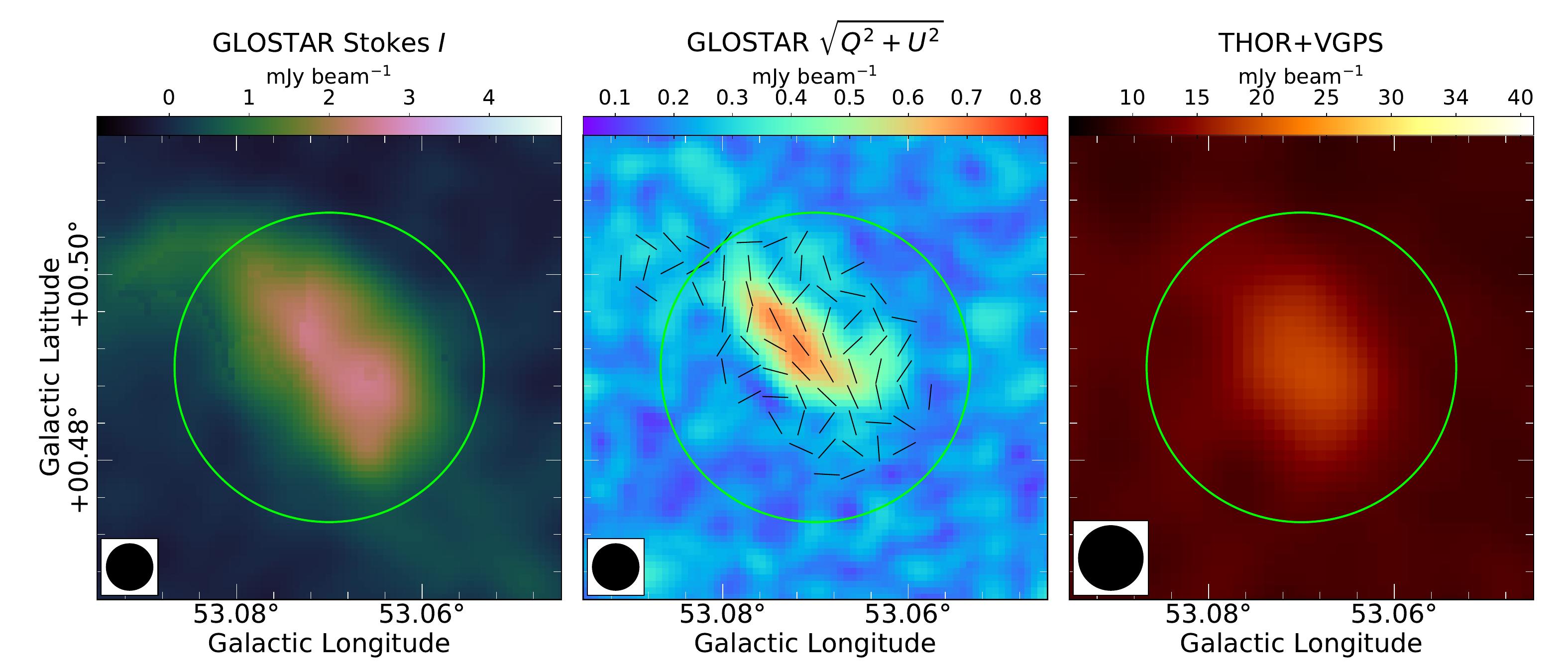}
    \caption{THOR SNR candidate G53.07$+$0.49 as seen in the GLOSTAR-VLA data (Stokes $I$, left, and linear polarization, middle), and the THOR+VGPS data on the right.}
    \label{fig:G53.07}
\end{figure*}

\citet{2017A&A...605A..58A} identified this object as a SNR candidate in
the THOR survey.  In the GLOSTAR-VLA data, we find a slightly elongated
structure (Fig.~\ref{fig:G53.07}).  The polarization data show that the
electric field vectors near the center roughly line up along the long axis
of this candidate.  It has a degree of polarization of $0.12 \pm 0.02$.  We
note  that this small angular size candidate is quite asymmetric.  If it
were indeed a SNR, it must be quite young or far and/or be expanding into
a region of the ISM with a large density gradient.  While we find that the
emission in the GLOSTAR-VLA data arises from the same position reported by
\citet{2017A&A...605A..58A}, \citet{2018ApJ...860..133D} report that the
peaks of flux density in low-frequency data obtained with the Westerbork
Synthesis Radio Telescope (WSRT) and the Low Frequency Array (LOFAR) have
a large offset ($2\arcmin$ to $3\arcmin$) from the VLA data, but this
could be the result of confusion with nearby sources due to the elongated
beam at lower frequencies \citep[see Fig.~3 of][]{2018ApJ...860..133D}.
Further observational campaigns at multiple frequencies sensitive to
various angular scales are needed to shed light on the nature of this candidate.

\subsubsection{Resolved SNR candidates}

\begin{figure*}
    \centering
    \includegraphics[width=0.48\textwidth]{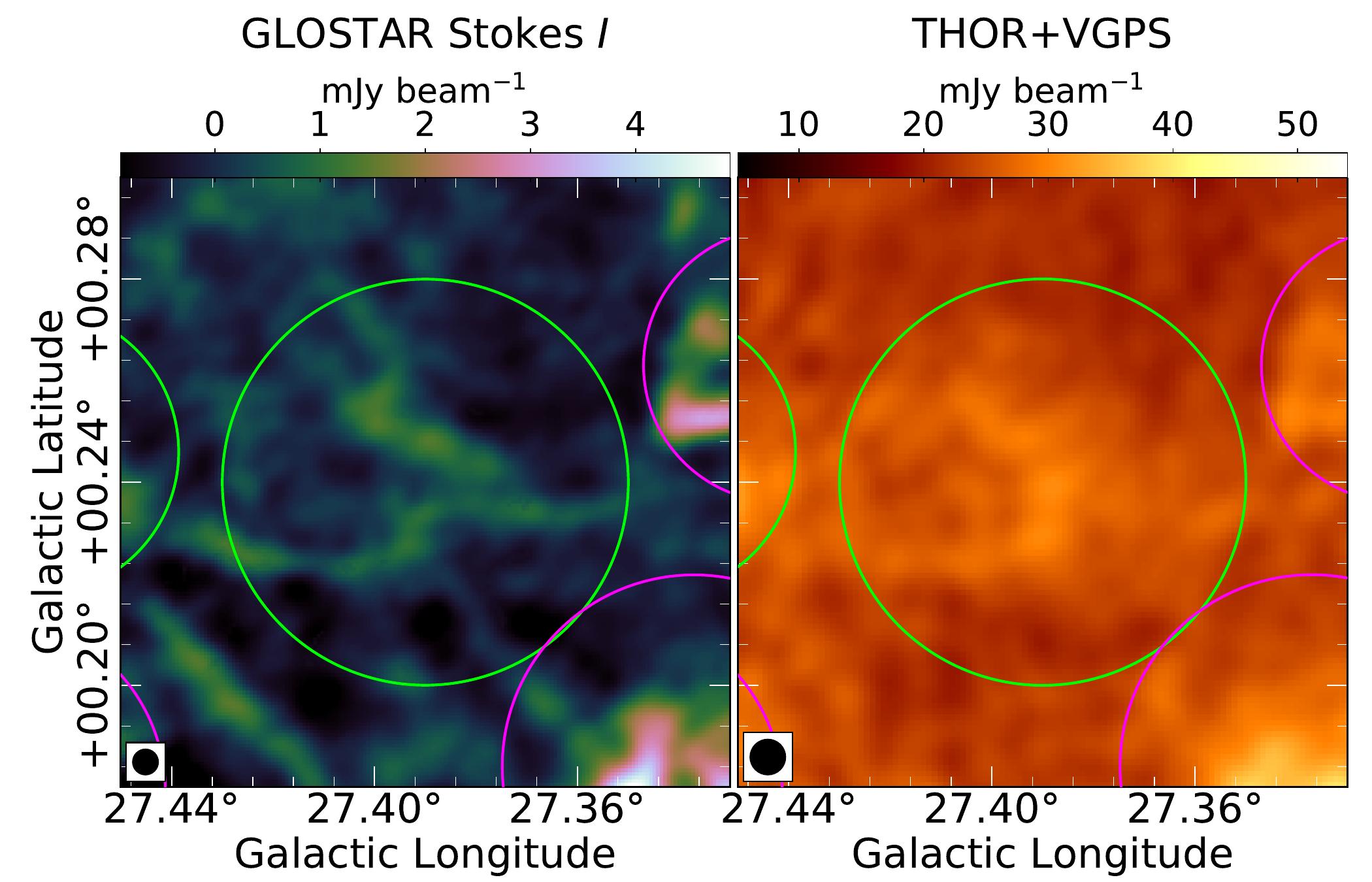}
    \includegraphics[width=0.48\textwidth]{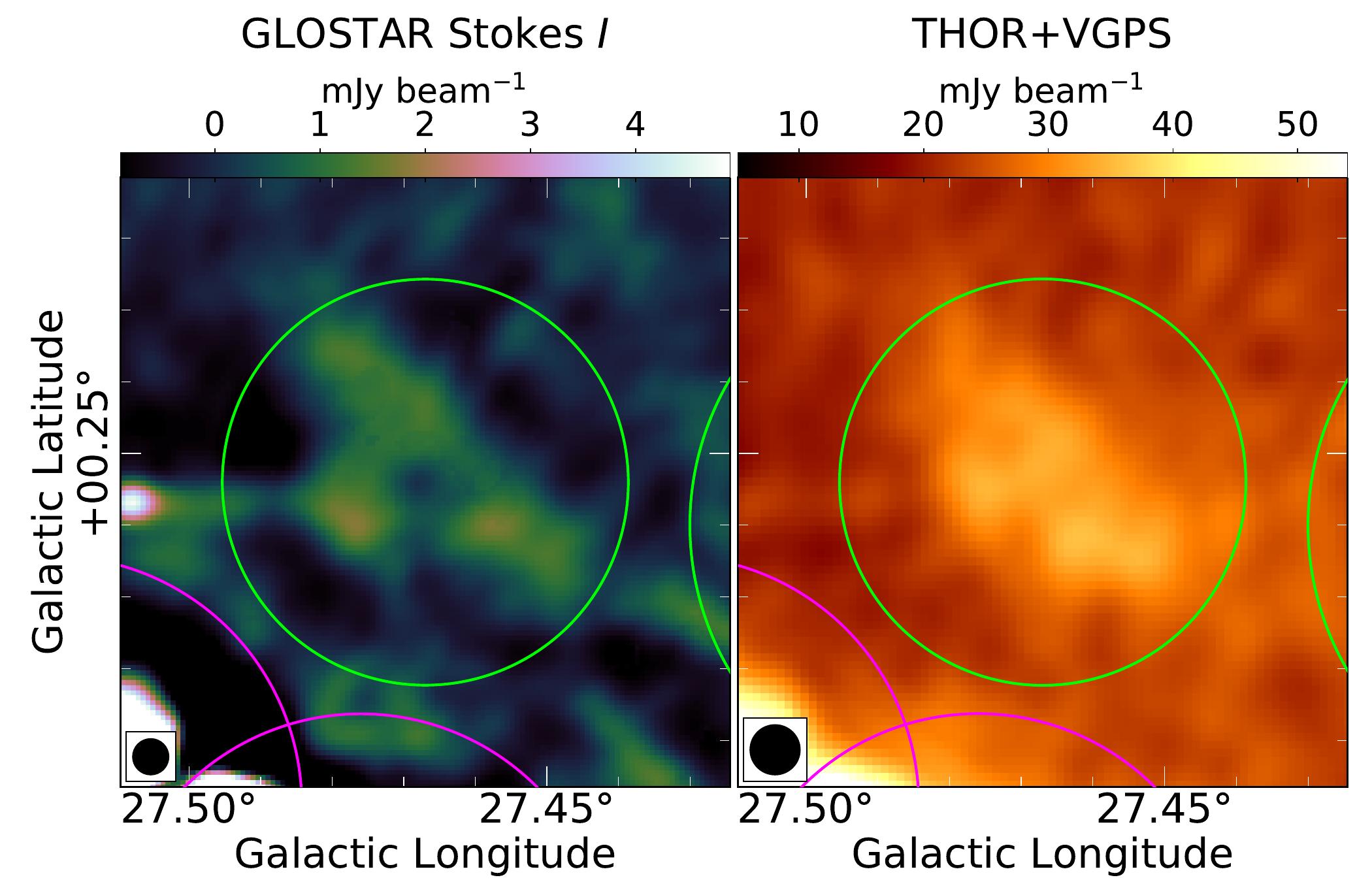}

    \includegraphics[width=0.48\textwidth]{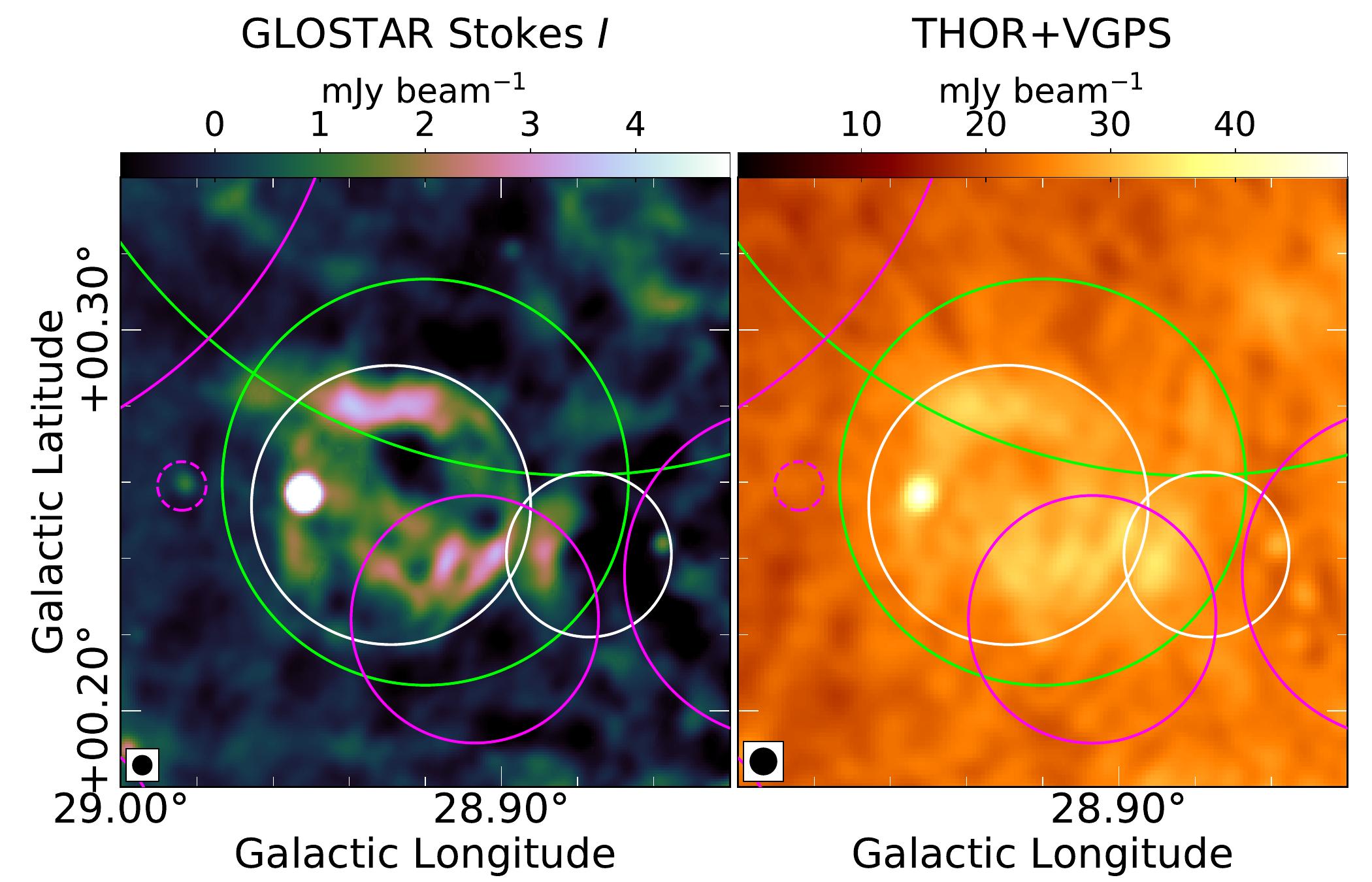}
    \includegraphics[width=0.48\textwidth]{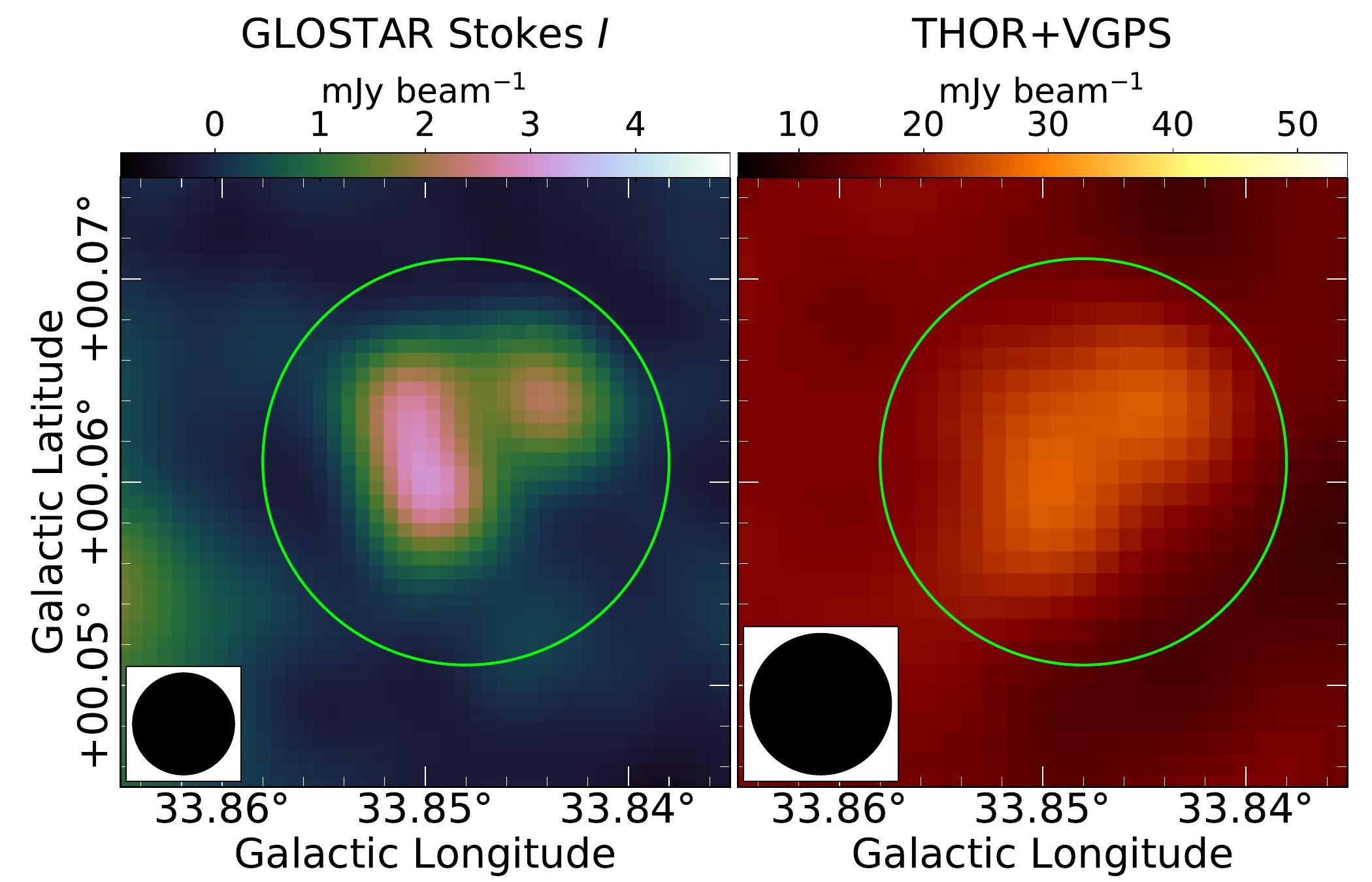}
    \caption{THOR SNR candidates resolved in the GLOSTAR-VLA data: G27.39$+$0.24 (top left), G27.47$+$0.25 (top right) G28.92$+$0.26 (bottom left), and G33.85$+$0.06 (bottom right).  The left panels are the GLOSTAR-VLA data, and the right panels are the THOR+VGPS data.  G27.39$+$0.24 and G27.47$+$0.25 are filaments mistaken for SNR emission, whereas G28.92$+$0.26 and G33.85$+$0.06 contain multiple distinct objects.}
    \label{fig:resolvedSNRs}
\end{figure*}

The THOR candidates G27.39$+$0.24, G27.47$+$0.25, G28.92$+$0.26,
and G33.85$+$0.06 are better resolved in the GLOSTAR-VLA data (Fig.
\ref{fig:resolvedSNRs}).  The filamentary structures of G27.39$+$0.24
and G27.47$+$0.25 that we observe in the GLOSTAR-VLA data
suggest that these may be nearby filaments that were
unresolved by THOR+VGPS data, rather than SNRs.
There seem to be two objects---possibly unrelated---in the extent of
G33.85$+$0.06.  Their sizes are comparable to the beam size; they are
more likely to be radio galaxies than SNRs.

In the region of the candidate G28.92$+$0.26, we observe a larger shell-shaped object centered at $l=28.93,~b=0.26$, and to its west, a smaller
object that resembles a partial shell (marked with two white circles in
Fig.~\ref{fig:resolvedSNRs}).  The bright compact object near
$l=28.95,~b=0.26$ has a thermal spectral index in the GLOSTAR-VLA data
($\alpha{\sim}0$) and was detected in MIPSGAL, implying thermal emission.
The other two shell-shaped objects are included in the list of GLOSTAR
SNR candidates (G028.877$+$0.241 and G028.929$+$0.254).

\section{Conclusions}
\label{sec:conclusions}

In the GLOSTAR-VLA data, we discover 80 new SNR candidates using the radio-MIR
anti-correlation property of SNRs.  In addition, 77 previously identified
candidates have also been detected.  We show that the degree of polarization
measured using the GLOSTAR-VLA data can reliably distinguish thermal and
nonthermal emission in many cases, in spite of the diffuse synchrotron
emission that permeates the ISM.  Following the positive polarization
measurements from the GLOSTAR-VLA data and favorable spectral index measurements
using data from lower frequency surveys, we are able to confirm six previously
identified candidates as SNRs (G26.75$+$0.73, G27.06$+$0.04, G28.78$-$0.44, G29.38+0.10, G51.04$+$0.07 and G51.26$+$0.11).
We were also able to measure significant polarization from three newly
discovered GLOSTAR SNR candidates, G005.989$+$0.019, G034.524$-$0.761, and G039.539$+$0.366.  Comparing our results with the predictions by \citet{1991ApJ...378...93L}, we find that over $50\%$ of SNRs in our survey region are yet to be discovered (\S\ref{subsubsec:nSNRs}).

The G19 SNR catalog contains 94 objects in the survey region of GLOSTAR,
and we detect all the objects
previously identified in radio continuum data, except G0.0$+$0.0
and G0.3$+$0.0, which lie in a very confused region.  We find that four
of these objects (G8.3$-$0.0, G10.5$-$0.0, G11.1$-$1.0 and G14.3$+$0.1)
are actually \ion{H}{ii}~regions mistaken for SNRs by cross-matching with the A14 catalog of
Galactic \ion{H}{ii}~regions.

The GLOSTAR-VLA data highlight the importance of resolution and
sensitivity in large-scale surveys:  we were able to detect almost all
radio SNRs in the survey due to the remarkable sensitivity, and the
higher resolution made it possible to reclassify several objects. The SNR
candidates G27.39$+$0.24 and G27.47$+$0.25 are filaments, and multiple
objects were identified in the candidates G18.76$-$0.07, G28.92$+$0.26,
G33.85$+$0.06, and in the G19 SNR G6.1$+$0.5.

The future addition of single dish data ---which are presently collected with the
Effelsberg 100 meter radio telescope--- to the GLOSTAR-VLA images will make
reliable spectral index measurements possible for extended objects.  This
should prove useful in confirming the SNR candidates.  If all
the detected candidates were confirmed as SNRs, it would nearly triple the
number of SNRs in the first quadrant of the Galaxy, bringing us closer to
the predicted number of SNRs in the Milky Way
\citep[${\sim}1000$;][]{1991ApJ...378...93L}.  Further deeper large-scale
surveys covering the entire Galactic plane should be able to rectify the
apparent deficiency of SNRs in the Galaxy.

\begin{acknowledgements}
    We thank the referee David Helfand for his valuable comments on the draft which helped improve the discussion and presentation of our work.
    We thank Miguel Araya for pointing out a naming inconsistency.
    HB acknowledges support from the European Research Council
    under the Horizon 2020 Framework Program via the ERC Consolidator
    Grant CSF-648505.  HB also acknowledges support from the Deutsche
    Forschungsgemeinschaft in the Collaborative Research Center (SFB 881)
    "The Milky Way System" (subproject B1).
    NR acknowledges Max-Planck-Gesellschaft for funding support through
    the Max Planck India Partner Group grant.
    This research has made use of NASA's Astrophysics Data System and
    the SIMBAD database.
    We have used the softwares Astropy \citep{2013A&A...558A..33A}, APLpy \citep{2012ascl.soft08017R},
    DS9 \citep{2003ASPC..295..489J} and Aladin \citep{2000A&AS..143...33B} at
    various stages of this research.
\end{acknowledgements}

\bibliographystyle{aa.bst}          
\bibliography{ref}

\begin{thebibliography}{75}
\expandafter\ifx\csname natexlab\endcsname\relax\def\natexlab#1{#1}\fi

\bibitem[{{Adams} {et~al.}(2013){Adams}, {Kochanek}, {Beacom}, {Vagins}, \&
  {Stanek}}]{2013ApJ...778..164A}
{Adams}, S.~M., {Kochanek}, C.~S., {Beacom}, J.~F., {Vagins}, M.~R., \&
  {Stanek}, K.~Z. 2013, \apj, 778, 164

\bibitem[{{Anderson} {et~al.}(2014){Anderson}, {Bania}, {Balser}, {Cunningham},
  {Wenger}, {Johnstone}, \& {Armentrout}}]{2014ApJS..212....1A}
{Anderson}, L.~D., {Bania}, T.~M., {Balser}, D.~S., {et~al.} 2014, \apjs, 212,
  1

\bibitem[{{Anderson} {et~al.}(2017){Anderson}, {Wang}, {Bihr}, {Rugel},
  {Beuther}, {Bigiel}, {Churchwell}, {Glover}, {Goodman}, {Henning}, {Heyer},
  {Klessen}, {Linz}, {Longmore}, {Menten}, {Ott}, {Roy}, {Soler}, {Stil}, \&
  {Urquhart}}]{2017A&A...605A..58A}
{Anderson}, L.~D., {Wang}, Y., {Bihr}, S., {et~al.} 2017, \aap, 605, A58

\bibitem[{{Araya}(2021)}]{2021arXiv210208851A}
{Araya}, M. 2021, arXiv e-prints, arXiv:2102.08851

\bibitem[{{Astropy Collaboration} {et~al.}(2013){Astropy Collaboration},
  {Robitaille}, {Tollerud}, {Greenfield}, {Droettboom}, {Bray}, {Aldcroft},
  {Davis}, {Ginsburg}, {Price-Whelan}, {Kerzendorf}, {Conley}, {Crighton},
  {Barbary}, {Muna}, {Ferguson}, {Grollier}, {Parikh}, {Nair}, {Unther},
  {Deil}, {Woillez}, {Conseil}, {Kramer}, {Turner}, {Singer}, {Fox}, {Weaver},
  {Zabalza}, {Edwards}, {Azalee Bostroem}, {Burke}, {Casey}, {Crawford},
  {Dencheva}, {Ely}, {Jenness}, {Labrie}, {Lim}, {Pierfederici}, {Pontzen},
  {Ptak}, {Refsdal}, {Servillat}, \& {Streicher}}]{2013A&A...558A..33A}
{Astropy Collaboration}, {Robitaille}, T.~P., {Tollerud}, E.~J., {et~al.} 2013,
  \aap, 558, A33

\bibitem[{{Bertin} \& {Arnouts}(1996)}]{1996A&AS..117..393B}
{Bertin}, E. \& {Arnouts}, S. 1996, \aaps, 117, 393

\bibitem[{{Beuther} {et~al.}(2016){Beuther}, {Bihr}, {Rugel}, {Johnston},
  {Wang}, {Walter}, {Brunthaler}, {Walsh}, {Ott}, {Stil}, {Henning},
  {Schierhuber}, {Kainulainen}, {Heyer}, {Goldsmith}, {Anderson}, {Longmore},
  {Klessen}, {Glover}, {Urquhart}, {Plume}, {Ragan}, {Schneider},
  {McClure-Griffiths}, {Menten}, {Smith}, {Roy}, {Shanahan}, {Nguyen-Luong}, \&
  {Bigiel}}]{2016A&A...595A..32B}
{Beuther}, H., {Bihr}, S., {Rugel}, M., {et~al.} 2016, \aap, 595, A32

\bibitem[{{Bonnarel} {et~al.}(2000){Bonnarel}, {Fernique}, {Bienaym{\'e}},
  {Egret}, {Genova}, {Louys}, {Ochsenbein}, {Wenger}, \&
  {Bartlett}}]{2000A&AS..143...33B}
{Bonnarel}, F., {Fernique}, P., {Bienaym{\'e}}, O., {et~al.} 2000, \aaps, 143,
  33

\bibitem[{{Brentjens} \& {de Bruyn}(2005)}]{2005A&A...441.1217B}
{Brentjens}, M.~A. \& {de Bruyn}, A.~G. 2005, \aap, 441, 1217

\bibitem[{{Brogan} {et~al.}(2006){Brogan}, {Gelfand}, {Gaensler}, {Kassim}, \&
  {Lazio}}]{2006ApJ...639L..25B}
{Brogan}, C.~L., {Gelfand}, J.~D., {Gaensler}, B.~M., {Kassim}, N.~E., \&
  {Lazio}, T.~J.~W. 2006, \apjl, 639, L25

\bibitem[{{Brose} {et~al.}(2020){Brose}, {Pohl}, {Sushch}, {Petruk}, \&
  {Kuzyo}}]{2020A&A...634A..59B}
{Brose}, R., {Pohl}, M., {Sushch}, I., {Petruk}, O., \& {Kuzyo}, T. 2020, \aap,
  634, A59

\bibitem[{Brunthaler {et~al.}(2020)Brunthaler, Menten, Dzib, Cotton, Wyrowski,
  Dokara, Gong, Medina, M{\"u}ller, Nguyen, Ortiz-Le{\'o}n, Reich, Rugel,
  Urquhart, Winkel, Yang, Beuther, Billington, Carrasco-Gonzalez, Csengeri,
  Murugeshan, Pandian, \& Roy}]{2021arXiv210600377B}
Brunthaler, A., Menten, K.~M., Dzib, S.~A., {et~al.} 2021, arXiv e-prints, arXiv:2106.00377

\bibitem[{{Cappellaro} {et~al.}(1993){Cappellaro}, {Turatto}, {Benetti},
  {Tsvetkov}, {Bartunov}, \& {Makarova}}]{1993A&A...273..383C}
{Cappellaro}, E., {Turatto}, M., {Benetti}, S., {et~al.} 1993, \aap, 273, 383

\bibitem[{{Carey} {et~al.}(2009){Carey}, {Noriega-Crespo}, {Mizuno}, {Shenoy},
  {Paladini}, {Kraemer}, {Price}, {Flagey}, {Ryan}, {Ingalls}, {Kuchar},
  {Pinheiro Gon{\c{c}}alves}, {Indebetouw}, {Billot}, {Marleau}, {Padgett},
  {Rebull}, {Bressert}, {Ali}, {Molinari}, {Martin}, {Berriman}, {Boulanger},
  {Latter}, {Miville-Deschenes}, {Shipman}, \& {Testi}}]{2009PASP..121...76C}
{Carey}, S.~J., {Noriega-Crespo}, A., {Mizuno}, D.~R., {et~al.} 2009, \pasp,
  121, 76

\bibitem[{{Castelletti} {et~al.}(2017){Castelletti}, {Supan}, {Petriella},
  {Giacani}, \& {Joshi}}]{2017A&A...602A..31C}
{Castelletti}, G., {Supan}, L., {Petriella}, A., {Giacani}, E., \& {Joshi},
  B.~C. 2017, \aap, 602, A31

\bibitem[{{Churchwell} {et~al.}(2009){Churchwell}, {Babler}, {Meade},
  {Whitney}, {Benjamin}, {Indebetouw}, {Cyganowski}, {Robitaille}, {Povich},
  {Watson}, \& {Bracker}}]{2009PASP..121..213C}
{Churchwell}, E., {Babler}, B.~L., {Meade}, M.~R., {et~al.} 2009, \pasp, 121,
  213

\bibitem[{{Clark} \& {Caswell}(1976)}]{1976MNRAS.174..267C}
{Clark}, D.~H. \& {Caswell}, J.~L. 1976, \mnras, 174, 267

\bibitem[{{Condon} {et~al.}(1998){Condon}, {Cotton}, {Greisen}, {Yin},
  {Perley}, {Taylor}, \& {Broderick}}]{1998AJ....115.1693C}
{Condon}, J.~J., {Cotton}, W.~D., {Greisen}, E.~W., {et~al.} 1998, \aj, 115,
  1693

\bibitem[{{Cotton}(2008)}]{2008PASP..120..439C}
{Cotton}, W.~D. 2008, \pasp, 120, 439

\bibitem[{{Cox} {et~al.}(1986){Cox}, {Kruegel}, \&
  {Mezger}}]{1986A&A...155..380C}
{Cox}, P., {Kruegel}, E., \& {Mezger}, P.~G. 1986, \aap, 155, 380

\bibitem[{{de Gasperin} {et~al.}(2018){de Gasperin}, {Intema}, \&
  {Frail}}]{2018MNRAS.474.5008D}
{de Gasperin}, F., {Intema}, H.~T., \& {Frail}, D.~A. 2018, \mnras, 474, 5008

\bibitem[{{Dokara} {et~al.}(2018){Dokara}, {Roy}, {Beuther}, {Anderson},
  {Rugel}, {Stil}, {Wang}, {Soler}, \& {Shanahan}}]{2018ApJ...866...61D}
{Dokara}, R., {Roy}, N., {Beuther}, H., {et~al.} 2018, \apj, 866, 61

\bibitem[{{Driessen} {et~al.}(2018){Driessen}, {Dom{\v{c}}ek}, {Vink},
  {Hessels}, {Arias}, \& {Gelfand }}]{2018ApJ...860..133D}
{Driessen}, L.~N., {Dom{\v{c}}ek}, V., {Vink}, J., {et~al.} 2018, \apj, 860,
  133

\bibitem[{{Dubner} \& {Giacani}(2015)}]{Dubner2015}
{Dubner}, G. \& {Giacani}, E. 2015, \aapr, 23, 3

\bibitem[{{Frail} {et~al.}(1994){Frail}, {Goss}, \&
  {Whiteoak}}]{1994ApJ...437..781F}
{Frail}, D.~A., {Goss}, W.~M., \& {Whiteoak}, J.~B.~Z. 1994, \apj, 437, 781

\bibitem[{{F{\"u}rst} \& {Reich}(2004)}]{2004mim..proc..141F}
{F{\"u}rst}, E. \& {Reich}, W. 2004, in The Magnetized Interstellar Medium, ed.
  B.~{Uyaniker}, W.~{Reich}, \& R.~{Wielebinski}, 141--146

\bibitem[{{F{\"u}rst} {et~al.}(1987){F{\"u}rst}, {Reich}, \&
  {Sofue}}]{1987A&AS...71...63F}
{F{\"u}rst}, E., {Reich}, W., \& {Sofue}, Y. 1987, \aaps, 71, 63

\bibitem[{{Gaensler} {et~al.}(2000){Gaensler}, {Dickel}, \&
  {Green}}]{2000ApJ...542..380G}
{Gaensler}, B.~M., {Dickel}, J.~R., \& {Green}, A.~J. 2000, \apj, 542, 380

\bibitem[{{Gaensler} {et~al.}(2001){Gaensler}, {Dickey}, {McClure-Griffiths},
  {Green}, {Wieringa}, \& {Haynes}}]{2001ApJ...549..959G}
{Gaensler}, B.~M., {Dickey}, J.~M., {McClure-Griffiths}, N.~M., {et~al.} 2001,
  \apj, 549, 959

\bibitem[{{Gaensler} \& {Slane}(2006)}]{2006ARA&A..44...17G}
{Gaensler}, B.~M. \& {Slane}, P.~O. 2006, \araa, 44, 17

\bibitem[{{Gao} {et~al.}(2019){Gao}, {Reich}, {Hou}, {Reich}, \&
  {Han}}]{2019A&A...623A.105G}
{Gao}, X.~Y., {Reich}, P., {Hou}, L.~G., {Reich}, W., \& {Han}, J.~L. 2019,
  \aap, 623, A105

\bibitem[{{Green} {et~al.}(2014){Green}, {Reeves}, \&
  {Murphy}}]{2014PASA...31...42G}
{Green}, A.~J., {Reeves}, S.~N., \& {Murphy}, T. 2014, \pasa, 31, e042

\bibitem[{{Green}(2019)}]{2019JApA...40...36G}
{Green}, D.~A. 2019, Journal of Astrophysics and Astronomy, 40, 36

\bibitem[{{H.~E.~S.~S. Collaboration} {et~al.}(2018){H.~E.~S.~S.
  Collaboration}, {Abdalla}, {Abramowski}, {Aharonian}, {Ait Benkhali},
  {Akhperjanian}, {Andersson}, {Ang{\"u}ner}, {Arrieta}, {Aubert}, \&
  et~al.}]{2018A&A...612A...2H}
{H.~E.~S.~S. Collaboration}, {Abdalla}, H., {Abramowski}, A., {et~al.} 2018,
  \aap, 612, A2

\bibitem[{{Hales} {et~al.}(2012){Hales}, {Murphy}, {Curran}, {Middelberg},
  {Gaensler}, \& {Norris}}]{2012MNRAS.425..979H}
{Hales}, C.~A., {Murphy}, T., {Curran}, J.~R., {et~al.} 2012, \mnras, 425, 979

\bibitem[{{Harvey-Smith} {et~al.}(2010){Harvey-Smith}, {Gaensler}, {Kothes},
  {Townsend}, {Heald}, {Ng}, \& {Green}}]{2010ApJ...712.1157H}
{Harvey-Smith}, L., {Gaensler}, B.~M., {Kothes}, R., {et~al.} 2010, \apj, 712,
  1157

\bibitem[{{Helfand} {et~al.}(2006){Helfand}, {Becker}, {White}, {Fallon}, \&
  {Tuttle}}]{2006AJ....131.2525H}
{Helfand}, D.~J., {Becker}, R.~H., {White}, R.~L., {Fallon}, A., \& {Tuttle},
  S. 2006, \aj, 131, 2525

\bibitem[{{Helfand} {et~al.}(1989){Helfand}, {Velusamy}, {Becker}, \&
  {Lockman}}]{1989ApJ...341..151H}
{Helfand}, D.~J., {Velusamy}, T., {Becker}, R.~H., \& {Lockman}, F.~J. 1989,
  \apj, 341, 151

\bibitem[{{Hurley-Walker} {et~al.}(2019{\natexlab{a}}){Hurley-Walker},
  {Filipovi{\'c}}, {Gaensler}, {Leahy}, {Hancock}, {Franzen}, {Offringa},
  {Callingham}, {Hindson}, {Wu}, {Bell}, {For}, {Johnston-Hollitt},
  {Kapi{\'n}ska}, {Morgan}, {Murphy}, {McKinley}, {Procopio}, {Staveley-Smith},
  {Wayth}, \& {Zheng}}]{2019PASA...36...45H}
{Hurley-Walker}, N., {Filipovi{\'c}}, M.~D., {Gaensler}, B.~M., {et~al.}
  2019{\natexlab{a}}, \pasa, 36, e045

\bibitem[{{Hurley-Walker} {et~al.}(2019{\natexlab{b}}){Hurley-Walker},
  {Gaensler}, {Leahy}, {Filipovi{\'c}}, {Hancock}, {Franzen}, {Offringa},
  {Callingham}, {Hindson}, {Wu}, {Bell}, {For}, {Johnston-Hollitt},
  {Kapi{\'n}ska}, {Morgan}, {Murphy}, {McKinley}, {Procopio}, {Staveley-Smith},
  {Wayth}, \& {Zheng}}]{2019PASA...36...48H}
{Hurley-Walker}, N., {Gaensler}, B.~M., {Leahy}, D.~A., {et~al.}
  2019{\natexlab{b}}, \pasa, 36, e048

\bibitem[{{Hurley-Walker} {et~al.}(2019{\natexlab{c}}){Hurley-Walker},
  {Hancock}, {Franzen}, {Callingham}, {Offringa}, {Hindson}, {Wu}, {Bell},
  {For}, {Gaensler}, {Johnston-Hollitt}, {Kapi{\'n}ska}, {Morgan}, {Murphy},
  {McKinley}, {Procopio}, {Staveley-Smith}, {Wayth}, \&
  {Zheng}}]{2019PASA...36...47H}
{Hurley-Walker}, N., {Hancock}, P.~J., {Franzen}, T.~M.~O., {et~al.}
  2019{\natexlab{c}}, \pasa, 36, e047

\bibitem[{{Iffrig} \& {Hennebelle}(2017)}]{2017A&A...604A..70I}
{Iffrig}, O. \& {Hennebelle}, P. 2017, \aap, 604, A70

\bibitem[{{Intema} {et~al.}(2017){Intema}, {Jagannathan}, {Mooley}, \&
  {Frail}}]{2017A&A...598A..78I}
{Intema}, H.~T., {Jagannathan}, P., {Mooley}, K.~P., \& {Frail}, D.~A. 2017,
  \aap, 598, A78

\bibitem[{{Joye} \& {Mandel}(2003)}]{2003ASPC..295..489J}
{Joye}, W.~A. \& {Mandel}, E. 2003, Astronomical Society of the Pacific
  Conference Series, Vol. 295, {New Features of SAOImage DS9}, ed. H.~E.
  {Payne}, R.~I. {Jedrzejewski}, \& R.~N. {Hook}, 489

\bibitem[{{Jun} \& {Norman}(1996)}]{1996ApJ...472..245J}
{Jun}, B.-I. \& {Norman}, M.~L. 1996, \apj, 472, 245

\bibitem[{{Kim} {et~al.}(2013){Kim}, {Koo}, \& {Moon}}]{2013ApJ...774....5K}
{Kim}, H.-J., {Koo}, B.-C., \& {Moon}, D.-S. 2013, \apj, 774, 5

\bibitem[{{Kothes} {et~al.}(2008){Kothes}, {Landecker}, {Reich}, {Safi-Harb},
  \& {Arzoumanian}}]{2008ApJ...687..516K}
{Kothes}, R., {Landecker}, T.~L., {Reich}, W., {Safi-Harb}, S., \&
  {Arzoumanian}, Z. 2008, \apj, 687, 516

\bibitem[{{Kothes} {et~al.}(2017){Kothes}, {Reich}, {Foster}, \&
  {Reich}}]{2017A&A...597A.116K}
{Kothes}, R., {Reich}, P., {Foster}, T.~J., \& {Reich}, W. 2017, \aap, 597,
  A116

\bibitem[{{Kothes} {et~al.}(2020){Kothes}, {Reich}, {Safi-Harb}, {Guest},
  {Reich}, \& {F{\"u}rst}}]{2020MNRAS.496..723K}
{Kothes}, R., {Reich}, W., {Safi-Harb}, S., {et~al.} 2020, \mnras, 496, 723

\bibitem[{{Lang} {et~al.}(2010){Lang}, {Wang}, {Lu}, \&
  {Clubb}}]{2010ApJ...709.1125L}
{Lang}, C.~C., {Wang}, Q.~D., {Lu}, F., \& {Clubb}, K.~I. 2010, \apj, 709, 1125

\bibitem[{{Lee} {et~al.}(2012){Lee}, {Koo}, \& {Lee}}]{2012JKAS...45..117L}
{Lee}, J.-W., {Koo}, B.-C., \& {Lee}, J.-E. 2012, Journal of Korean
  Astronomical Society, 45, 117

\bibitem[{{Li} {et~al.}(1991){Li}, {Wheeler}, {Bash}, \&
  {Jefferys}}]{1991ApJ...378...93L}
{Li}, Z., {Wheeler}, J.~C., {Bash}, F.~N., \& {Jefferys}, W.~H. 1991, \apj,
  378, 93

\bibitem[{{Liszt}(2009)}]{2009A&A...508.1331L}
{Liszt}, H.~S. 2009, \aap, 508, 1331

\bibitem[{{Lockman}(1989)}]{1989ApJS...71..469L}
{Lockman}, F.~J. 1989, \apjs, 71, 469

\bibitem[{{Lockman} {et~al.}(1996){Lockman}, {Pisano}, \&
  {Howard}}]{1996ApJ...472..173L}
{Lockman}, F.~J., {Pisano}, D.~J., \& {Howard}, G.~J. 1996, \apj, 472, 173

\bibitem[{{Medina} {et~al.}(2019){Medina}, {Urquhart}, {Dzib}, {Brunthaler},
  {Cotton}, {Menten}, {Wyrowski}, {Beuther}, {Billington}, {Carrasco-Gonzalez},
  {Csengeri}, {Gong}, {Hofner}, {Nguyen}, {Ortiz-Le{\'o}n}, {Ott}, {Pandian},
  {Roy}, {Sarkar}, {Wang}, \& {Winkel}}]{2019A&A...627A.175M}
{Medina}, S. N.~X., {Urquhart}, J.~S., {Dzib}, S.~A., {et~al.} 2019, \aap, 627,
  A175

\bibitem[{{Mertsch} \& {Sarkar}(2013)}]{2013JCAP...06..041M}
{Mertsch}, P. \& {Sarkar}, S. 2013, \jcap, 2013, 041

\bibitem[{{Milne}(1987)}]{1987AuJPh..40..771M}
{Milne}, D.~K. 1987, Australian Journal of Physics, 40, 771

\bibitem[{{Milne} \& {Dickel}(1974{\natexlab{a}})}]{1974AuJPh..27..549M}
{Milne}, D.~K. \& {Dickel}, J.~R. 1974{\natexlab{a}}, Australian Journal of
  Physics, 27, 549

\bibitem[{{Milne} \& {Dickel}(1974{\natexlab{b}})}]{1974IAUS...60..335M}
{Milne}, D.~K. \& {Dickel}, J.~R. 1974{\natexlab{b}}, in Galactic Radio
  Astronomy, ed. F.~J. {Kerr} \& S.~C. {Simonson}, Vol.~60, 335

\bibitem[{{Petriella}(2019)}]{2019A&A...626A..65P}
{Petriella}, A. 2019, \aap, 626, A65

\bibitem[{{Pinheiro Gon{\c{c}}alves} {et~al.}(2011){Pinheiro Gon{\c{c}}alves},
  {Noriega-Crespo}, {Paladini}, {Martin}, \& {Carey}}]{2011AJ....142...47P}
{Pinheiro Gon{\c{c}}alves}, D., {Noriega-Crespo}, A., {Paladini}, R., {Martin},
  P.~G., \& {Carey}, S.~J. 2011, \aj, 142, 47

\bibitem[{{Ranasinghe} \& {Leahy}(2018)}]{2018AJ....155..204R}
{Ranasinghe}, S. \& {Leahy}, D.~A. 2018, \aj, 155, 204

\bibitem[{{Robitaille} \& {Bressert}(2012)}]{2012ascl.soft08017R}
{Robitaille}, T. \& {Bressert}, E. 2012, {APLpy: Astronomical Plotting Library
  in Python}

\bibitem[{{Simmons} \& {Stewart}(1985)}]{1985A&A...142..100S}
{Simmons}, J.~F.~L. \& {Stewart}, B.~G. 1985, \aap, 142, 100

\bibitem[{{Stupar} \& {Parker}(2011)}]{2011MNRAS.414.2282S}
{Stupar}, M. \& {Parker}, Q.~A. 2011, \mnras, 414, 2282

\bibitem[{{Sun} {et~al.}(2007){Sun}, {Han}, {Reich}, {Reich}, {Shi},
  {Wielebinski}, \& {F{\"u}rst}}]{2007A&A...463..993S}
{Sun}, X.~H., {Han}, J.~L., {Reich}, W., {et~al.} 2007, \aap, 463, 993

\bibitem[{{Sun} {et~al.}(2011){Sun}, {Reich}, {Reich}, {Xiao}, {Gao}, \&
  {Han}}]{2011A&A...536A..83S}
{Sun}, X.~H., {Reich}, P., {Reich}, W., {et~al.} 2011, \aap, 536, A83

\bibitem[{{Supan} {et~al.}(2018){Supan}, {Castelletti}, {Peters}, \&
  {Kassim}}]{2018A&A...616A..98S}
{Supan}, L., {Castelletti}, G., {Peters}, W.~M., \& {Kassim}, N.~E. 2018, \aap,
  616, A98

\bibitem[{{Tammann} {et~al.}(1994){Tammann}, {Loeffler}, \&
  {Schroeder}}]{1994ApJS...92..487T}
{Tammann}, G.~A., {Loeffler}, W., \& {Schroeder}, A. 1994, \apjs, 92, 487

\bibitem[{{Ueno} {et~al.}(2006){Ueno}, {Yamauchi}, {Bamba}, {Yamaguchi},
  {Koyama}, \& {Ebisawa}}]{2006IAUS..230..333U}
{Ueno}, M., {Yamauchi}, S., {Bamba}, A., {et~al.} 2006, in IAU Symposium, Vol.
  230, Populations of High Energy Sources in Galaxies, ed. E.~J.~A. {Meurs} \&
  G.~{Fabbiano}, 333--337

\bibitem[{{Wardle} \& {Kronberg}(1974)}]{1974ApJ...194..249W}
{Wardle}, J.~F.~C. \& {Kronberg}, P.~P. 1974, \apj, 194, 249

\bibitem[{{Whiteoak} \& {Green}(1996)}]{1996A&AS..118..329W}
{Whiteoak}, J.~B.~Z. \& {Green}, A.~J. 1996, \aaps, 118, 329

\bibitem[{{Wilson} {et~al.}(2013){Wilson}, {Rohlfs}, \&
  {H{\"u}ttemeister}}]{2013tra..book.....W}
{Wilson}, T.~L., {Rohlfs}, K., \& {H{\"u}ttemeister}, S. 2013, {Tools of Radio
  Astronomy}

\bibitem[{{Yamauchi} {et~al.}(2016){Yamauchi}, {Sumita}, \&
  {Bamba}}]{2016PASJ...68S...6Y}
{Yamauchi}, S., {Sumita}, M., \& {Bamba}, A. 2016, \pasj, 68, S6

\end{thebibliography}

\begin{appendix}
\section{GLOSTAR-VLA and GLIMPSE images of newly identified SNR candidates}
\label{ims:newcandSNRs}

The images of newly identified SNR candidates from the GLOSTAR-VLA data
are shown here.  The marking
scheme is the same as in the text: red circles mark G19 SNRs,
solid and dashed magenta circles mark confirmed and candidate \ion{H}{ii}~regions from
A14 catalog, green circles mark previously identified SNR candidates, and
white circles mark new GLOSTAR SNR candidates.  The beam of GLOSTAR-VLA
data is shown in the bottom left corner in black.

{\centering
\noindent\includegraphics[width=0.47\textwidth]{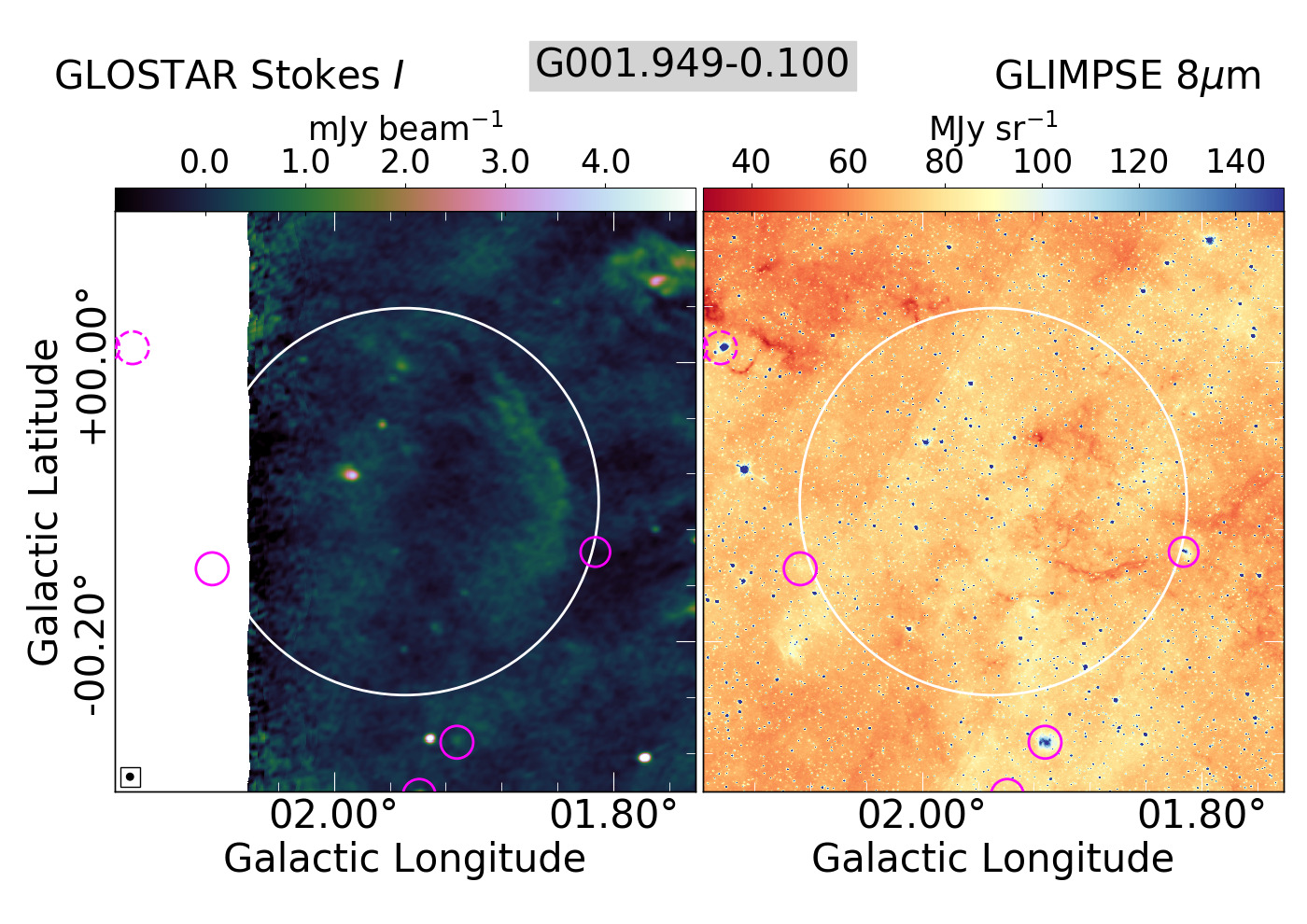}\\
\noindent\includegraphics[width=0.47\textwidth]{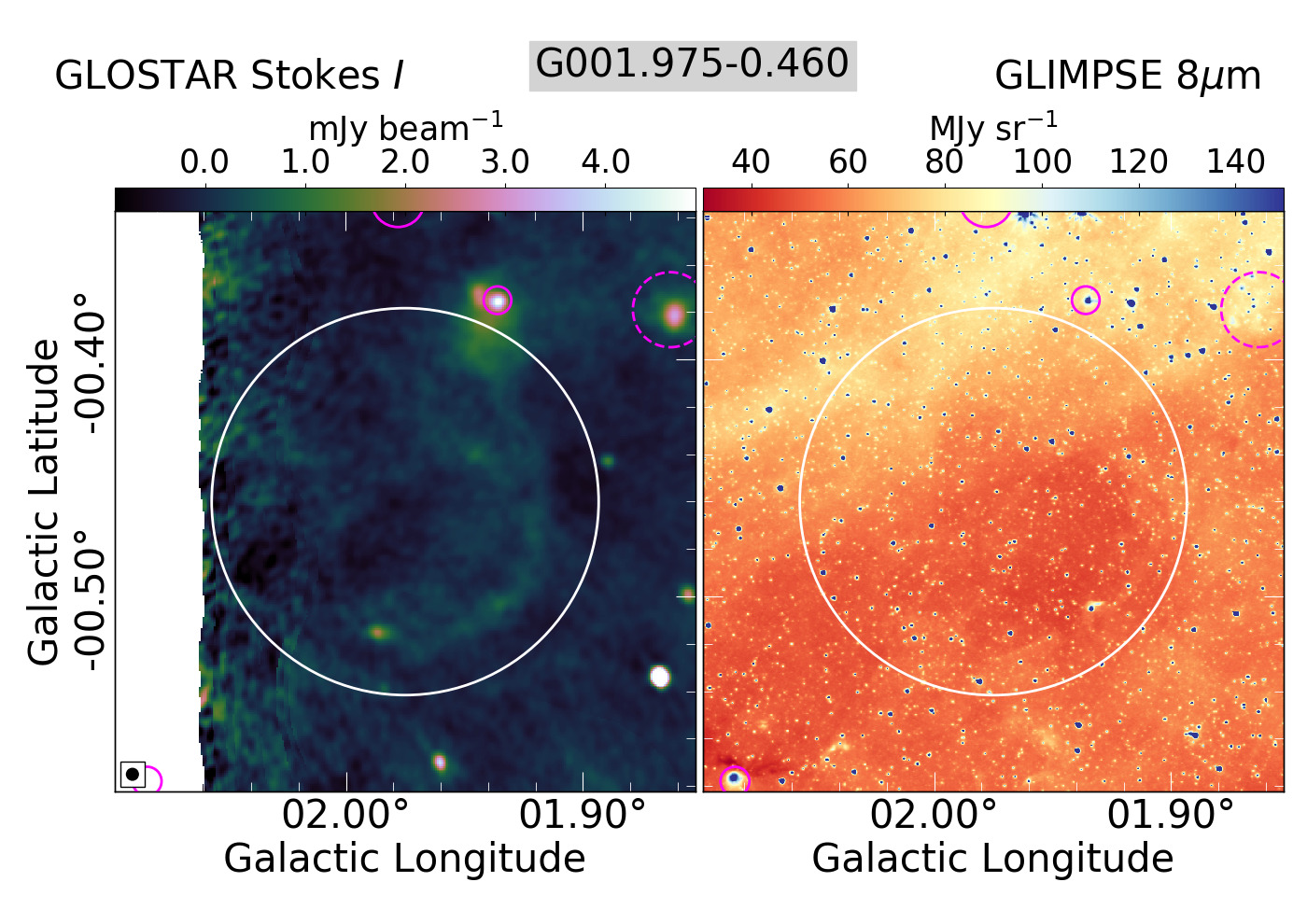}\\
\noindent\includegraphics[width=0.47\textwidth]{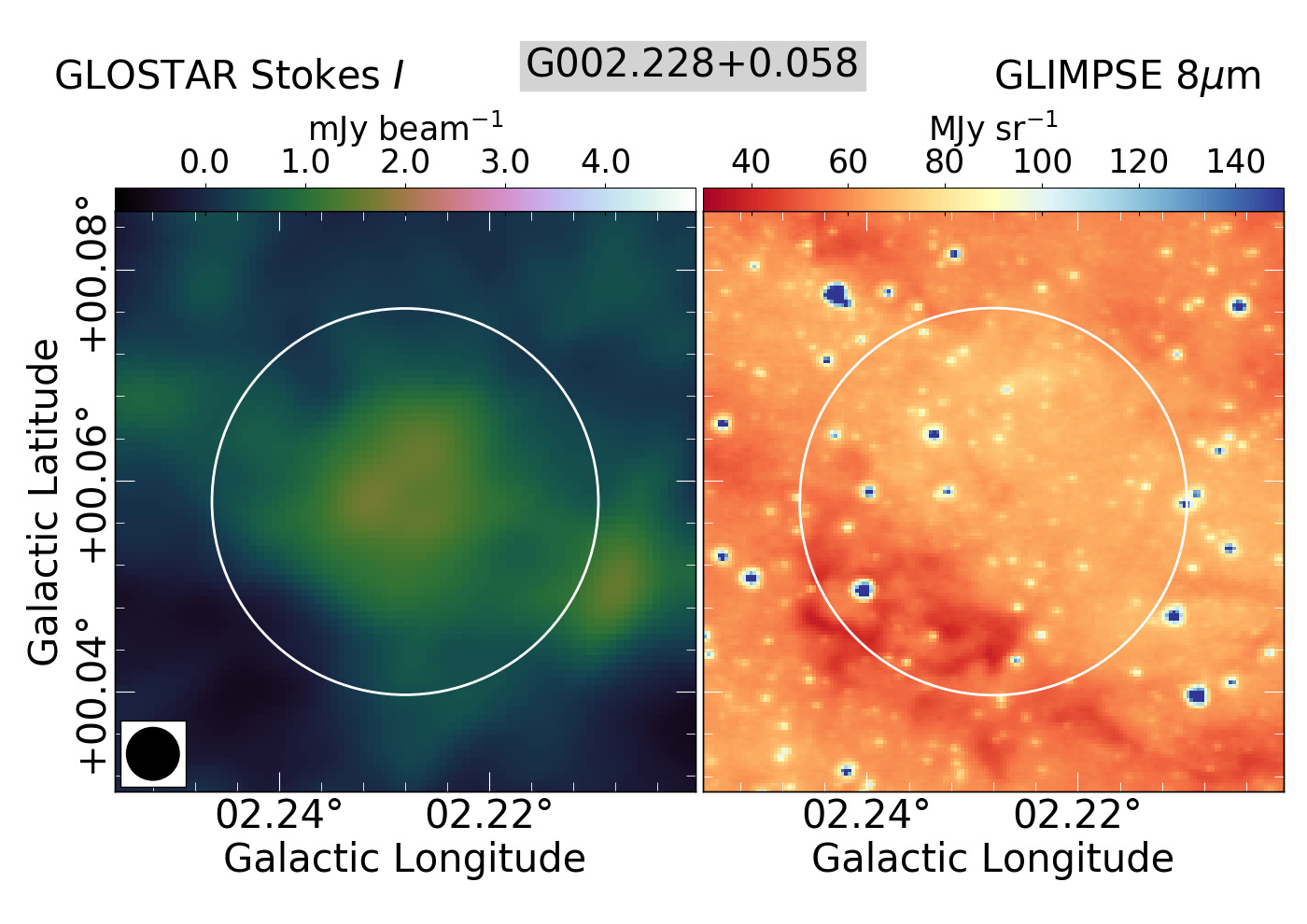}\\
\noindent\includegraphics[width=0.47\textwidth]{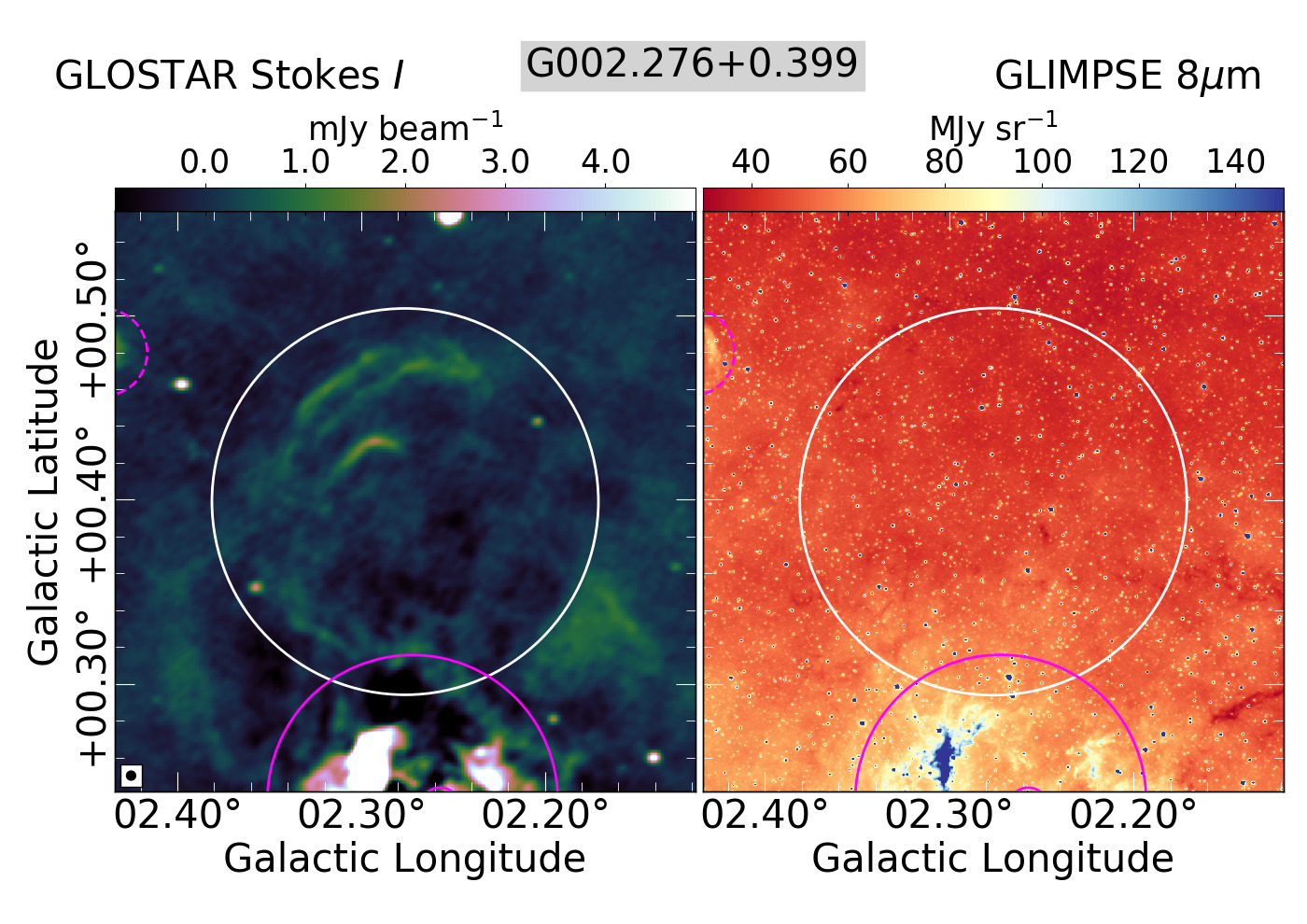}\\
\noindent\includegraphics[width=0.47\textwidth]{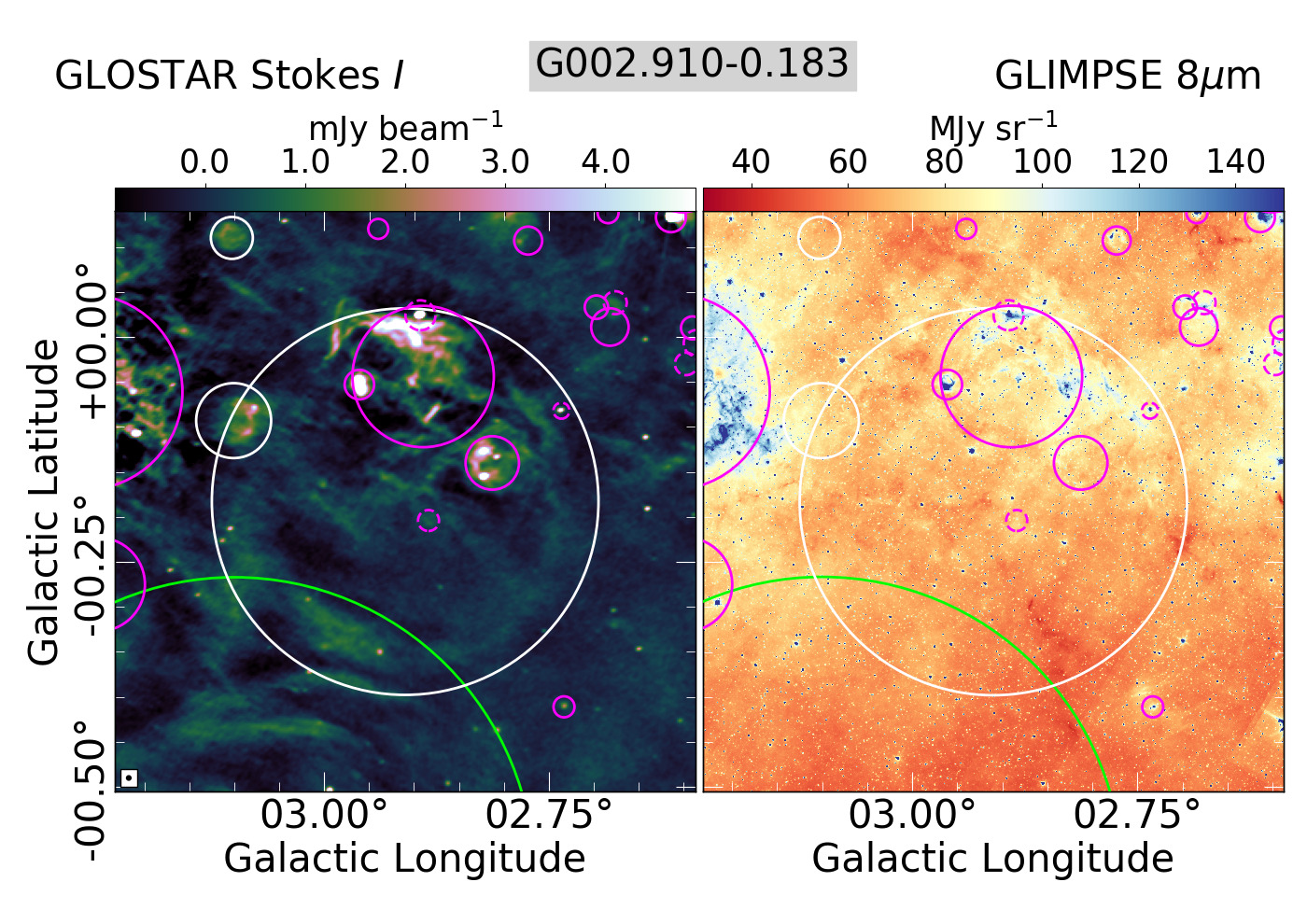}\\
\noindent\includegraphics[width=0.47\textwidth]{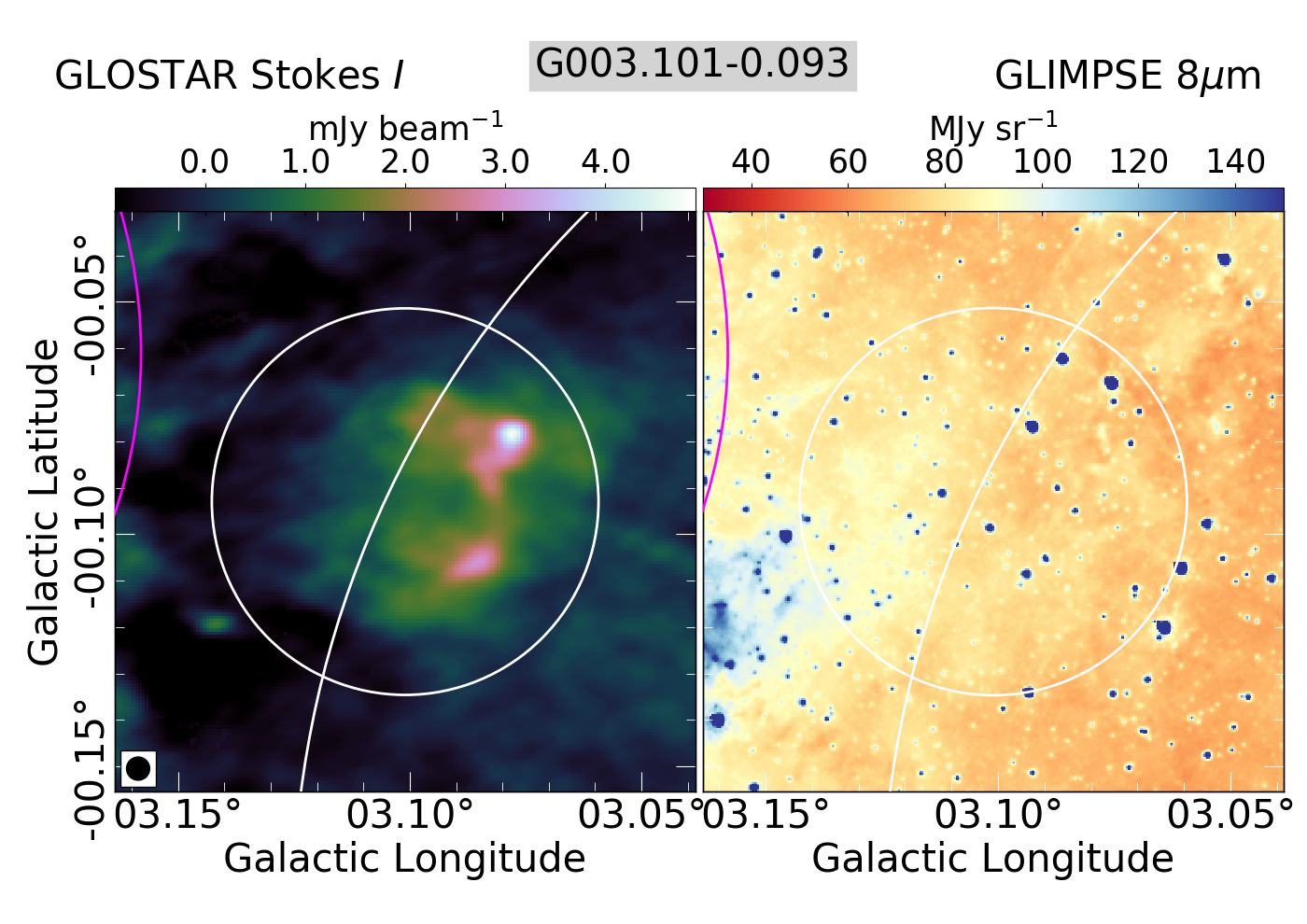}\\
\noindent\includegraphics[width=0.47\textwidth]{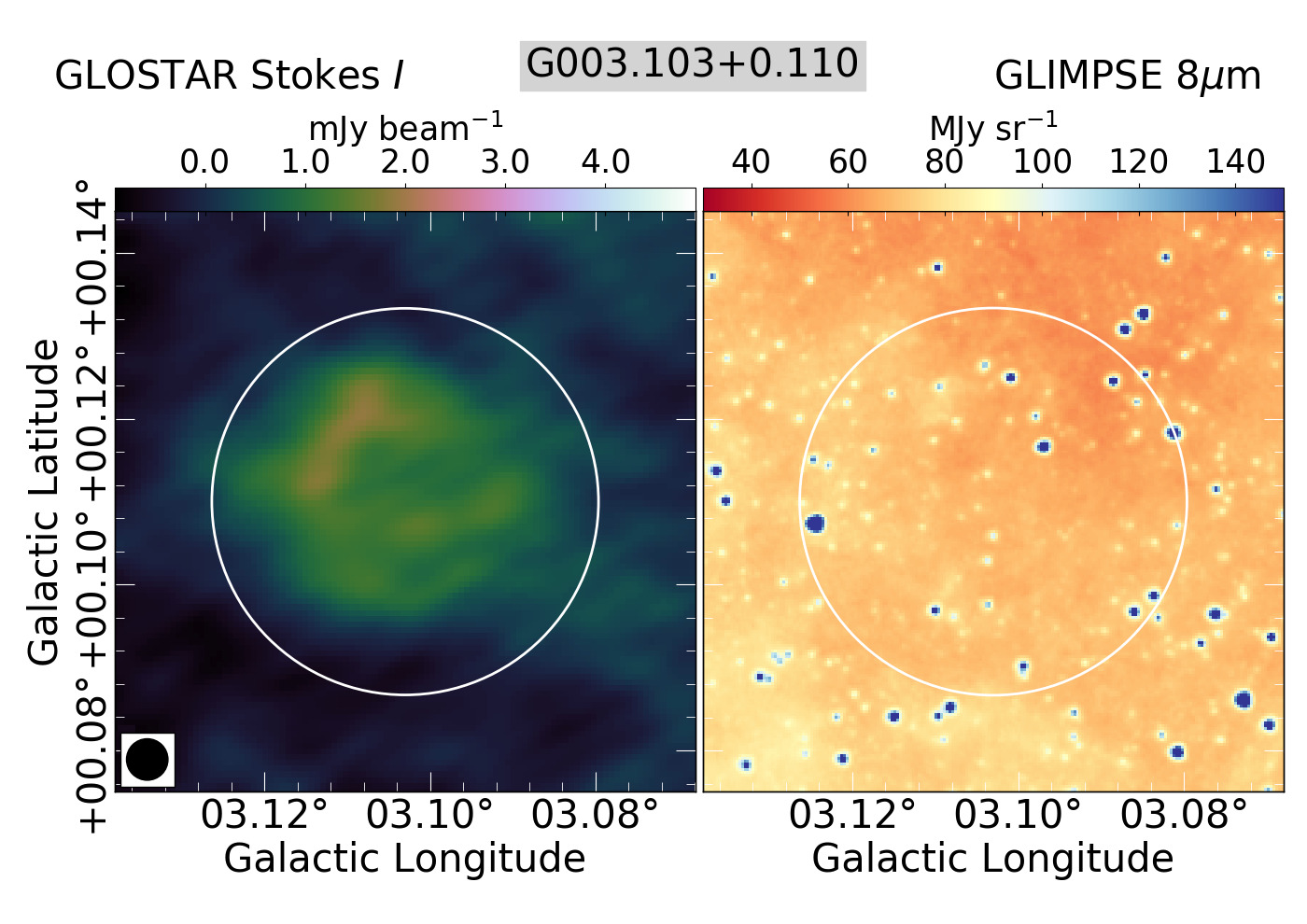}\\
\noindent\includegraphics[width=0.47\textwidth]{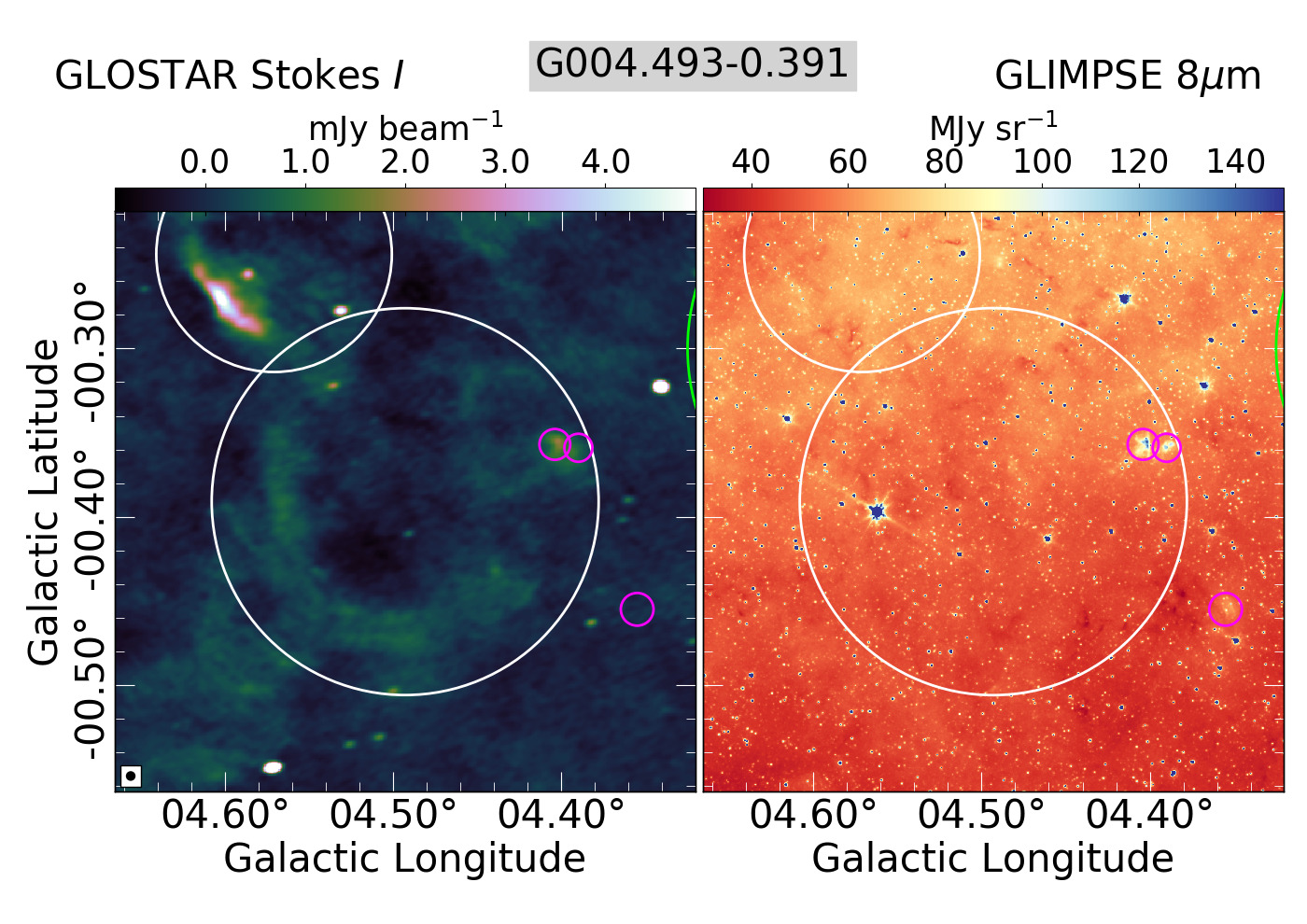}\\
\noindent\includegraphics[width=0.47\textwidth]{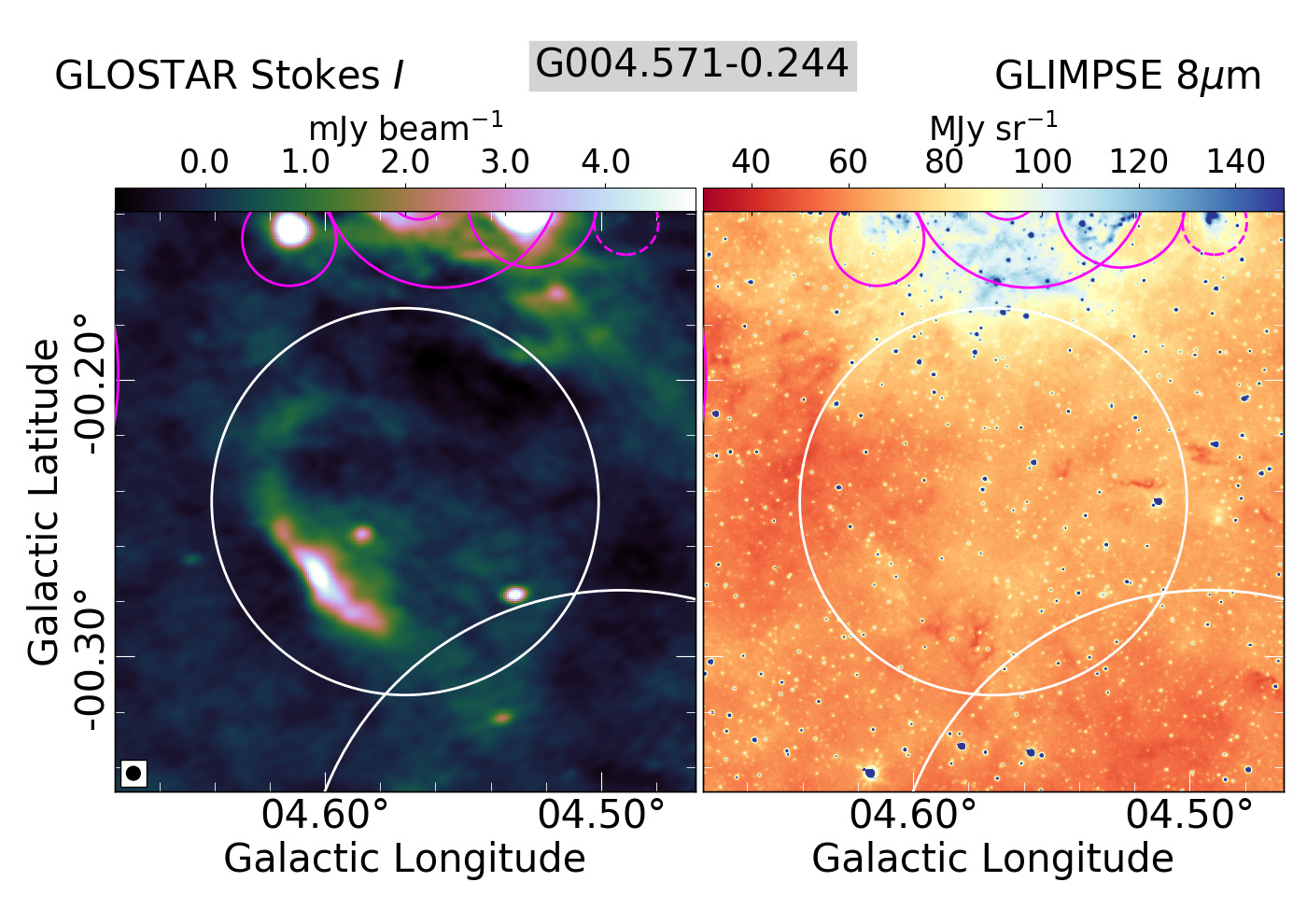}\\
\noindent\includegraphics[width=0.47\textwidth]{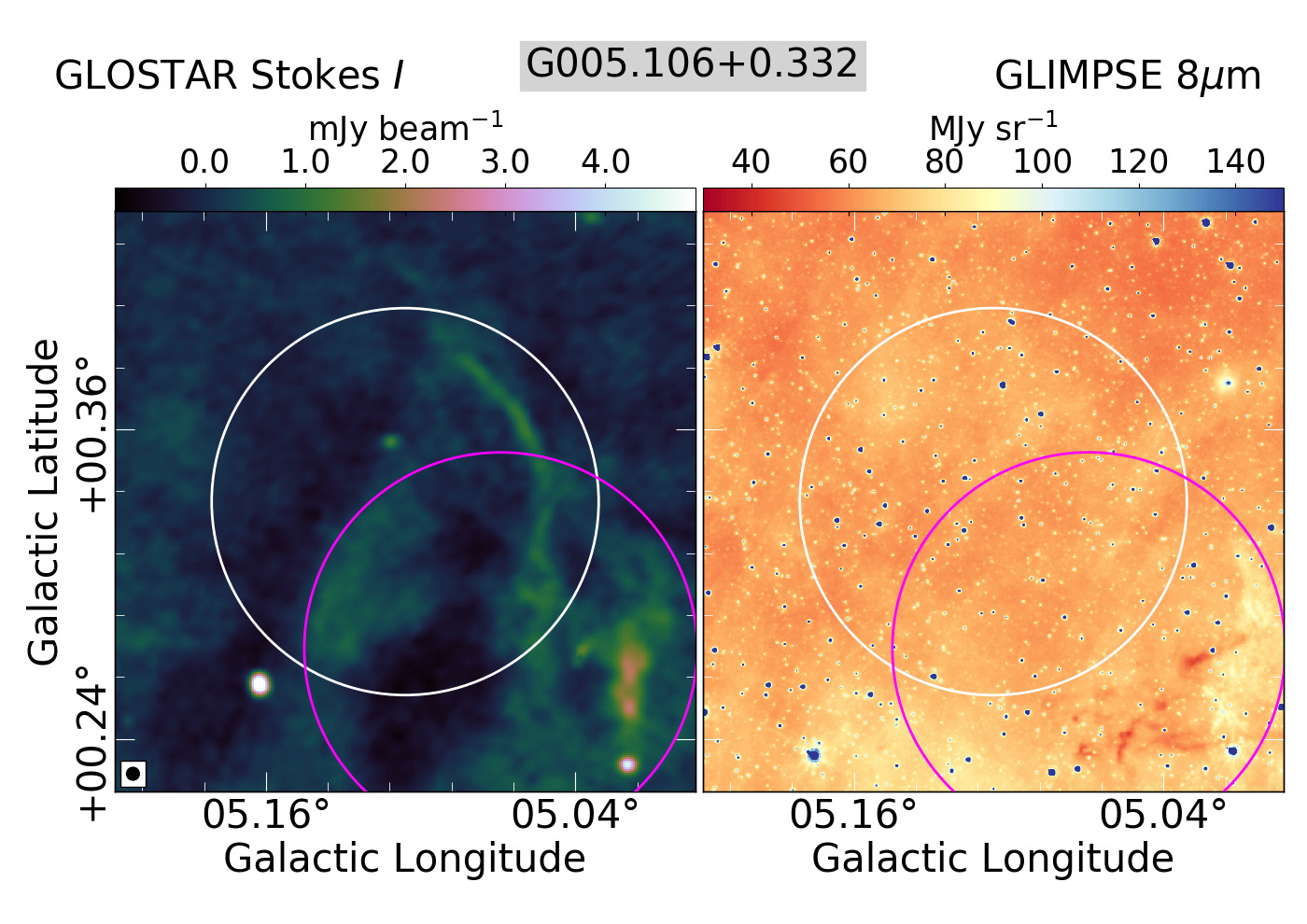}\\
\noindent\includegraphics[width=0.47\textwidth]{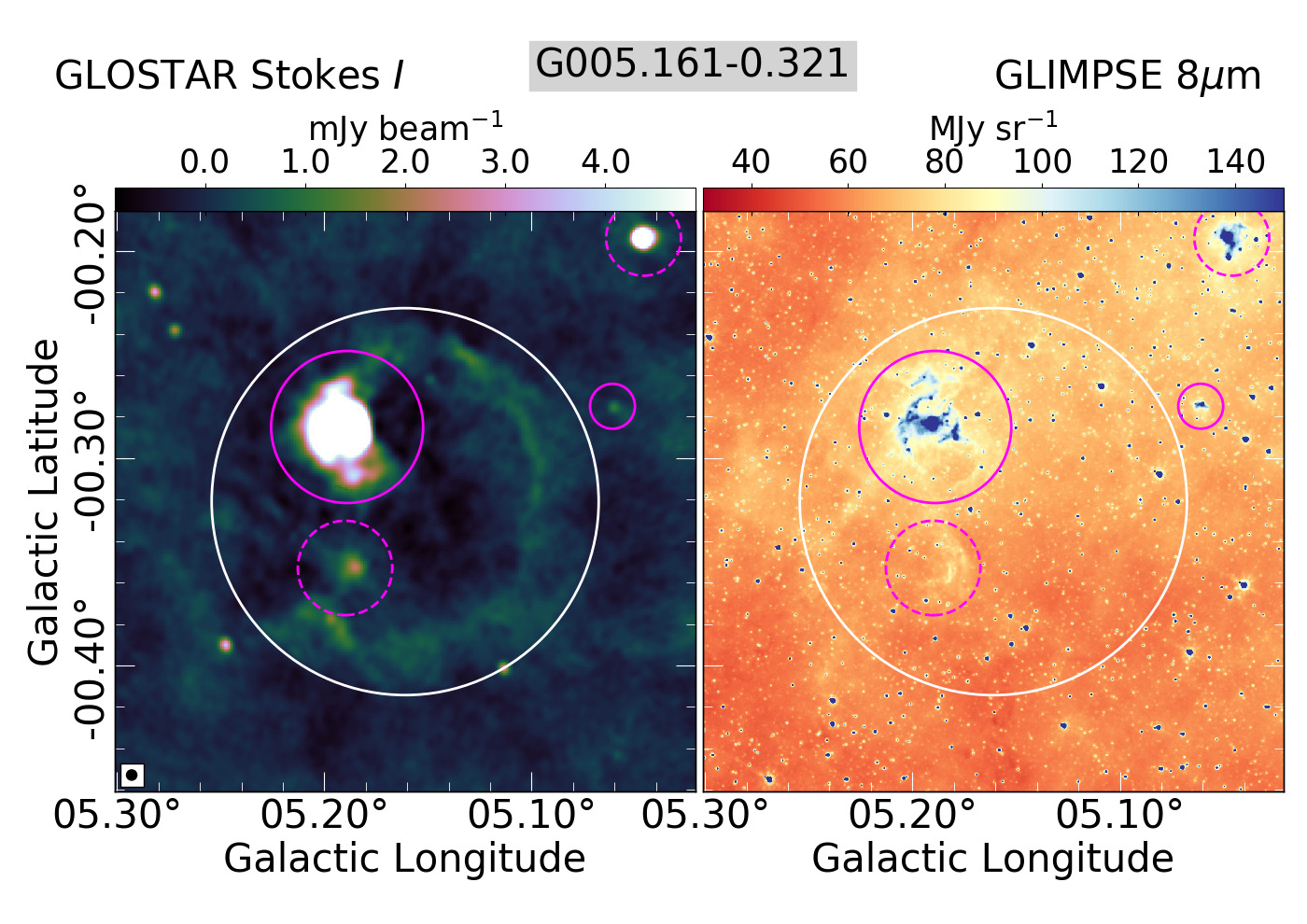}\\
\noindent\includegraphics[width=0.47\textwidth]{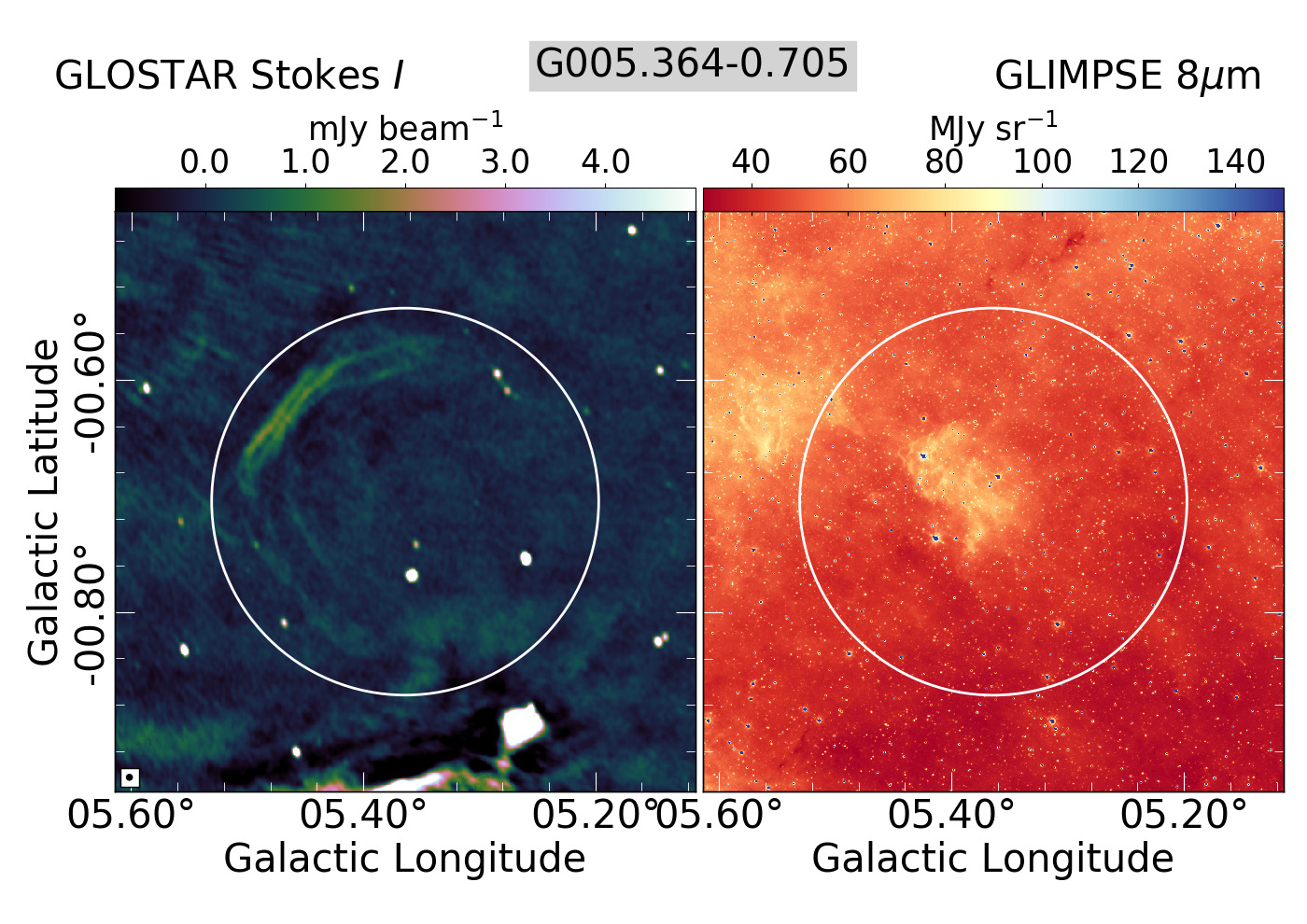}\\
\noindent\includegraphics[width=0.47\textwidth]{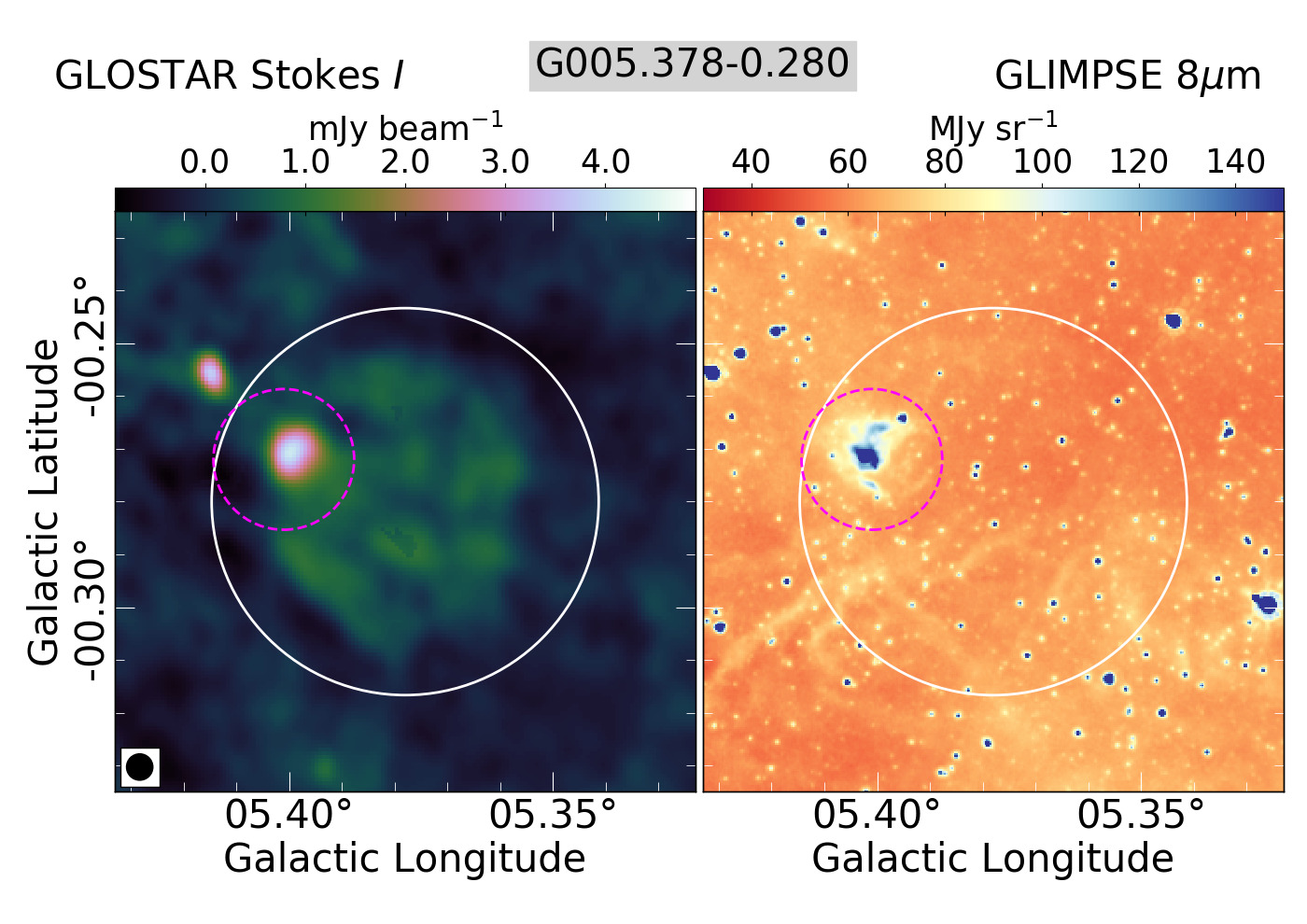}\\
\noindent\includegraphics[width=0.47\textwidth]{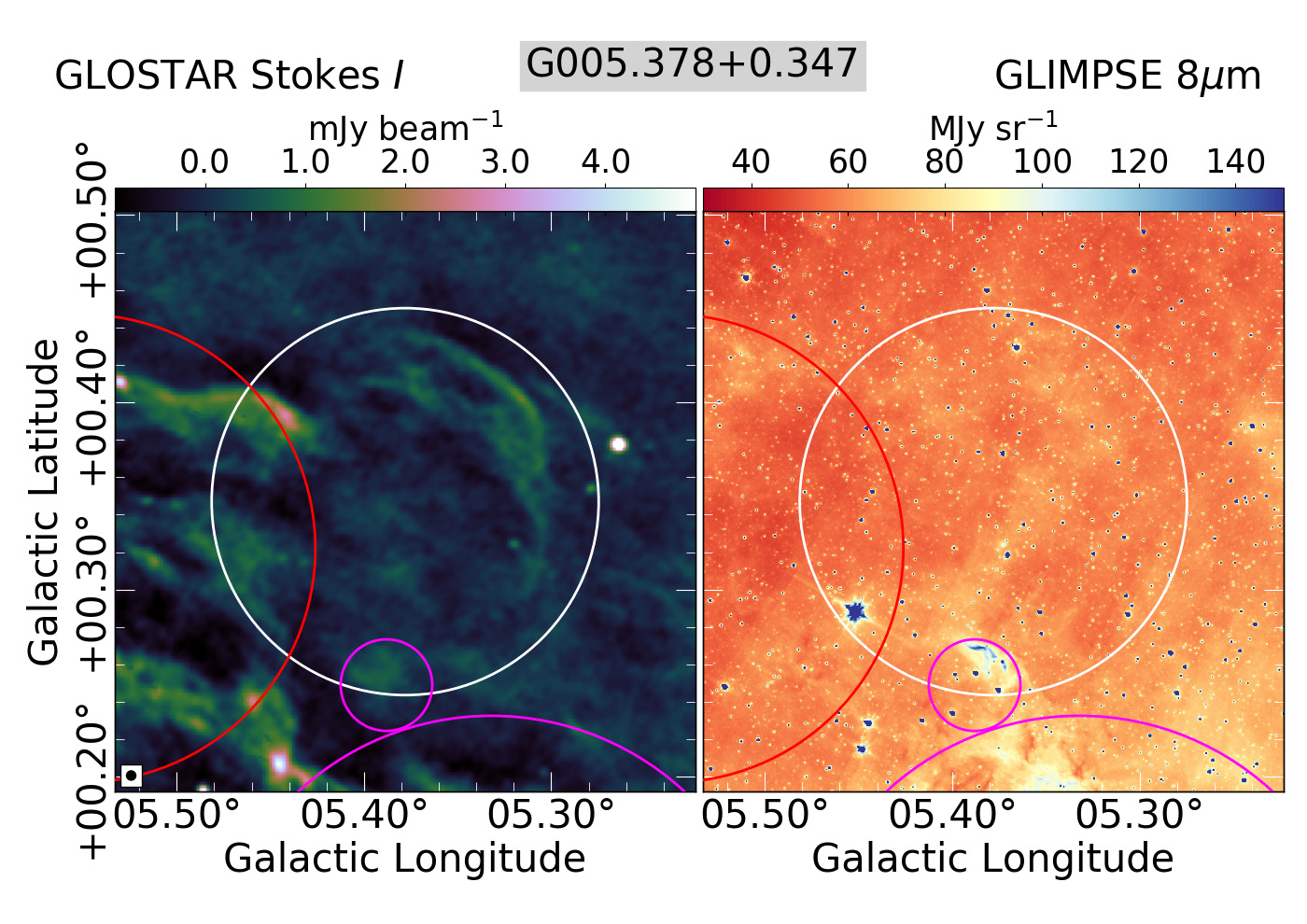}\\
\noindent\includegraphics[width=0.47\textwidth]{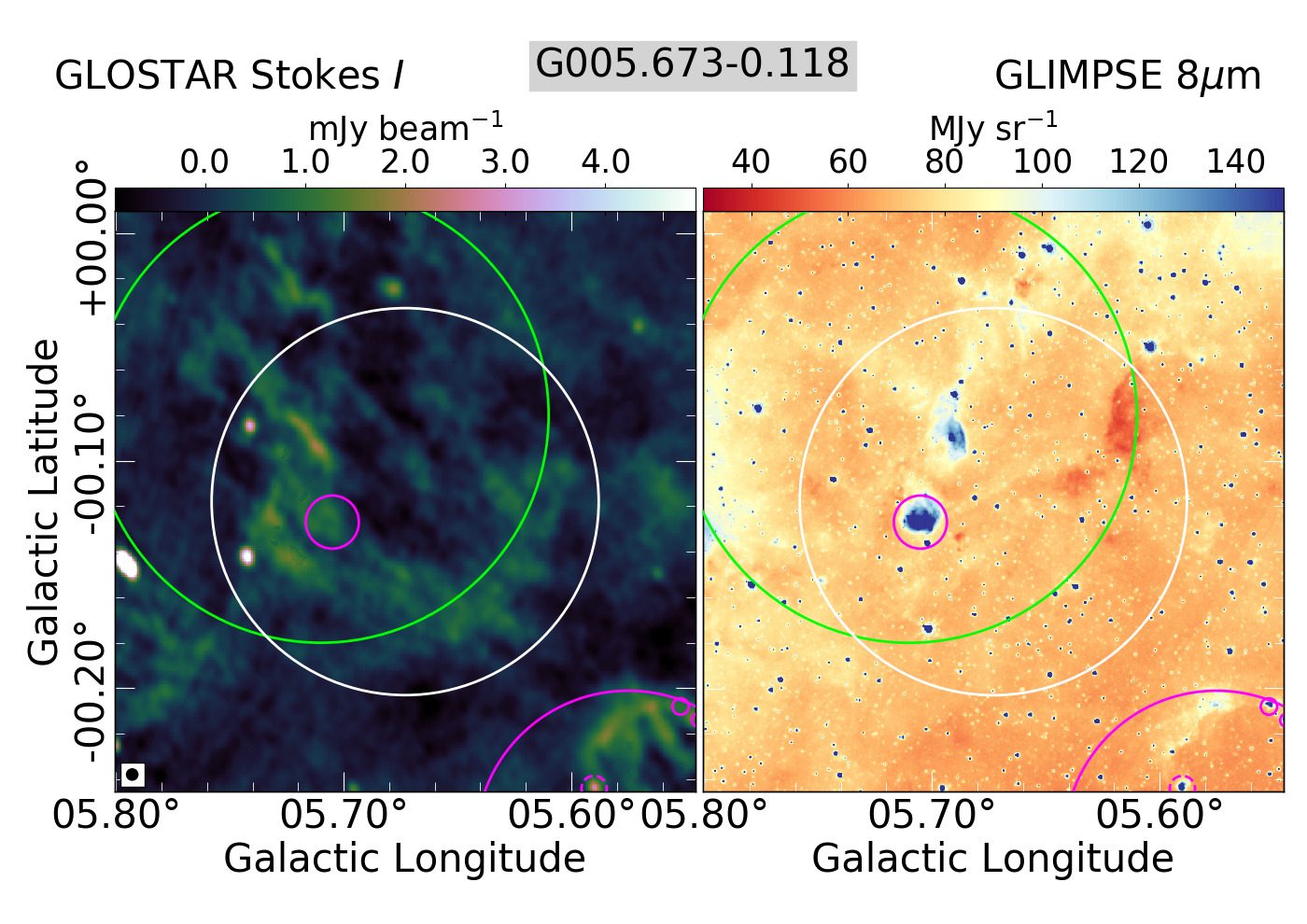}\\
\noindent\includegraphics[width=0.47\textwidth]{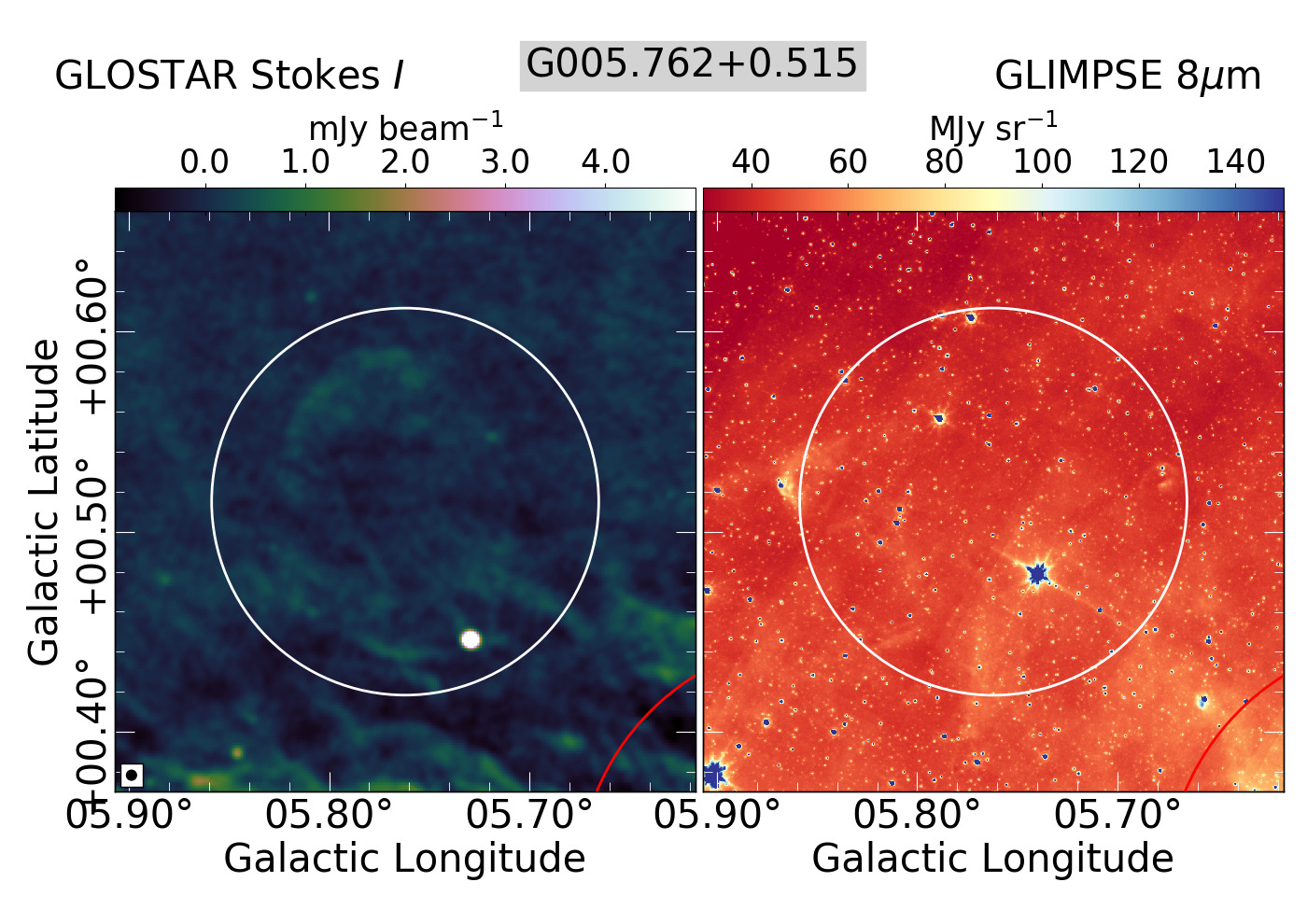}\\
\noindent\includegraphics[width=0.47\textwidth]{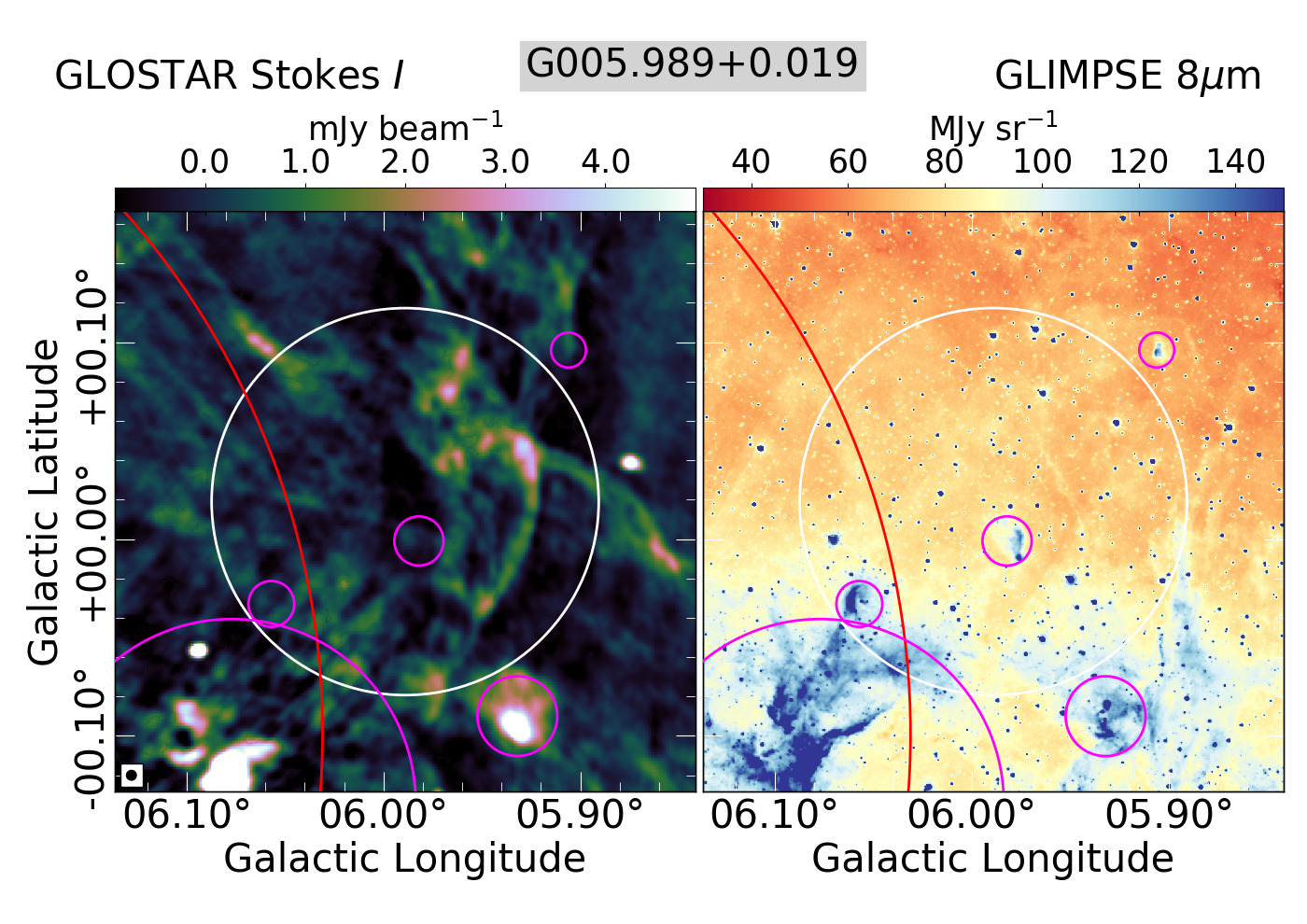}\\
\noindent\includegraphics[width=0.47\textwidth]{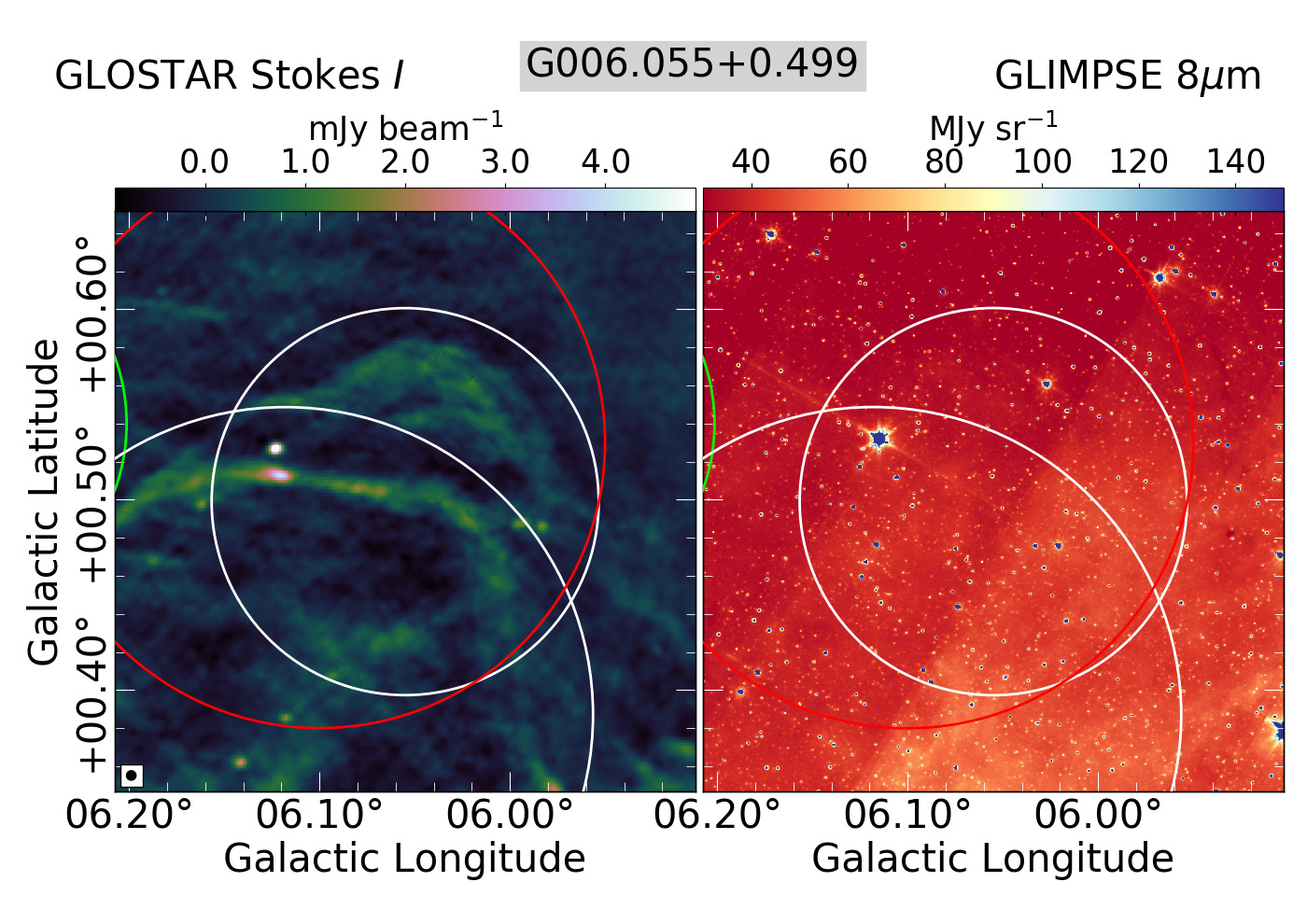}\\
\noindent\includegraphics[width=0.47\textwidth]{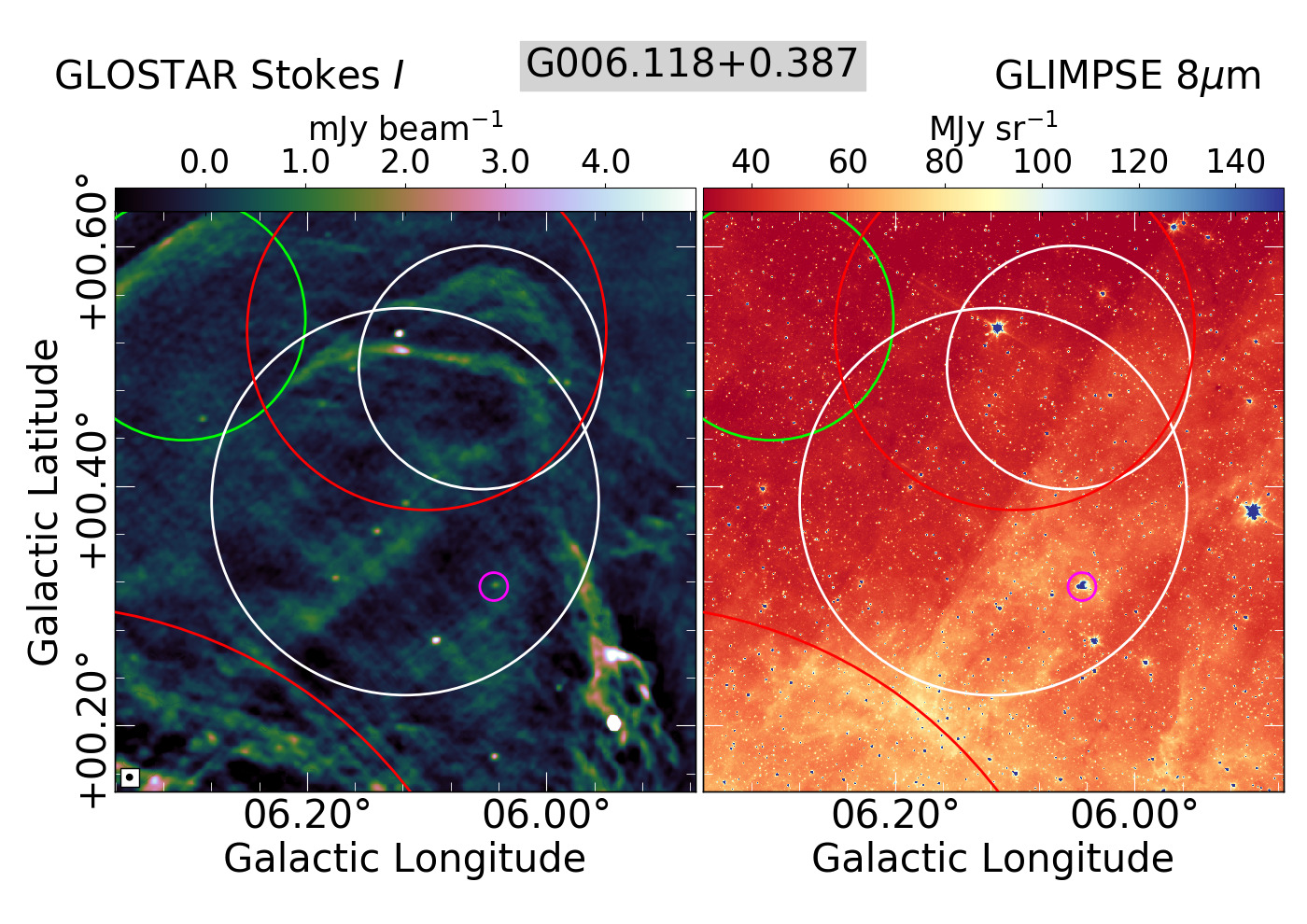}\\
\noindent\includegraphics[width=0.47\textwidth]{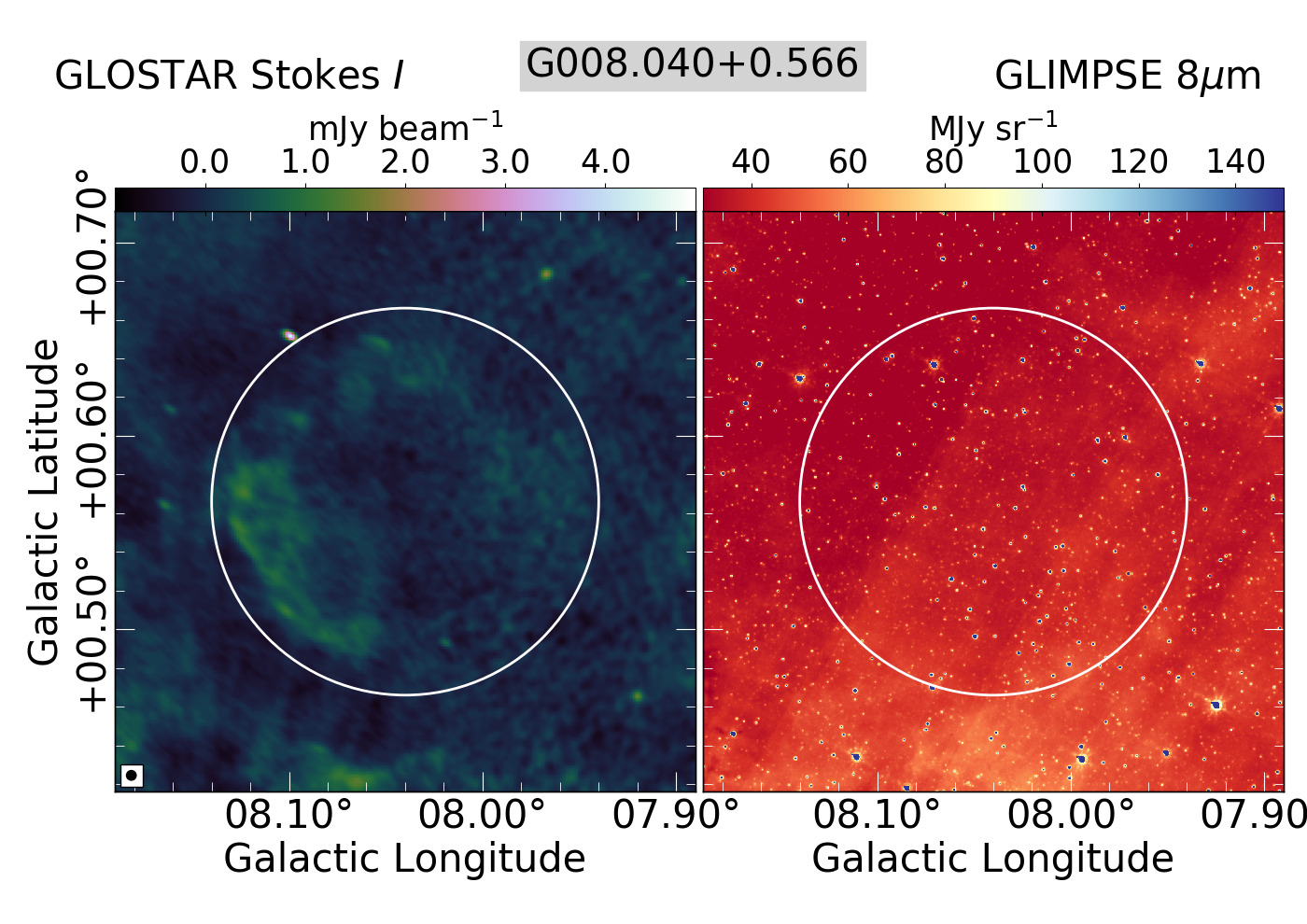}\\
\noindent\includegraphics[width=0.47\textwidth]{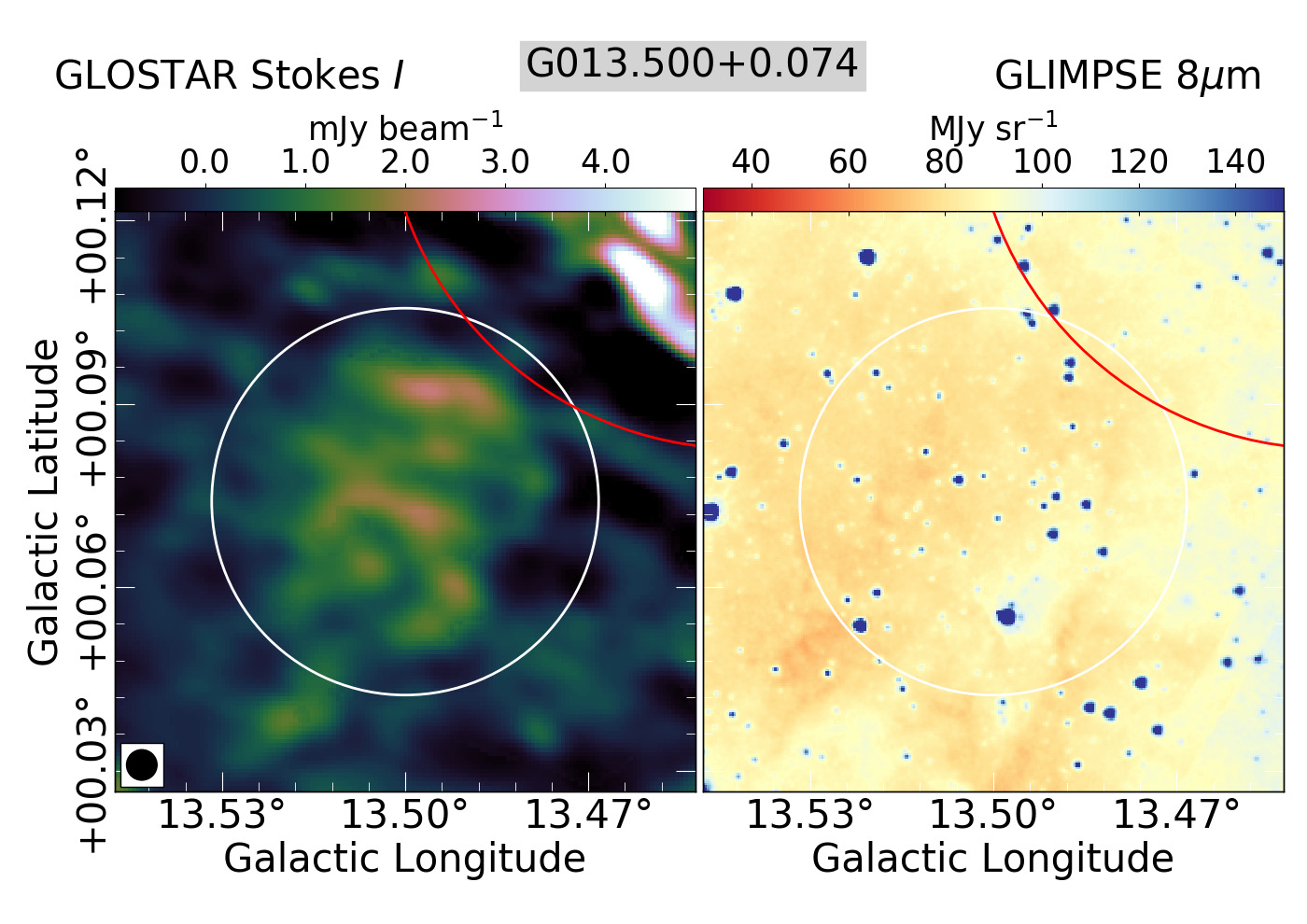}\\
\noindent\includegraphics[width=0.47\textwidth]{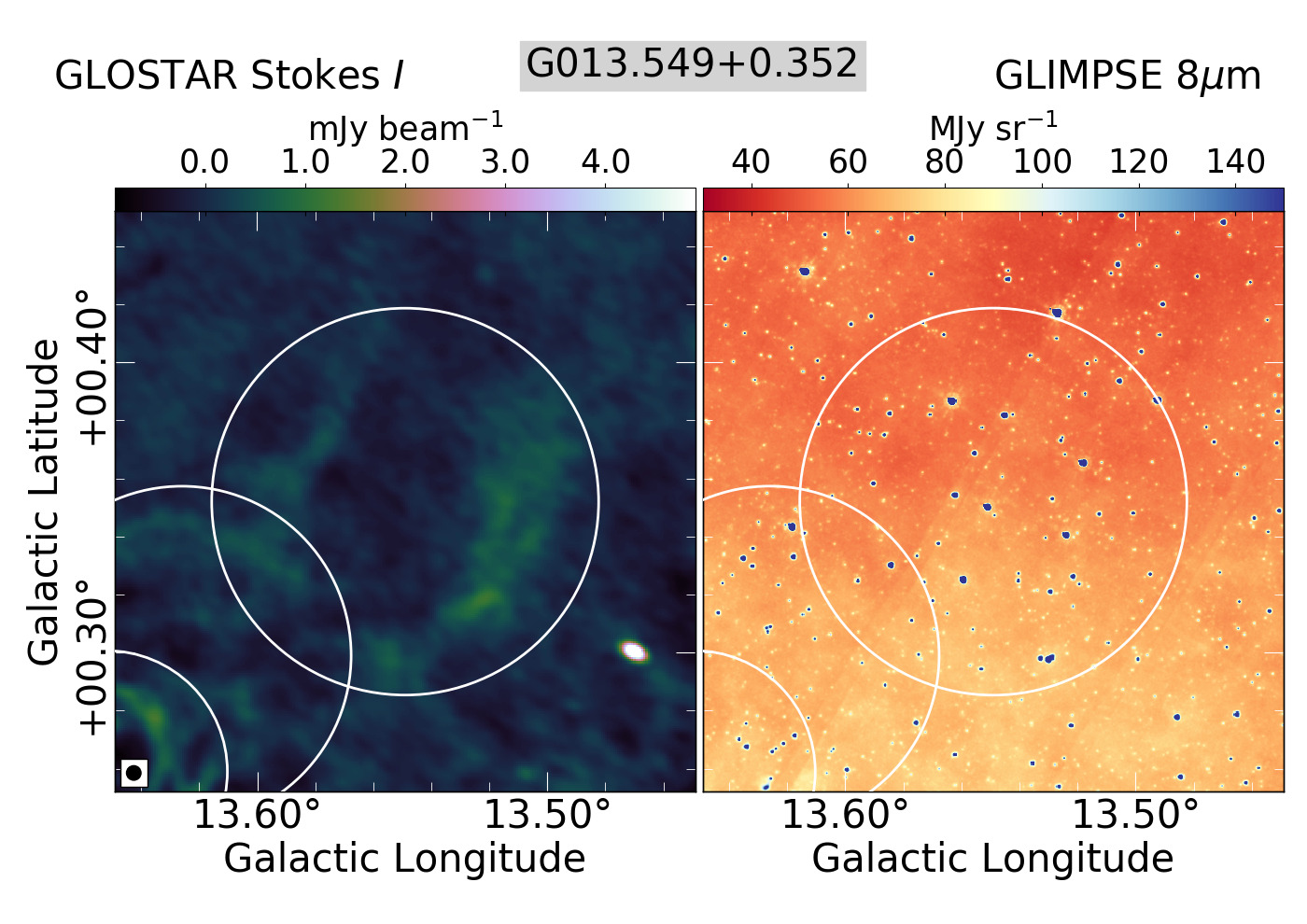}\\
\noindent\includegraphics[width=0.47\textwidth]{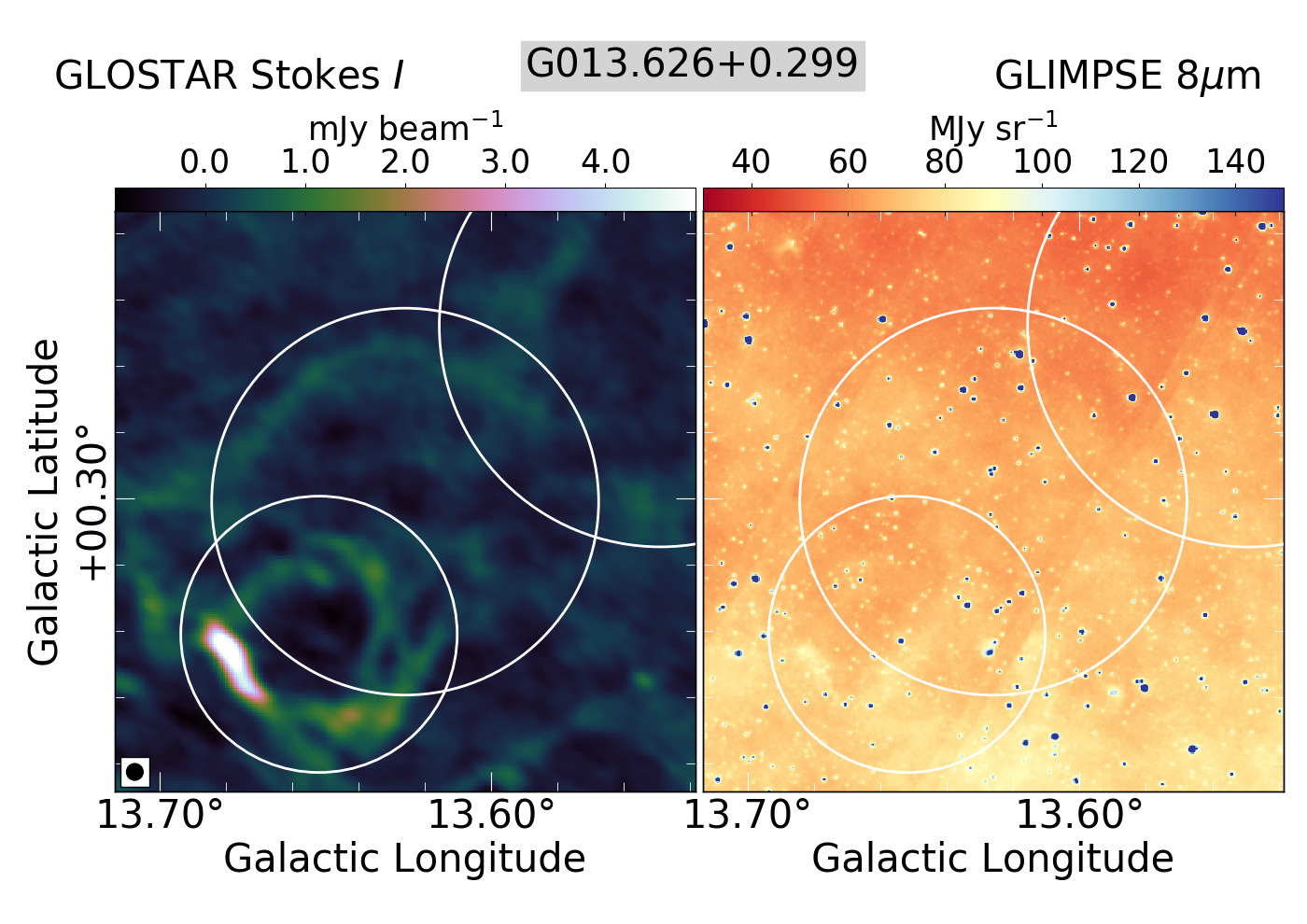}\\
\noindent\includegraphics[width=0.47\textwidth]{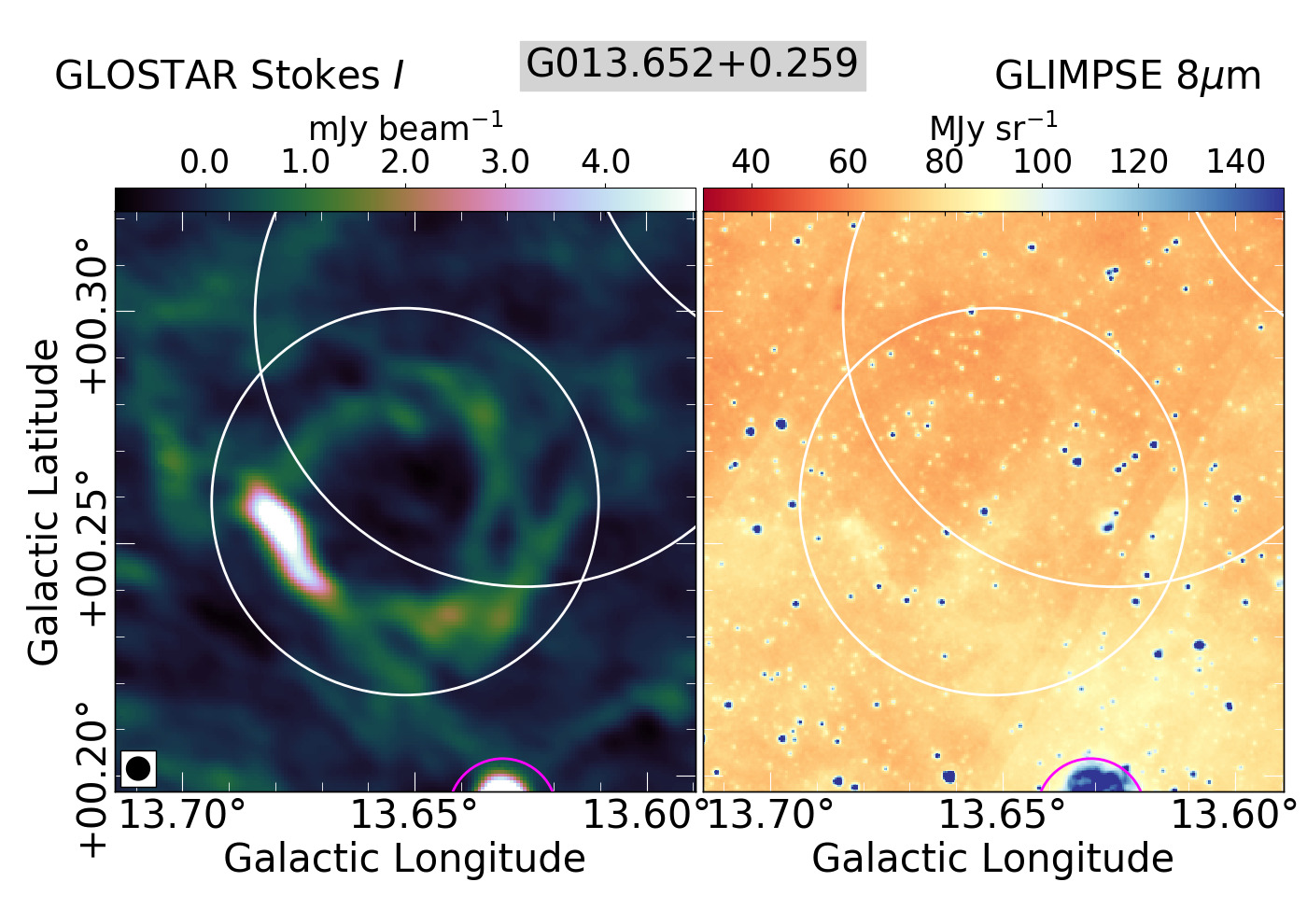}\\
\noindent\includegraphics[width=0.47\textwidth]{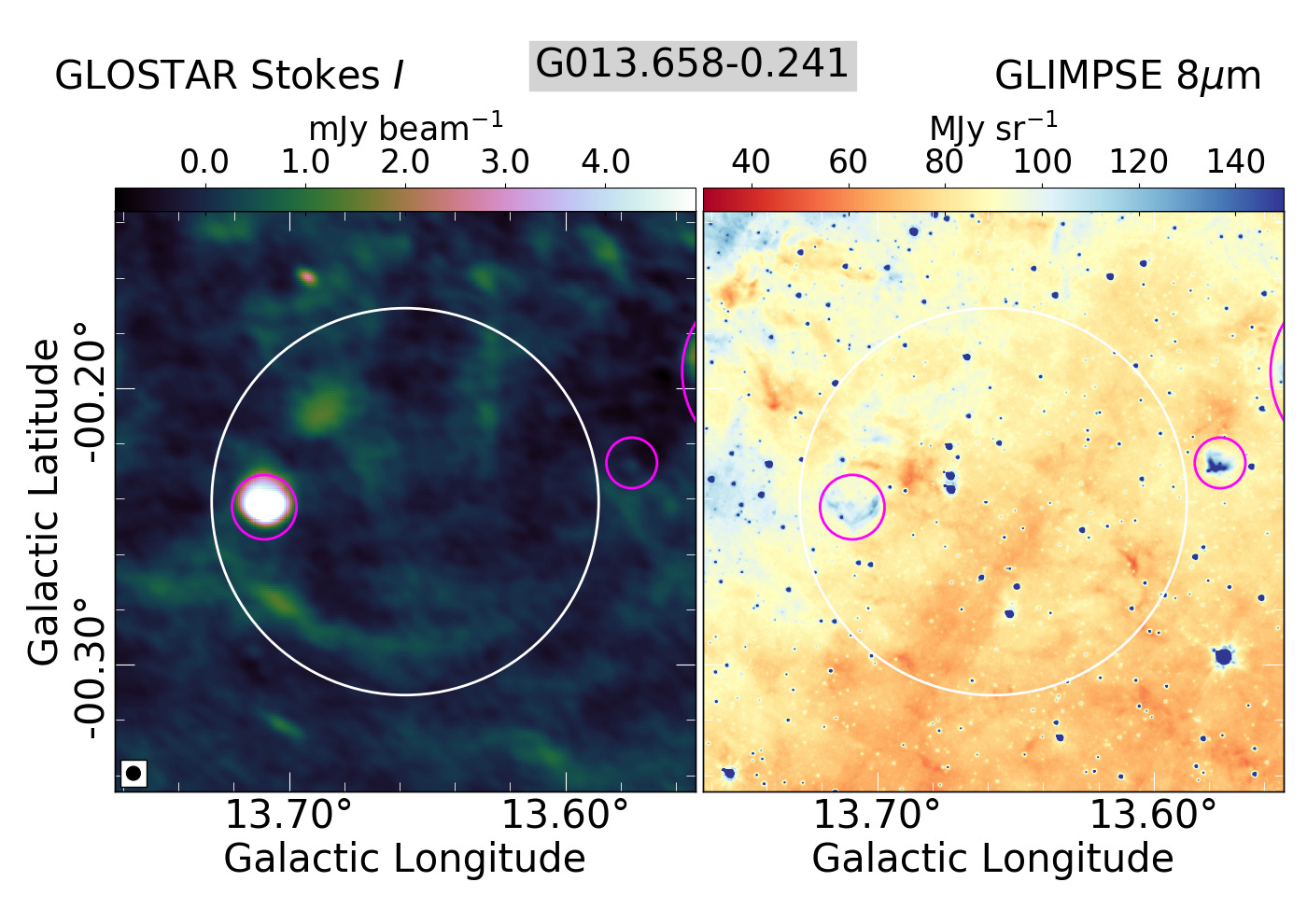}\\
\noindent\includegraphics[width=0.47\textwidth]{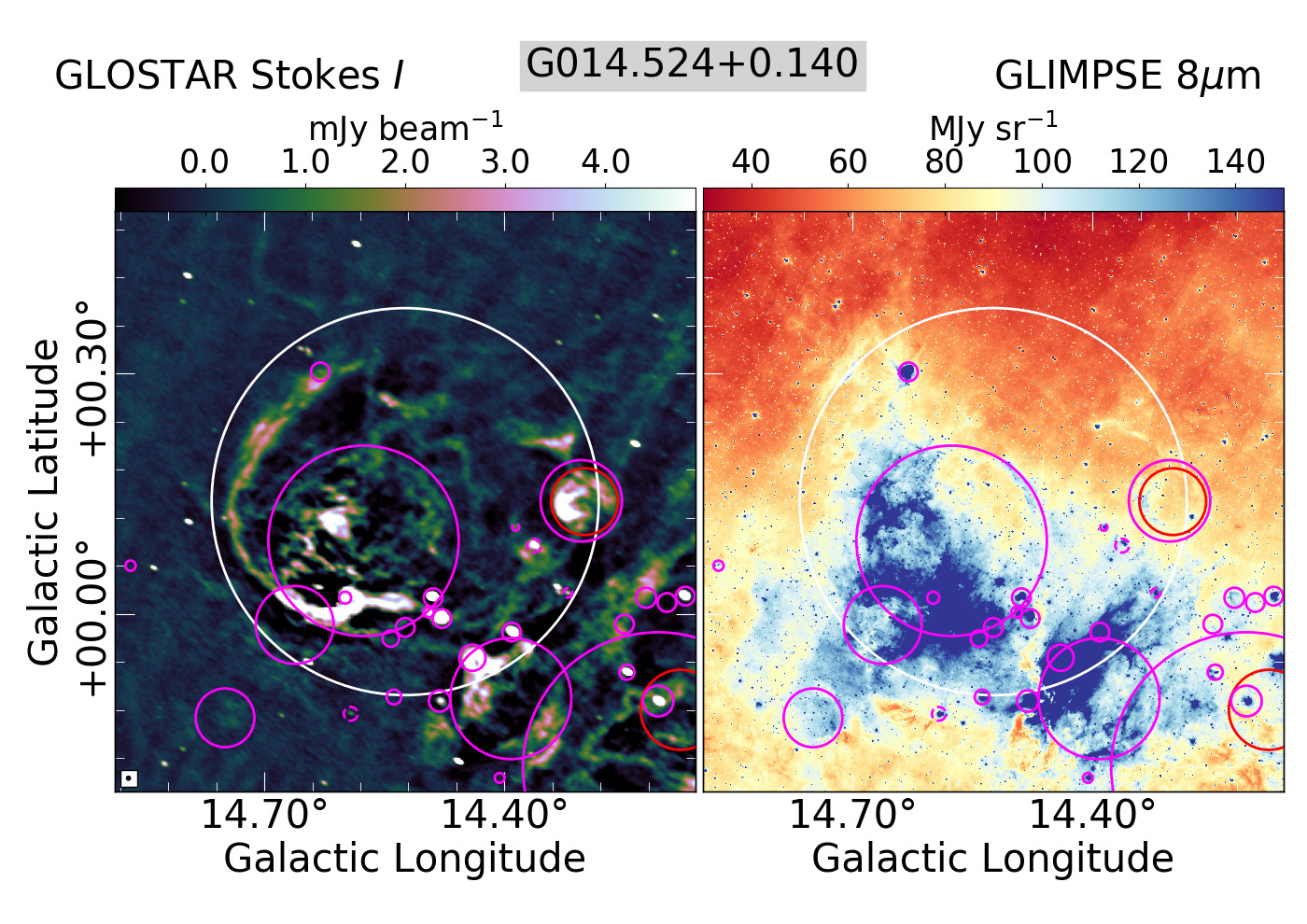}\\
\noindent\includegraphics[width=0.47\textwidth]{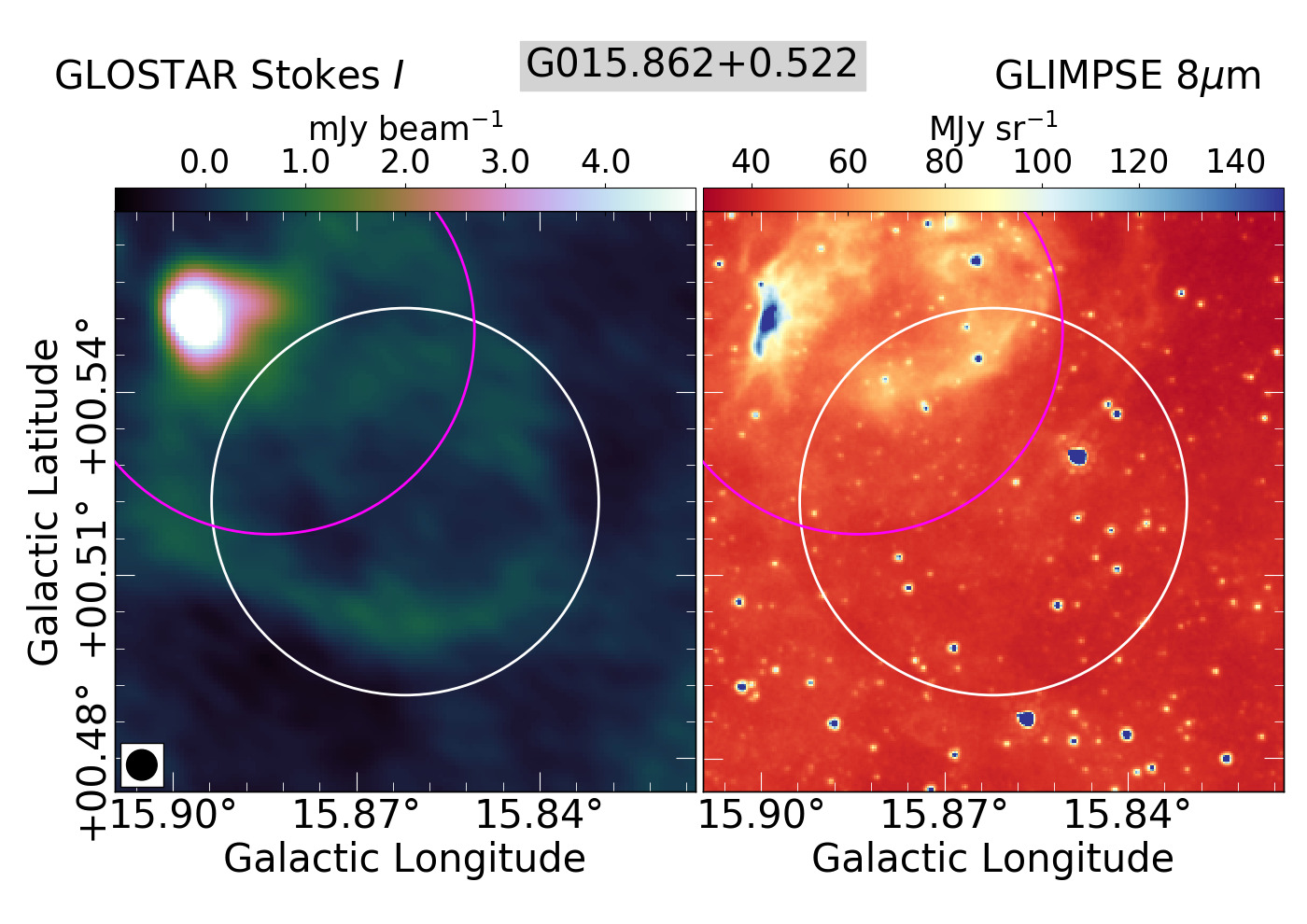}\\
\noindent\includegraphics[width=0.47\textwidth]{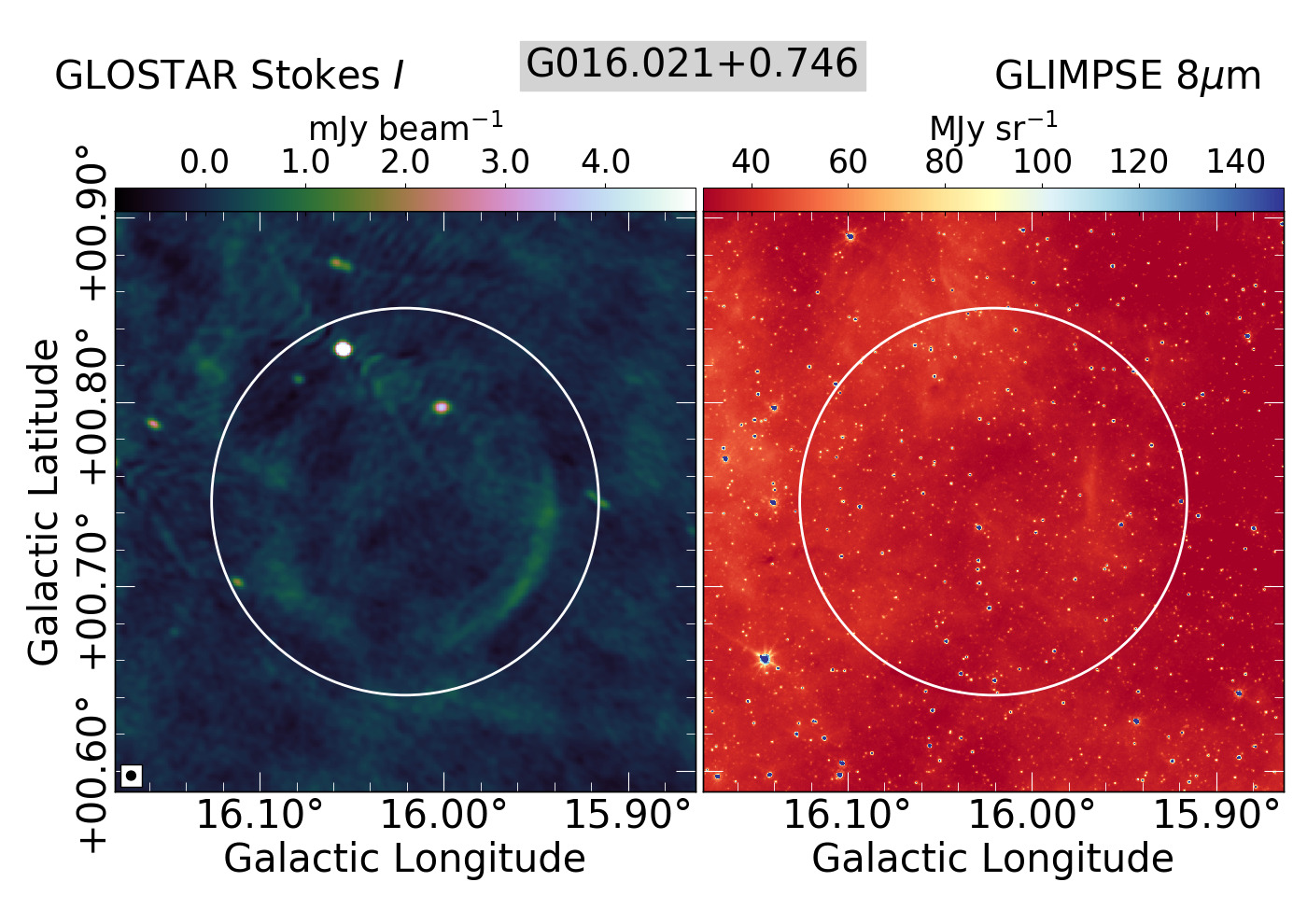}\\
\noindent\includegraphics[width=0.47\textwidth]{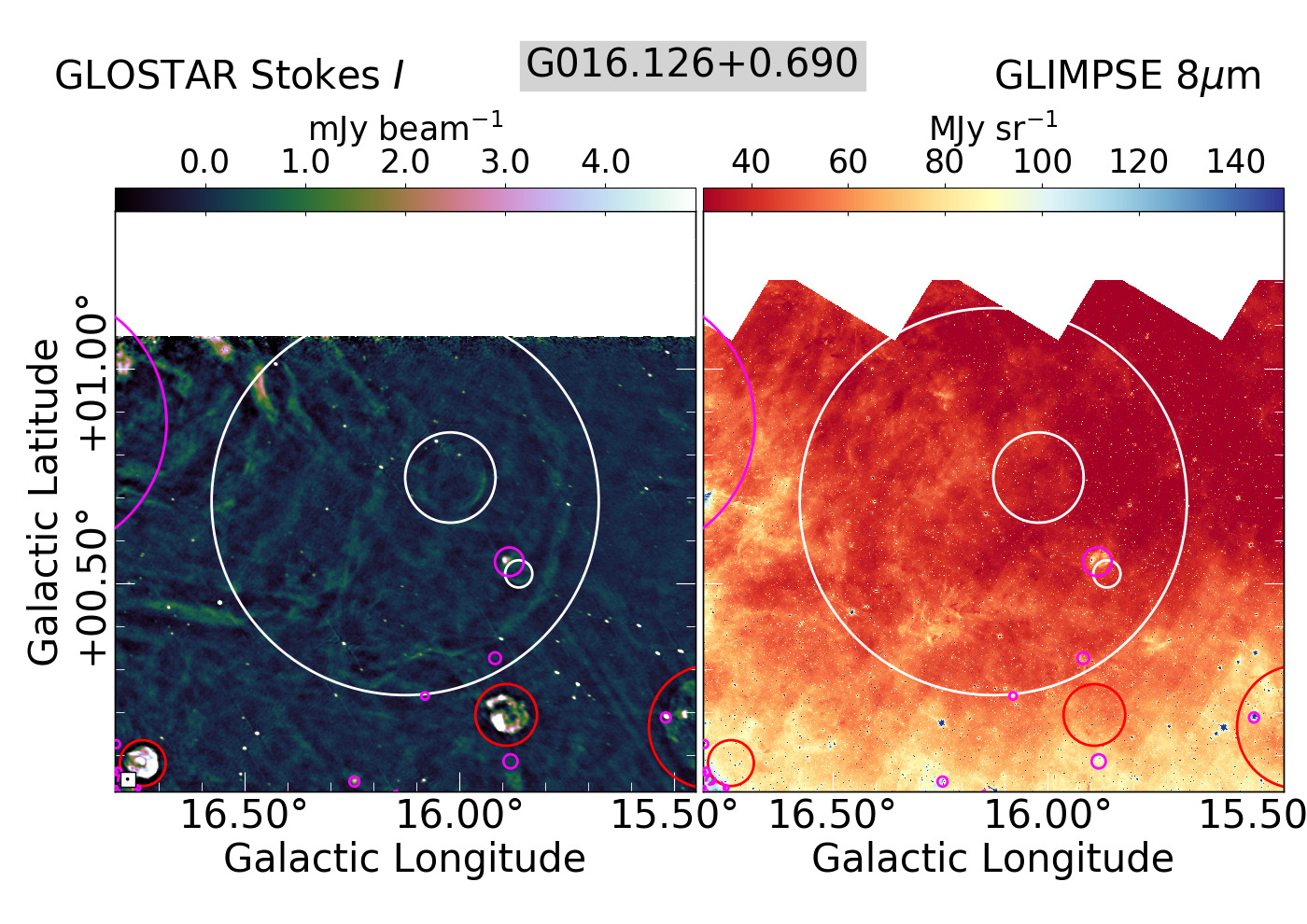}\\
\noindent\includegraphics[width=0.47\textwidth]{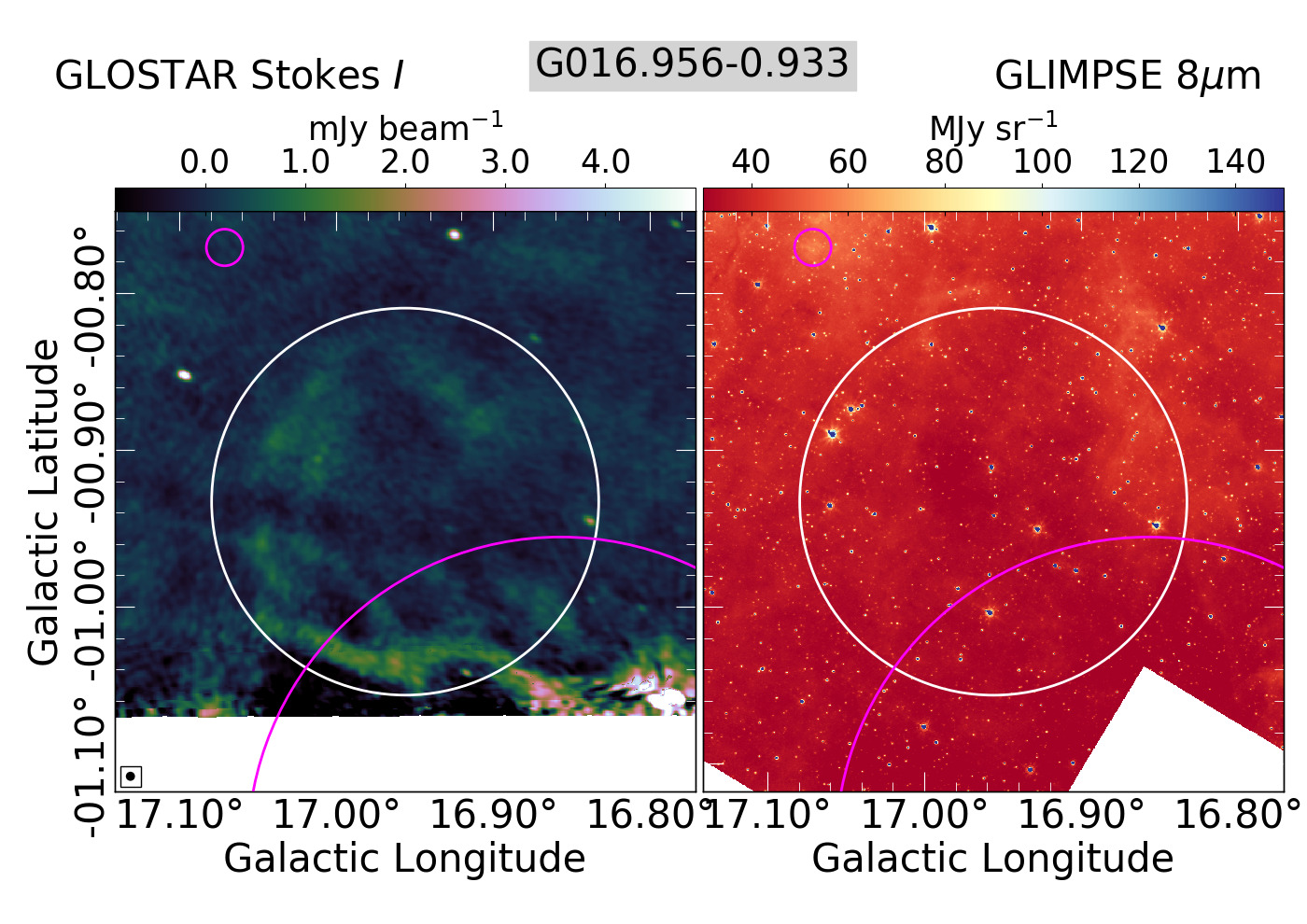}\\
\noindent\includegraphics[width=0.47\textwidth]{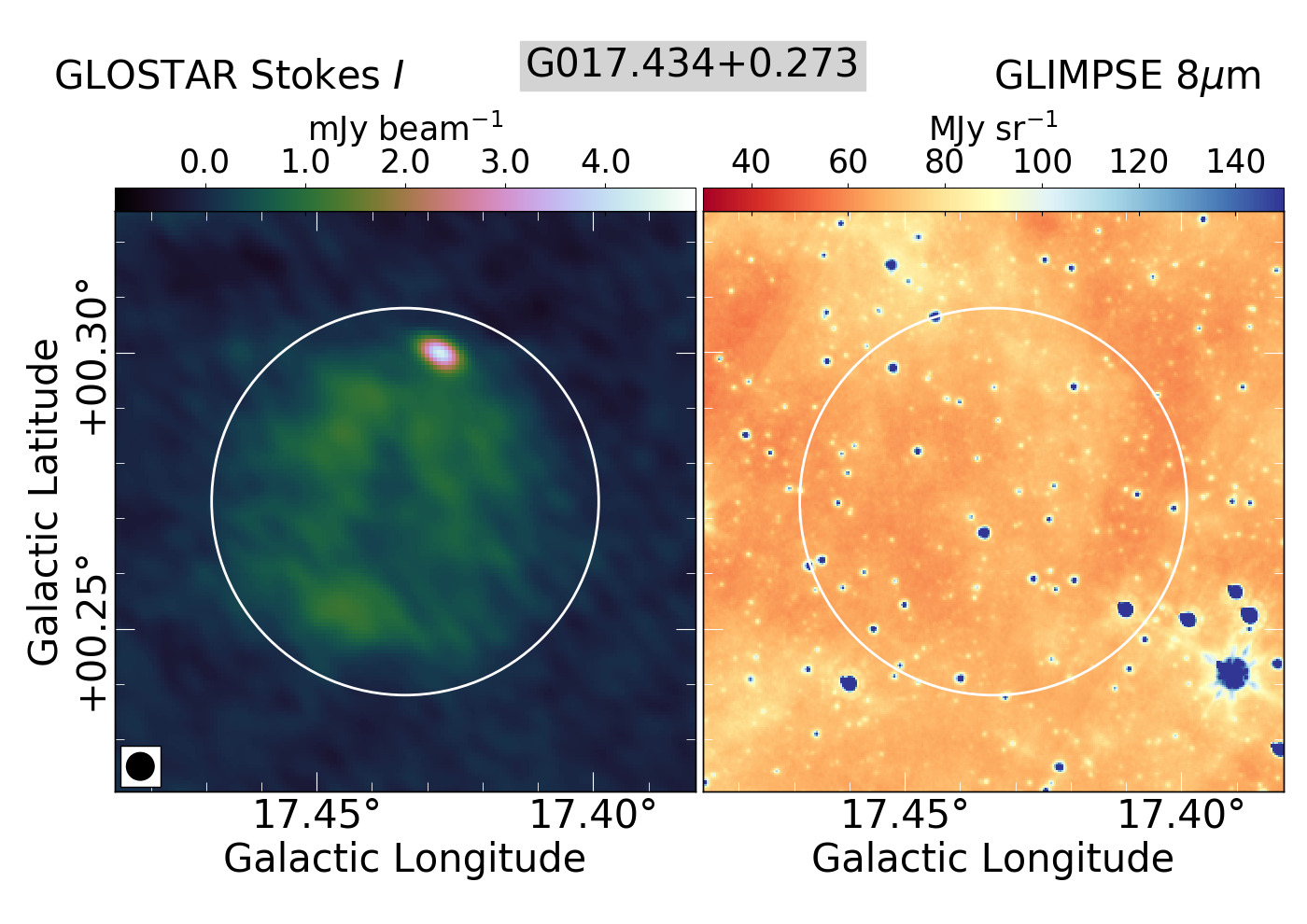}\\
\noindent\includegraphics[width=0.47\textwidth]{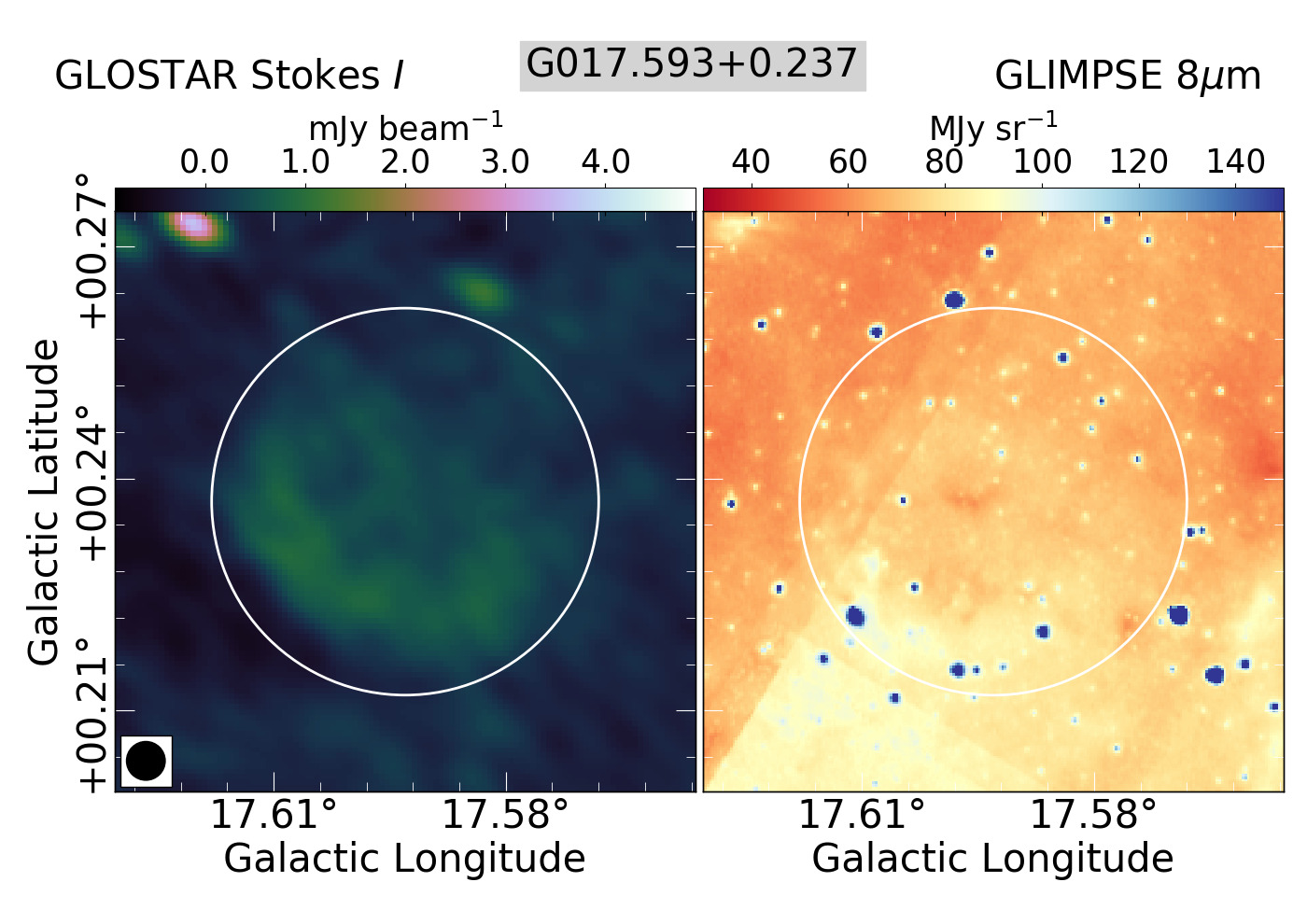}\\
\noindent\includegraphics[width=0.47\textwidth]{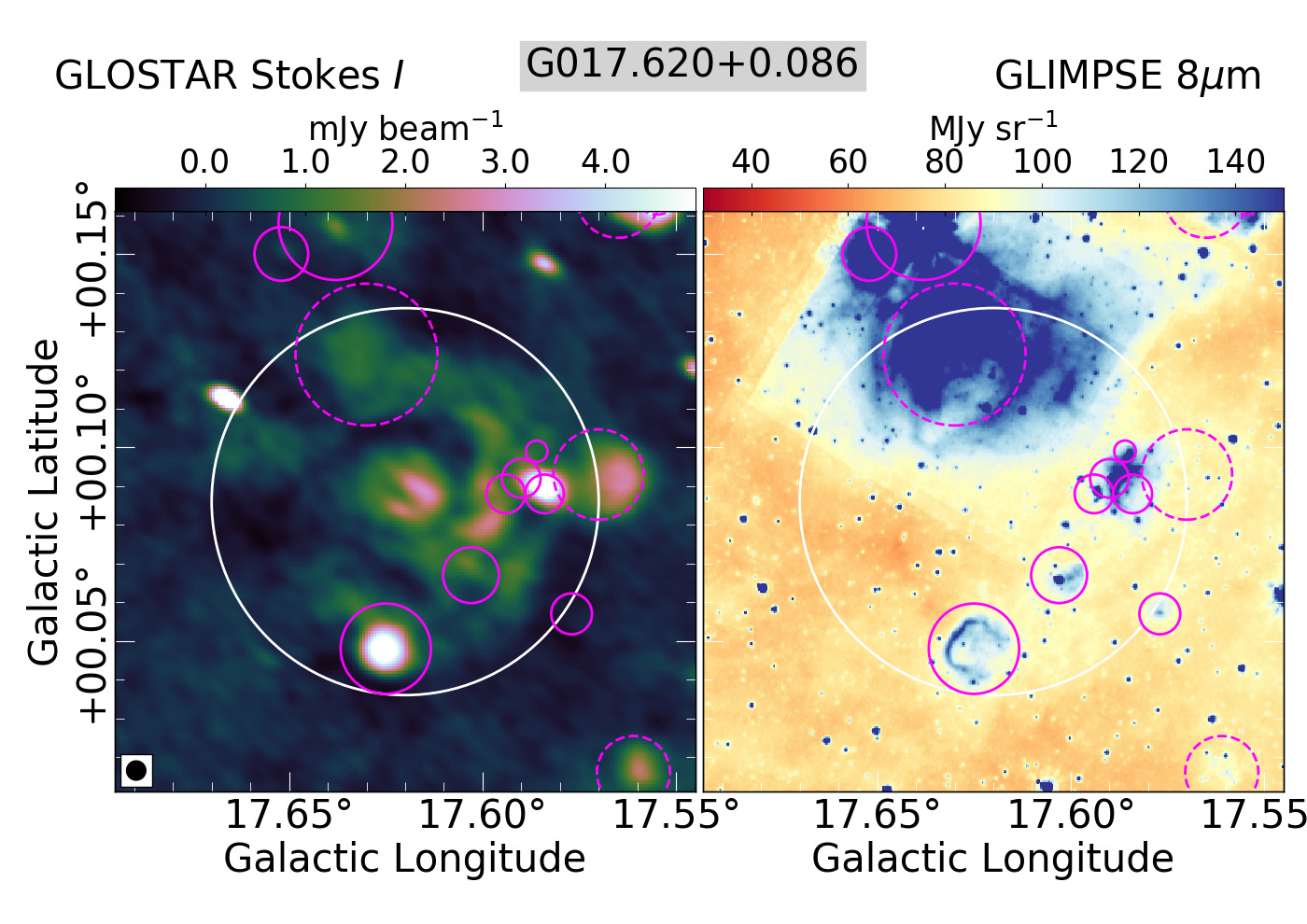}\\
\noindent\includegraphics[width=0.47\textwidth]{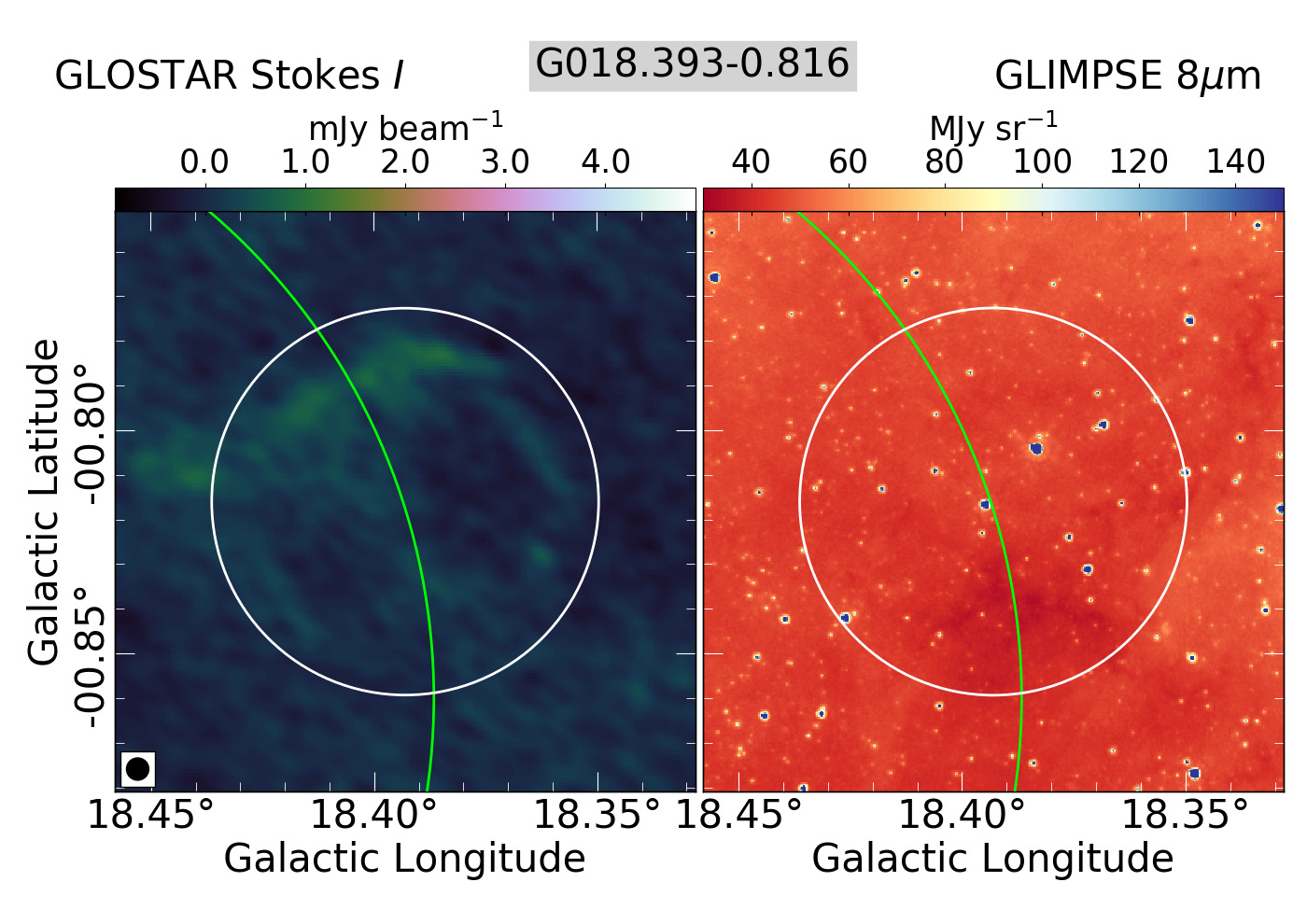}\\
\noindent\includegraphics[width=0.47\textwidth]{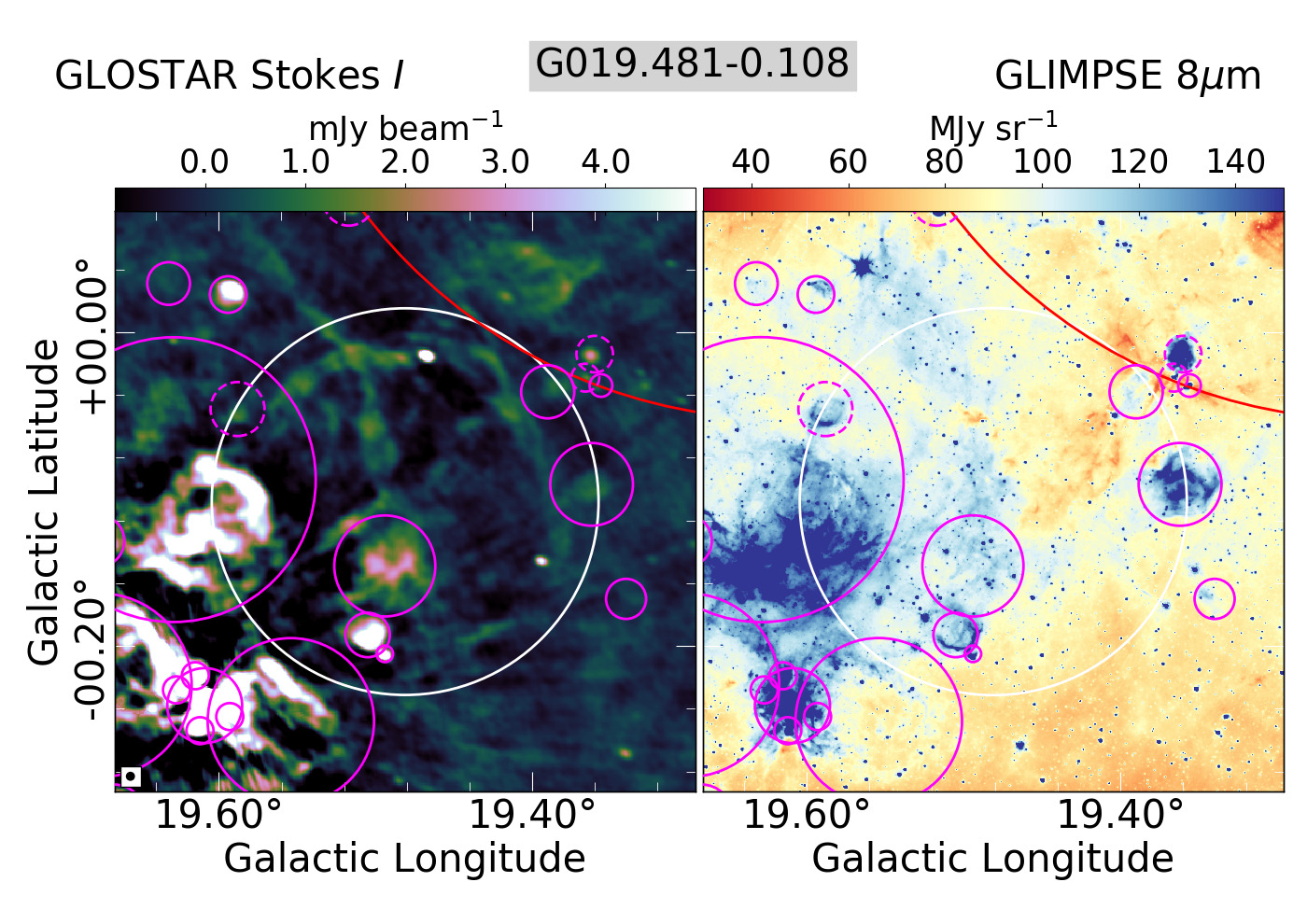}\\
\noindent\includegraphics[width=0.47\textwidth]{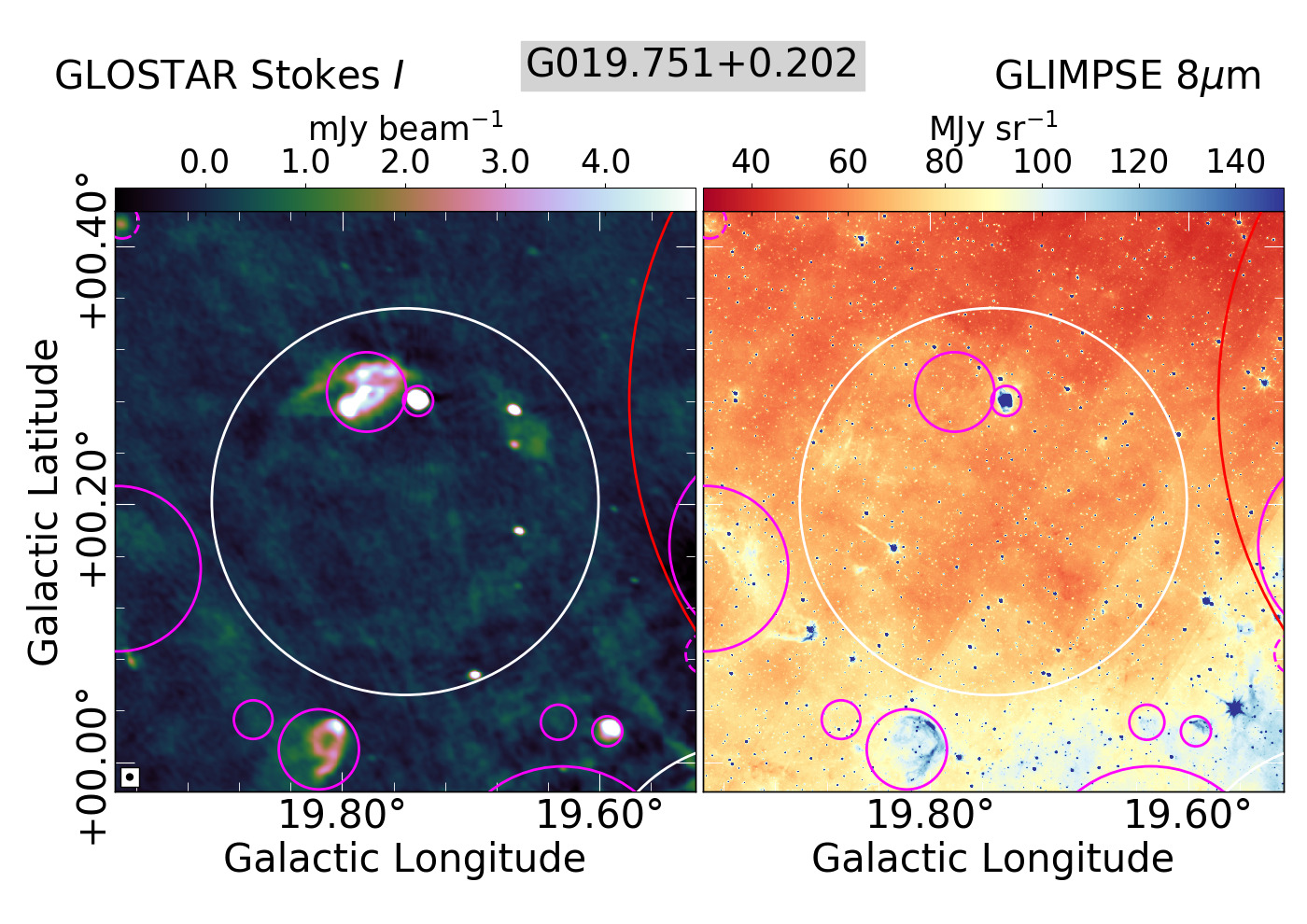}\\
\noindent\includegraphics[width=0.47\textwidth]{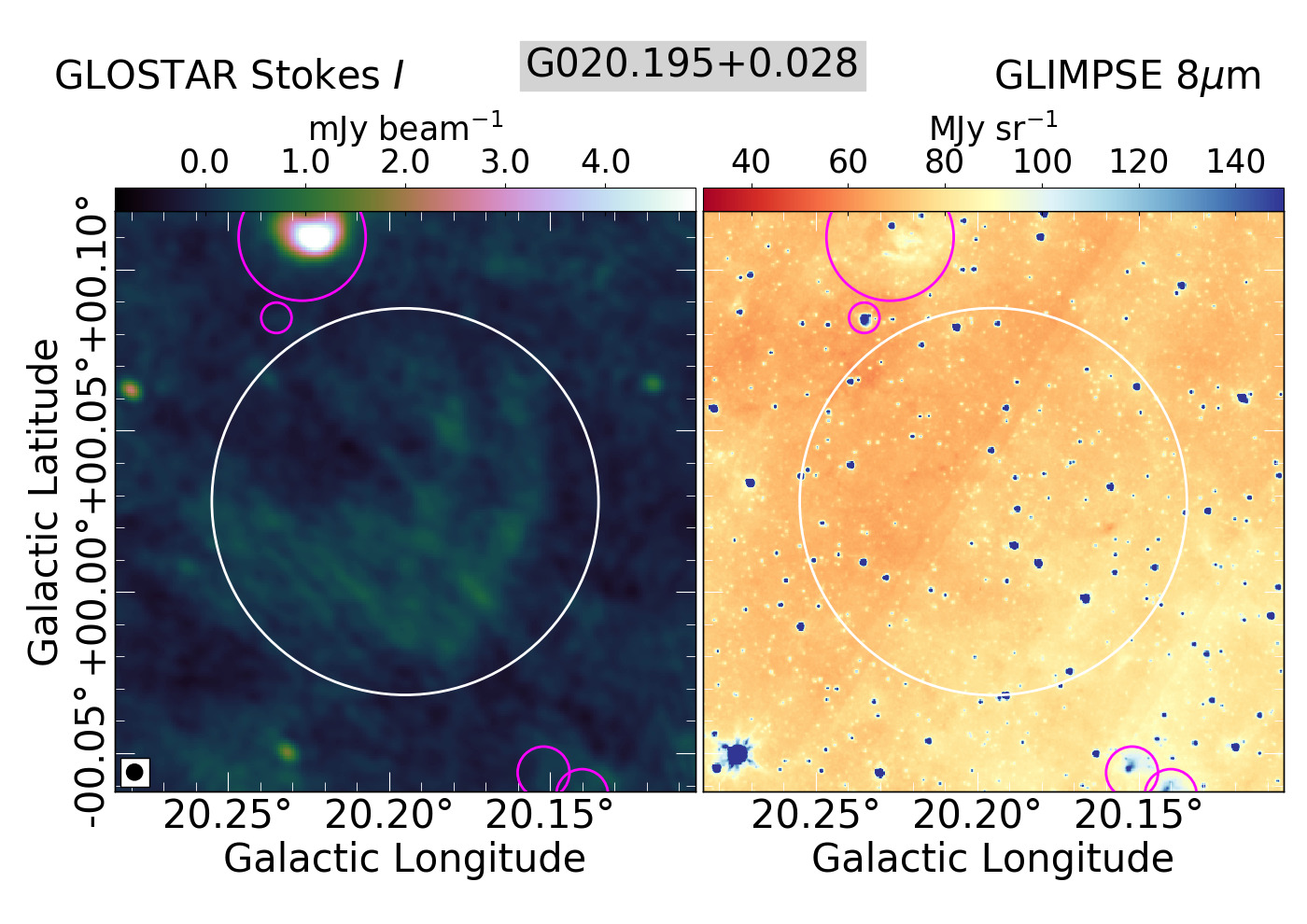}\\
\noindent\includegraphics[width=0.47\textwidth]{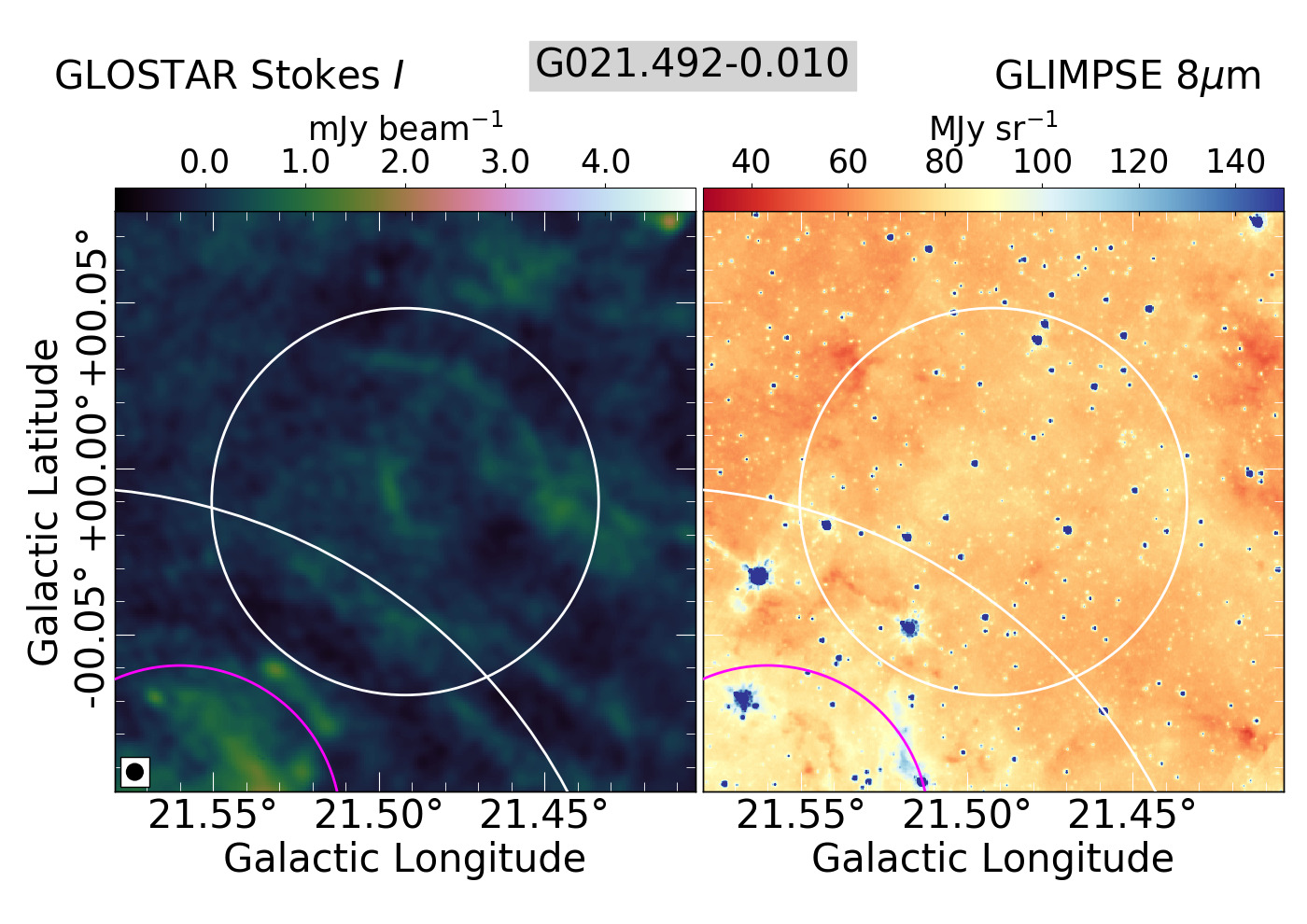}\\
\noindent\includegraphics[width=0.47\textwidth]{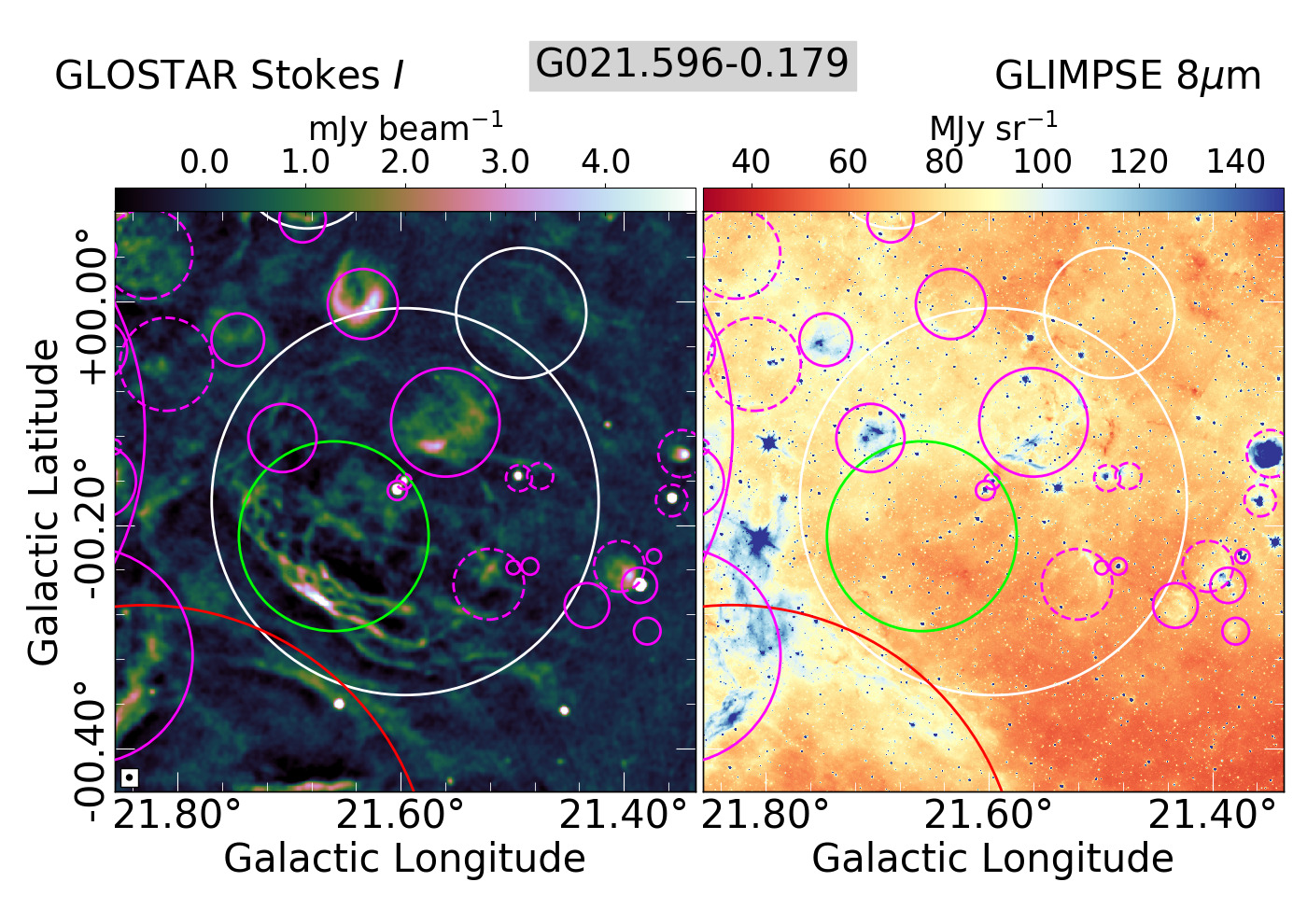}\\
\noindent\includegraphics[width=0.47\textwidth]{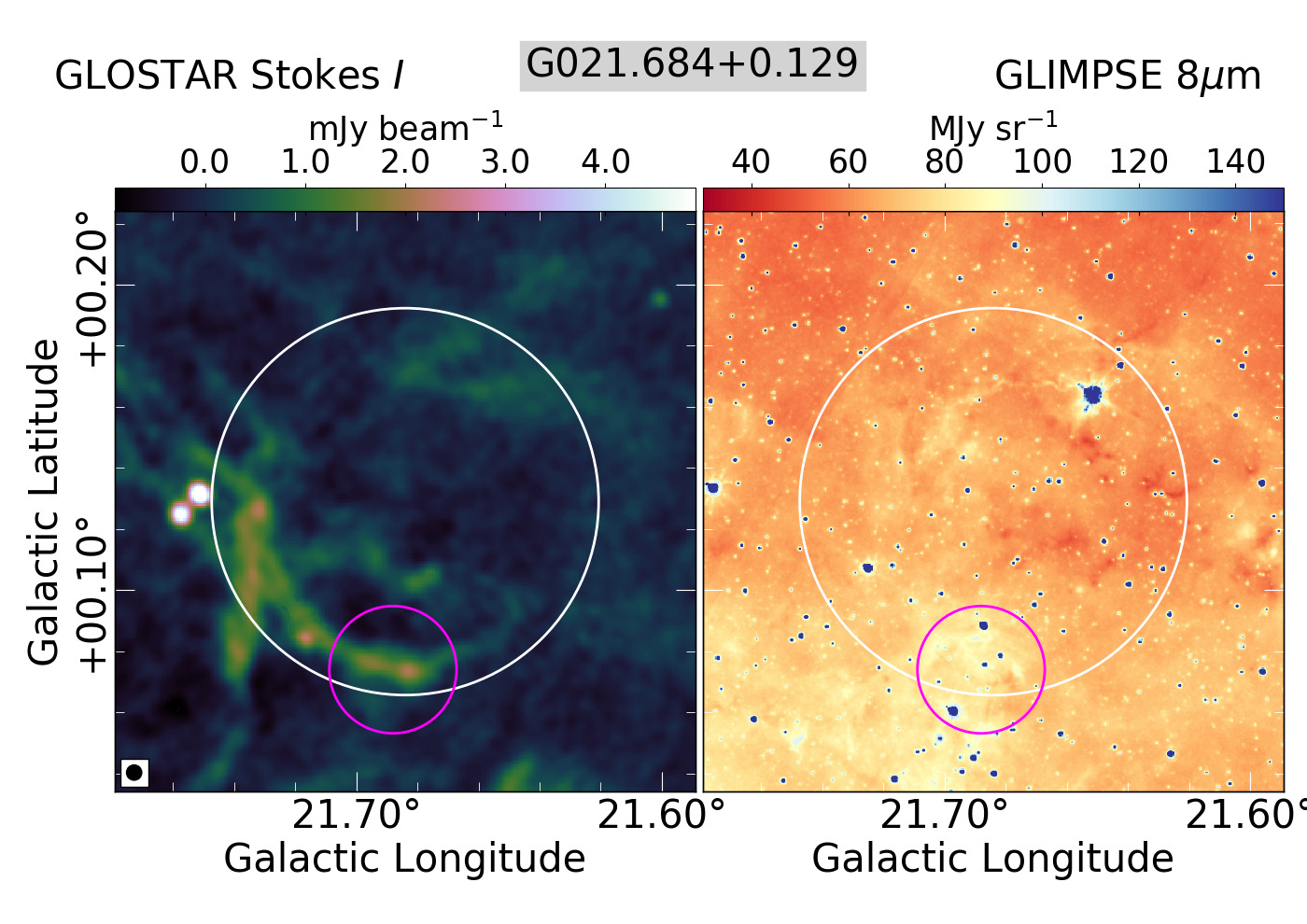}\\
\noindent\includegraphics[width=0.47\textwidth]{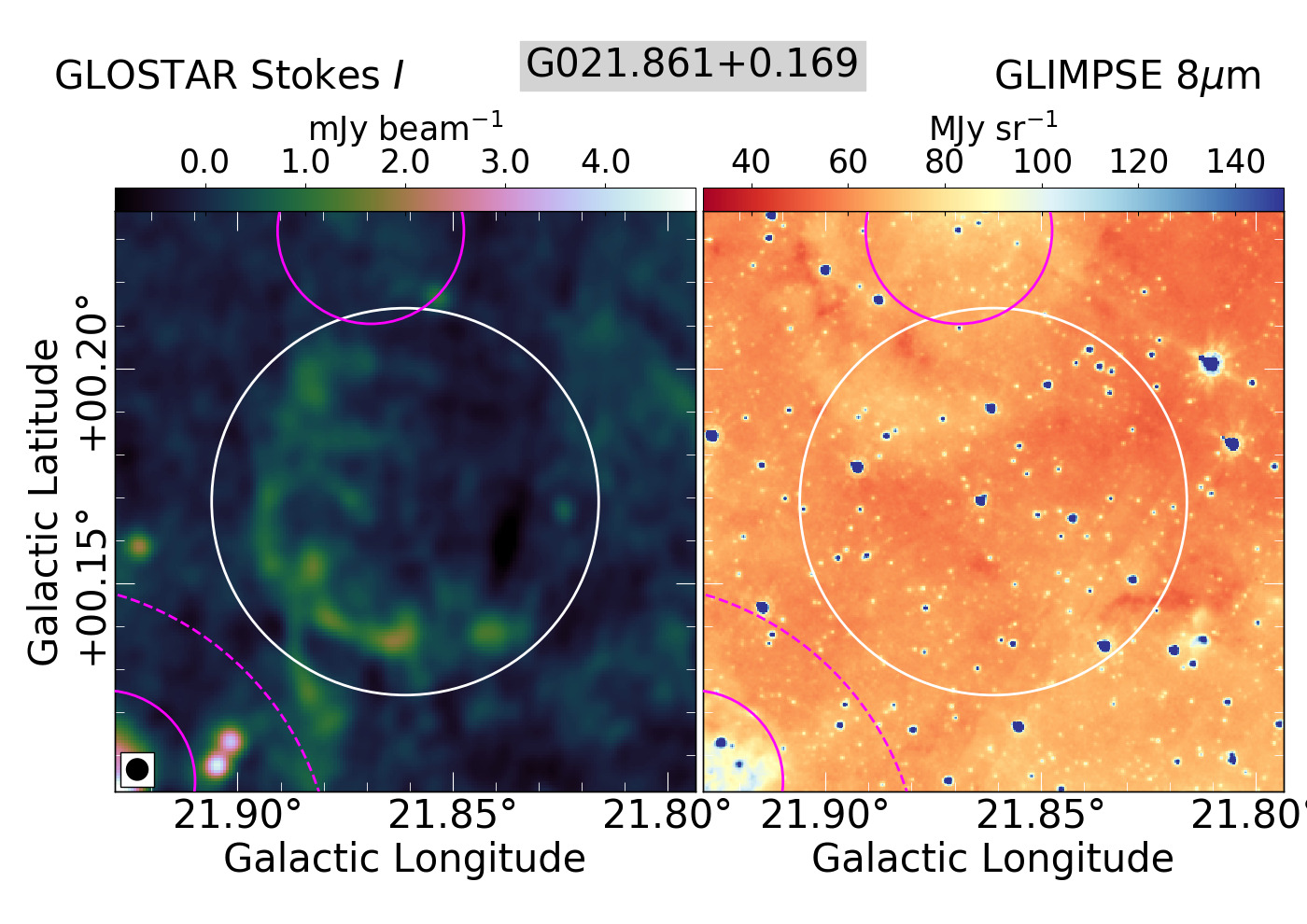}\\
\noindent\includegraphics[width=0.47\textwidth]{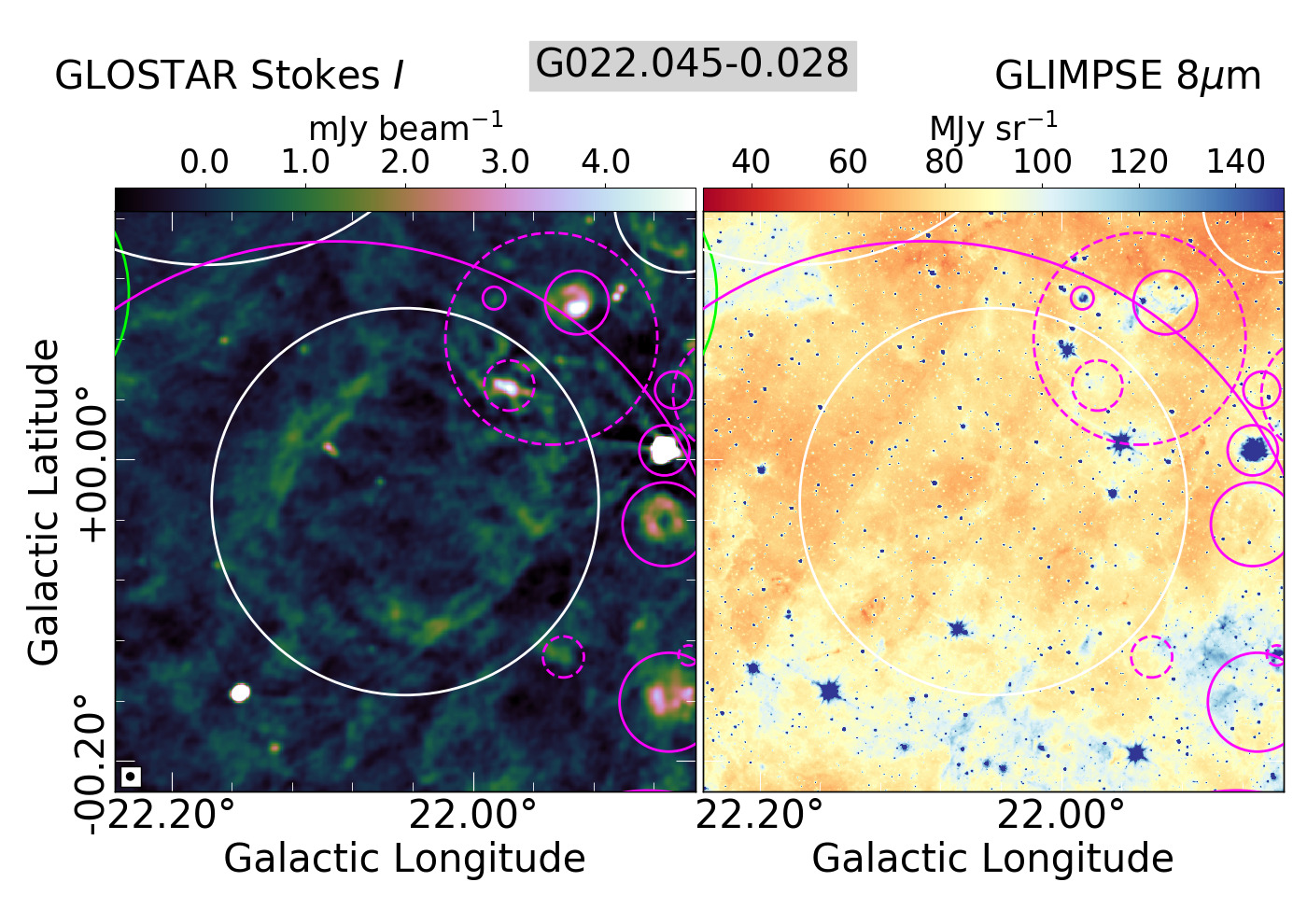}\\
\noindent\includegraphics[width=0.47\textwidth]{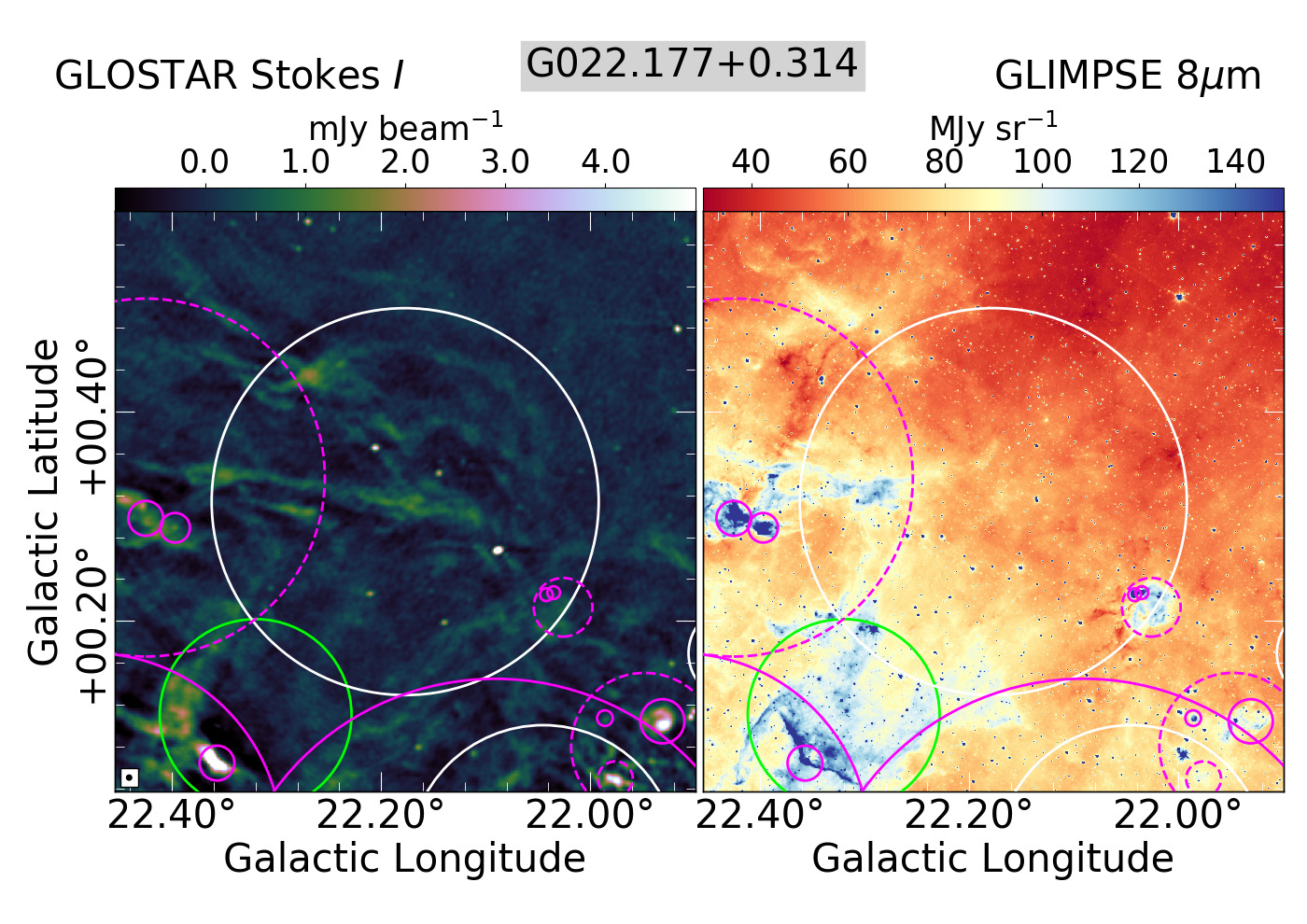}\\
\noindent\includegraphics[width=0.47\textwidth]{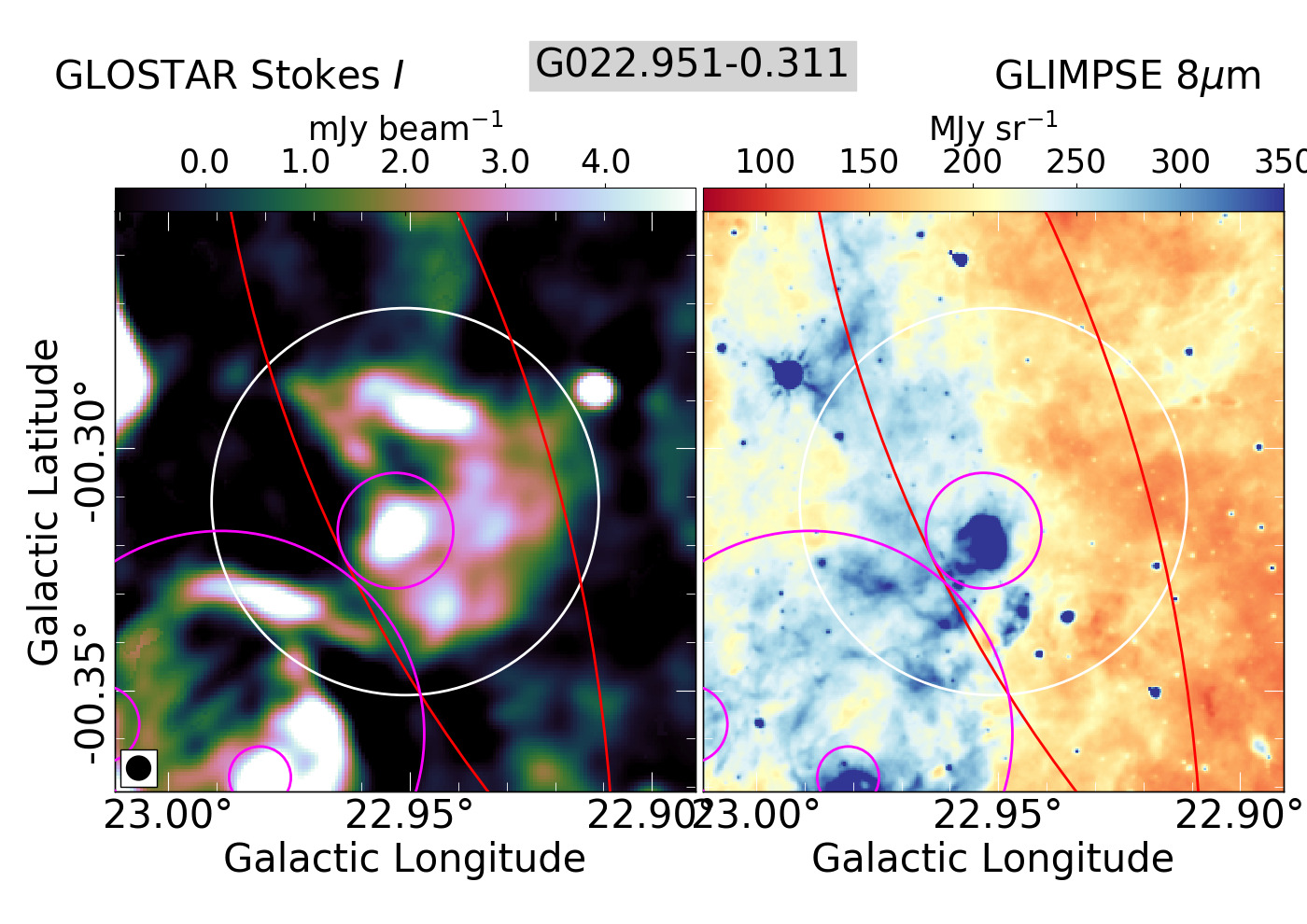}\\
\noindent\includegraphics[width=0.47\textwidth]{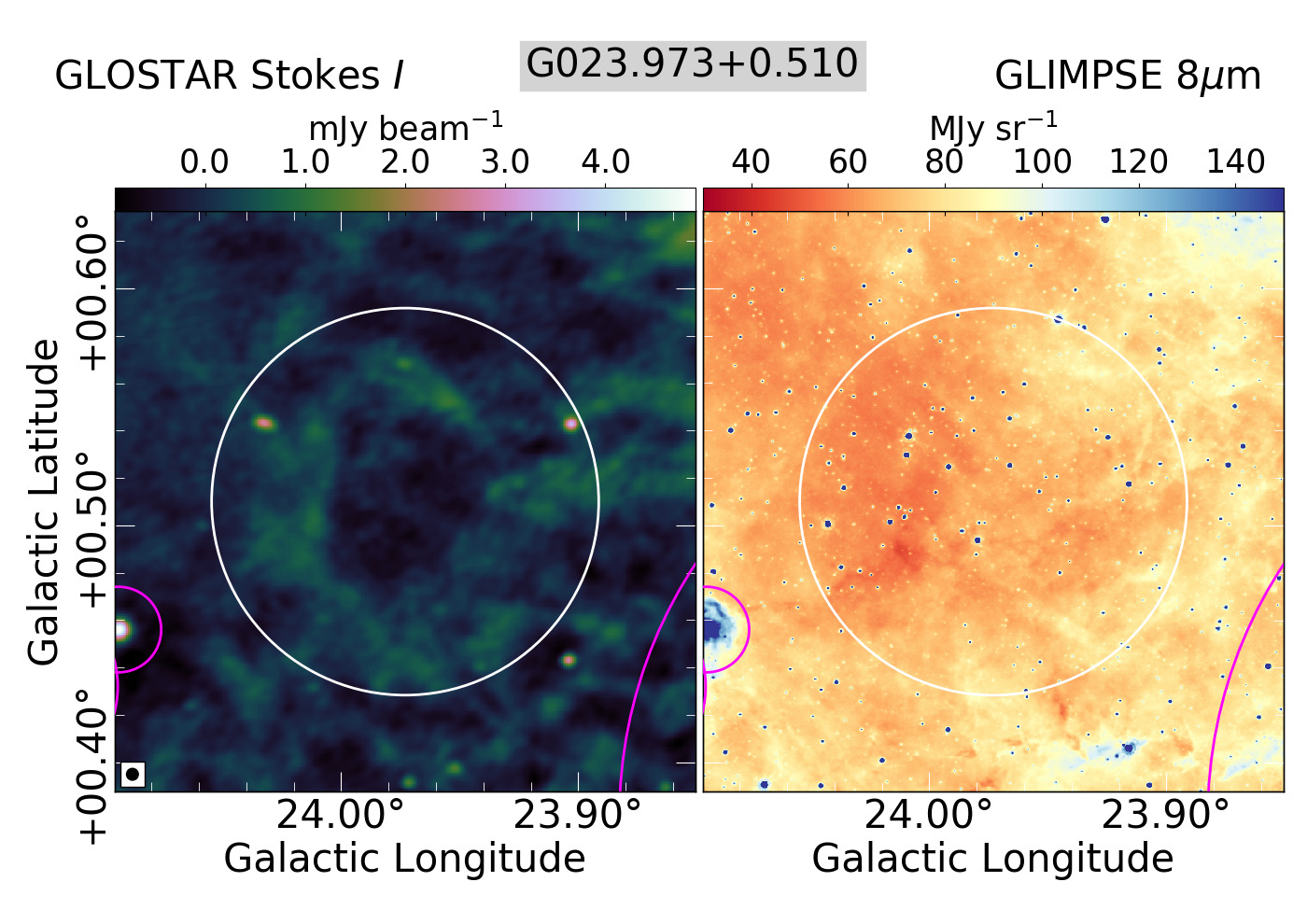}\\
\noindent\includegraphics[width=0.47\textwidth]{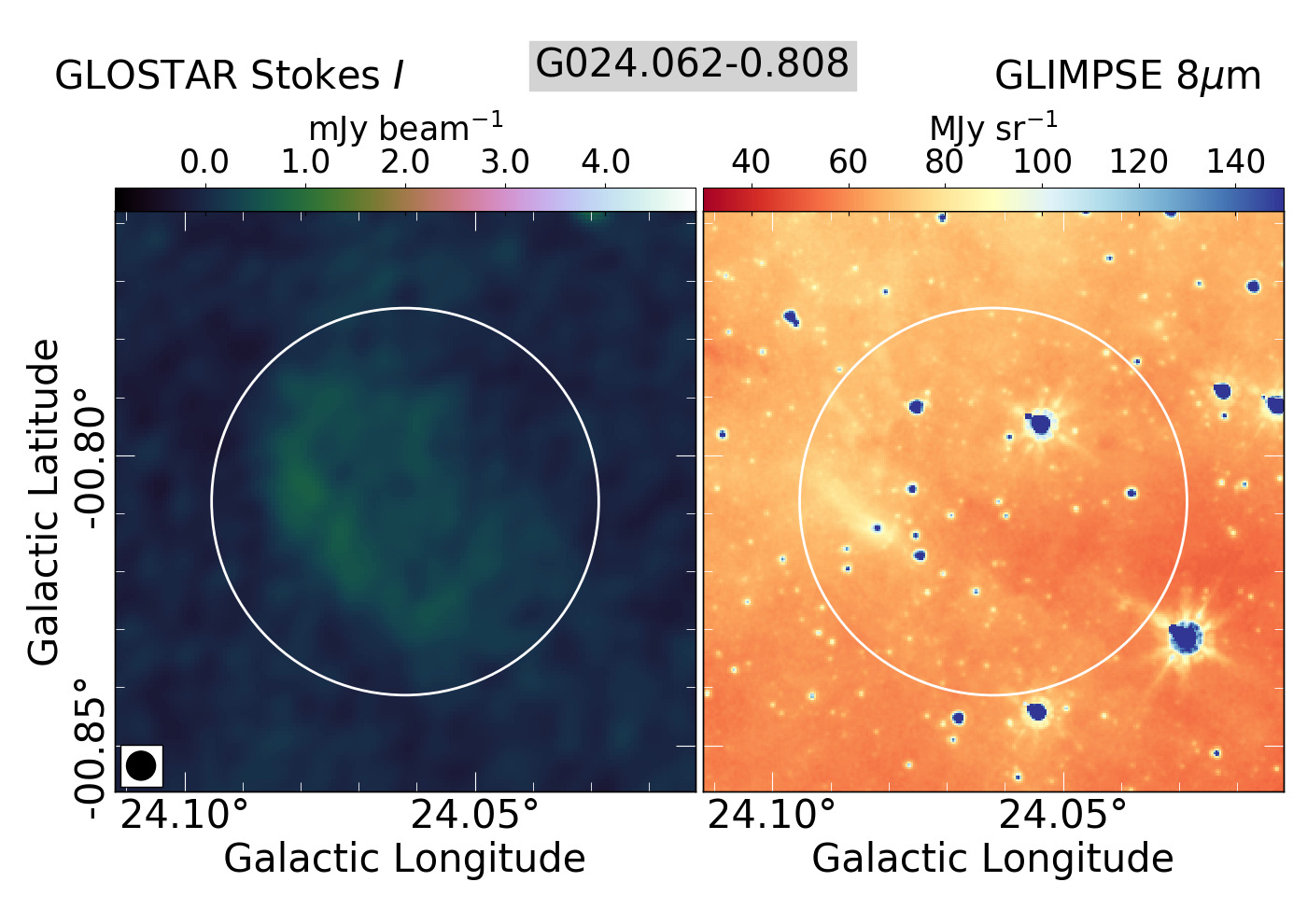}\\
\noindent\includegraphics[width=0.47\textwidth]{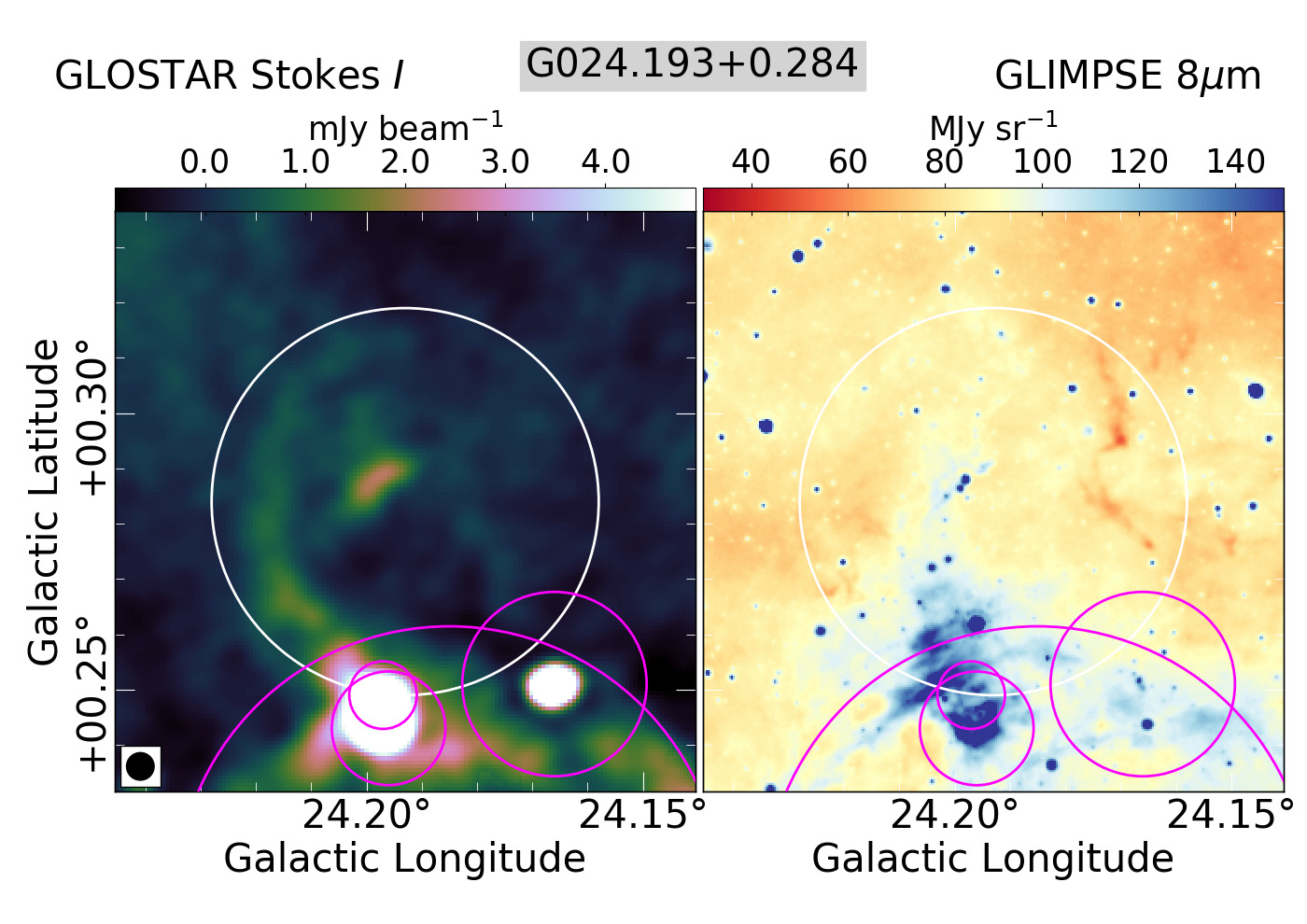}\\
\noindent\includegraphics[width=0.47\textwidth]{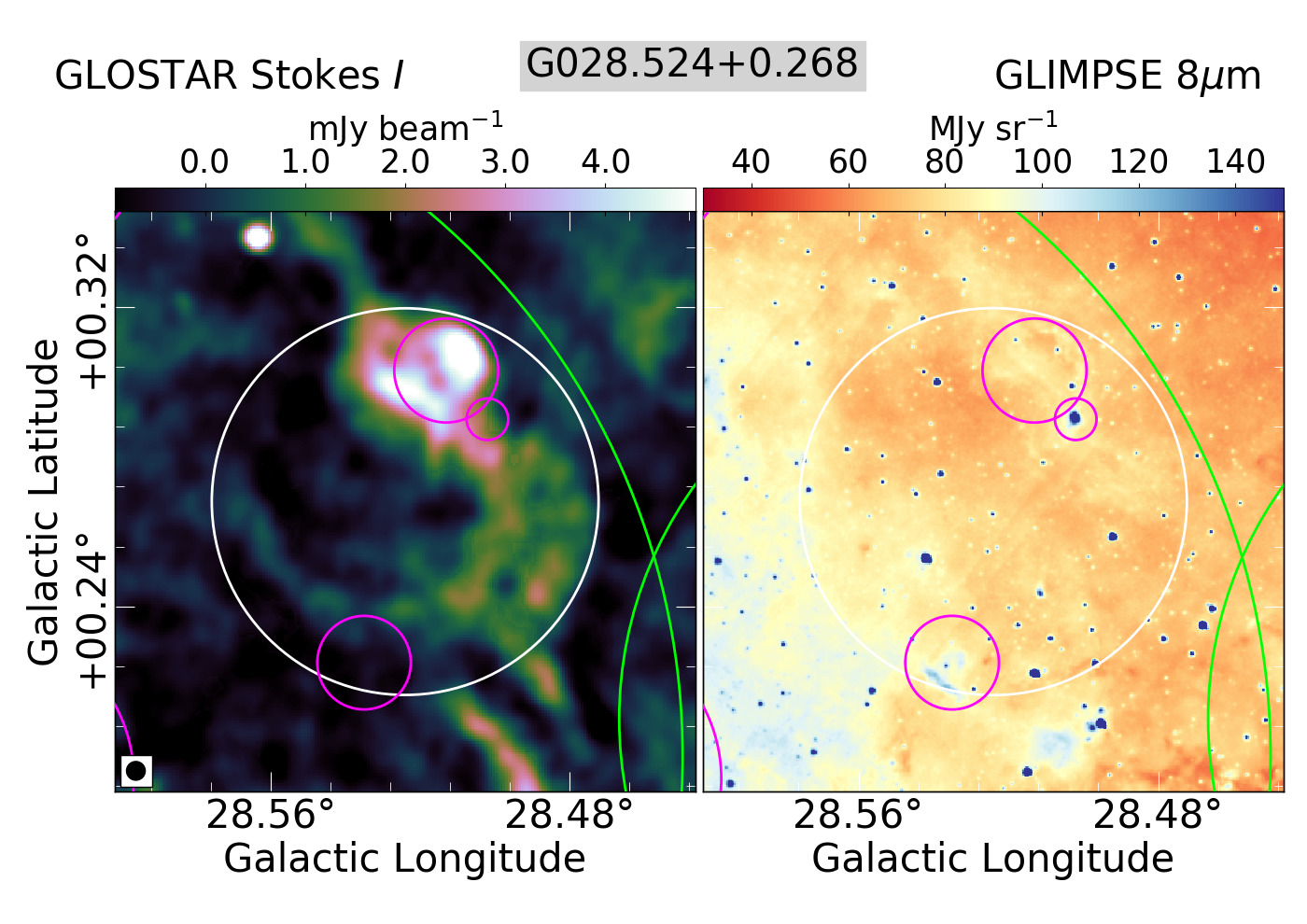}\\
\noindent\includegraphics[width=0.47\textwidth]{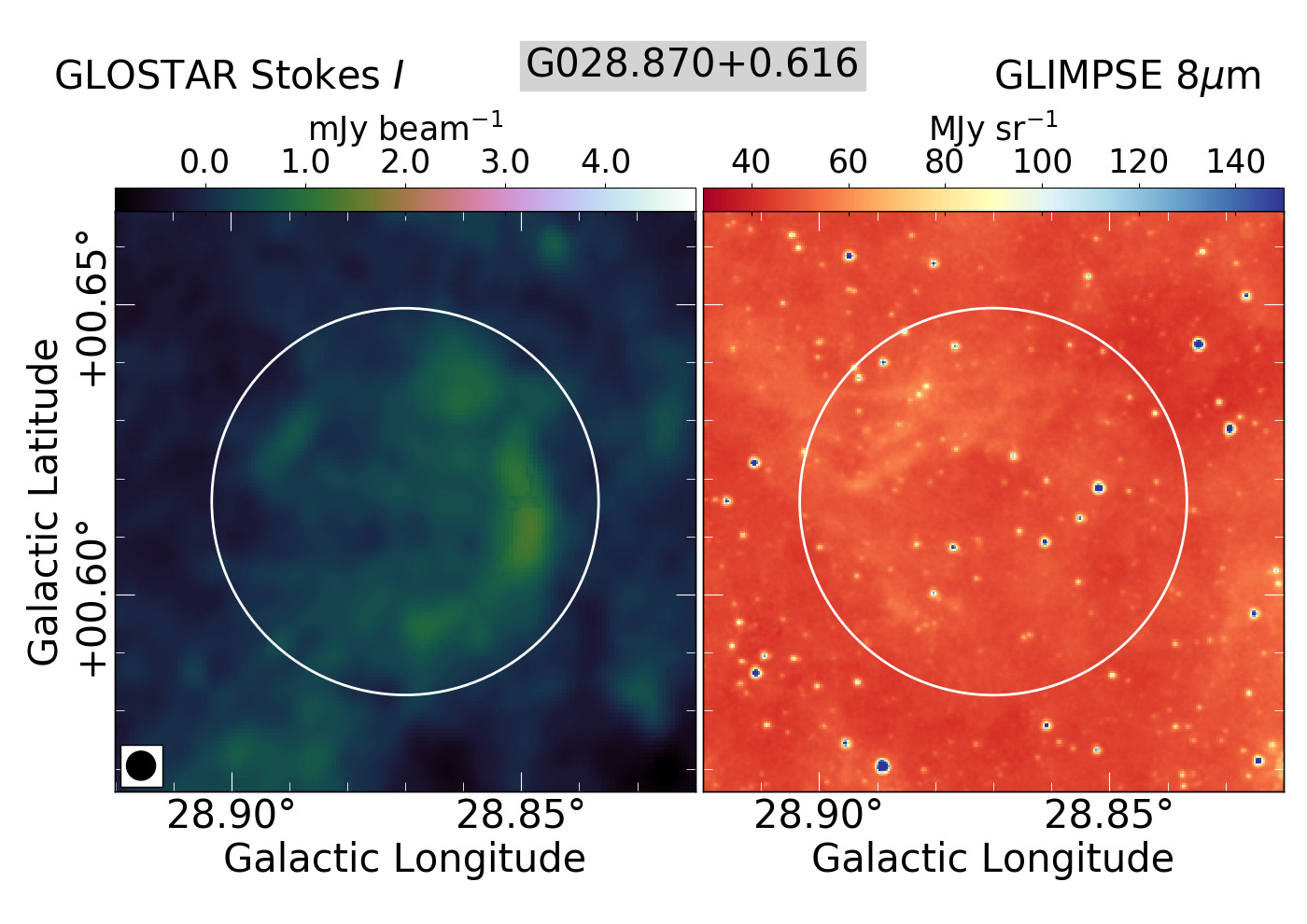}\\
\noindent\includegraphics[width=0.47\textwidth]{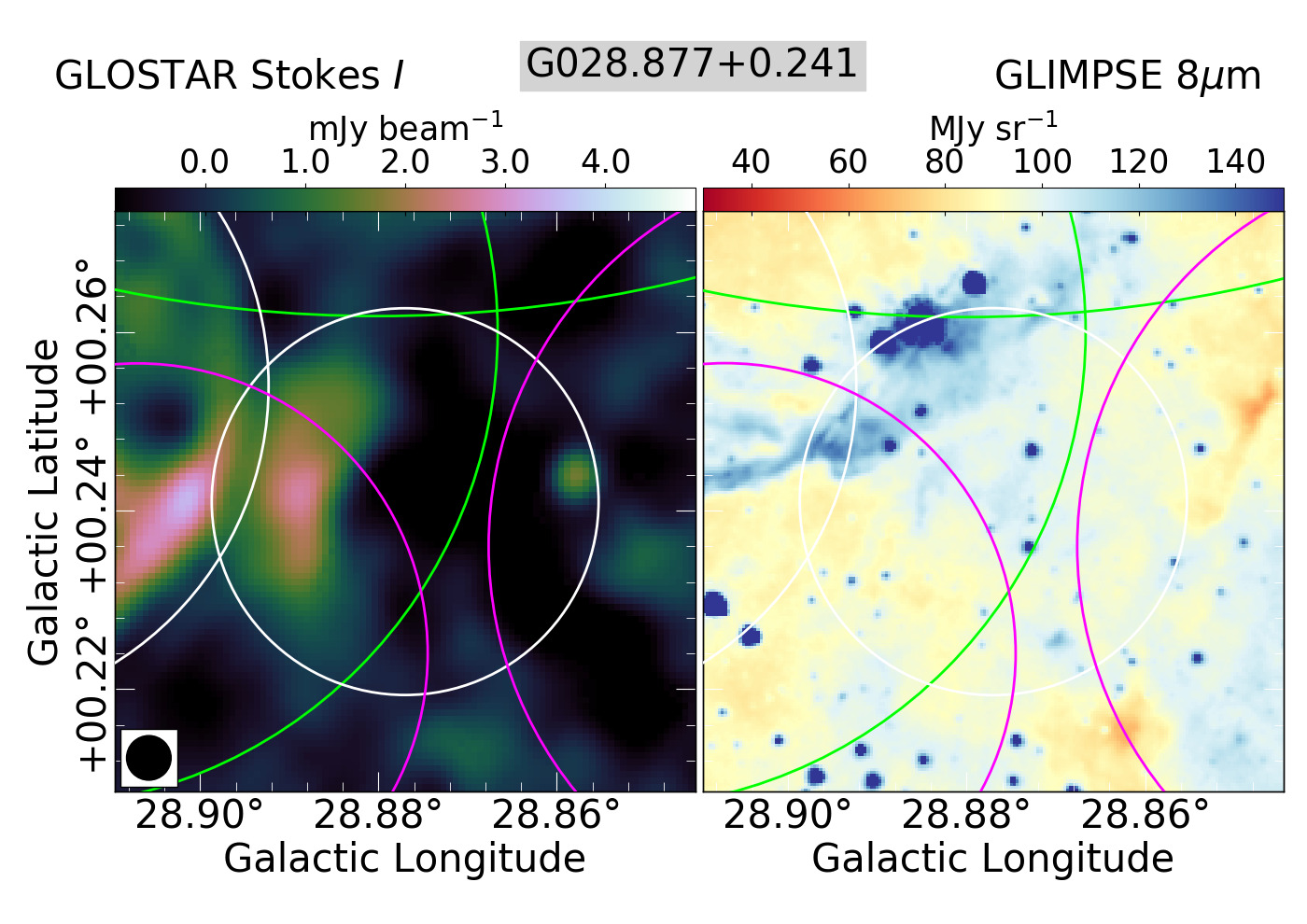}\\
\noindent\includegraphics[width=0.47\textwidth]{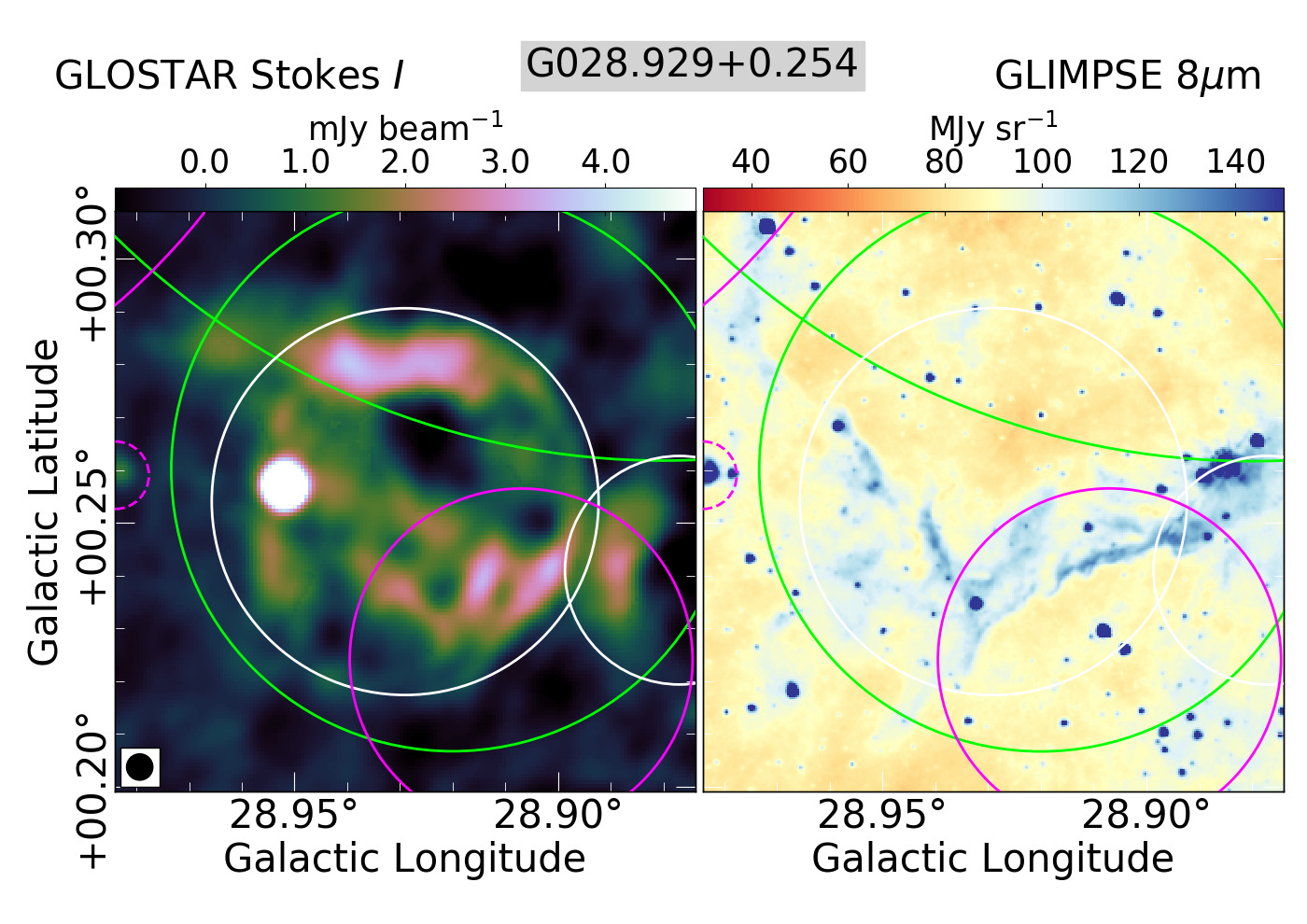}\\
\noindent\includegraphics[width=0.47\textwidth]{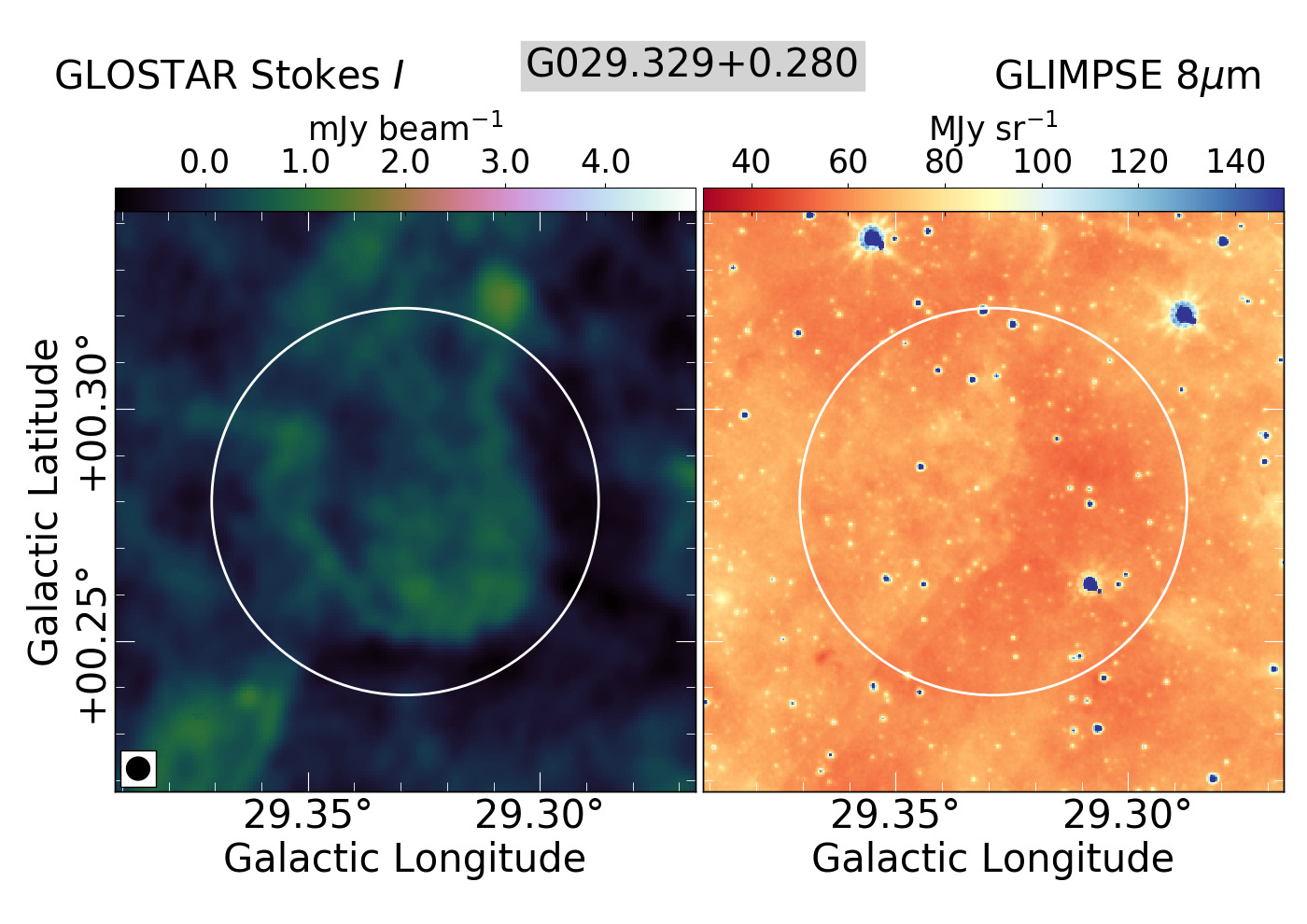}\\
\noindent\includegraphics[width=0.47\textwidth]{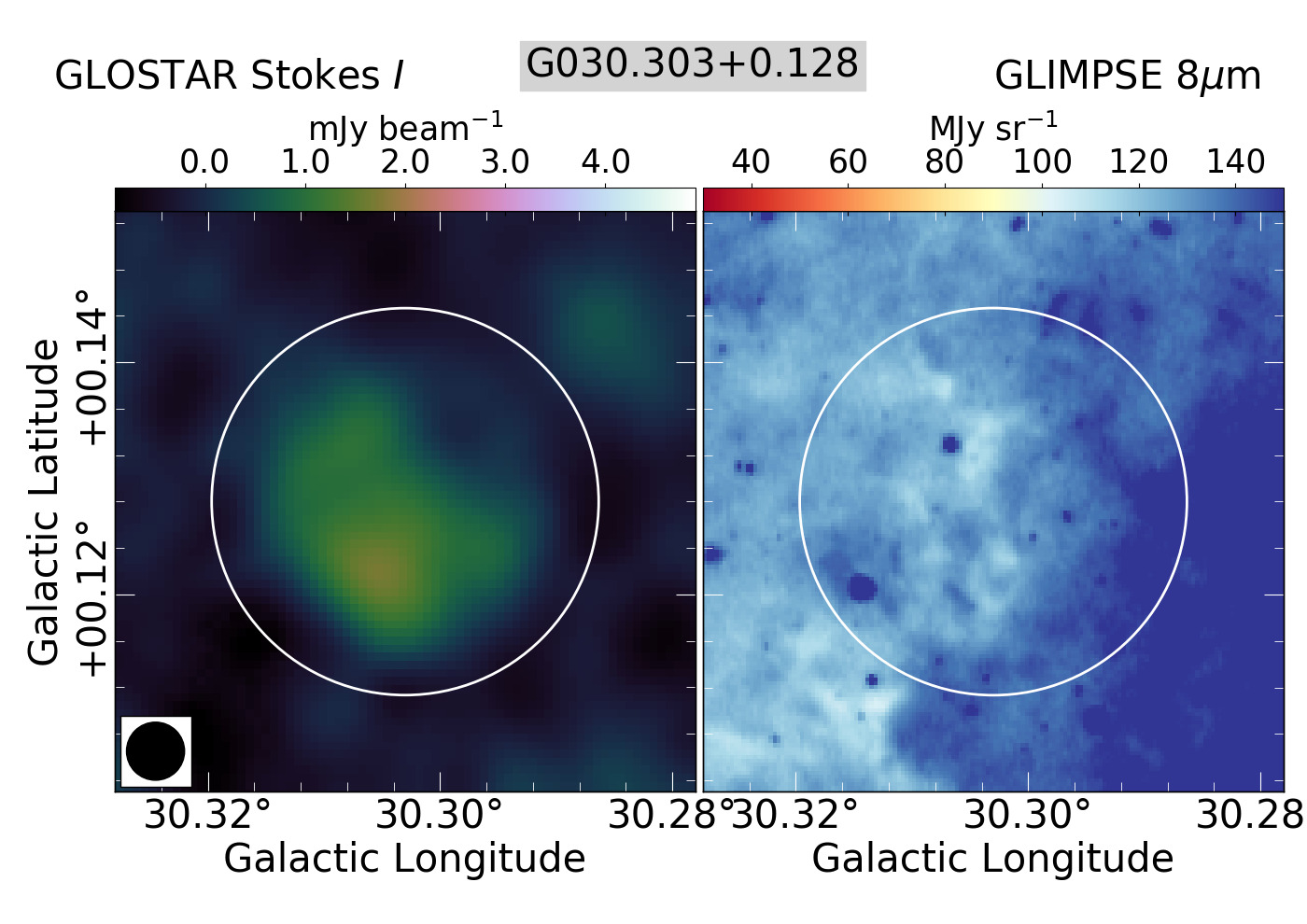}\\
\noindent\includegraphics[width=0.47\textwidth]{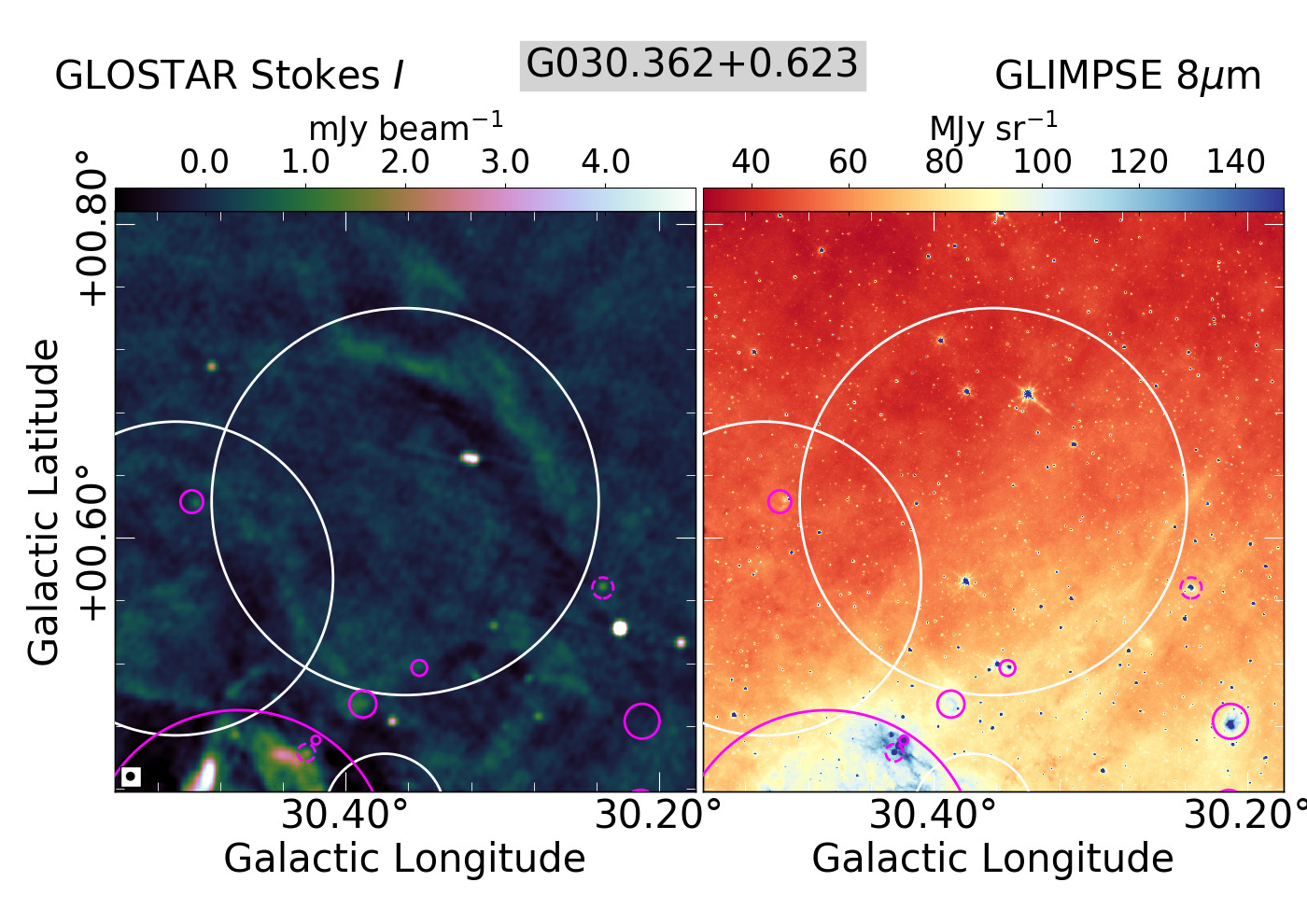}\\
\noindent\includegraphics[width=0.47\textwidth]{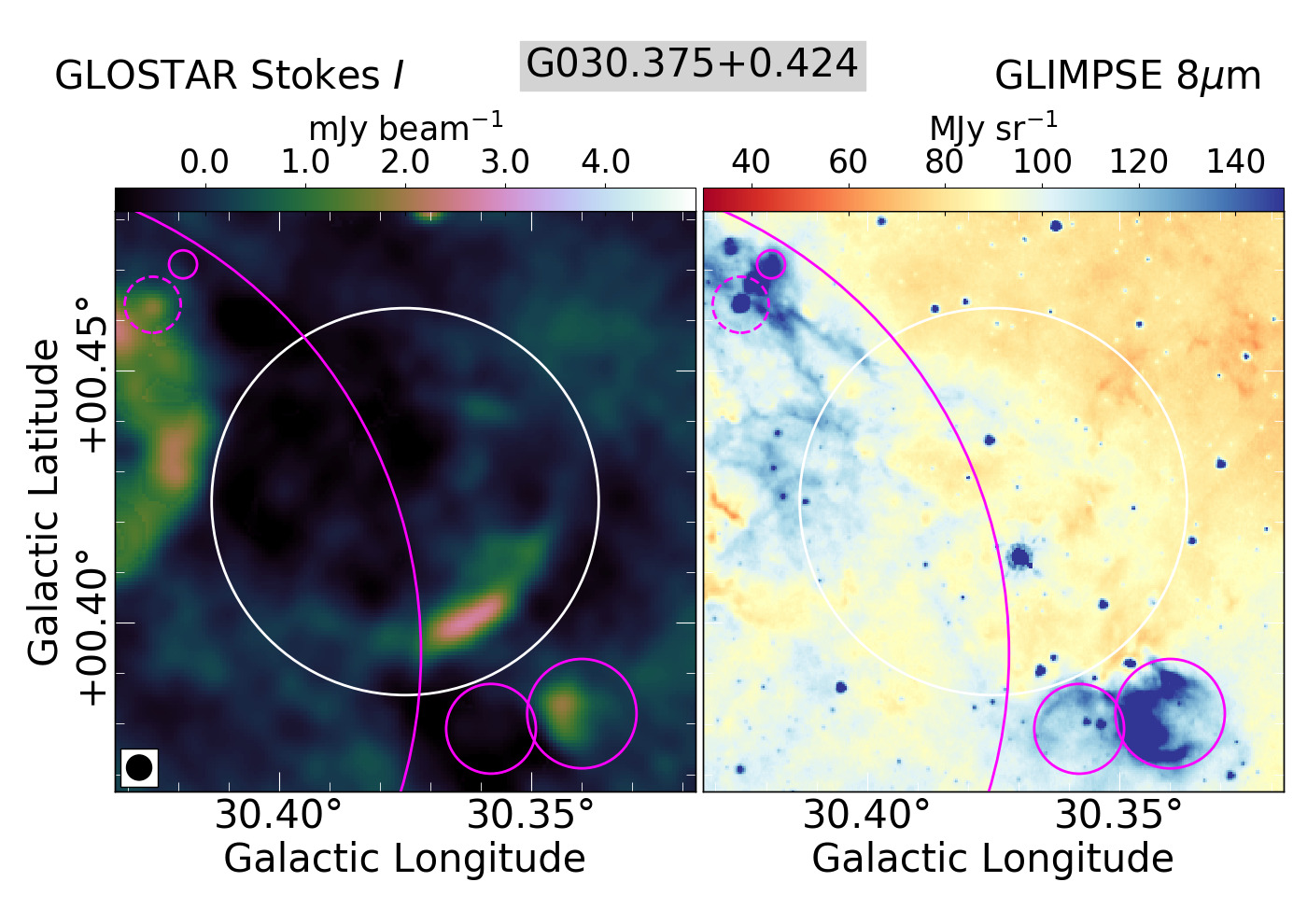}\\
\noindent\includegraphics[width=0.47\textwidth]{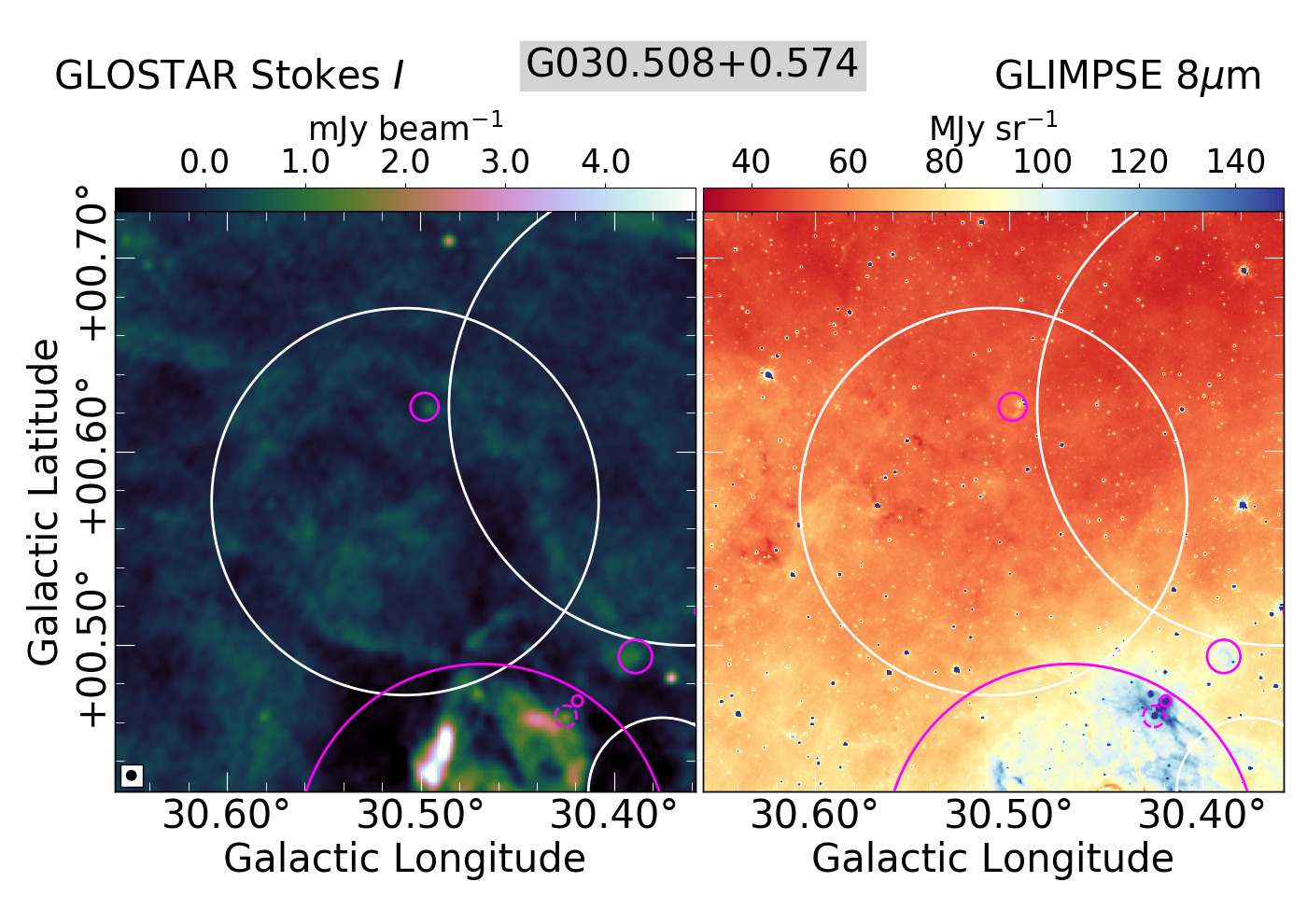}\\
\noindent\includegraphics[width=0.47\textwidth]{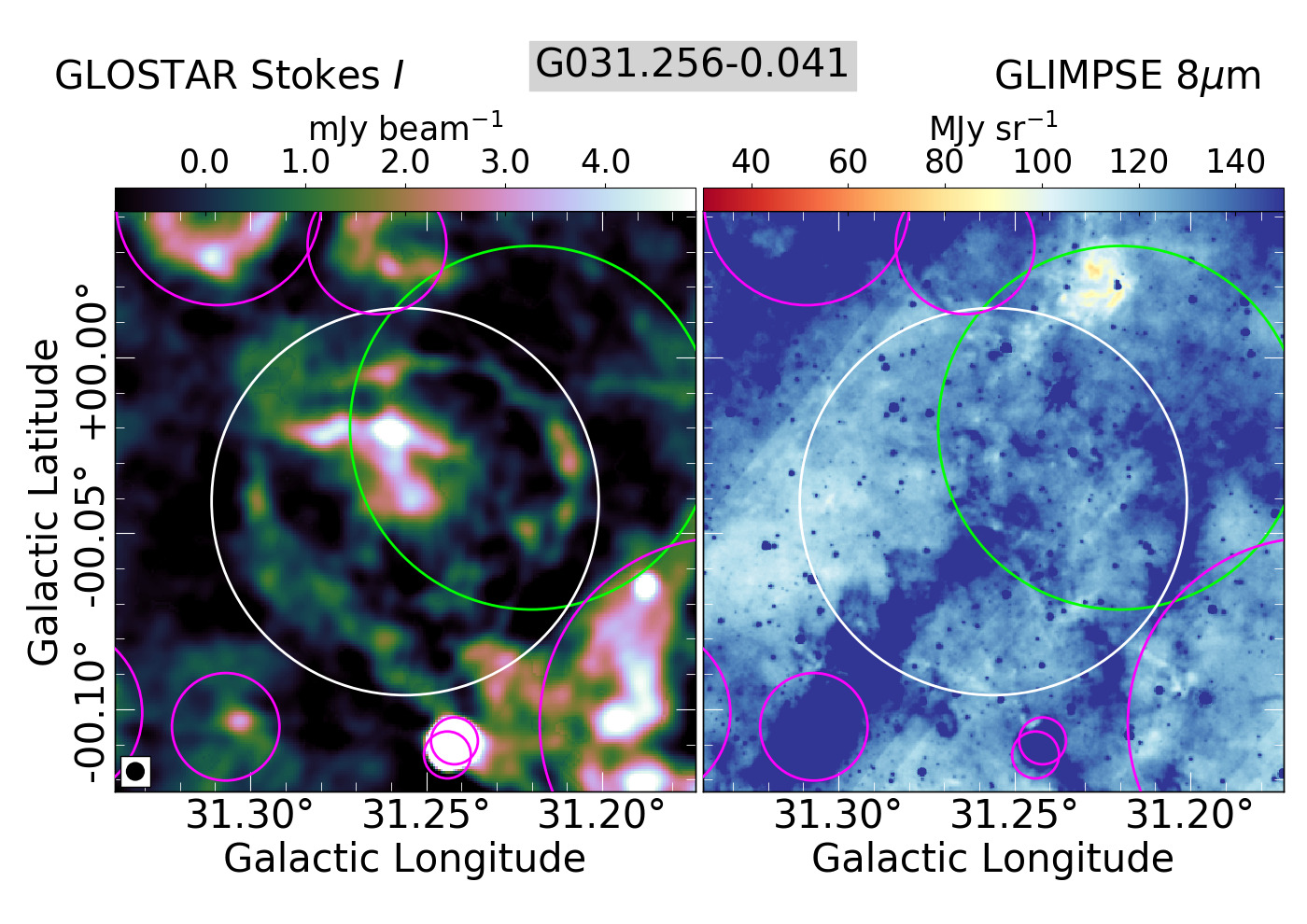}\\
\noindent\includegraphics[width=0.47\textwidth]{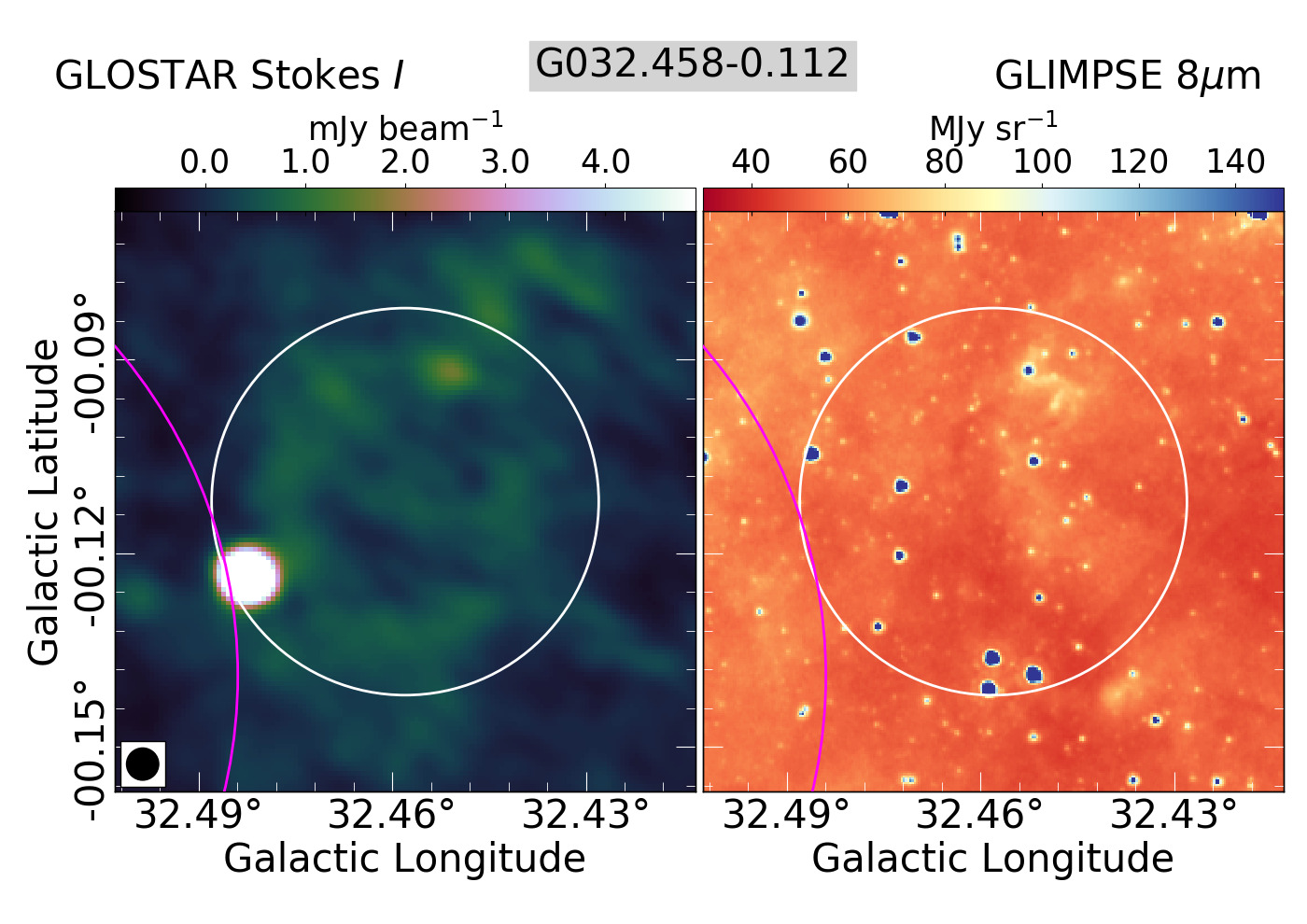}\\
\noindent\includegraphics[width=0.47\textwidth]{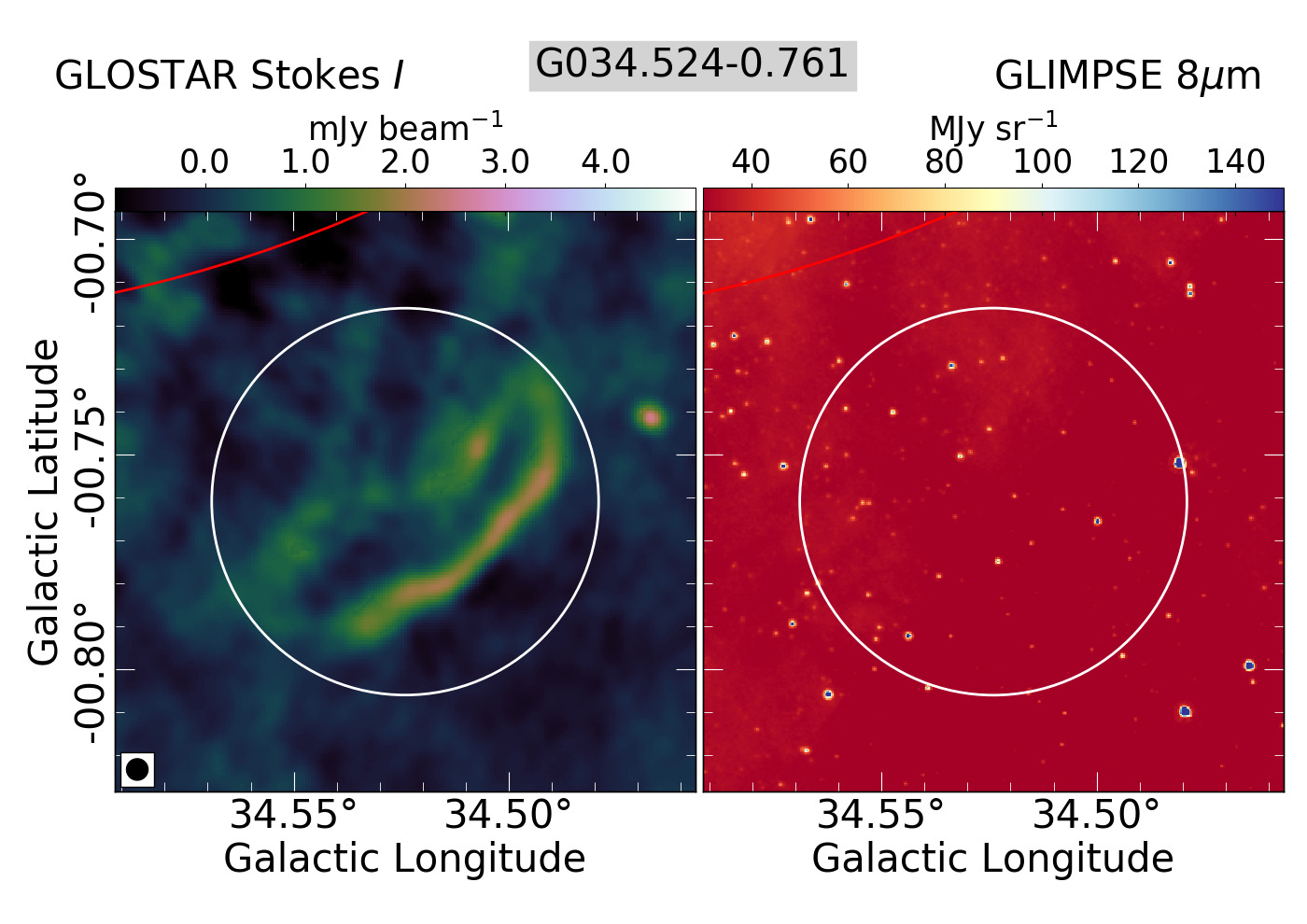}\\
\noindent\includegraphics[width=0.47\textwidth]{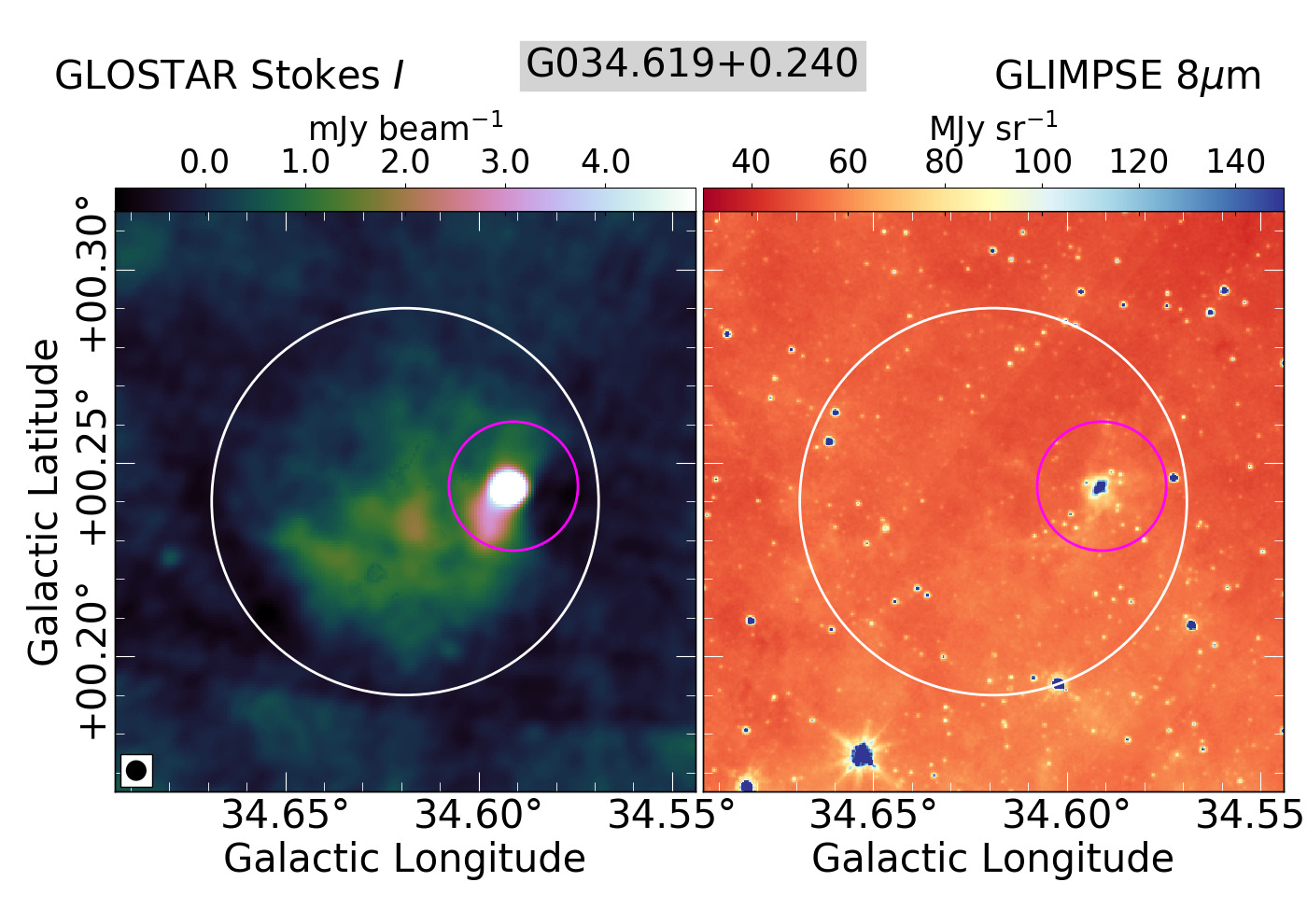}\\
\noindent\includegraphics[width=0.47\textwidth]{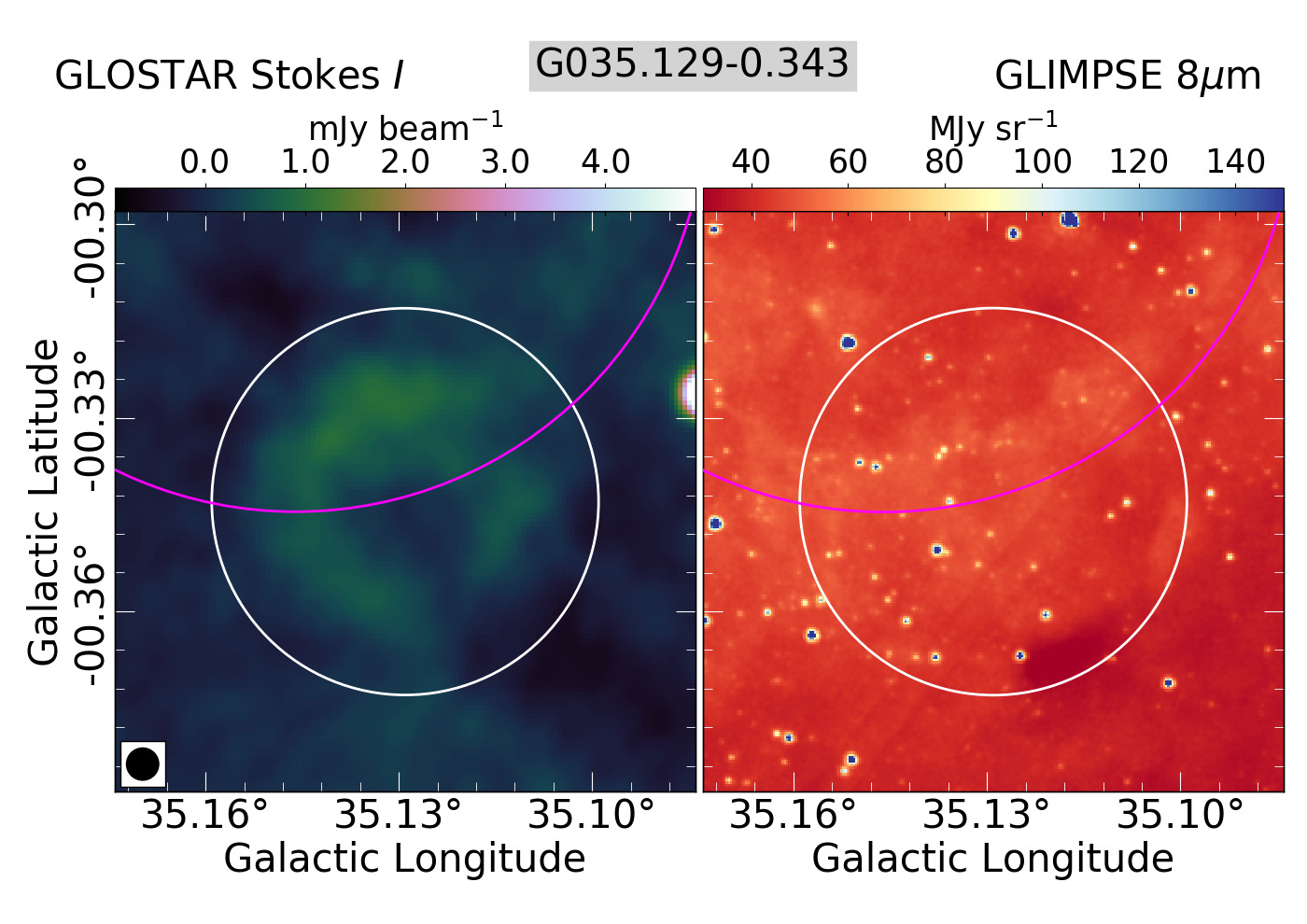}\\
\noindent\includegraphics[width=0.47\textwidth]{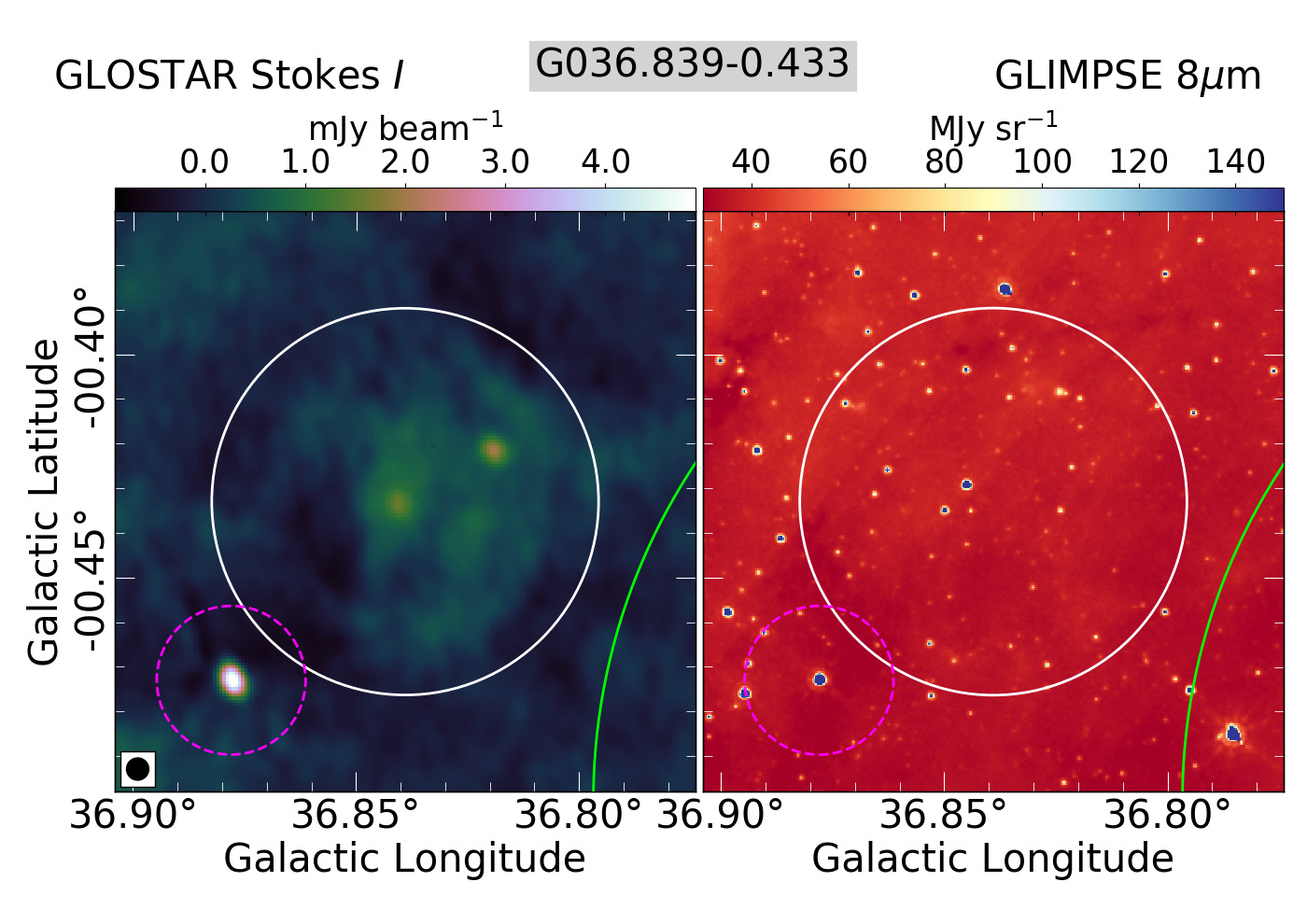}\\
\noindent\includegraphics[width=0.47\textwidth]{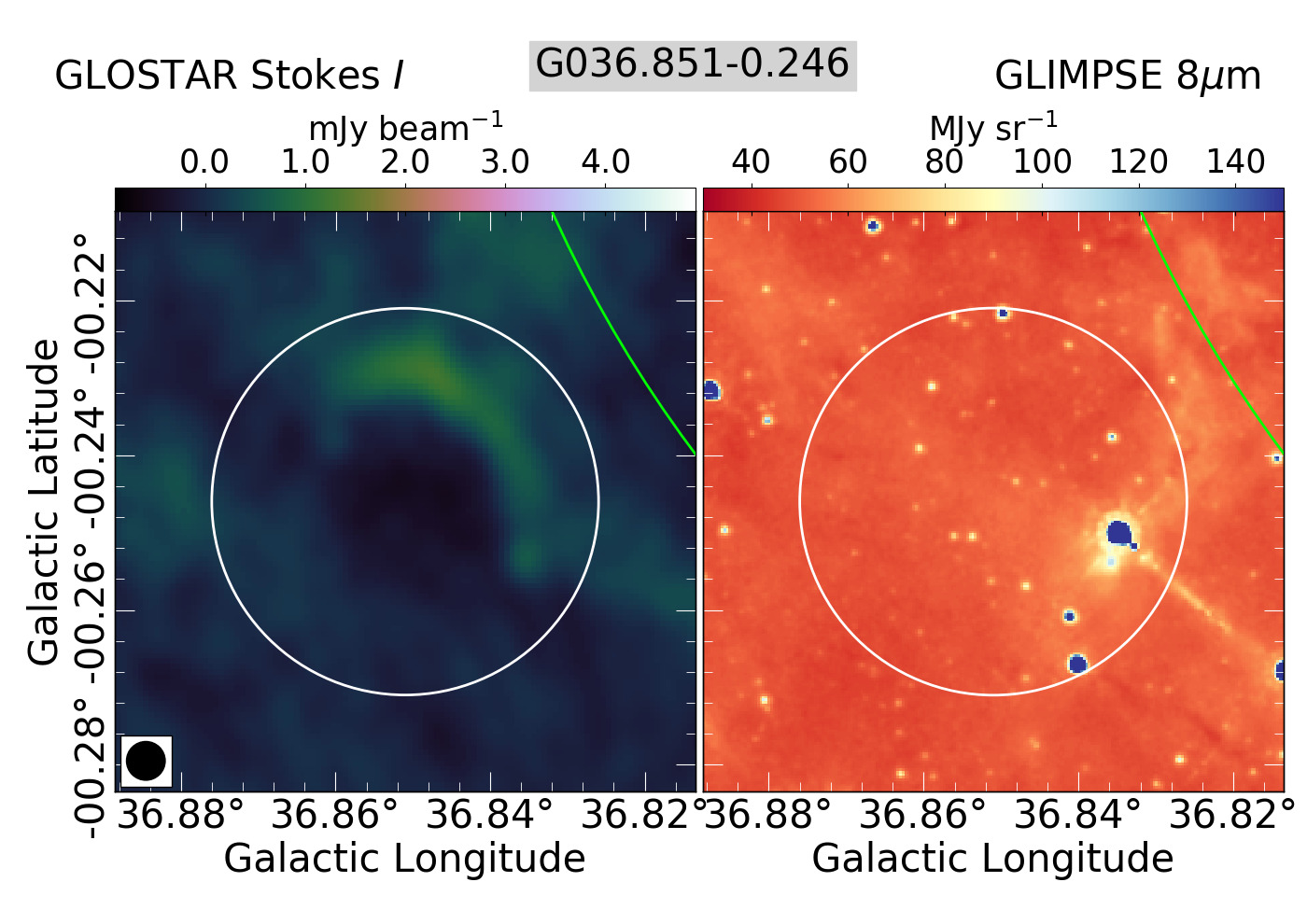}\\
\noindent\includegraphics[width=0.47\textwidth]{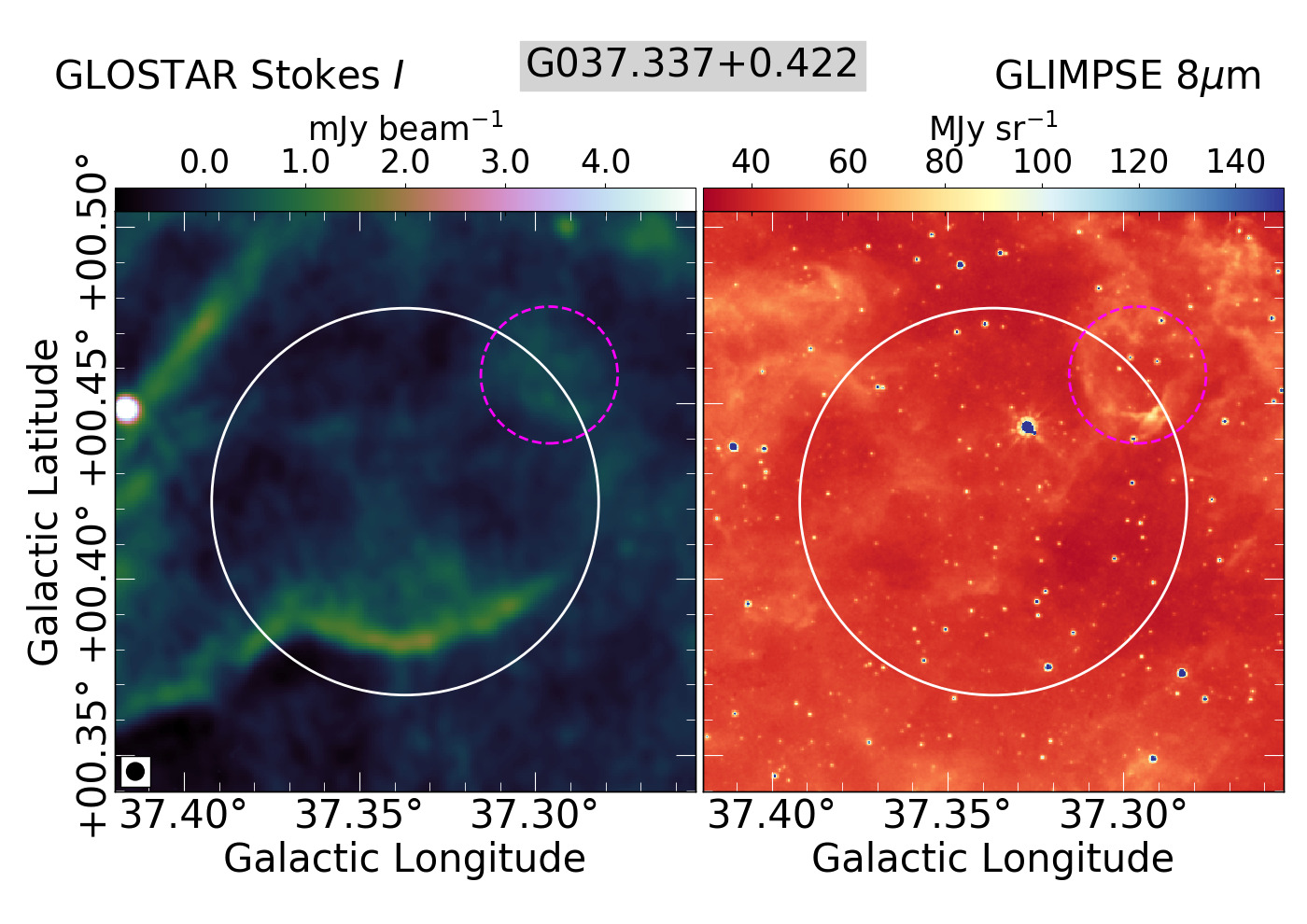}\\
\noindent\includegraphics[width=0.47\textwidth]{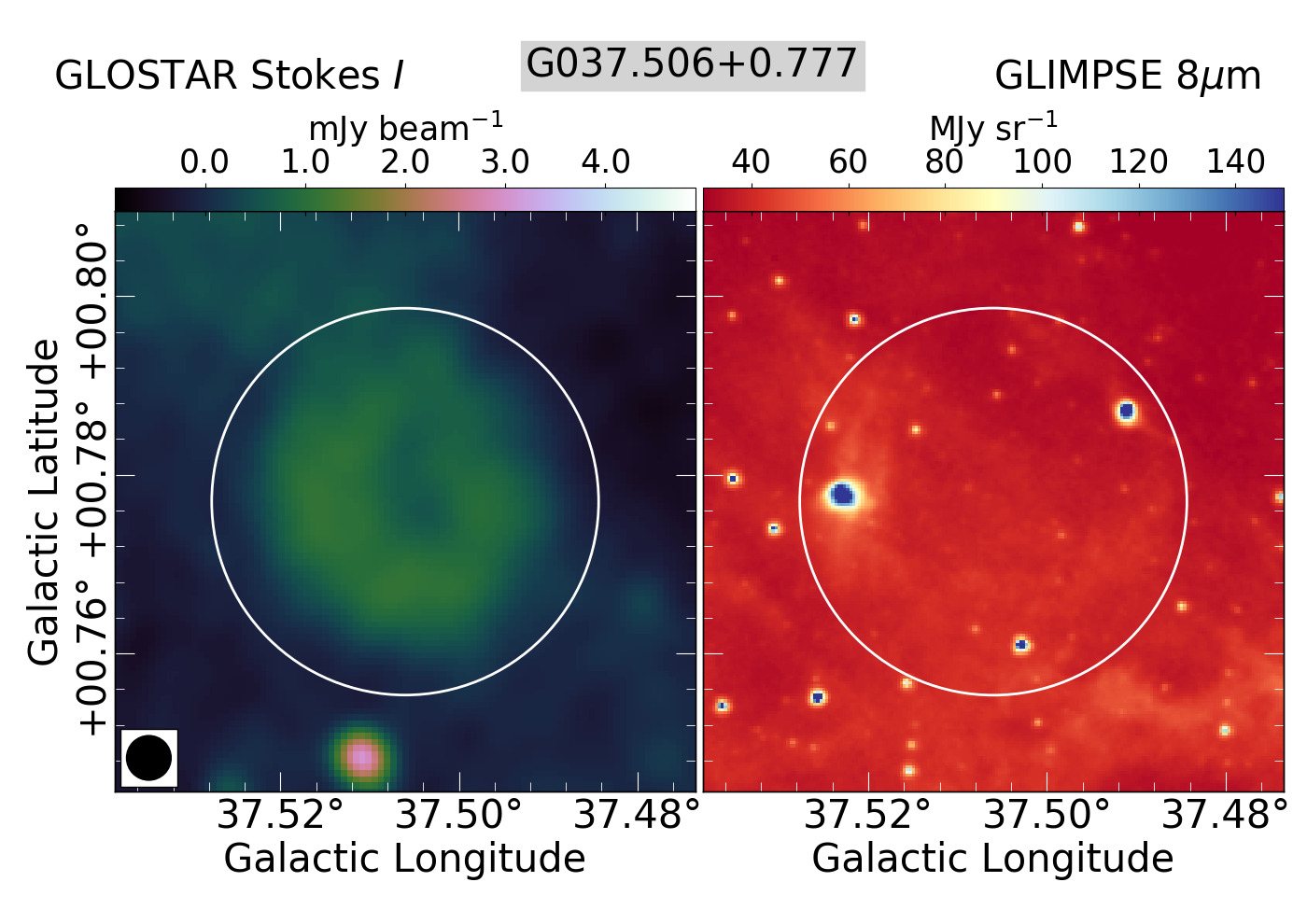}\\
\noindent\includegraphics[width=0.47\textwidth]{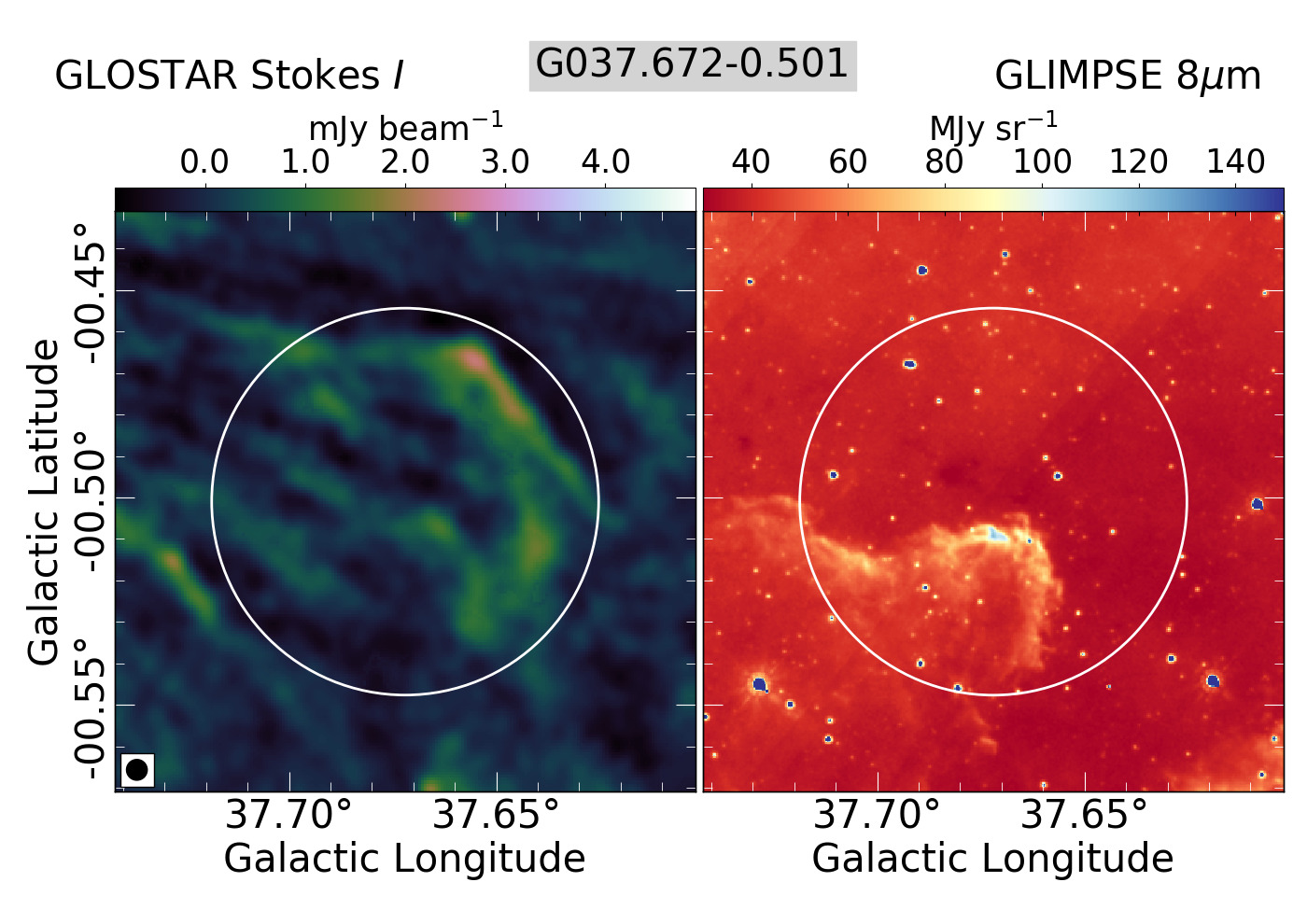}\\
\noindent\includegraphics[width=0.47\textwidth]{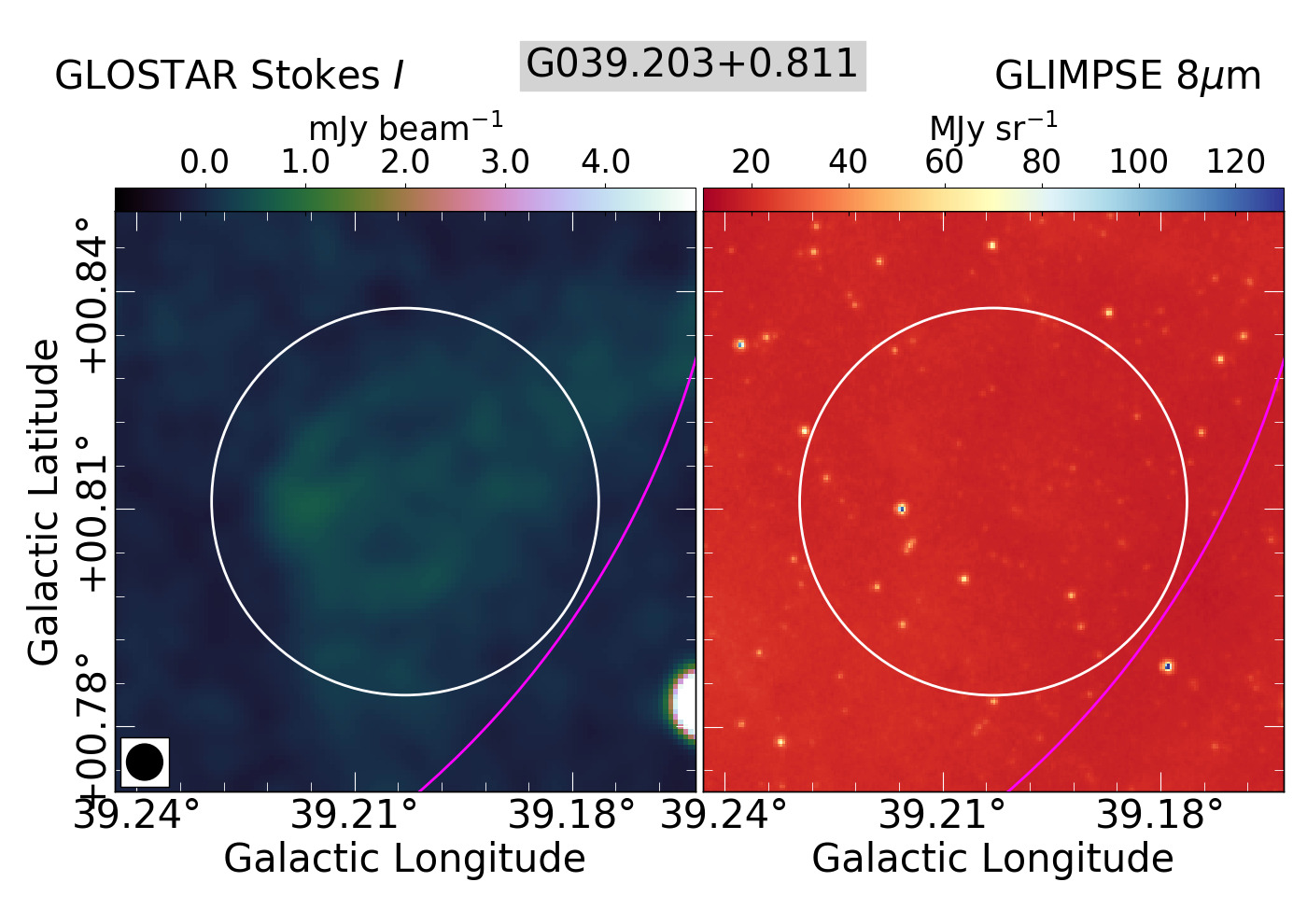}\\
\noindent\includegraphics[width=0.47\textwidth]{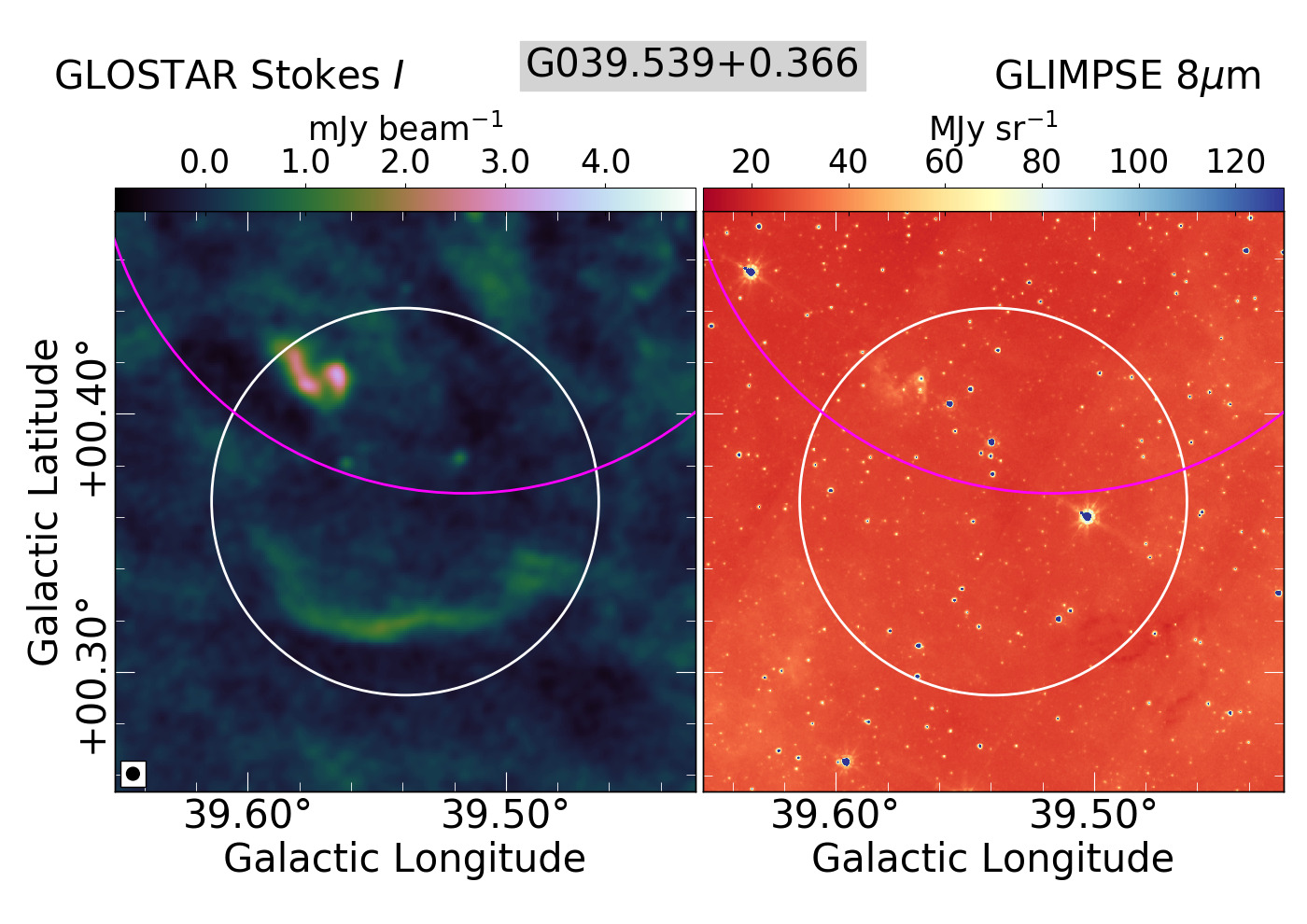}\\
\noindent\includegraphics[width=0.47\textwidth]{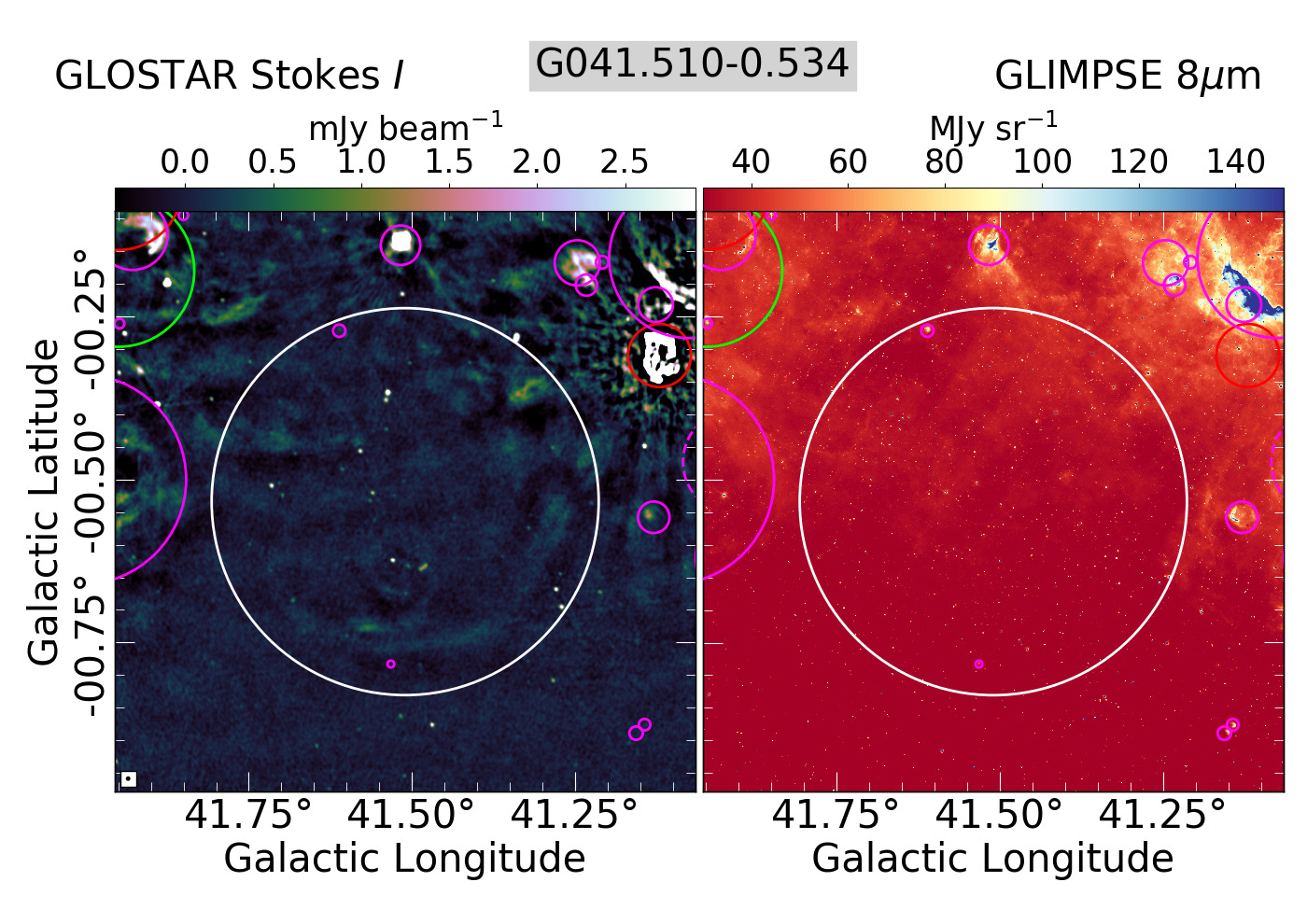}\\
\noindent\includegraphics[width=0.47\textwidth]{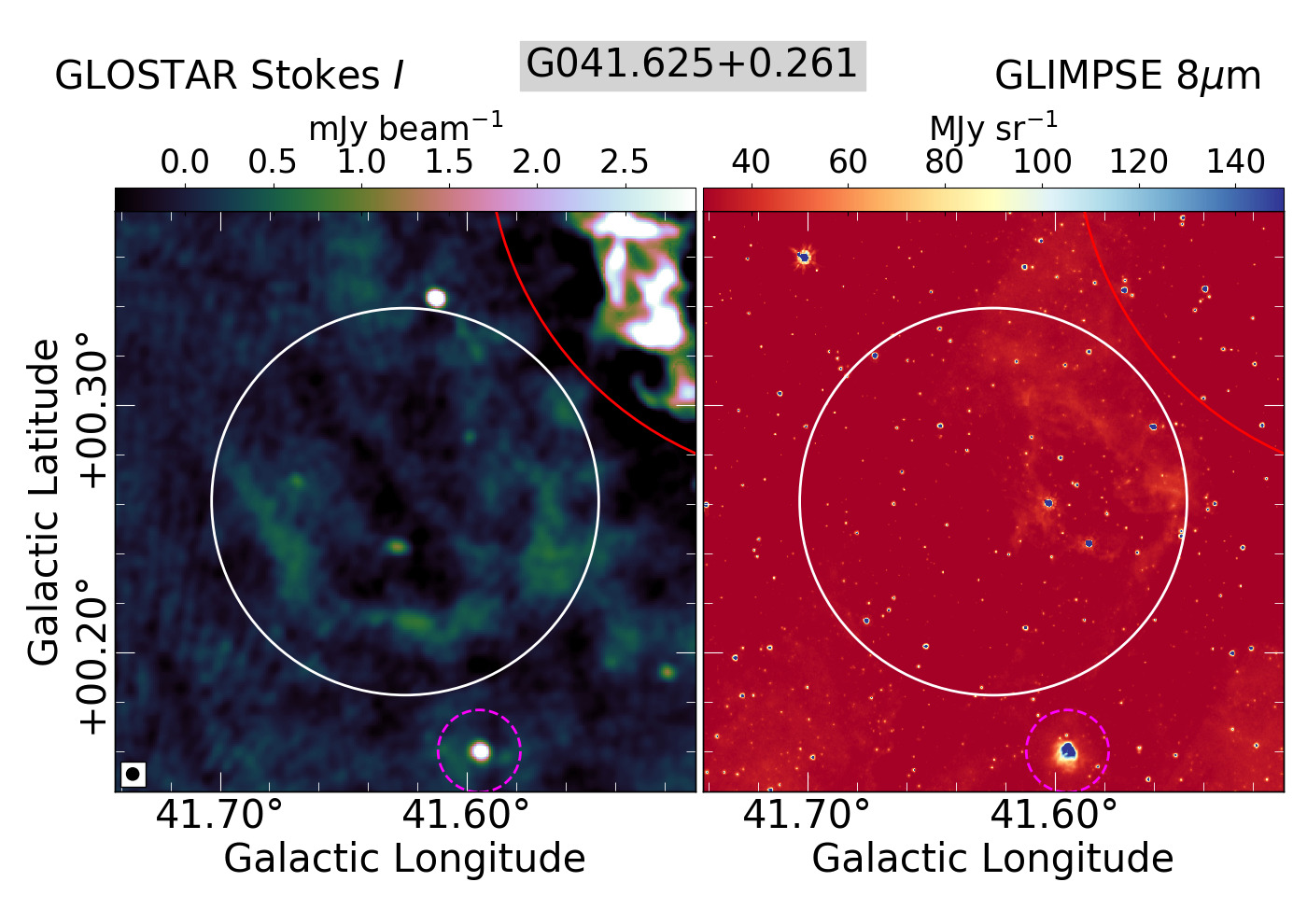}\\
\noindent\includegraphics[width=0.47\textwidth]{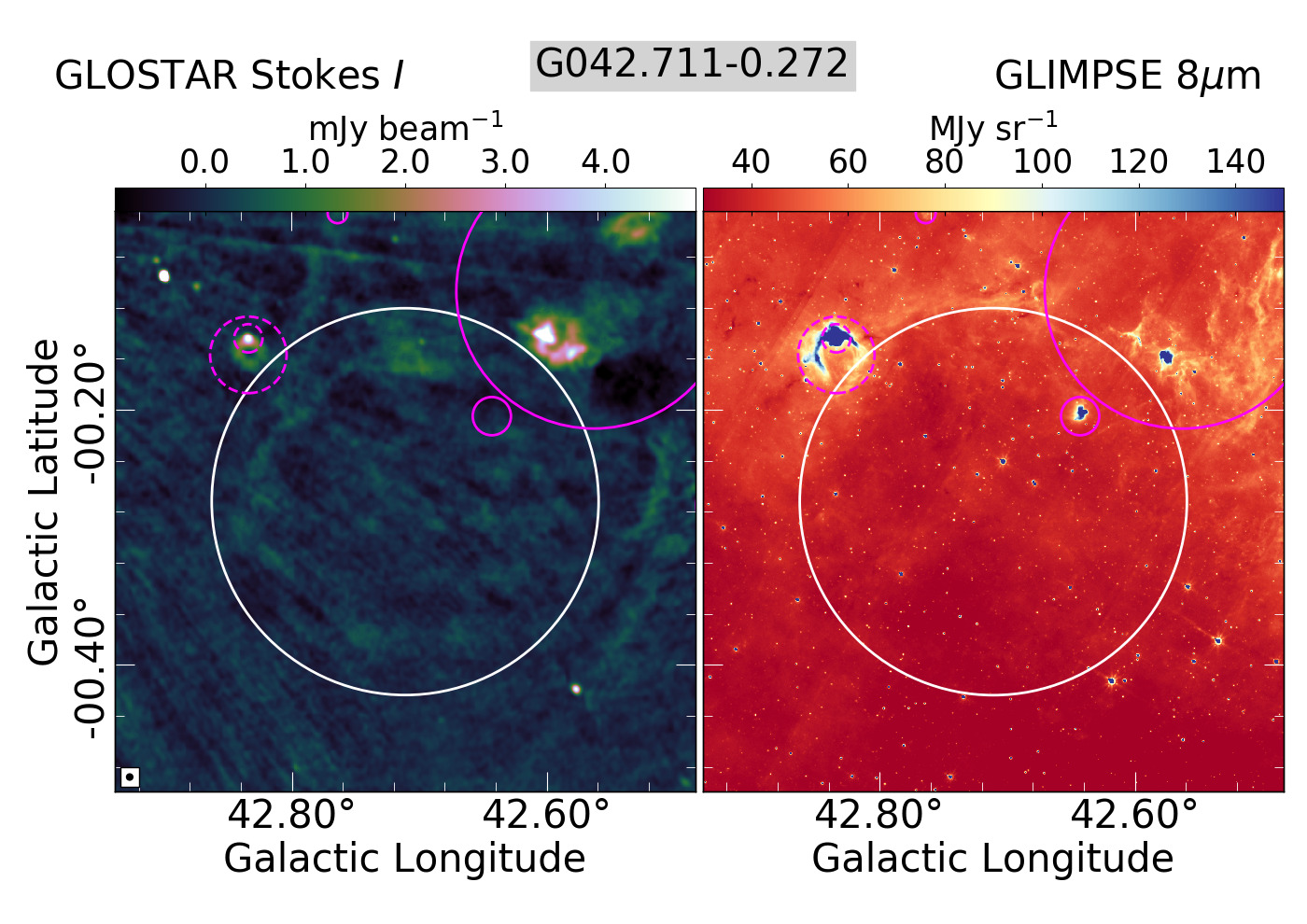}\\
\noindent\includegraphics[width=0.47\textwidth]{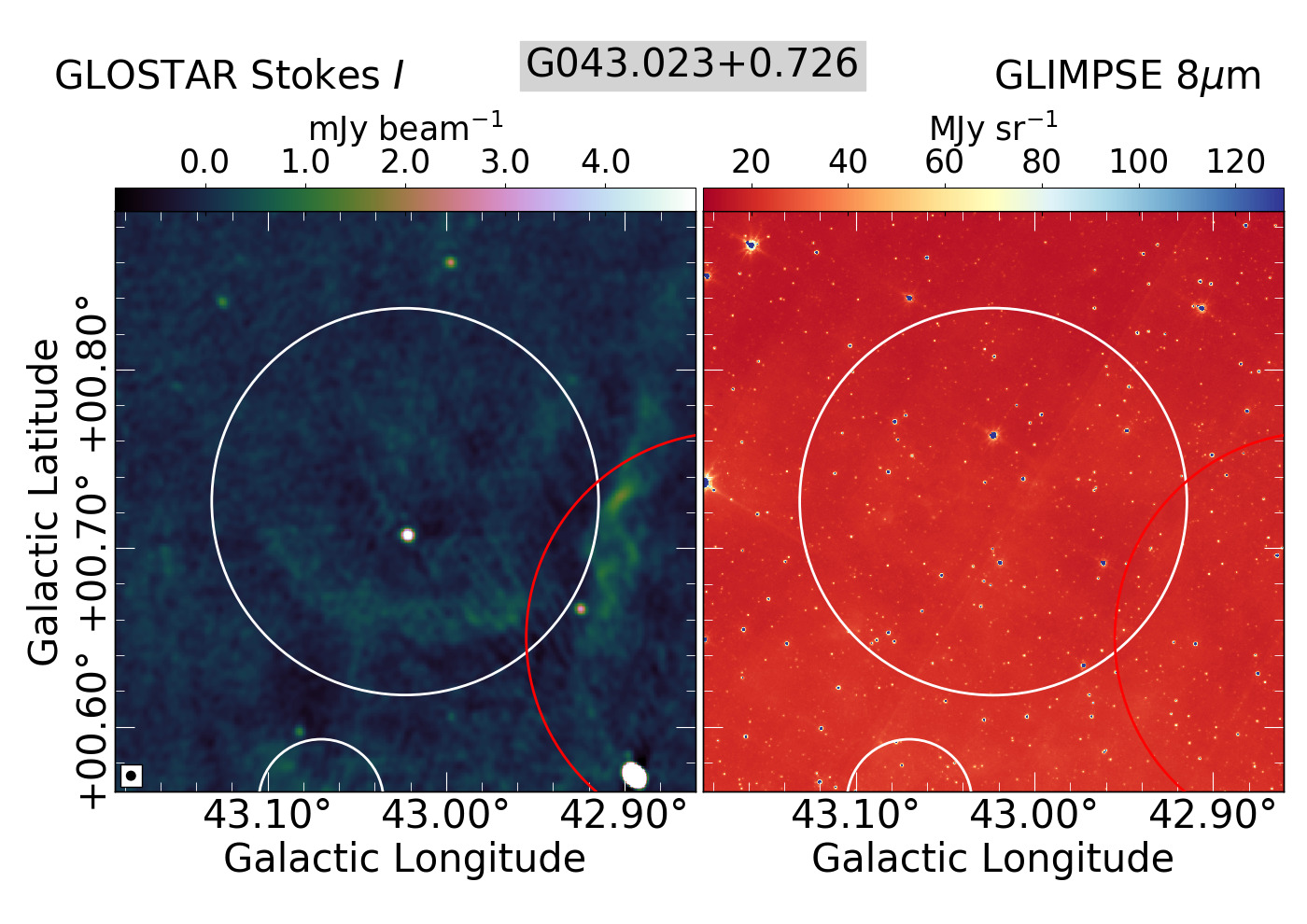}\\
\noindent\includegraphics[width=0.47\textwidth]{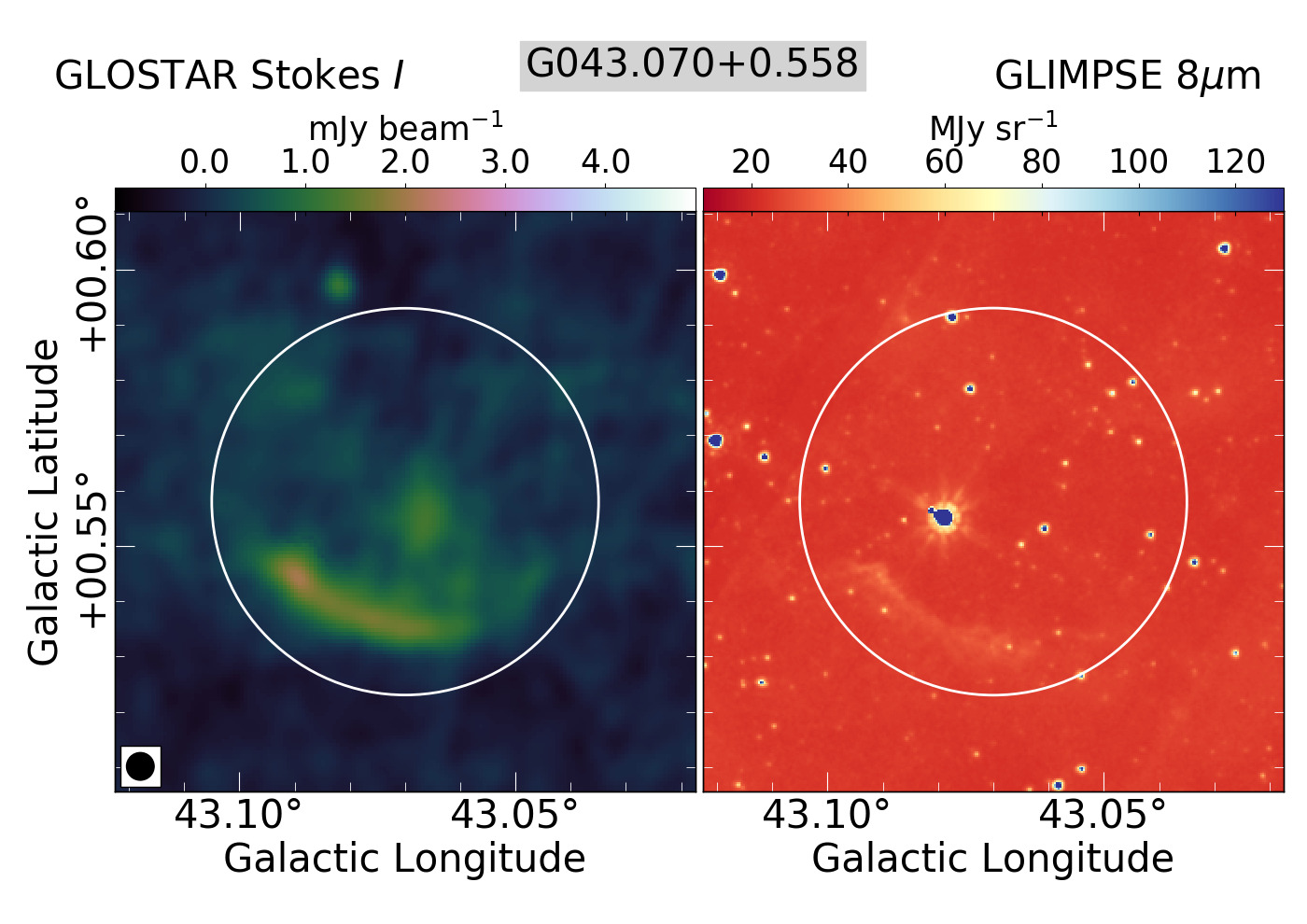}\\
\noindent\includegraphics[width=0.47\textwidth]{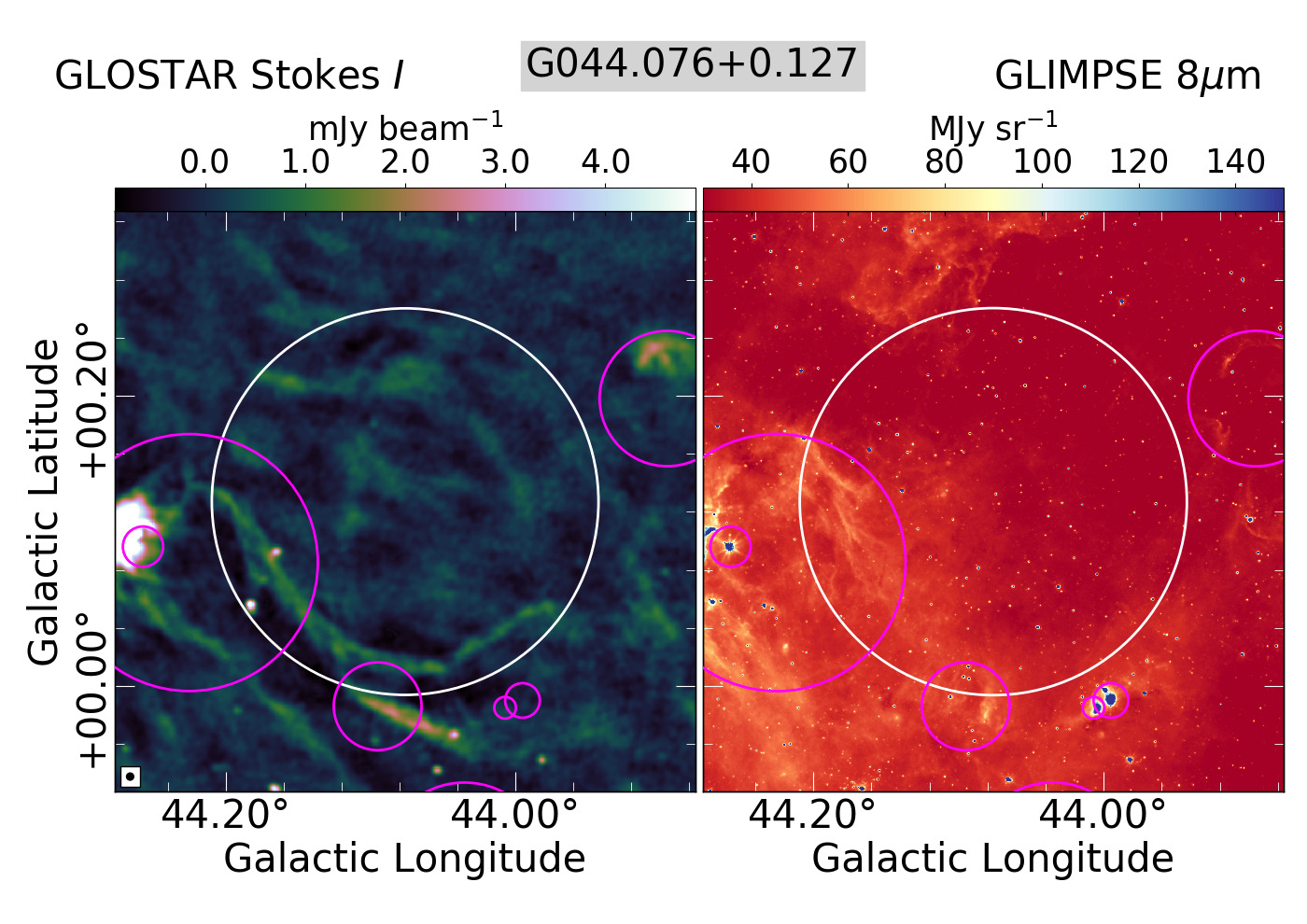}\\
\noindent\includegraphics[width=0.47\textwidth]{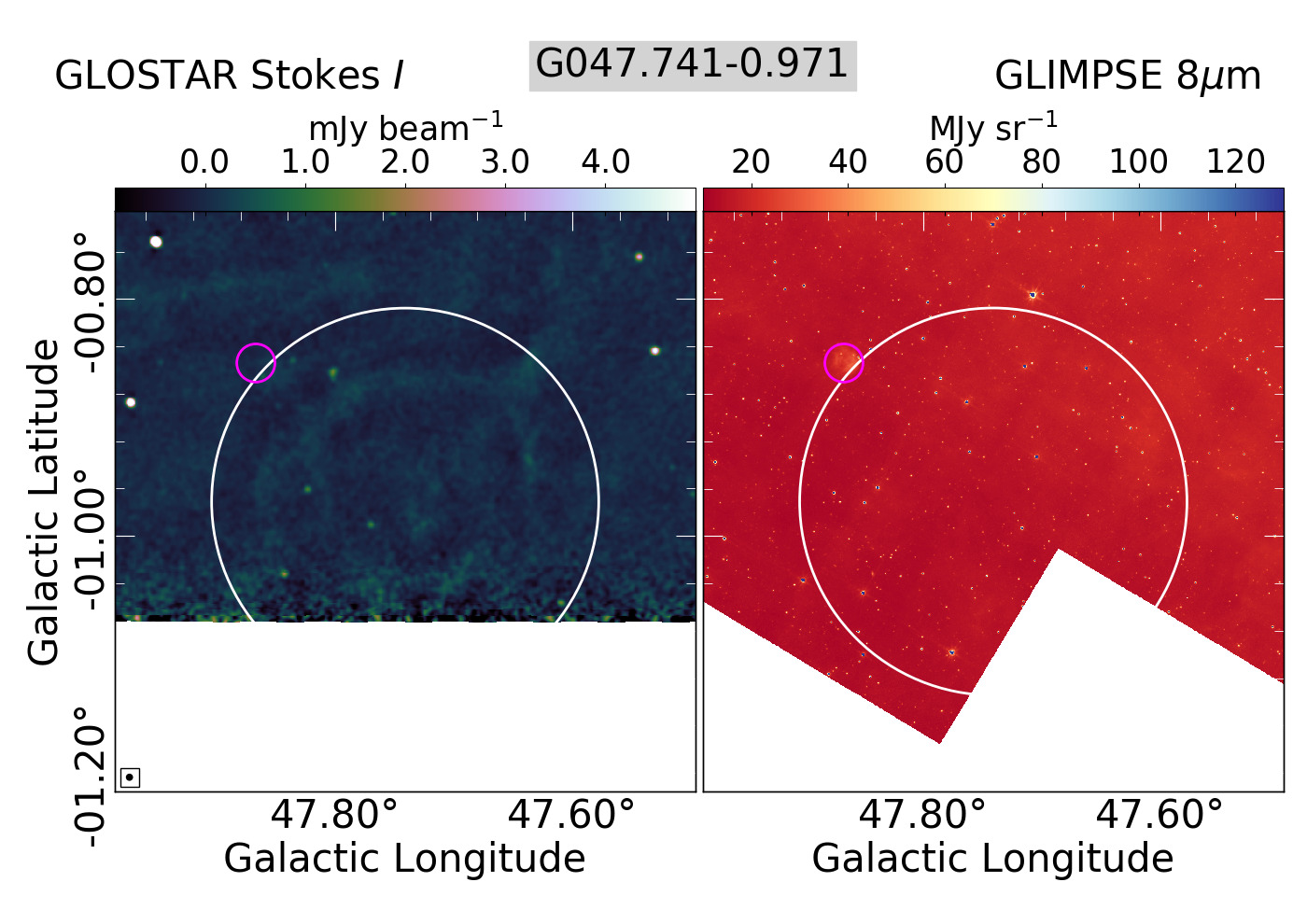}\\
\noindent\includegraphics[width=0.47\textwidth]{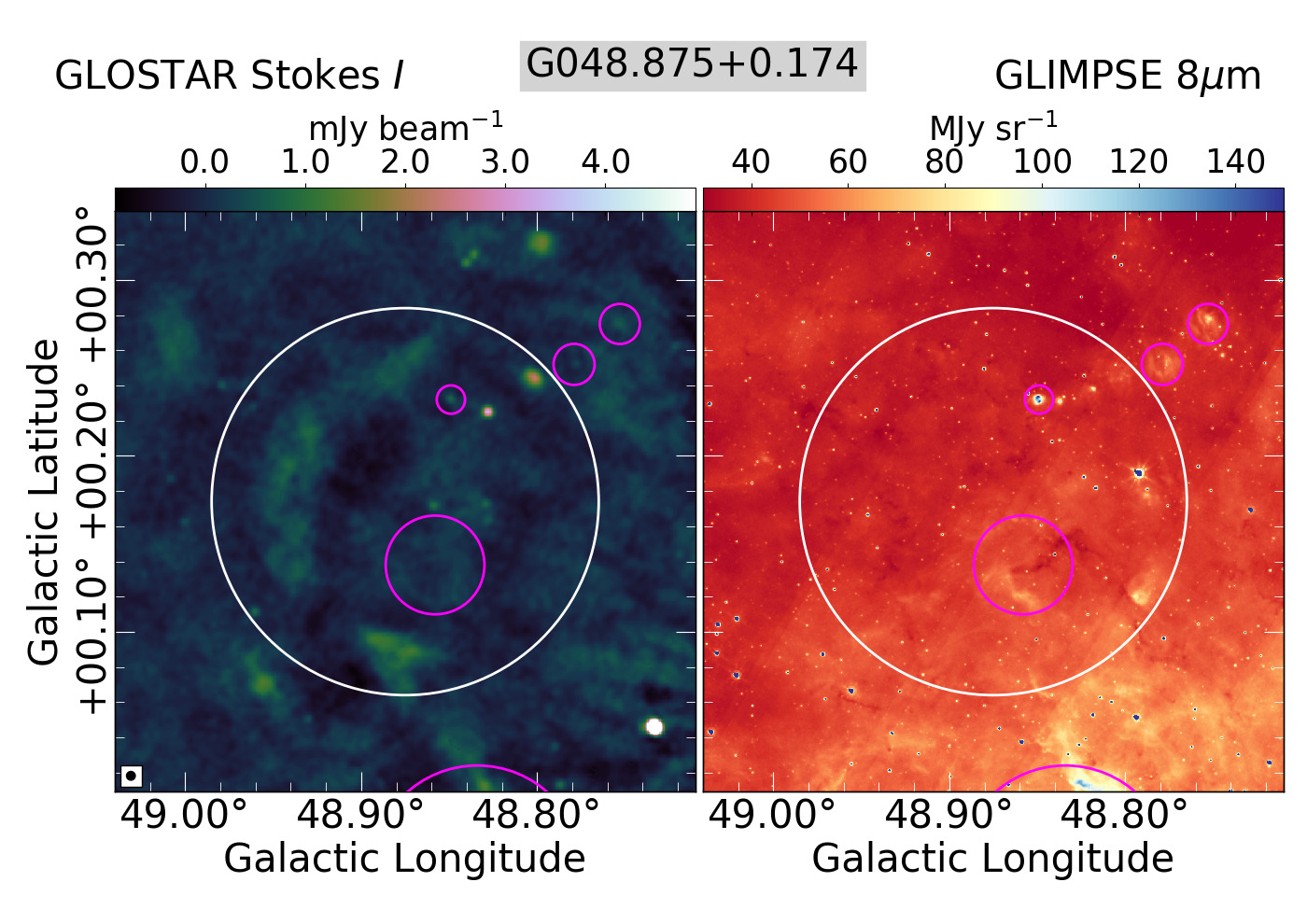}\\
\noindent\includegraphics[width=0.47\textwidth]{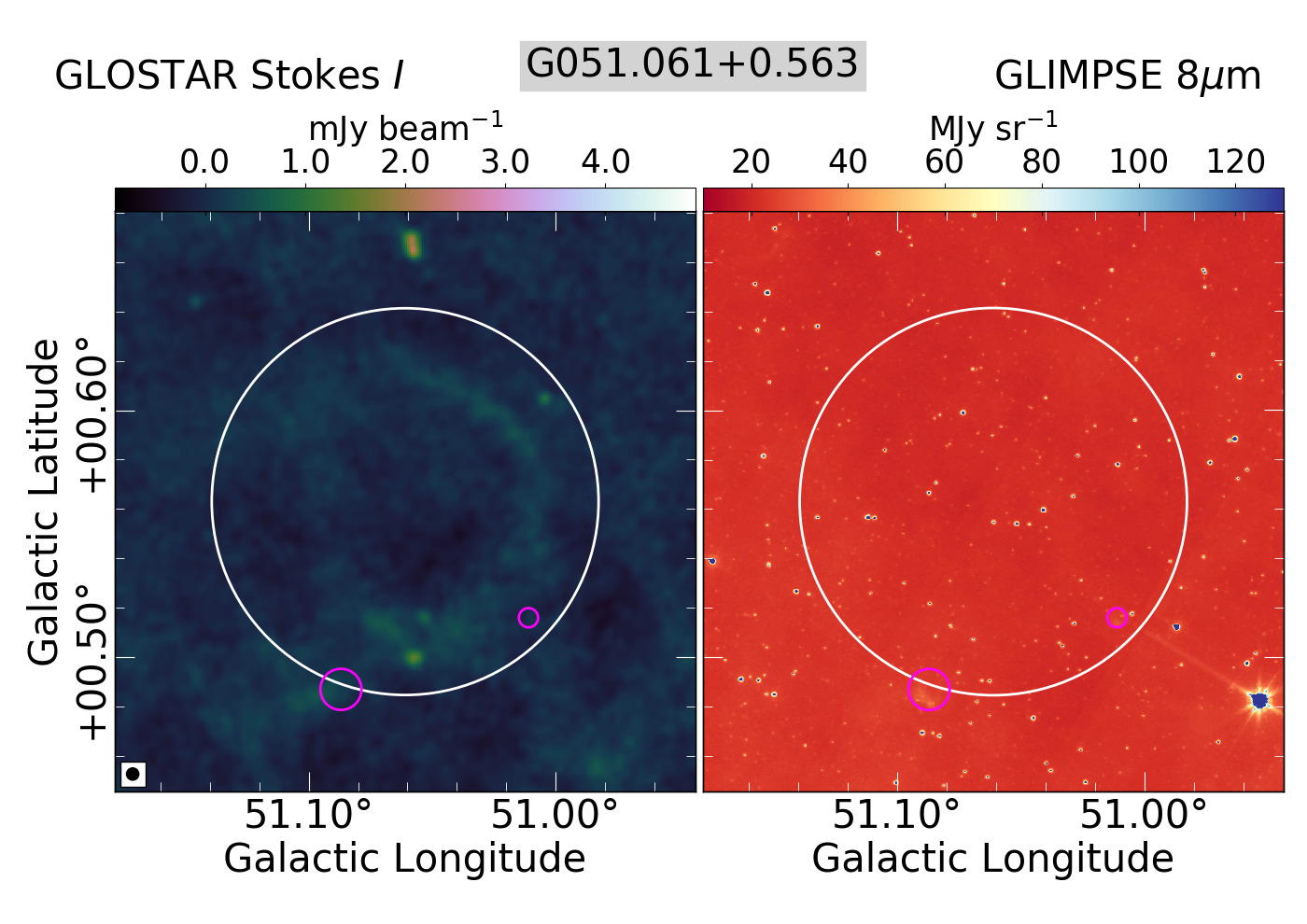}\\
\noindent\includegraphics[width=0.47\textwidth]{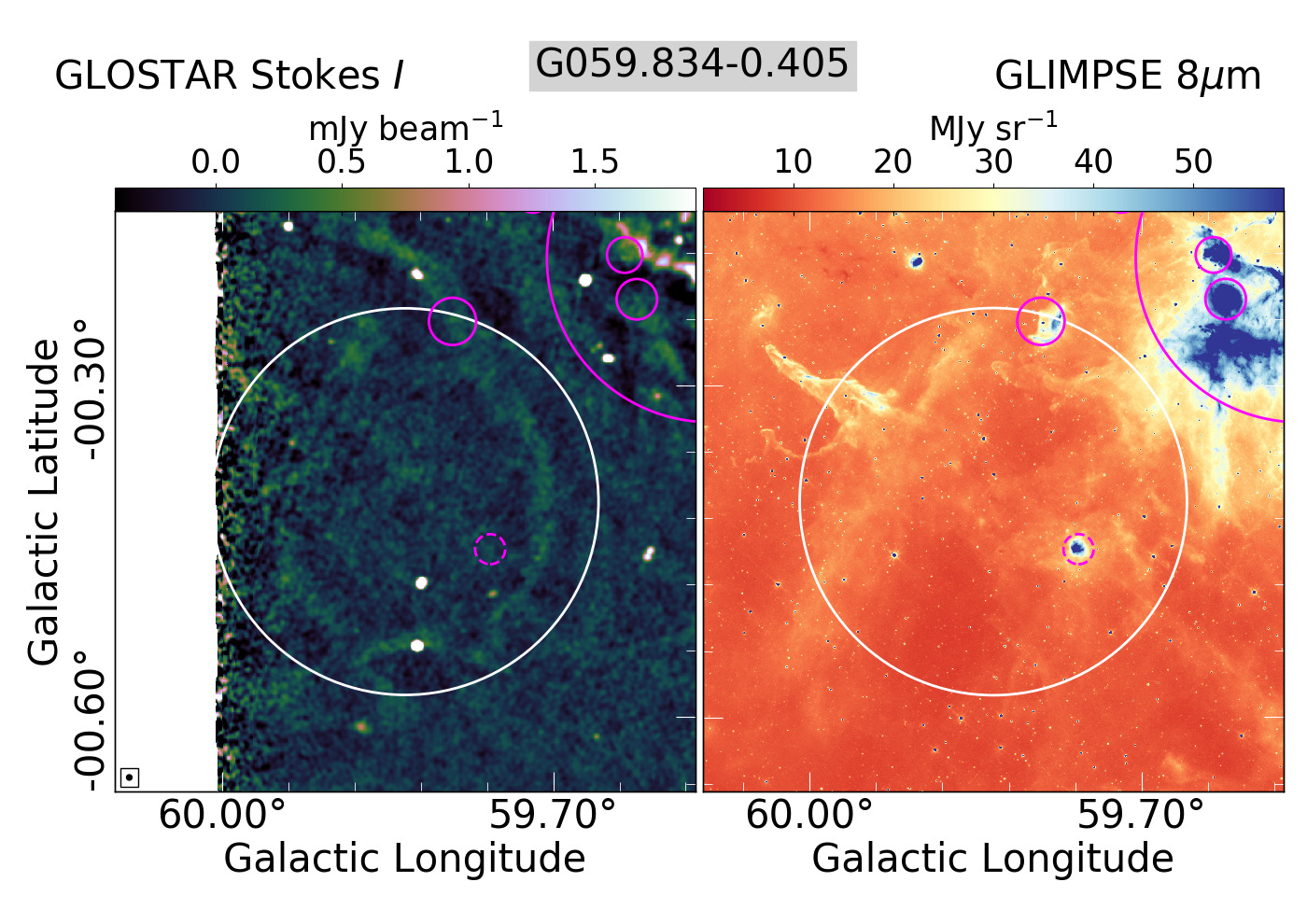}\\
\noindent\includegraphics[width=0.47\textwidth]{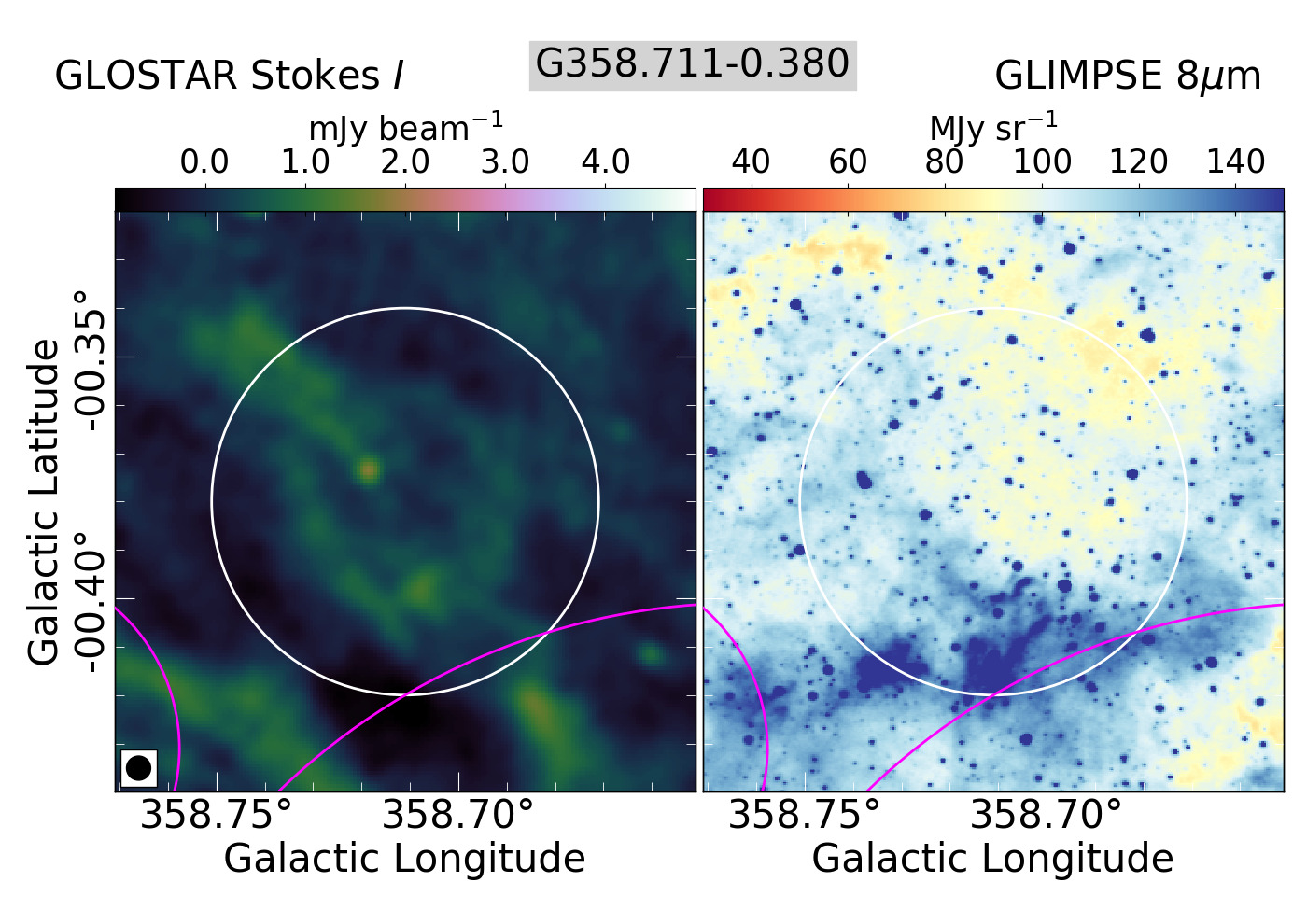}\\
\noindent\includegraphics[width=0.47\textwidth]{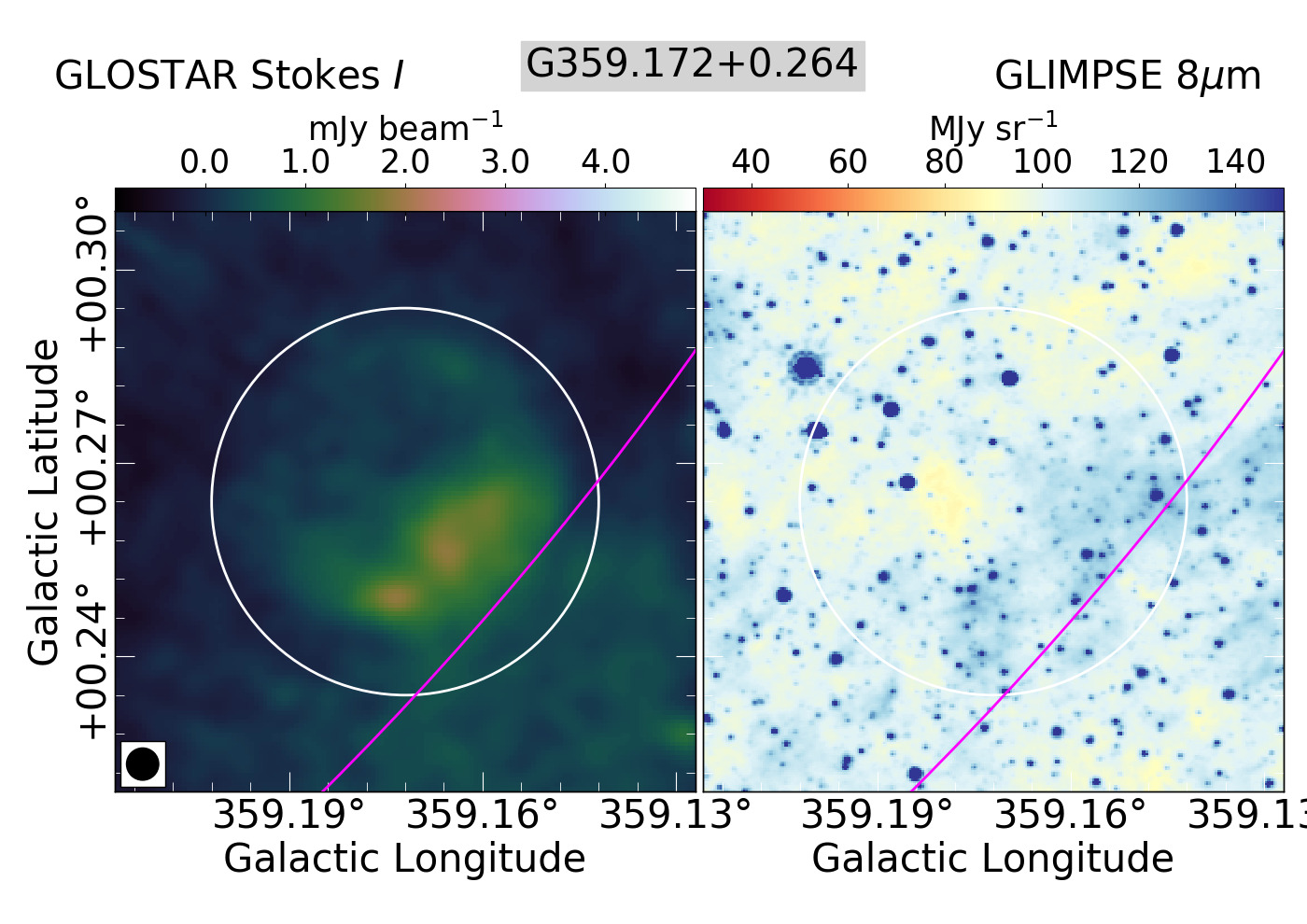}\\
}

\end{appendix}

\clearpage
\onecolumn

\setcounter{table}{1}
\begin{longtable}{ l r r r c c r r r }
\caption{\label{table:G19SNRs} G19 SNRs identified in the GLOSTAR-VLA images.  We note that the quoted Stokes $I$ flux densities should be considered as only lower limits due to missing short spacing data (see \S\ref{subsec:GLOSTARdata} and \S\ref{subsubsec:missflux}).} \\
\hline\hline
    Name & GLong\tablefootmark{a} & GLat\tablefootmark{a} & Radius     & Type\tablefootmark{b} & Remarks & $S_{5.8\mathrm{GHz}}$ & $L_{5.8\mathrm{GHz}}$ & $p_{5.8\mathrm{GHz}}$  \\
         & ($\degr$)              & ($\degr$)            & ($\arcmin$) &                       &         & (Jy)                  & (Jy)                  &  \\
\hline
\endfirsthead
\caption{continued.}\\
\hline\hline
    Name & GLong\tablefootmark{a} & GLat\tablefootmark{a} & Radius      & Type\tablefootmark{b} & Remarks & $S_{5.8\mathrm{GHz}}$ & $L_{5.8\mathrm{GHz}}$ & $p_{5.8\mathrm{GHz}}$  \\
         & ($\degr$)              & ($\degr$)             & ($\arcmin$) &                       &         & (Jy)                  & (Jy)                  &  \\
\hline
\endhead
\hline
\endfoot
G0.0$+$0.0 & $000.000$ & $000.000$ & $1.8$ & S & Not detected & - & - & - \\
G0.3$+$0.0 & $000.300$ & $000.000$ & $7.5$ & S & Not detected & - & - & - \\
G0.9$+$0.1 & $000.874$ & $000.074$ & $4.4$ & C &  & 2.320 $\pm$ 0.019 & 0.0525 $\pm$ 0.0243 & 0.023 $\pm$ 0.010 \\
G1.0$-$0.1 & $001.011$ & $-000.174$ & $5.5$ & S &  & - & - & - \\
G1.4$-$0.1 & $001.454$ & $-000.135$ & $6.2$ & S &  & 0.093 $\pm$ 0.010 & 0.0092 $\pm$ 0.0049 & 0.099 $\pm$ 0.054 \\
G1.9$+$0.3 & $001.869$ & $000.325$ & $1.3$ & S &  & 0.316 $\pm$ 0.001 & 0.0165 $\pm$ 0.0029 & 0.052 $\pm$ 0.009 \\
G3.7$-$0.2 & $003.786$ & $-000.284$ & $8.4$ & S &  & 0.120 $\pm$ 0.013 & 0.0229 $\pm$ 0.0045 & 0.192 $\pm$ 0.043 \\
G3.8$+$0.3 & $003.860$ & $000.392$ & $13.1$ & S? &  & 0.066 $\pm$ 0.012 & 0.0005 $\pm$ 0.0001 & 0.007 $\pm$ 0.002 \\
G5.5$+$0.3 & $005.551$ & $000.322$ & $7.5$ & S & & - & - & - \\
G6.1$+$0.5 & $006.100$ & $000.530$ & $9.0$ & S & Multiple objects\footnote{GLOSTAR SNR candidates G006.055$+$0.499, G006.118$+$0.387} & 0.112 $\pm$ 0.011 & 0.0175 $\pm$ 0.0064 & 0.156 $\pm$ 0.059 \\
G6.4$-$0.1 & $006.431$ & $-000.099$ & $24.0$ & C &  & - & - & - \\
G6.5$-$0.4 & $006.530$ & $-000.470$ & $13.4$ & S &  & - & - & - \\
G7.0$-$0.1 & $007.071$ & $-000.092$ & $9.9$ & S &  & - & - & - \\
G7.2$+$0.2 & $007.201$ & $000.187$ & $5.6$ & S &  & 0.128 $\pm$ 0.011 & $<$0.0010 & $<$0.008 \\
G8.3$-$0.0 & $008.310$ & $-000.090$ & $2.5$ & S & H II region & 0.449 $\pm$ 0.005 & 0.0105 $\pm$ 0.0039 & 0.023 $\pm$ 0.009 \\
G8.7$-$0.1 & $008.776$ & $-000.143$ & $22.5$ & S? &  & - & - & - \\
G8.9$+$0.4 & $008.840$ & $000.460$ & $18.0$ & S &  & 0.051 $\pm$ 0.018 & 0.0075 $\pm$ 0.0028 & 0.145 $\pm$ 0.075 \\
G9.7$+$0.0 & $009.684$ & $-000.077$ & $6.1$ & S &  & 0.164 $\pm$ 0.008 & 0.0108 $\pm$ 0.0017 & 0.065 $\pm$ 0.011 \\
G9.8$+$0.6 & $009.776$ & $000.575$ & $7.9$ & S &  & 0.323 $\pm$ 0.008 & 0.0028 $\pm$ 0.0009 & 0.009 $\pm$ 0.003 \\
G9.9$-$0.8 & $009.963$ & $-000.813$ & $7.4$ & S &  & - & - & - \\
G10.5$-$0.0 & $010.590$ & $-000.040$ & $3.0$ & S & H II region & 0.185 $\pm$ 0.007 & $<$0.0008 & $<$0.004 \\
G11.0$+$0.0 & $011.032$ & $-000.056$ & $4.6$ & S &  & 0.318 $\pm$ 0.011 &  - & - \\
G11.1$-$1.0 & $011.196$ & $-001.055$ & $8.5$ & S & H II region & - & - & - \\
G11.1$-$0.7 & $011.148$ & $-000.730$ & $6.5$ & S &  & 0.783 $\pm$ 0.007 & 0.0129 $\pm$ 0.0055 & 0.016 $\pm$ 0.007 \\
G11.1$+$0.1 & $011.186$ & $000.124$ & $6.0$ & S &  & 0.018 $\pm$ 0.015 & $<$0.0006 & $<$0.033 \\
G11.2$-$0.3 & $011.181$ & $-000.347$ & $2.8$ & C &  & 1.942 $\pm$ 0.005 & 0.1943 $\pm$ 0.0873 & 0.100 $\pm$ 0.045 \\
G11.4$-$0.1 & $011.386$ & $-000.065$ & $5.2$ & S? &  & 0.762 $\pm$ 0.009 & 0.0863 $\pm$ 0.0454 & 0.113 $\pm$ 0.060 \\
G11.8$-$0.2 & $011.890$ & $-000.210$ & $2.0$ & S &  & - & - & - \\
G12.0$-$0.1 & $011.962$ & $-000.090$ & $5.0$ & ? &  & 0.150 $\pm$ 0.009 &  - & - \\
G12.2$+$0.3 & $012.261$ & $000.303$ & $3.6$ & S &  & 0.096 $\pm$ 0.003 & 0.0045 $\pm$ 0.0007 & 0.046 $\pm$ 0.008 \\
G12.5$+$0.2 & $012.584$ & $000.228$ & $3.9$ & C? &  & 0.024 $\pm$ 0.005 & $<$0.0004 & $<$0.016 \\
G12.7$+$0.0 & $012.714$ & $-000.006$ & $3.3$ & S &  & 0.035 $\pm$ 0.008 &  - & - \\
G12.8$+$0.0 & $012.822$ & $-000.016$ & $1.7$ & C? &  & - & - & - \\
G13.5$+$0.2 & $013.445$ & $000.141$ & $3.5$ & S? &  & 0.436 $\pm$ 0.009 &  - & - \\
G14.1$-$0.1 & $014.180$ & $-000.120$ & $3.0$ & S &  & 0.005 $\pm$ 0.010 & $<$0.0004 & $<$0.075 \\
G14.3$+$0.1 & $014.300$ & $000.140$ & $2.5$ & S & H II region & 0.367 $\pm$ 0.008 & 0.0012 $\pm$ 0.0002 & 0.003 $\pm$ 0.001 \\
G15.4$+$0.1 & $015.416$ & $000.166$ & $8.6$ & C? &  & 0.176 $\pm$ 0.006 & 0.0006 $\pm$ 0.0002 & 0.003 $\pm$ 0.001 \\
G15.9$+$0.2 & $015.891$ & $000.194$ & $4.3$ & S? &  & 0.595 $\pm$ 0.004 & 0.0186 $\pm$ 0.0066 & 0.031 $\pm$ 0.011 \\
G16.0$-$0.5 & $016.003$ & $-000.491$ & $12.3$ & S &  & 0.054 $\pm$ 0.010 & 0.0014 $\pm$ 0.0002 & 0.026 $\pm$ 0.006 \\
G16.4$-$0.5 & $016.410$ & $-000.550$ & $6.5$ & S &  & - & - & - \\
G16.7$+$0.1 & $016.736$ & $000.082$ & $3.2$ & C &  & 0.725 $\pm$ 0.003 & 0.0167 $\pm$ 0.0037 & 0.023 $\pm$ 0.005 \\
G17.0$+$0.0 & $017.012$ & $-000.037$ & $3.2$ & S &  & - & - & - \\
G17.4$-$0.1 & $017.479$ & $-000.115$ & $5.0$ & S &  & 0.049 $\pm$ 0.004 & $<$0.0004 & $<$0.007 \\
G18.1$-$0.1 & $018.150$ & $-000.172$ & $4.6$ & S &  & 0.237 $\pm$ 0.007 & 0.0148 $\pm$ 0.0016 & 0.063 $\pm$ 0.007 \\
G18.6$-$0.2 & $018.618$ & $-000.284$ & $3.4$ & S &  & 0.374 $\pm$ 0.010 & 0.0134 $\pm$ 0.0052 & 0.036 $\pm$ 0.014 \\
G18.8$+$0.3 & $018.773$ & $000.390$ & $10.3$ & S &  & 2.245 $\pm$ 0.013 & 0.4520 $\pm$ 0.0627 & 0.201 $\pm$ 0.028 \\
G18.9$-$1.1 & $018.900$ & $-001.100$ & $16.5$ & C? &  & - & - & - \\
G19.1$+$0.2 & $019.240$ & $000.281$ & $20.2$ & S &  & - & - & - \\
G20.0$-$0.2 & $019.990$ & $-000.190$ & $6.8$ & F &  & 0.758 $\pm$ 0.010 & 0.1434 $\pm$ 0.0102 & 0.189 $\pm$ 0.014 \\
G21.0$-$0.4 & $021.030$ & $-000.470$ & $5.6$ & S &  & 0.046 $\pm$ 0.006 & 0.0022 $\pm$ 0.0004 & 0.048 $\pm$ 0.010 \\
G21.5$-$0.9 & $021.500$ & $-000.885$ & $1.4$ & C &  & 5.775 $\pm$ 0.002 & 0.6240 $\pm$ 0.1007 & 0.108 $\pm$ 0.017 \\
G21.6$-$0.8 & $021.640$ & $-000.820$ & $8.1$ & S &  & 0.027 $\pm$ 0.006 & 0.0070 $\pm$ 0.0019 & 0.256 $\pm$ 0.089 \\
G21.8$-$0.6 & $021.830$ & $-000.530$ & $15.5$ & S &  & 1.834 $\pm$ 0.018 & 0.3581 $\pm$ 0.0550 & 0.195 $\pm$ 0.030 \\
G22.7$-$0.2 & $022.710$ & $-000.200$ & $16.9$ & S? &  & 0.663 $\pm$ 0.025 & 0.1555 $\pm$ 0.0169 & 0.235 $\pm$ 0.027 \\
G23.3$-$0.3 & $023.251$ & $-000.397$ & $20.6$ & S &  & 1.838 $\pm$ 0.016 & 0.1367 $\pm$ 0.0267 & 0.074 $\pm$ 0.015 \\
G24.7$+$0.6 & $024.550$ & $000.650$ & $15.2$ & C? &  & 0.922 $\pm$ 0.018 & 0.1173 $\pm$ 0.0195 & 0.127 $\pm$ 0.021 \\
G24.7$-$0.6 & $024.860$ & $-000.660$ & $13.1$ & S? &  & 0.698 $\pm$ 0.014 & 0.1744 $\pm$ 0.0191 & 0.250 $\pm$ 0.028 \\
G27.4$+$0.0 & $027.390$ & $-000.010$ & $3.1$ & S &  & 0.780 $\pm$ 0.008 & 0.0181 $\pm$ 0.0044 & 0.023 $\pm$ 0.006 \\
G27.8$+$0.6 & $027.700$ & $000.630$ & $23.4$ & F &  & - & - & - \\
G28.6$-$0.1 & $028.610$ & $-000.110$ & $5.3$ & S &  & 1.170 $\pm$ 0.012 & 0.0637 $\pm$ 0.0126 & 0.054 $\pm$ 0.011 \\
G29.6$+$0.1 & $029.560$ & $000.110$ & $3.3$ & S &  & 0.085 $\pm$ 0.005 & 0.0027 $\pm$ 0.0006 & 0.032 $\pm$ 0.007 \\
G29.7$-$0.3 & $029.710$ & $-000.240$ & $2.7$ & C &  & 2.027 $\pm$ 0.006 & 0.0887 $\pm$ 0.0376 & 0.044 $\pm$ 0.019 \\
G30.7$+$1.0 & $030.700$ & $001.000$ & $12.0$ & S? &  & - & - & - \\
G31.5$-$0.6 & $031.540$ & $-000.660$ & $10.7$ & S? &  & 0.254 $\pm$ 0.011 & 0.0006 $\pm$ 0.0001 & 0.002 $\pm$ 0.001 \\
G31.9$+$0.0 & $031.870$ & $000.020$ & $4.5$ & S &  & 1.546 $\pm$ 0.009 & 0.1218 $\pm$ 0.0280 & 0.079 $\pm$ 0.018 \\
G32.1$-$0.9 & $032.130$ & $-000.960$ & $21.5$ & C? & Not detected & - & - & - \\
G32.4$+$0.1 & $032.420$ & $000.110$ & $4.4$ & S &  & - & - & - \\
G32.8$-$0.1 & $032.790$ & $-000.040$ & $11.5$ & S? &  & - & - & - \\
G33.2$-$0.6 & $033.180$ & $-000.570$ & $9.2$ & S &  & 0.160 $\pm$ 0.008 & 0.0027 $\pm$ 0.0007 & 0.017 $\pm$ 0.004 \\
G33.6$+$0.1 & $033.670$ & $000.030$ & $6.7$ & S &  & 1.222 $\pm$ 0.013 & 0.3606 $\pm$ 0.0987 & 0.295 $\pm$ 0.081 \\
G34.7$-$0.4 & $034.660$ & $-000.400$ & $19.2$ & C &  & - & - & - \\
G35.6$-$0.4 & $035.590$ & $-000.440$ & $8.6$ & S? &  & 0.296 $\pm$ 0.009 & 0.1060 $\pm$ 0.0299 & 0.359 $\pm$ 0.102 \\
G36.6$-$0.7 & $036.590$ & $-000.810$ & $7.0$ & S? &  & - & - & - \\
G39.2$-$0.3 & $039.220$ & $-000.320$ & $4.5$ & C &  & 1.029 $\pm$ 0.006 & 0.2832 $\pm$ 0.0279 & 0.275 $\pm$ 0.027 \\
G40.5$-$0.5 & $040.520$ & $-000.510$ & $12.5$ & S &  & 0.050 $\pm$ 0.007 & 0.0058 $\pm$ 0.0007 & 0.115 $\pm$ 0.022 \\
G41.1$-$0.3 & $041.120$ & $-000.310$ & $2.9$ & S &  & 1.330 $\pm$ 0.004 & 0.1016 $\pm$ 0.0176 & 0.076 $\pm$ 0.013 \\
G41.5$+$0.4 & $041.450$ & $000.410$ & $8.5$ & S? &  & 0.670 $\pm$ 0.007 & 0.0268 $\pm$ 0.0049 & 0.040 $\pm$ 0.007 \\
G42.0$-$0.1 & $041.950$ & $-000.050$ & $5.9$ & S? &  & - & - & - \\
G42.8$+$0.6 & $042.840$ & $000.650$ & $6.9$ & S &  & 0.037 $\pm$ 0.005 &  - & - \\
G43.3$-$0.2 & $043.270$ & $-000.190$ & $3.2$ & S &  & 3.335 $\pm$ 0.009 &  - & - \\
G45.7$-$0.4 & $045.591$ & $-000.352$ & $13.7$ & S &  & 0.149 $\pm$ 0.011 & 0.0221 $\pm$ 0.0058 & 0.148 $\pm$ 0.040 \\
G46.8$-$0.3 & $046.770$ & $-000.280$ & $10.0$ & S &  & 0.935 $\pm$ 0.009 & 0.3954 $\pm$ 0.0936 & 0.423 $\pm$ 0.100 \\
G49.2$-$0.7 & $049.170$ & $-000.540$ & $19.2$ & S? &  & 0.757 $\pm$ 0.037 & 0.1887 $\pm$ 0.0165 & 0.249 $\pm$ 0.025 \\
G53.4$+$0.0 & $053.410$ & $000.030$ & $5.0$ & S &  & 0.027 $\pm$ 0.009 & $<$0.0002 & $<$0.009 \\
G54.1$+$0.3 & $054.100$ & $000.300$ & $6.0$ & C? &  & - & - & - \\
G54.4$-$0.3 & $054.500$ & $-000.280$ & $25.0$ & S &  & 0.646 $\pm$ 0.022 & 0.2005 $\pm$ 0.0342 & 0.310 $\pm$ 0.054 \\
G55.0$+$0.3 & $054.810$ & $-000.090$ & $36.5$ & S &  & - & - & - \\
G57.2$+$0.8 & $057.240$ & $000.820$ & $6.7$ & S? &  & 0.182 $\pm$ 0.004 & 0.0147 $\pm$ 0.0021 & 0.081 $\pm$ 0.012 \\
G59.5$+$0.1 & $059.590$ & $000.110$ & $9.0$ & S &  & 0.031 $\pm$ 0.007 & 0.0002 $\pm$ 0.0001 & 0.007 $\pm$ 0.003 \\
G358.1$+$1.0 & $358.100$ & $001.000$ & $10.0$ & S &  & - & - & - \\
G358.5$-$0.9 & $358.606$ & $-000.971$ & $9.6$ & S &  & - & - & - \\
G359.0$-$0.9 & $358.940$ & $-000.940$ & $17.1$ & S &  & - & - & - \\
G359.1$+$0.9 & $359.097$ & $000.982$ & $6.0$ & S &  & 0.070 $\pm$ 0.005 &  - & - \\
G359.1$-$0.5 & $359.118$ & $-000.503$ & $13.8$ & S &  & 0.823 $\pm$ 0.030 & 0.0787 $\pm$ 0.0118 & 0.096 $\pm$ 0.015 \\
\end{longtable}
\tablefoot{
Measurements of total flux density, $S_{5.8\mathrm{GHz}}$, linearly polarized flux density, $L_{5.8\mathrm{GHz}}$, and degree of polarization, $p=L_{5.8\mathrm{GHz}}/S_{5.8\mathrm{GHz}}$, are explained in \S \ref{sec:methods}.\\
\tablefoottext{a}{Typical uncertainties in the coordinates of the positions are $\lesssim 20\arcsec$.}\\
\tablefoottext{b}{S---shell, C---composite, F---filled, ?---uncertain (taken from \citealt{2019JApA...40...36G}).
}
}

\clearpage

\begin{longtable}{ l r c c r r r }
\caption{\label{table:newcandSNRs} Newly identified SNR candidates in the GLOSTAR-VLA data.  We note that the quoted Stokes $I$ flux densities should be considered as only lower limits due to missing short spacing data (see \S\ref{subsec:GLOSTARdata} and \S\ref{subsubsec:missflux}).} \\
\hline\hline
    Name\tablefootmark{a} & Radius      & Type\tablefootmark{b} & Detected in\tablefootmark{c} & $S_{5.8\mathrm{GHz}}$ & $L_{5.8\mathrm{GHz}}$ & $p_{5.8\mathrm{GHz}}$ \\
    ($l\degr+b\degr$)     & ($\arcmin$) &                       &         & (Jy)                  & (Jy)                  & \\
\hline
\endfirsthead
\caption{continued.}\\
\hline\hline
    Name\tablefootmark{a} & Radius      & Type\tablefootmark{b} & Detected in\tablefootmark{c} & $S_{5.8\mathrm{GHz}}$ & $L_{5.8\mathrm{GHz}}$ & $p_{5.8\mathrm{GHz}}$ \\
    ($l\degr+b\degr$)     & ($\arcmin$) &                       &         & (Jy)                  & (Jy)                  & \\
\hline
\endhead
\hline
\endfoot
G001.949$-$0.100 & $8.3$ & S & G150 & 0.010 $\pm$ 0.007 & $<$0.0007 & $<$0.063 \\
G001.975$-$0.460 & $4.9$ & S &  & 0.002 $\pm$ 0.004 & $<$0.0008 & $<$0.476 \\
G002.228$+$0.058 & $1.1$ & F &  & 0.026 $\pm$ 0.007 & $<$0.0006 & $<$0.024 \\
G002.276$+$0.399 & $6.3$ & S & G150 & 0.051 $\pm$ 0.007 & 0.0006 $\pm$ 0.0002 & 0.013 $\pm$ 0.004 \\
G002.910$-$0.183 & $12.9$ & S &  & - & - & - \\
G003.101$-$0.093 & $2.5$ & S &  & 0.051 $\pm$ 0.006 & $<$0.0008 & $<$0.016 \\
G003.103$+$0.110 & $1.4$ & F &  & 0.017 $\pm$ 0.003 & $<$0.0006 & $<$0.037 \\
G004.493$-$0.391 & $6.9$ & S & G150 & 0.026 $\pm$ 0.008 & $<$0.0006 & $<$0.023 \\
G004.571$-$0.244 & $4.2$ & S & G150 & 0.097 $\pm$ 0.007 & $<$0.0006 & $<$0.006 \\
G005.106$+$0.332 & $4.5$ & S & M20,M90,G150 & 0.024 $\pm$ 0.005 & $<$0.0005 & $<$0.019 \\
G005.161$-$0.321 & $5.6$ & S &  & 0.030 $\pm$ 0.005 & 0.0006 $\pm$ 0.0000 & 0.018 $\pm$ 0.003 \\
G005.364$-$0.705 & $10.0$ & S &  & 0.082 $\pm$ 0.007 & $<$0.0006 & $<$0.008 \\
G005.378$-$0.280 & $2.2$ & F &  & 0.017 $\pm$ 0.002 &  - & - \\
G005.378$+$0.347 & $6.2$ & S & M20,M90,G150 & 0.007 $\pm$ 0.006 & 0.0017 $\pm$ 0.0003 & 0.237 $\pm$ 0.208 \\
G005.673$-$0.118 & $5.1$ & S & M20,M90,G150 & 0.037 $\pm$ 0.014 & $<$0.0006 & $<$0.017 \\
G005.762$+$0.515 & $5.8$ & S &  & - & - & - \\
G005.989$+$0.019 & $5.9$ & S & M20,M90,G150 & 0.086 $\pm$ 0.012 & 0.0147 $\pm$ 0.0014 & 0.171 $\pm$ 0.029 \\
G006.055$+$0.499 & $6.1$ & S & M20,M90,G150 \footnote{From G19 SNR G6.1$+$0.5} & - & - & - \\
G006.118$+$0.387 & $9.7$ & S? & M20,M90,G150 \footnote{From G19 SNR G6.1$+$0.5} & - & - & - \\
G008.040$+$0.566 & $6.0$ & S & M20,M90,G150 & 0.032 $\pm$ 0.006 & $<$0.0006 & $<$0.017 \\
G013.500$+$0.074 & $1.9$ & F & M20,M90,G150 & 0.026 $\pm$ 0.005 &  - & - \\
G013.549$+$0.352 & $4.0$ & S &  & 0.008 $\pm$ 0.005 & $<$0.0004 & $<$0.047 \\
G013.626$+$0.299 & $3.5$ & S &  & - & - & - \\
G013.652$+$0.259 & $2.5$ & S & M20,M90,G150 & 0.046 $\pm$ 0.004 & $<$0.0005 & $<$0.011 \\
G013.658$-$0.241 & $4.2$ & S &  & 0.013 $\pm$ 0.005 & $<$0.0004 & $<$0.029 \\
G014.524$+$0.140 & $14.5$ & S? & M20,M90,G150 & - & - & - \\
G015.862$+$0.522 & $1.9$ & S & M20 & 0.005 $\pm$ 0.002 & $<$0.0003 & $<$0.063 \\
G016.021$+$0.746 & $6.3$ & S & TV,M20,M90,G150 & 0.014 $\pm$ 0.005 & 0.0002 $\pm$ 0.0000 & 0.018 $\pm$ 0.006 \\
G016.126$+$0.690 & $27.0$ & S &  & - & - & - \\
G016.956$-$0.933 & $7.4$ & S &  & 0.153 $\pm$ 0.006 &  - & - \\
G017.434$+$0.273 & $2.1$ & F? & M20,G150 & 0.050 $\pm$ 0.002 & $<$0.0005 & $<$0.009 \\
G017.593$+$0.237 & $1.5$ & S & M20 & 0.010 $\pm$ 0.001 & $<$0.0005 & $<$0.048 \\
G017.620$+$0.086 & $3.0$ & C & M20,M90,G150 & - & - & - \\
G018.393$-$0.816 & $2.6$ & S & M20,M90,G150 & 0.009 $\pm$ 0.002 & $<$0.0004 & $<$0.044 \\
G019.481$-$0.108 & $7.4$ & S & M20,M90,G150 & - & - & - \\
G019.751$+$0.202 & $9.0$ & S & TV,M20 & - & - & - \\
G020.195$+$0.028 & $3.6$ & F? & TV,G150 & 0.004 $\pm$ 0.004 & $<$0.0004 & $<$0.109 \\
G021.492$-$0.010 & $3.5$ & C? &  & 0.004 $\pm$ 0.004 & $<$0.0003 & $<$0.097 \\
G021.596$-$0.179 & $10.4$ & S & TV,M20 & 0.228 $\pm$ 0.013 & $<$0.0004 & $<$0.002 \\
G021.684$+$0.129 & $3.8$ & S & TV,M20 & 0.060 $\pm$ 0.005 & 0.0003 $\pm$ 0.0001 & 0.004 $\pm$ 0.002 \\
G021.861$+$0.169 & $2.7$ & S & TV,M20 & 0.014 $\pm$ 0.004 & $<$0.0003 & $<$0.025 \\
G022.045$-$0.028 & $7.7$ & S & TV,M20,G150 \footnote{possibly related to X-ray SNR candidate G22.00+0.00 \citep{2006IAUS..230..333U,2016PASJ...68S...6Y}} & 0.073 $\pm$ 0.014 & $<$0.0004 & $<$0.005 \\
G022.177$+$0.314 & $11.1$ & S & TV,M20 & - & - & - \\
G022.951$-$0.311 & $2.4$ & S & TV,M20,G150 & - & - & - \\
G023.973$+$0.510 & $4.9$ & S & TV & 0.002 $\pm$ 0.007 & $<$0.0003 & $<$0.151 \\
G024.062$-$0.808 & $2.0$ & F & M20 & 0.015 $\pm$ 0.001 & $<$0.0003 & $<$0.018 \\
G024.193$+$0.284 & $2.1$ & S & M20 & 0.008 $\pm$ 0.003 & $<$0.0004 & $<$0.044 \\
G028.524$+$0.268 & $3.1$ & S & TV,M20,G150 & - & - & - \\
G028.870$+$0.616 & $2.0$ & S & TV & 0.008 $\pm$ 0.003 & $<$0.0003 & $<$0.036 \\
G028.877$+$0.241 & $1.3$ & S & TV,M20 \footnote{From THOR SNR candidate G28.92$+$0.26} & 0.010 $\pm$ 0.003 & $<$0.0004 & $<$0.034 \\
G028.929$+$0.254 & $2.2$ & S & TV,M20,G150 \footnote{From THOR SNR candidate G28.92$+$0.26} & 0.092 $\pm$ 0.006 & $<$0.0003 & $<$0.004 \\
G029.329$+$0.280 & $2.5$ & S & TV,M20 & - & - & - \\
G030.303$+$0.128 & $1.0$ & F &  & 0.002 $\pm$ 0.003 & $<$0.0003 & $<$0.199 \\
G030.362$+$0.623 & $7.4$ & S &  & 0.017 $\pm$ 0.008 & 0.0002 $\pm$ 0.0001 & 0.011 $\pm$ 0.006 \\
G030.375$+$0.424 & $2.3$ & S & TV,M20 & 0.010 $\pm$ 0.004 & $<$0.0003 & $<$0.029 \\
G030.508$+$0.574 & $6.0$ & S &  & 0.001 $\pm$ 0.008 & $<$0.0003 & $<$0.204 \\
G031.256$-$0.041 & $3.3$ & C & TV,M20,M90 & 0.086 $\pm$ 0.013 &  - & - \\
G032.458$-$0.112 & $1.8$ & S &  & 0.002 $\pm$ 0.003 & $<$0.0003 & $<$0.194 \\
G034.524$-$0.761 & $2.7$ & S & TV,G150 & 0.046 $\pm$ 0.005 & 0.0030 $\pm$ 0.0006 & 0.066 $\pm$ 0.014 \\
G034.619$+$0.240 & $3.0$ & F & TV & - & - & - \\
G035.129$-$0.343 & $1.8$ & S &  & 0.007 $\pm$ 0.002 & $<$0.0003 & $<$0.049 \\
G036.839$-$0.433 & $2.6$ & F &  & 0.010 $\pm$ 0.003 & $<$0.0003 & $<$0.034 \\
G036.851$-$0.246 & $1.5$ & S &  & 0.002 $\pm$ 0.002 & $<$0.0003 & $<$0.160 \\
G037.337$+$0.422 & $3.3$ & S & TV,M20 & 0.026 $\pm$ 0.004 & $<$0.0002 & $<$0.009 \\
G037.506$+$0.777 & $1.3$ & F & TV & 0.024 $\pm$ 0.001 & $<$0.0002 & $<$0.010 \\
G037.672$-$0.501 & $2.8$ & S & TV & 0.024 $\pm$ 0.005 & $<$0.0004 & $<$0.018 \\
G039.203$+$0.811 & $1.6$ & S &  & 0.009 $\pm$ 0.001 & 0.0004 $\pm$ 0.0002 & 0.047 $\pm$ 0.019 \\
G039.539$+$0.366 & $4.5$ & S & TV,M20,G150 & 0.024 $\pm$ 0.003 & 0.0014 $\pm$ 0.0004 & 0.060 $\pm$ 0.019 \\
G041.510$-$0.534 & $17.8$ & S & TV,G150 & - & - & - \\
G041.625$+$0.261 & $4.7$ & S &  & 0.003 $\pm$ 0.005 & $<$0.0003 & $<$0.082 \\
G042.711$-$0.272 & $9.1$ & S & G150 & 0.026 $\pm$ 0.010 & $<$0.0003 & $<$0.013 \\
G043.023$+$0.726 & $6.5$ & S &  & 0.007 $\pm$ 0.004 &  - & - \\
G043.070$+$0.558 & $2.1$ & S & TV & 0.032 $\pm$ 0.002 & $<$0.0003 & $<$0.010 \\
G044.076$+$0.127 & $8.0$ & S &  & 0.086 $\pm$ 0.010 & 0.0005 $\pm$ 0.0001 & 0.006 $\pm$ 0.002 \\
G047.741$-$0.971 & $9.8$ & S &  & - & - & - \\
G048.875$+$0.174 & $6.6$ & S & G150 & 0.026 $\pm$ 0.004 & $<$0.0003 & $<$0.011 \\
G051.061$+$0.563 & $4.7$ & S & TV,G150 & 0.010 $\pm$ 0.003 & $<$0.0002 & $<$0.024 \\
G059.834$-$0.405 & $10.5$ & S &  & - & - & - \\
G358.711$-$0.380 & $2.4$ & S &  & 0.003 $\pm$ 0.004 & $<$0.0005 & $<$0.151 \\
G359.172$+$0.264 & $1.8$ & F &  & 0.018 $\pm$ 0.004 & $<$0.0007 & $<$0.039 \\
\end{longtable}
\tablefoot{
Measurements of total flux density, $S_{5.8\mathrm{GHz}}$, linearly polarized flux density, $L_{5.8\mathrm{GHz}}$, and degree of polarization, $p=L_{5.8\mathrm{GHz}}/S_{5.8\mathrm{GHz}}$, are explained in \S \ref{sec:methods}.\\
\tablefoottext{a}{Typical uncertainties in the coordinates of the positions are $\lesssim 20\arcsec$.}\\
\tablefoottext{b}{S---shell, C---composite, F---filled, ?---uncertain.}\\
\tablefoottext{c}{The candidates that have at least a faint counterpart in lower frequency surveys have been marked: 20 cm THOR+VGPS---TV, 150 cm GLEAM---G150, 90 cm MAGPIS---M90, 20 cm MAGPIS---M20.}\\
}


\begin{landscape}
\begin{longtable}{ l r r r c c c r r r }
\caption{\label{tab:prevcandSNRs} Previously identified SNR candidates with counterparts in the GLOSTAR-VLA data.  We note that the quoted Stokes $I$ flux densities should be considered as only lower limits due to missing short spacing data (see \S\ref{subsec:GLOSTARdata} and \S\ref{subsubsec:missflux}).}\\
\hline\hline
    Name & GLong\tablefootmark{a} & GLat\tablefootmark{a} & Radius      & Type\tablefootmark{b} & Remarks & References & $S_{5.8\mathrm{GHz}}$ & $L_{5.8\mathrm{GHz}}$ & $p_{5.8\mathrm{GHz}}$ \\
         & ($\degr$)              & $\degr$)              & ($\arcmin$) &                       &         &            & (Jy)                  & (Jy)                  & \\
\hline
\endfirsthead
\caption{continued.}\\
\hline\hline
    Name & GLong\tablefootmark{a} & GLat\tablefootmark{a} & Radius      & Type\tablefootmark{b} & Remarks & References & $S_{5.8\mathrm{GHz}}$ & $L_{5.8\mathrm{GHz}}$ & $p_{5.8\mathrm{GHz}}$ \\
         & ($\degr$)              & $\degr$)              & ($\arcmin$) &                       &         &            & (Jy)                  & (Jy)                  & \\
\hline
\endhead
\hline
\endfoot
G3.10$-$0.60 & $003.100$ & $-000.600$ & $20$ & F &  & 1 & - & - & - \\
G4.20$-$0.30 & $004.200$ & $-000.300$ & $7.5$ & F &  & 2 & - & - & - \\
G5.71$-$0.08 & $005.710$ & $-000.080$ & $6$ & F & Related to G005.673$-$0.084? & 3 & - & - & - \\
G6.31$+$0.54 & $006.303$ & $000.540$ & $6.1$ & S? &  & 3 & 0.089 $\pm$ 0.007 & 0.0140 $\pm$ 0.0036 & 0.158 $\pm$ 0.042 \\
G6.4500$-$0.5583 & $006.445$ & $-000.552$ & $3.1$ & S & Part of SNR G6.5$-$0.4? & 3, 4 & 0.053 $\pm$ 0.009 & $<$0.0010 & $<$0.019 \\
G8.8583$-$0.2583 & $008.858$ & $-000.258$ & $2$ & S & Part of SNR G8.7$-$0.1? & 4 & - & - & - \\
G11.5500$+$0.3333 & $011.550$ & $000.333$ & $2.25$ & - & A14 H II region & 4 & - & - & - \\
G15.51$-$0.15 & $015.495$ & $-000.154$ & $6.5$ & C & PWN? & 3, $*$ & 0.166 $\pm$ 0.008 & 0.0062 $\pm$ 0.0020 & 0.037 $\pm$ 0.012 \\
G17.80$-$0.02 & $017.800$ & $-000.020$ & $4.4$ & S &  & 5 & - & - & - \\
G18.45$-$0.42 & $018.450$ & $-000.420$ & $7.6$ & S &  & 5 & - & - & - \\
G18.53$-$0.86 & $018.530$ & $-000.860$ & $8.6$ & ? &  & 5, 6 & - & - & - \\
G18.76$-$0.07 & $018.760$ & $-000.073$ & $0.8$ & ? &  & 4, 5, $*$ & 0.084 $\pm$ 0.002 & 0.0071 $\pm$ 0.0011 & 0.084 $\pm$ 0.013 \\
G19.13$+$0.90 & $019.130$ & $000.900$ & $12$ & ? &  & 3 & 0.148 $\pm$ 0.008 & 0.0069 $\pm$ 0.0025 & 0.046 $\pm$ 0.017 \\
G19.75$-$0.69 & $019.750$ & $-000.690$ & $13.2$ & F &  & 5, 6 & - & - & - \\
G19.96$-$0.33 & $019.960$ & $-000.330$ & $5.9$ & C? &  & 5 & 0.002 $\pm$ 0.008 & $<$0.0004 & $<$0.198 \\
G20.26$-$0.86 & $020.260$ & $-000.860$ & $7.5$ & F? &  & 5 & 0.144 $\pm$ 0.009 & $<$0.0004 & $<$0.003 \\
G21.66$-$0.21 & $021.660$ & $-000.210$ & $5.1$ & F? & Larger structure\footnote{GLOSTAR SNR candidate G021.596$-$0.179} & 5, 6 & - & - & - \\
G22.32$+$0.11 & $022.320$ & $000.110$ & $5.5$ & S & & 4, 5 & - & - & - \\
G23.11$+$0.19 & $023.110$ & $000.190$ & $12.1$ & S & SNR & 5, 6, 7 & - & - & - \\
G23.85$-$0.18 & $023.855$ & $-000.180$ & $2.7$ & S &  & 5 & 0.062 $\pm$ 0.006 &  - & - \\
G25.49$+$0.01 & $025.490$ & $000.010$ & $7.4$ & S & & 5, 8 & - & - & - \\
G26.04$-$0.42 & $026.040$ & $-000.420$ & $13.5$ & F? &  & 5 & - & - & - \\
G26.13$+$0.13 & $026.130$ & $000.130$ & $11.3$ & S &  & 5 & - & - & - \\
G26.53$+$0.07 & $026.530$ & $000.070$ & $11.2$ & C? &  & 5 & - & - & - \\
G26.75$+$0.73 & $026.750$ & $000.730$ & $5.3$ & S & SNR & 5, $*$ & 0.033 $\pm$ 0.004 & 0.0229 $\pm$ 0.0127 & 0.702 $\pm$ 0.398 \\
G27.06$+$0.04 & $027.060$ & $000.040$ & $7.5$ & S & SNR\footnote{Flux densities measured only for the arc} & 4, 5, 9, $*$ & 0.317 $\pm$ 0.008 & 0.0320 $\pm$ 0.0034 & 0.101 $\pm$ 0.011 \\
G27.18$+$0.30 & $027.180$ & $000.305$ & $0.9$ & F & PWN? & 5, $*$ & 0.048 $\pm$ 0.001 & $<$0.0004 & $<$0.008 \\
G27.24$-$0.14 & $027.240$ & $-000.140$ & $6.1$ & S & Likely thermal & 5 & - & - & - \\
G27.39$+$0.24 & $027.390$ & $000.240$ & $2.4$ & x & Filaments & 5, $*$ & 0.005 $\pm$ 0.005 &  - & - \\
G27.47$+$0.25 & $027.467$ & $000.246$ & $1.7$ & x & Filaments & 5, $*$ & 0.018 $\pm$ 0.003 &  - & - \\
G27.78$-$0.33 & $027.780$ & $-000.330$ & $3.7$ & S &  & 5 & 0.051 $\pm$ 0.002 & $<$0.0003 & $<$0.005 \\
G28.21$+$0.02 & $028.210$ & $000.020$ & $2.5$ & F & & 5 & - & - & - \\
G28.22$-$0.09 & $028.216$ & $-000.087$ & $1.7$ & F &  & 5 & 0.035 $\pm$ 0.003 &  - & - \\
G28.33$+$0.06 & $028.330$ & $000.060$ & $3.2$ & F &  & 5 & - & - & - \\
G28.36$+$0.21 & $028.360$ & $000.210$ & $6.4$ & S &  & 4, 5, 6 & 0.084 $\pm$ 0.016 & 0.0023 $\pm$ 0.0007 & 0.027 $\pm$ 0.009 \\
G28.56$+$0.00 & $028.564$ & $000.000$ & $1.5$ & S &  & 4, 5, 9 & 0.573 $\pm$ 0.005 & 0.0162 $\pm$ 0.0051 & 0.028 $\pm$ 0.009 \\
G28.64$+$0.20 & $028.640$ & $000.200$ & $11.4$ & S & & 4, 5 & - & - & - \\
G28.78$-$0.44 & $028.780$ & $-000.436$ & $6.6$ & S & SNR & 4, 5, 6, $*$ & 0.054 $\pm$ 0.007 & 0.0260 $\pm$ 0.0057 & 0.485 $\pm$ 0.125 \\
G28.88$+$0.41 & $028.880$ & $000.410$ & $8.9$ & F &  & 5 & 0.108 $\pm$ 0.020 & $<$0.0003 & $<$0.003 \\
G28.92$+$0.26 & $028.920$ & $000.260$ & $3.2$ & S & Multiple objects\footnote{GLOSTAR SNR candidates G028.929$+$0.254, G028.877$+$0.241} & 5 & - & - & - \\
G29.38$+$0.10 & $029.380$ & $000.100$ & $5.1$ & C & PWN + radio galaxy & 4, 5, 9, 10, 11, 12, $*$ & 0.578 $\pm$ 0.012 & 0.0898 $\pm$ 0.0085 & 0.155 $\pm$ 0.015 \\
G29.41$-$0.18 & $029.410$ & $-000.180$ & $7.5$ & S &  & 5 & 0.007 $\pm$ 0.014 & $<$0.0004 & $<$0.054 \\
G31.22$-$0.02 & $031.220$ & $-000.020$ & $3.1$ & C &  & 5 & - & - & - \\
G31.93$+$0.16 & $031.936$ & $000.172$ & $2.4$ & ? & Filaments? & 5 & 0.014 $\pm$ 0.003 &  - & - \\
G32.22$-$0.21 & $032.220$ & $-000.210$ & $3.1$ & F &  & 5 & 0.115 $\pm$ 0.005 &  - & - \\
G32.37$-$0.51 & $032.370$ & $-000.510$ & $12$ & S? &  & 5 & - & - & - \\
G33.85$+$0.06 & $033.848$ & $000.061$ & $0.6$ & F? & Unlikely to be a SNR & 5, $*$ & 0.006 $\pm$ 0.001 &  - & - \\
G34.93$-$0.24 & $034.933$ & $-000.244$ & $8.1$ & S &  & 5 & - & - & - \\
G36.66$-$0.50 & $036.660$ & $-000.500$ & $8.2$ & F? &  & 5 & - & - & - \\
G36.68$-$0.14 & $036.680$ & $-000.140$ & $10$ & S &  & 5 & 0.206 $\pm$ 0.012 & 0.0004 $\pm$ 0.0001 & 0.002 $\pm$ 0.000 \\
G36.90$+$0.49 & $036.902$ & $000.488$ & $3.8$ & F? &  & 5 & - & - & - \\
G37.62$-$0.22 & $037.622$ & $-000.220$ & $1.6$ & S &  & 5 & 0.026 $\pm$ 0.006 & 0.0008 $\pm$ 0.0002 & 0.030 $\pm$ 0.010 \\
G37.88$+$0.32 & $037.880$ & $000.320$ & $11.4$ & S &  & 5 & - & - & - \\
G38.17$+$0.09 & $038.170$ & $000.090$ & $14.7$ & F? &  & 5 & - & - & - \\
G38.62$-$0.24 & $038.620$ & $-000.240$ & $2.5$ & F? &  & 5 & - & - & - \\
G38.68$-$0.43 & $038.680$ & $-000.430$ & $4.3$ & S? &  & 5 & - & - & - \\
G38.83$-$0.01 & $038.834$ & $-000.013$ & $1.3$ & S? &  & 5 & 0.033 $\pm$ 0.001 & $<$0.0003 & $<$0.009 \\
G39.56$-$0.32 & $039.560$ & $-000.320$ & $8.5$ & F? &  & 5 & - & - & - \\
G41.95$-$0.18 & $041.950$ & $-000.180$ & $7$ & S &  & 5 & - & - & - \\
G42.62$+$0.14 & $042.620$ & $000.140$ & $3.3$ & S? &  & 5 & 0.140 $\pm$ 0.003 & 0.0003 $\pm$ 0.0000 & 0.002 $\pm$ 0.000 \\
G46.18$-$0.02 & $046.158$ & $-000.020$ & $6.7$ & F &  & 5 & 0.045 $\pm$ 0.004 & 0.0002 $\pm$ 0.0000 & 0.005 $\pm$ 0.001 \\
G46.54$-$0.03 & $046.540$ & $-000.026$ & $6.2$ & S &  & 5 & 0.036 $\pm$ 0.008 & $<$0.0003 & $<$0.007 \\
G47.15$+$0.73 & $047.150$ & $000.730$ & $0.8$ & C? &  & 5 & 0.009 $\pm$ 0.000 & $<$0.0002 & $<$0.024 \\
G51.04$+$0.07 & $051.040$ & $000.070$ & $3.8$ & F & SNR & 5, 9, 13, 14, 15, $*$ & 0.100 $\pm$ 0.011 & 0.0070 $\pm$ 0.0014 & 0.070 $\pm$ 0.016 \\
G51.21$+$0.11 & $051.209$ & $000.113$ & $14.9$ & S & Multiple objects\footnote{complex of SNRs G51.04$+$0.07 and G51.26$+$0.11} & 5, 9, 13, 14, 15, $*$ & - & - & - \\
G51.26$+$0.11 & $051.260$ & $000.110$ & $11.3$ & S & SNR & 5, 9, 13, 14, 15, $*$ & 1.054 $\pm$ 0.015 & 0.0655 $\pm$ 0.0128 & 0.062 $\pm$ 0.012 \\
G52.37$-$0.70 & $052.370$ & $-000.700$ & $17.7$ & F? & Likely thermal & 5, 14 & - & - & - \\
G53.07$+$0.49 & $053.070$ & $000.490$ & $1$ & F &  & 5, 14, $*$ & 0.030 $\pm$ 0.001 & 0.0035 $\pm$ 0.0007 & 0.118 $\pm$ 0.022 \\
G53.84$-$0.75 & $053.840$ & $-000.750$ & $18.7$ & F &  & 5, 14 & - & - & - \\
G54.11$+$0.25 & $054.110$ & $000.250$ & $7.2$ & C &  & 5, 14, $*$ & 0.418 $\pm$ 0.013 & 0.0343 $\pm$ 0.0069 & 0.082 $\pm$ 0.017 \\
G56.56$-$0.75 & $056.560$ & $-000.750$ & $11.6$ & F &  & 5 & - & - & - \\
G57.12$+$0.35 & $057.120$ & $000.350$ & $14.1$ & S &  & 5 & - & - & - \\
G58.70$-$0.31 & $058.700$ & $-000.310$ & $4.4$ & F &  & 5 & 0.008 $\pm$ 0.002 & 0.0002 $\pm$ 0.0000 & 0.025 $\pm$ 0.006 \\
G59.46$+$0.83 & $059.460$ & $000.830$ & $4.5$ & F &  & 5 & 0.006 $\pm$ 0.002 & $<$0.0002 & $<$0.039 \\
G358.70$+$0.70 & $358.745$ & $000.810$ & $13$ & S &  & 1 & - & - & - \\
G359.07$+$0.02 & $359.070$ & $000.020$ & $11$ & S? &  & 16, 17, 18 & - & - & - \\
G359.56$-$0.08 & $359.560$ & $-000.080$ & $1$ & S? &  & 19 & - & - & - \\
\end{longtable}
\tablefoot{
The measurements of total flux density, $S_{5.8\mathrm{GHz}}$, linearly polarized flux density, $L_{5.8\mathrm{GHz}}$, and degree of polarization, $p=L_{5.8\mathrm{GHz}}/S_{5.8\mathrm{GHz}}$, are explained in \S \ref{sec:methods}.\\
\tablefoottext{a}{Typical uncertainties in the coordinates of the positions are $\lesssim 20\arcsec$.}\\
\tablefoottext{b}{S---shell, C---composite, F---filled, ?---uncertain.}
\tablebib{
(1)~\citet{1994MNRAS.270..847G}; (2)~\citet{2001ESASP.459..109T};
(3)~\citet{2006ApJ...639L..25B}; (4)~\citet{2006AJ....131.2525H};
(5)~\citet{2017A&A...605A..58A}; (6)~\citet{2019PASA...36...45H};
(7)~\citet{2019ApJ...885..129M}; (8)~\citet{2003ApJ...589..253B};
(9)~\citet{2018ApJ...866...61D}; (10)~\citet{2019PASA...36...48H};
(11)~\citet{2017A&A...602A..31C}; (12)~\citet{2019A&A...626A..65P};
(13)~\citet{2014A&A...565A...6S};
(14)~\citet{2018ApJ...860..133D}; (15)~\citet{2018A&A...616A..98S};
(16)~\citet{2000AJ....119..207L}; (17)~\citet{2010PASJ...62..971N};
(18)~\citet{2015MNRAS.453..172P}; (19)~\citet{2007A&A...462.1065M};
($*$)~Discussed in the text of this paper.
}}
\end{landscape}

\end{document}